\documentclass[9pt]{amsart}

\usepackage{latexsym,pifont}
\usepackage{epsfig}
\usepackage{rotating}

\usepackage{graphicx}

\usepackage{color}

\usepackage{ifpdf}
\ifpdf
\usepackage{epstopdf}
\usepackage{hyperref}
\else
\usepackage[hypertex]{hyperref}
\fi

\usepackage{amsfonts,amsmath,amssymb,mathrsfs,extarrows,MnSymbol}
\usepackage{tikzsymbols}
\usepackage[all]{xy}

\xyoption{knot}

\usepackage{tikz}
\usetikzlibrary{matrix,arrows}

\usepackage[makeroom]{cancel}

\newtheorem{theorem}{Theorem}[section]
\newtheorem*{theorem*}{Theorem}
\newtheorem{proposition}[theorem]{Proposition}

\newtheorem{remark}[theorem]{Remark}
\newtheorem{example}[theorem]{Example}
\newtheorem{definition}[theorem]{Definition}
\newtheorem{corollary}[theorem]{Corollary}
\newtheorem{lemma}[theorem]{Lemma}

\newtheorem{propanition}[theorem]{Proposition/Definition}

\newtheorem{conjecture}[theorem]{Conjecture}
\newtheorem{convention}[theorem]{Convention}

\newcommand{\alxydim}[2]{\begin{aligned}\xymatrix#1{#2}\end{aligned}}

\newcommand{\brem}{\begin{remark}}
	\newcommand{\erem}{\end{remark}}
\newcommand{\beg}{\begin{example}}
	\newcommand{\eeg}{\end{example}}
\newcommand{\bedef}{\begin{definition}}
	\newcommand{\exdef}{\end{definition}}
\newcommand{\berop}{\begin{proposition}}
	\newcommand{\eerop}{\end{proposition}}
\newcommand{\belem}{\begin{lemma}}
	\newcommand{\elem}{\end{lemma}}
\newcommand{\bethe}{\begin{theorem}}
	\newcommand{\ethe}{\end{theorem}}
\newcommand{\becor}{\begin{corollary}}
	\newcommand{\ecor}{\end{corollary}}
\newcommand{\beroof}{\noindent\begin{proof}}
	\newcommand{\eroof}{\end{proof}}
\newcommand{\becon}{\begin{convention}}
	\newcommand{\econ}{\end{convention}}
	\newcommand{\efact}{\begin{flushright}$\checkmark$\end{flushright}\end{Fact}}
\newcommand{\becj}{\begin{conjecture}}
	\newcommand{\ecj}{\begin{flushright}$\boxtimes$\end{flushright}\end{conjecture}}

\newcommand{\barr}{\begin{array}}
	\newcommand{\earr}{\end{array}}
\newcommand{\ben}{\begin{enumerate}}
	\newcommand{\een}{\end{enumerate}}
\newcommand{\bit}{\begin{itemize}}
	\newcommand{\eit}{\end{itemize}}

\newcommand{\qq}{\begin{eqnarray}}
	\newcommand{\qqq}{\end{eqnarray}}

\newcommand{\nn}{\nonumber}

\newcommand{\ovl}[1]{\overline{#1}}
\newcommand{\unl}[1]{\underline{#1}}

\newcommand{\Reqref}[1]{Eq.\,\eqref{#1}}

\newcommand\void[1]{}

\newcommand{\tx}[1]{\textrm{#1}} 

\newcommand{\gt}[1]{\mathfrak{#1}}

\def\cA{\mathcal{A}}

\def\cC{\mathcal{C}}
\def\cD{\mathcal{D}}
\def\cE{\mathcal{E}}

\def\cG{\mathcal{G}}

\def\cI{\mathcal{I}}

\def\cK{\mathcal{K}}
\def\ceL{\mathcal{L}}
\def\cM{\mathcal{M}}

\def\cO{\mathcal{O}}
\def\xcP{\mathcal{P}}

\def\cS{\mathcal{S}}

\def\cU{\mathcal{U}}

\def\cW{\mathcal{W}}

\def\xcA{\mathscr{A}}

\def\xcD{\mathscr{D}}
\def\xcE{\mathscr{E}}
\def\xcF{\mathscr{F}}
\def\xcG{\mathscr{G}}

\def\xcI{\mathscr{I}}

\def\xcL{\mathscr{L}}

\def\xcO{\mathscr{O}}
\def\xcP{\mathscr{P}}

\def\xcR{\mathscr{R}}

\def\xcV{\mathscr{V}}
\def\xcW{\mathscr{W}}

\def\bbB{\mathbf{B}}

\def\t{\mathbf{t}}

\def\bB{{\mathbb{B}}}
\def\bC{{\mathbb{C}}}

\def\bH{{\mathbb{H}}}

\def\bN{{\mathbb{N}}}

\def\bR{{\mathbb{R}}}
\def\bS{{\mathbb{S}}}

\def\bZ{{\mathbb{Z}}}

\def\a{\alpha}
\def\b{\beta}
\def\g{\gamma}
\def\G{\Gamma}
\def\d{\delta}
\def\D{\Delta}
\def\ep{\epsilon}
\def\vep{\varepsilon}

\def\la{\lambda}
\def\La{\Lambda}
\def\om{\omega}
\def\Om{\Omega}

\def\si{\sigma}
\def\Si{\Sigma}

\def\t{\tau}

\def\Ups{\Upsilon}
\def\z{\zeta}

\def\egt{\gt{e}}
\def\Egt{\gt{E}}

\def\ggt{\gt{g}}

\def\Igt{\gt{I}}

\def\Vgt{\gt{V}}

\newcommand{\sfd}{{\mathsf d}}

\newcommand{\sfi}{{\mathsf i}}
\newcommand{\sfI}{{\mathsf I}}

\newcommand{\sfJ}{{\mathsf J}}

\newcommand{\sfM}{{\mathsf M}}

\newcommand{\sfS}{{\mathsf S}}

\newcommand{\sfT}{{\mathsf T}}

\newcommand{\txA}{{\rm A}}

\newcommand{\txB}{{\rm B}}

\newcommand{\txd}{{\rm d}}

\newcommand{\txg}{{\rm g}}
\newcommand{\txG}{{\rm G}}

\newcommand{\txH}{{\rm H}}

\newcommand{\txm}{{\rm m}}

\newcommand{\txT}{{\rm T}}

\newcommand{\txV}{{\rm V}}

\def\vC{\check{C}}
\def\Cv{\v{C}}

\def\vZ{\check{Z}}

\def\vd{\check{\d}}

\def\exp{{\rm exp}}
\def\id{{\rm id}}
\newcommand{\pr}{{\rm pr}}

\def\too{\longrightarrow}
\def\ev{{\rm ev}}

\newcommand{\grpd}[2]{\hspace{-5pt}\alxydim{@C=.5cm@R=1.cm}{ #1 \ar@<.25ex>[r] \ar@<-.25ex>[r] & #2}\hspace{-3pt}}

\def\obj{{\rm Ob}}

\def\Hom{{\rm Hom}}
\def\morf{{\rm Mor}}
\def\1morf{1{\rm -Mor}}
\def\2morf{2{\rm -Mor}}
\def\dim{{\rm dim}}
\def\im{{\rm im}}
\def\ker{{\rm ker}}

\newcommand{\Id}{{\rm Id}}
\def\Inv{{\rm Inv}}

\def\bgrb{\gt{BGrb}}

\newcommand{\Gr}{{\rm {\bf Gr}}}
\newcommand{\BisGr}{{\rm Bisec({\bf Gr})}}

\newcommand{\Set}{{\rm {\bf Set}}}

\newcommand{\Top}{{\rm {\bf Top}}}
\newcommand{\Man}{{\rm {\bf Man}}}

\def\p{\partial}

\newcommand{\Diff}{{\rm Diff}}

\def\emb{\hookrightarrow}

\def\curv{{\rm curv}}
\def\Hol{{\rm Hol}}
\def\Bun{{\rm {\bf Bun}}}
\newcommand{\breP}{{\rm \breve{P}}}

\def\bd1{{\boldsymbol{1}}}
\def\brd0{{\boldsymbol{0}}}

\def\ad{{\rm ad}}
\def\Ad{{\rm Ad}}

\def\Lie{{\rm Lie}}

\newcommand{\uj}{{\rm U}(1)}

\newcommand\MCR{\theta_{\rm R}}

\def\x{\times}
\def\ox{\otimes}

\def\lx{{\hspace{-0.04cm}\ltimes\hspace{-0.05cm}}}

\newcommand{\fibx}[2]{\hspace{-2pt}{}_{#1}\hspace{-3pt}\x_{#2}\hspace{-2pt}}
\def\ract{\vartriangleleft}
\def\lact{\vartriangleright}
\def\mact{\blacktriangleleft}
\def\mlact{\blacktriangleright}
\def\ulact{\,\unl{\vartriangleright}\,}
\def\must{\stackrel{!}{=}}

\def\rstr{\big\vert}

\newcommand{\corr}[1]{\left\langle #1 \right\rangle}

\newcommand\colo{\hspace{-2pt}:}

\numberwithin{equation}{section} 

\newcount\hour\newcount\minute
\hour=\time \divide\hour by60 \minute=\time
{\multiply\hour by60 \global\advance\minute by-\hour}
\edef\militarytime{\number\hour:\ifnum\minute<10 0\fi\number\minute}

\begin{document}

\title{Making Poisson $\sigma$ again\\
	--- an \'etude on groupoidal gauging,\ anomalies,\ gerbes,\ and defects}

\author[R.R. Suszek]{\vspace{45pt}Rafa\l ~R.\ ~Suszek}
\address{Katedra Metod Matematycznych Fizyki,\ Wydzia\l ~Fizyki Uniwersytetu Warszawskiego,\ ul.\ Pasteura 5,\ PL-02-093 Warszawa,\ Poland} 
\email{suszek@fuw.edu.pl}
	
\begin{abstract}
The Poisson $\sigma$-model (PSM) over a closed worldsheet $\Sigma$ is reinterpreted as a theory of connections on a trivial principaloid bundle $\xcP_{\rm triv}\equiv\Si\x\xcG\to\Sigma$ with the symplectic structure groupoid $\,\grpd{(\xcG,\om)}{(M,\Pi)}$ integrating the cotangent algebroid over the Poisson target $(M,\Pi)$,\ and that of sections of the shadow bundle $\xcF_{\rm triv}\equiv\xcP_{\rm triv}/\xcG=\Si\x M\to\Si$---as defined in 2503.09886 [math.DG].\ The flat directions of the PSM are explained as linearisations of automorphisms of $\xcP_{\rm triv}$,\ modelled on smooth maps from $\Si$ into the group $\bB(\xcG,\om)$ of lagrangean bisections of $\xcG$,\ which induce Poisson automorphisms of $(M,\Pi)$.\ The gauge anomaly of the ensuing field theory is derived and demonstrated to vanish for prequantisable $\om$.\ A general anomaly-cancellation mechanism is subsequently proposed,\ which couples the bare model---rewritten as the holonomy of a connection-dependent trivial gerbe on $\xcF_{\rm triv}$---to a background gerbe over $M$ endowed with an $\om$-twisted $\xcG$-equivariant structure.\ The coupling takes the form of the tensor product of the two gerbes and the extended PSM is defined as the holonomy of the product gerbe.\ An essential generalisation of the theory is then proposed---to shadows $\xcF$ of principaloid bundles $\xcP$ of \emph{arbitrary} topology,\ in keeping with the interpretation of the gauging as an effective configurational descent to the orbistack $M//\xcG$.\ The generalisation uses local product gerbes depending on \emph{local} connections over trivialisations of $\xcF$.\ The $\xcG$-equivariant structure is proven to ensure a 2-stacky descent of the local gerbes to a global one over $\xcF$.\ Its holonomy is taken as the definition of the generalised PSM.\ A complete cohomological discussion of obstructions against the descent is provided.\ Finally,\ a reformulation of the generalised PSM as a field theory with a \emph{topological} gauge-symmetry defect is given,\ in which the $\xcG$-equivariant structure is employed---alongside the transition cocycle of $\xcP$---as the datum of the defect.
\end{abstract}

\maketitle
	
\tableofcontents
	
\section{Introduction}

The gauge principle underlies much of modern physical modelling---from the Standard Model of fundamental interactions,\ through the 2$d$ $\si$-models of string theory on homogeneous spaces,\ all the way to topological field theory,\ such as the 3$d$ Chern--Simons theory.\ It comes with a neat geometric assignment of field-theoretic types:\ Gauge fields are identified with connections on bundles of frames (aka gauge bundles),\ and matter fields are represented by global sections of the latter's quotients with respect to a principal action of the symmetry model\footnote{We intentionally abstain from specifying the symmetry model at this stage.\ The classic one,\ largely regarded as synonymous with the notion of symmetry,\ is a group acting smoothly on a manifold,\ {\it i.e.},\ an action groupoid.\ In what follows,\ we stumble upon generalisations of this classic case,\ given by arbitrary Lie groupoids.} on their total spaces,\ representing a global change of frame.\ A `minimal' coupling between the two field species is captured by the covariant derivative,\ and gauge symmetry is implemented by those automorphisms of the gauge bundles---locally presented as families of \emph{diffeomorphisms} of their typical fibres,\ indexed smoothly by points in the spacetime---which are equivariant with respect to the global change of frame.\ The geometric framework naturally accomodates dynamics---typically driven by the curvature of the corresponding connections---of both:\ gauge fields and probes charged under these fields.\ The topological couplings arising in the latter setting,\ also in a categorified (`higher gauge') form,\ employ differential characters \cite{Alvarez:1984es,Gawedzki:1987ak},\ which pave the way to prequantisation of the dynamics and a rigorous description of Aharonov--Bohm-type interference phenomena.

This elaborate geometric structure reflects the deeper idea behind the gauging,\ to wit:\ an effective descent of dynamics to the space of orbits of the symmetry within the typical fibre of the configuration bundle of the physical system,\ {\it i.e.},\ within the space of internal degrees of freedom of the theory.\ The space of orbits need not admit a smooth structure,\ and so it is not---in general---directly accessible to field-theoretic constructions,\ but it does admit a homotopy model---the classifying space of the symmetry-action groupoid,\ given by the geometric realisation of the groupoid's simplicial nerve.\ The homotopy model serves as the base of a universal gauge bundle\footnote{The bundle is called the universal principal bundle of the symmetry-action groupoid,\ see,\ {\it e.g.},\ \cite{Arias:2011}.},\ reproducing its physical avatar upon pullback \emph{to the associated configuration bundle},\ the latter two fibering over the spacetime of the physical theory.\ Furthermore,\ as shown by Dupont \cite{Dupont:1976},\ it has a well-defined model of differential cohomology---the Bott--Shulman-Stasheff  cohomology (and its Beilinson--Deligne-type resolutions)---which suggests itself as a natural source of extensions of cohomological structures on the configuration fibre of the theory to be employed in the gauging.\ This becomes relevant whenever such structures enter the definition of the dynamics,\ as is the case,\ {\it e.g.},\ for the Polyakov--Alvarez--Gaw\c{e}dzki (PAG) $\si$-model of charged loop mechanics in external higher gauge fields.\ The existence and status of the universal bundles and their relation to their physical counterparts---the gauge bundles---over the spacetime,\ indicates the necessity to incorporate \emph{all} isoclasses of the latter in a consistent gauging,\ and---indeed---violating this rule has been demonstrated to lead to pathologies in the quantum r\'egime,\ see:\ \cite[Sec.\,4]{Gawedzki:2010rn} for a discussion and further references to the literature.

Lifting classical gauge symmetries to the quantum theory may encounter obstructions,\ which go under the name of gauge anomalies.\ These are both ubiquitous and naturally quantifiable in the presence of (background) cohomological structures over the configuration bundle---de Rham cocycles and their geometrisations (aka $n$-gerbes and related higher-geometric objects)---which couple to charges carried by the internal degrees of freedom of the theory \cite{Gawedzki:2010rn}.\ Such structures are themselves associated with an independent (higher) gauge symmetry,\ and making the two gauge symmetries compatible with each other---typically through equivariantisation of the original cohomological structure relative to the symmetry to be gauged,\ see:\ {\it ibidem},\ and also \cite{Gawedzki:2003pm,Gawedzki:2007uz,Gawedzki:2008um,Runkel:2008gr,Gawedzki:2012fu,Suszek:2012ddg,Suszek:2026col}---requires the vanishing of certain obstruction classes in (equivariant) cohomology.\ This,\ then,\ becomes a source of superselection criteria for admissible backgrounds of the dynamics of interest,\ distinguishing those which allow for an effective descent to the space of orbits of the symmetry under gauging,\ effectively represented by descent to the quotient configuration bundles of arbitrary topology.

Thus,\ altogether,\ a fully fledged non-anomalous realisation of the gauge principle endows the physical theory not merely with a tightly knitted geometric structure,\ but also with a unique conceptual status---that of a model of dynamics admitting coherent configurational descent.\ Accordingly,\ it becomes highly desirable to uncover or reconstruct such a realisation whenever a gauge symmetry manifests,\ {\it e.g.},\ through the emergence of a nontrivial kernel of the presymplectic form of the theory in its lagrangean formulation \cite{Gawedzki:1972ms}.\medskip

There is an important topological field theory which has long been known to exhibit \emph{infinitesimal} gauge symmetries---that is null directions of the associated presymplectic form---without an {\it a priori} integrated (configurational) model and so also without the corresponding gauge bundle and gauge field.\ The field theory at issue is the Poisson $\si$-model (PSM) of Ikeda \cite{Ikeda:1993fh},\ Schaller and Strobl \cite{Schaller:1994es}.\ As established by Cattaneo and Felder in \cite{Cattaneo:1999fm},\ the field theory becomes---upon quantisation in Feynman's path-integral scheme---of key relevance to Kontsevich's programme of deformation quantisation of Poisson manifolds \cite{Kontsevich:1997vb},\ and so it seems worthwhile---in the light of our hitherto discussion---to identify and `reconstruct' rigorously the complete gauge geometry and field content behind the elementary manifestation of symmetry.\ This is the subject matter of the research presented below,\ developing and elaborating observations which arose as a spin-off from the Author's study of the gauging of groupoidal symmetries in the dynamical PAG $\si$-model,\ reported in \cite{Suszek:2026col}.

In its basic formulation,\ the PSM is a lagrangean theory of vector-bundle morphisms $\vartheta\colo\txT\Si\to\txT^*M$,\ covering smooth base components $X\colo\Si\to M$,\ from the tangent bundle $\txT\Si$ over a compact oriented two-dimensional `spacetime' manifold $\Si$ without boundary---called the worldsheet---to the cotangent bundle $\txT^*M$ over a Poisson manifold $(M,\Pi)$.\ The Poisson bivector $\Pi$ endows $\txT^*M$ with a canonical structure of a Lie algebroid \cite{Coste:1987},\ and critical points of the PSM action functional correspond to Lie-algebroid morphisms from the tangent algebroid over the worldsheet to the contangent algebroid over the target space $M$ \cite{Bojowald:2004wu}.\ The action functional has flat directions represented by smoothly $\Si$-indexed families $\ep_\mu(\si)\,\sfd(x^\mu\circ X)(\si),\ \si\in\Si$ of (de Rham-)closed sections of the restricted bundle $\txT^*M\rstr_{X(\Si)}$,\ and the transformation law for components $\vartheta_\mu(\si)\in\txT^*_\si\Si$ of $\vartheta$ in a local coordinate basis $\sfd(x^\mu\circ X)(\si)$ near $X(\si)$,
\qq\nn
\theta_\mu\longmapsto\vartheta_\mu+\sfd\ep_\mu-\bigl(\p_\mu\Pi^{\la\nu}\circ X\bigr)\,\ep_\la\,\vartheta_\nu\,,
\qqq
bears a suggestive formal resemblance to the linerisation of the usual affine transformation law for a Lie algebra-valued gauge field in a field theory in which a \emph{Lie-group} symmetry has been gauged---with one crucial caveat:\ The finite-dimensional tangent algebra of a Lie group is replaced by the infinite-dimensional Lie algebra of (compactly supported) sections $\G_{\rm c}(\txT^*M)$ of the cotangent Lie algebroid,\ and in the stead of a standard Lie algebra-valued gauge field we find a 1-form on the worldsheet $\Si$ taking values in that Lie algebroid $\txT^*M$ \emph{over the target manifold} $M$,\ further restricted to the worldsheet's image $X(\Si)\subset M$.\ On one hand,\ these observations document the failure of the \emph{standard} model of gauge field theory to capture the field and symmetry content of the PSM.\ On the other hand,\ they provide a clear-cut indication of a natural generalisation of this standard model which holds out a prospect of elucidating the structure described,\ and with it---the model itself:\ Such a generalisation would have the symmetry model---and so also the typical fibre of the gauge bundle---given by a Lie groupoid that integrates the cotangent Lie algebroid,\ and gauge transformations locally modelled on smoothly $\Si$-indexed families of elements of a Lie group\footnote{Taking the same Lie group also as the typical fibre of the gauge bundle would lead to an infinite-dimensional principal bundle (in general),\ and so this logical possibility shall not be pursued.} that integrates $\G_{\rm c}(\txT^*M)$ and maps homomorphically to $\Diff(M)$.\ Neither integration is guaranteed to exist (see,\ {\it e.g.},\ \cite{Crainic:2003,Schmeding:2015bis}),\ but whenever the former exists,\ then so does the latter\footnote{The veracity of the converse implication remains an open question since \cite{Schmeding:2015bis}.\ The Author is grateful to Alexander Schmeding for clarifying this point.},\ see:\ \cite{Rybicki:2002ALG,Schmeding:2015bis}.\ Under such circumstances,\ the integrating groupoid $\xcG$ is equipped with an extra differential structure---a multiplicative symplectic form $\om$ on its arrow manifold,\ which turns $\xcG$ into a symplectic groupoid in the sense of Weinstein {\it et al.} \cite{Weinstein:1987,Coste:1987}.\ The Lie group of $\G_{\rm c}(\txT^*M)$ is then identified with the group of bisections of $\xcG$,\ the latter being those global sections of its source map which induce diffeomorphisms of the Poisson object manifold $M$ under post-composition with the target map.\ The existence of the integrating symplectic groupoid $\grpd{(\xcG,\om)}{(M,\Pi)}$,\ with its group of bisections $\bB(\xcG)$,\ is the base assumption and the point of departure of the investigation undertaken in the present work.\medskip

Basic geometric objects---gauge bundles with compatible connections,\ alongside matter configuration bundles---behind the gauge principle for symmetries modelled on an arbitrary Lie groupoid\footnote{We refer the Reader to the Appendices for the notation used in the paper.} $(M,\xcG,s,t,\Id,\Inv,\txm\equiv .)$ were introduced by Strobl and the Author in \cite{Strobl:2025}.\ They were subsequently employed by the Author in the gauging of such symmetries in the PAG $\si$-model\footnote{See also:\ \cite{Suszek:2025,Filipek:2025MSc} for preliminary accounts.}  \cite{Suszek:2026col},\ drawing heavily on a former joint work with Gaw\c{e}dzki and Waldorf \cite{Gawedzki:2010rn,Gawedzki:2012fu} on the gauging of group-modelled symmetries in that model---this is the broad field-theoretic context in which some of the results reported herein were first stumbled upon.

The relevant gauge bundles,\ dubbed principaloid $\xcG$-bundles in \cite{Strobl:2025},\ are fibre-bundle objects 
\qq\nn
\alxydim{@C=.75cm@R=1cm}{ \xcP \ar@{->>}[d]_{\pi_\xcP} \ar[rd]^{\mu} & & \xcG \ar@{=>}[ld] \ar@/_1.pc/[ll]_{\xcR} \\ \Si & M & }\,,
\qqq
in the category of right $\xcG$-modules with the trivial $\xcG$-action on the base $\Si$,\ and the one on the total space $\xcP$,\ denoted as  $\xcR$,\ modelled---by $\xcG$-equivariant local trivialisations $\xcP\t_i\colo\pi_\xcP^{-1}(O_i)\xrightarrow{\cong}O_i\x\xcG$---on the canonical right action of the structure groupoid $\grpd{\xcG}{M}$ on its arrow manifold $\xcG$,\ realised through the right arrow adjunction $r\colo\xcG\fibx{s}{t}\xcG\to\xcG,\ (g,h)\mapsto g.h$.\ These $\xcG$-modules have smooth (Godement) quotients $\xcP/\xcG=\xcF\xrightarrow{\pi_\xcF}\Si$,\ called shadow bundles,\ which play the r\^ole of configuration bundles of the ensuing gauge field theory,\ with the typical fibre $M$ recovered in local trivialisations $\xcF\t_i\colo\pi_\xcF^{-1}(O_i)\xrightarrow{\cong}O_i\x M$.\ Moreover,\ the gauge bundles admit $\xcG$-invariant Ehresmann connections,\ and the corresponding projector fields $\Theta\colo\txT\xcP\to\ker\,\txT\pi_\xcP$ encode,\ in the local presentation:
\qq\nn
\Theta_i(\si,g)\equiv\xcP\t_i^{-1\,*}\Theta(\si,g)=\id_{\txT_g\xcG}-\txT r_g\circ A_i\bigl(\si,t(g)\bigr)\,,\quad(\si,g)\in O_i\x\xcG\,,
\qqq
gauge fields $A_i\in\G(\pr_1^*\txT^*O_i\ox\pr_2^*\xcE)$ over $O_i\x t(\xcG)\equiv O_i\x M$ with values in the tangent Lie algebroid $\xcE=\Id^*\ker\,\txT s$ of the structure groupoid.\ Vertical automorphisms of $\xcP$ which are equivariant with respect to the defining action $\xcR$,\ and hence are locally modelled on smoothly $O_i$-indexed families of bisections of $\xcG$ realised through the canonical left action $L\colo\bB(\xcG)\to\Diff(\xcG),\ \b\mapsto m_\xcG(\b\circ t(\cdot),\cdot)$ of $\bB(\xcG)$ on $\xcG$,\ compose the gauge group ${\rm Gau}(\xcP)$ of the principaloid $\xcG$-bundle $\xcP$.\ The group is homomorphically mapped to the group of (vertical) automorphisms of the shadow bundle,\ $\xcF_*\colo{\rm Gau}(\xcP)\to{\rm Aut}_{\Bun(\Si)/\Si}(\xcF)$,\ and thus gives rise to the shadow gauge group $\im\,\xcF_*\equiv{\rm Gau}(\xcF)$,\ with local model $O_i\to\bB(\xcG)\xrightarrow{t\circ}\Diff(M)$.\ Gauge automorphisms $\Phi\in{\rm Gau}(\xcP)$ become realised naturally on matter fields:\ $\G(\xcF)\ni\phi$ as $\phi\mapsto\xcF_*(\Phi)\circ\phi=:\phi^\Phi$,\ and on projector fields $\Theta$ as $\Theta\mapsto\txT\Phi\circ\Theta\circ\txT\Phi^{-1}$,\ the latter realisation inducing `standard' affine transformations of the gauge fields $A_i$.\ It is in this geometric framework  that we seek to reinterpret the classic Poisson $\si$-model upon internalising the concepts and constructions of \cite{Strobl:2025} in the category of symplectic groupoids.\medskip

Our reconstruction begins in Sec.\,\ref{sec:PsiM},\ where---after recalling the definition of the topological field theory of interest---we match the PSM field types with those present on a trivial principaloid $(\xcG,\om)$-bundle $\Si\x\xcG\equiv\xcP_{\rm triv}$,\ endowed with a trivial projector field $\Theta_{\rm triv}(\si,g)=\id_{\txT_g\xcG}-\txT r_g\circ A(\si,t(g)),\ A\in\G(\pr_1^*\txT^*\Si\ox\pr_2^*\xcE)$,\ and on its shadow $\Si\x M\equiv\xcF_{\rm triv}$,\ with global sections $\phi\in\G(\xcF_{\rm triv})\cong C^\infty(\Si,M)$.\ The matching:
\qq\nn
(X,\vartheta)=\bigl(\pr_2\circ\phi,\kappa_\xcE\circ A(\phi)\bigr)\,,
\qqq
which invokes the canonical isomorphism of Lie algebroids $\kappa_\xcE\colo\xcE\xrightarrow{\cong}\txT^*M$ \cite{Coste:1987},\ is accompanied by a rewriting of the PSM action functional in terms of tensor fields on the principaloid bundle and its typical fibre: 
\qq\nn
-S_{\rm PSM}[X,\vartheta]=\int_{\phi(\Si)}\om[\Theta_{\rm triv}]=:\widetilde S{}_{PSM}[\phi,\Theta_{\rm triv}]\,,\qquad\quad\om[\Theta_{\rm triv}]:=(\id_\Si\x\Id)^*\bigl(\pr_2^*\om\circ(\Theta_{\rm triv},\Theta_{\rm triv})\bigr)\in\Om^2(\xcF_{\rm triv})\,.
\qqq
The rewriting paves the way to a hands-on analysis,\ carried out in Sec.\,\ref{sub:anom-kill-first},\ of the behaviour of the reinterpreted field theory under the well-defined gauge transformations of its fields $(\phi,A)$ engendered by automorphisms $\Phi\colo\xcP_{\rm triv}\ni(\si,g)\mapsto(\si,L_{h(\si)}(g))\in\xcP_{\rm triv}$ with $h\colo\Si\to\bB(\xcG)$.\ The analysis---making ample use of the nontrivial $\bB(\xcG)$-equivariance properties of the multiplicative form $\om$ encoded by its Spencer data,\ as introduced and studied at length by Crainic {\it et al.}\ in \cite{Crainic:2015msp}---demonstrates the necessity of an $\om$-holonomic reduction of the structure group of the principaloid bundle:
\qq\nn
\bB(\xcG)\searrow\bB(\xcG,\om)
\qqq
to the lagrangean subgroup
\qq\nn
\bB(\xcG,\om)=\left\{\ \b\in\bB(\xcG) \quad\vert\quad \b^*\om=0 \ \right\}\,,
\qqq
which then becomes a model for the reduced gauge groups:\ ${\rm Gau}_\om(\xcP_{\rm triv})\ni\Phi$ and ${\rm Gau}_\om(\xcF_{\rm triv})\equiv\xcF_*({\rm Gau}_\om(\xcP_{\rm triv}))$.\ The reduction is proven to be---\emph{in general}---insufficient to ensure the ${\rm Gau}_\om(\xcP_{\rm triv})$-invariance of the field theory:\ The latter exhibits a lagrangean anomaly \eqref{eq:gtrafo-classic},\ which takes the simple form
\qq\nn
\cA_{\rm DF}^{\rm PSM}\bigl[\phi^\Phi,A^\Phi]=\cA_{\rm DF}^{\rm PSM}[\phi,A]\cdot\exp\left(-\sfi\int_{\phi(\Si)}\widetilde h{}^*\om\right)\,,\qquad\qquad\widetilde h\colo\xcF_{\rm triv}\too\xcG,\ (\si,m)\longmapsto h(\si)(m)
\qqq
in the Dirac picture
\qq\nn
\cA_{\rm DF}^{\rm PSM}[\phi,A]=\exp\left(\sfi\,\widetilde S{}_{PSM}[\phi,\Theta_{\rm triv}]\right)\,.
\qqq
The derivation of the lagrangean anomaly takes the discussion of the reinterpreted PSM to the next level as it provides a natural criterion of ((pre)quantum-mechanical) consistency on the space of PSM configuration backgrounds $\grpd{(\xcG,\om)}{(M,\Pi)}$:\ It distinguishes those backgrounds for which the anomaly vanishes identically,\ that is---the prequantisable ones,\ with
\qq\label{eq:preq-constr}
{\rm Per}(\om)\subset 2\pi\bZ\,,
\qqq
{\it i.e.},\ those for which there exist principal $\bC^\x$-bundles 
\qq\label{eq:preqbndl} 
\alxydim{@C=1.25cm@R=1.cm}{ \bC^\x \ar@{^{(}~>}[r] & (P_\om,\cA) \ar[d]_{\pi_{P_\om}} \\ & (\xcG,\om) }
\qqq
with (principal) connection $\cA\in\Om^1(P_\om)$ of curvature $\om$,\ such that $\pi_{P_\om}^*\om=\sfd\cA$.\ This is the first key quantitative \emph{prediction} of our investigation.\medskip

An alternative to \emph{avoiding} the lagrangean anomaly through the imposition of the superselection rule \eqref{eq:preq-constr} on admissible configuration backgrounds $\grpd{(\xcG,\om)}{(M,\Pi)}$ is conceived in Sec.\,\ref{sec:coaug-cancel}.\ It amounts to augmenting the geometric background by an extra cohomological component that \emph{cancels} the anomaly through a topological coupling to the PSM matter field.\ The choice of an augmentation which prevents the leakage of the symplectic charge due to the anomaly in the original physical system is suggested by the embedding of the cohomological datum $\om$ of the latter system in a cochain complex better suited to the present considerations of an effective configurational descent of the field theory to the orbistack $M//\xcG$ (or,\ more adequately,\ $M//\bB(\xcG,\om)$) implemented by the gauging,\ to wit:\ the diagonal complex of the BSS bicomplex mentioned at the beginning of this Introduction.\ In it,\ the standard de Rham cohomology is combined with the simplicial cohomology of the nerve $N_\bullet(\grpd{\xcG}{M})\equiv\xcG_\bullet$ of the symplectic groupoid\footnote{The Reader is urged to consult App.\,\ref{app:simplicial} for a precise definition.}.\ Thus,\ the symplectic form is regarded as a $(2,1)$-component of a BSS 3-cocycle\footnote{3-cocycles of this form have been considered in the literature,\ see,\ {\it e.g.},\ the early works of Cattaneo and Xu \cite{Cattaneo:2004} and those of Bursztyn {\it et al.}\ \cite{Bursztyn:2004} on twisted presymplectic groupoids.} $(H_M,\om,0,0)\in Z^3_{\rm dR}(M)\oplus\Om^2(\xcG_1)\oplus\Om^1(\xcG_2)\oplus\Om^0(\xcG_3)$ (for $\xcG_n=\xcG\fibx{s}{t}\xcG\fibx{s}{t}\cdots\fibx{s}{t}\xcG\subset\xcG^{\x n},\ n>0$),\ and it is a geometrisation of the complete BSS 3-cocycle that is considered,\ instead of that of its $(2,1)$-component in isolation.\ The geometrisation assumes the form of an $\om$-twisted $\xcG$-equivariant structure on Murray's abelian gerbe $\cG$ of curvature $H_M$ \cite{Murray:1994db,Murray:1999ew} over the Poisson manifold $(M,\Pi)$,\ as defined in \cite{Gawedzki:2010rn}.\ Besides the gerbe $\cG$,\ the structure is composed of distinguished cells of the monoidal bicategory of gerbes over the stratified manifold $\xcG_\bullet$:\ a gerbe 1-isomorphism over $\xcG_1$:
\qq\nn
\Upsilon\colo t^*\cG\xrightarrow{\ \cong\ }s^*\cG\ox\cI_\om\,,
\qqq
written in terms of the trivial gerbe $\cI_\om$ associated with the 2-form $\om$,\ and a gerbe 2-isomorphism over $\xcG_2$:
\qq\nn
\g\colo\bigl(\pr_2^*\Upsilon\ox\Id_{\cI_{\pr_1^*\om}}\bigr)\circ\pr_1^*\Upsilon\overset{\cong}{\Longrightarrow}\txm^*\Upsilon\,,
\qqq
the latter being subject to a coherence (`associativity') constraint over $\cG_3$,\ see:\ App.\,\ref{app:gerbarium} for details.

While the idea may seem non-obvious in the present context,\ it is actually entirely natural from the vantage point offered by the 2$d$ PAG $\si$-model \cite{Gawedzki:2010rn,Suszek:2026col} with a rigid Lie-groupoidal symmetry $\grpd{\xcG}{M}$ gauged,\ which was the key source of the requisite intuition.\ In that motivating field-theoretic setting,\ the ontic priority between the geometric structures is reversed:\ One presupposes the existence of a gerbe $\cG$,\ determining the $\si$-model's  coupling in the form of the corresponding Cheeger--Simons differential character with curvature $H_M$ (aka the surface holonomy) \cite{Alvarez:1984es,Gawedzki:1987ak},\ and then discovers the necessity to equip it with a $\xcG$-equivariant structure in an attempt to gauge the $\si$-model's groupoidal symmetry.\ The structure is generically twisted by a multiplicative 2-form $\om$---a fact determined by the contraction of the gerbe's curvature $H_M$ with the fundamental vector field of the action of $\xcG$ on the target space $M$,\ captured by $\sfd\om$---and so the BSS extension $(H_M,\om,0,0)$ of $H_M$ arises.\ Its aforementioned geometrisation ensures the cancellation of the surface holonomy of the pullback gerbe $\pr_2^*\cG$ over the configuration bundle $\xcF_{\rm triv}$ under its shadow gauge transformations from the \emph{holonomically reduced} gauge group $\bB(\xcG,\om)$.\ The source of the cancellation---the anomaly counterterm---is a multiplicative correction to the holonomy of $\pr_2^*\cG$ of the form $\cA_{\rm DF}^{\rm PSM}[\phi,A]$.\ This observation,\ originally made in \cite{Suszek:2026col},\ sheds  a new light on the geometric status of the PSM,\ and leads directly to the postulated anomaly-cancellation mechanism based on the $\xcG$-equivariant extension
\qq\label{eq:Gext}
\om\nearrow(H_M,\om,0,0)\nearrow(\cG,\Ups,\g)\,,
\qqq 
followed by a (co-)augmentation
\qq\label{eq:coaug}
\om[\Theta_{\rm triv}]\nearrow\pr_2^*\cG\ox\cI_{\om[\Theta_{\rm triv}]}=:\cG[\Theta_{\rm triv}]\,,
\qqq
which yields the desired ${\rm Gau}_\om(\xcP_{\rm triv})$-invariant Dirac--Feynman amplitude
\qq\label{eq:ADFtPsiG}
\cA_{\rm DF}^{tP\si[\cG]}[\phi,A]=\Hol_{\pr_2^*\cG}\bigl(\phi(\Si)\bigr)\cdot\cA_{\rm DF}^{\rm PSM}[\phi,A]\,.
\qqq
This is the second key quantitative result of our investigation,\ presented in detail in Sec.\,\ref{sec:coaug-cancel}.\ This result is substantially enriched by the following observations,\ which bring the section to a close:\ The gauge invariance for the \emph{trivial} gauge bundle $\xcP_{\rm triv}\equiv\Si\x\xcG$ calls only for the 1-isomorphism $\Ups$,\ and the ensuing coaugmentation of the original (reinterpreted) PSM subsumes the conservative scenario---the one in which the anomaly vanishes owing to the prequantisability of $\om$---as its trivial case $\cG\equiv\cI_0$.\ Furthermore,\ it gives us complete control over corrections to the Euler--Lagrange equations of the field theory,\ which are proportional to the curvature $H_M$ of the coaugmenting gerbe.\ In this manner,\ the mechanics of the field theory---and so also its Lie-algebroidal interpretation---can be preserved,\ or twisted \`a la Klim\v cik and Strobl \cite{Klimcik:2002}.\medskip

As explained at the beginning of the present Introduction,\ the logic of the configurational descent to the orbistack $M//\xcG$ (resp.\ $M//\bB(\xcG,\om)$) requires that the (reinterpreted) PSM be defined on principaloid and shadow bundles of \emph{arbitrary} topology.\ The exhaustive analysis of the gauging of Lie-group symmetries in the PAG $\si$-model in \cite{Gawedzki:2010rn,Gawedzki:2012fu},\ extended to symmetries modelled on arbitrary Lie groupoids in \cite{Suszek:2026col},\ shows that the extension \eqref{eq:Gext} and the subsequent coaugmentation \eqref{eq:coaug} open an avenue to a methodical generalisation of the model \eqref{eq:ADFtPsiG} beyond the trivial case considered in the literature.\ The generalisation---worked out in Sec.\,\ref{sec:coaug-PSM-def} along the lines of \cite{Suszek:2026col}---bases on the local trivialisability of the gauge bundles and exploits the structure of a 2-stack on the smooth site (relative to the Grothendieck sieve determined by open covers) carried by gerbes \cite{Stevenson:2000wj}.\ The latter structure is an analogon of the 1-stack structure carried by principal $\bC^\x$-bundles,\ which was described by Brylinski in \cite{Brylinski:1993ab}.\ Its coherent application in the present field-theoretic setting calls for the fully fledged $\om$-twisted $\xcG$-equivariant structure on the target gerbe.\ In the previously encountered trivial case $\cG=\cI_0$,\ it boils down to a multiplicative structure\footnote{Prequantisations of this distinguished type have been considered in the mathematical literature in the context of quantisation of Poisson structures,\ see,\ {\it e.g.},\ \cite{Weinstein:1991,Crainic:2004,Crainic:2007,Laurent-Gengoux:2007,Hawkins:2008}.} on the prequantum bundle \eqref{eq:preqbndl} from the conservative anomaly-evasion scenario,\ derived in Sec.\,\ref{sub:geom-cohom-equiv}.\ Our discussion of the extension culminates with a definition of the gerbe-coaugmented PSM---the third key quantitative result of our investigation---which is none other than an equivariantised character for the coaugmenting gerbe $\cG$,\ originally introduced in \cite{Suszek:2025,Filipek:2025MSc,Suszek:2026col}.\ The intrinsically higher-cohomological definition of the field theory affords a quantitative analysis of gauge anomalies in terms of the cohomology of the \v Cech--Deligne--Dupont tricomplex that resolves the BSS bicomplex.\ Here,\ the anomalies are identified---in direct analogy with \cite{Gawedzki:2010rn}---with topological obstructions against the existence of the $\xcG$-equivariant gerbe,\ and---as such---they reflect the clash of the two gauge systems:\ the 1-gauge system $(\xcP,\Theta,\xcF)$ and the 2-gauge system $(\cG,\Ups,\g)$,\ a mechanism reminiscent of the standard emergence of anomalies in the quantum field theory in the presence of competing symmetries \cite{Adler:1969,Bell:1969,tHooft:1980},\ see:\ \cite{Bertlmann:2000} for an at-length modern account.\medskip

The vast generalisation of the (reinterpreted) classic PSM to arbitrary gauge and matter bundles attained in Sec.\,\ref{sec:coaug-PSM-def},\ while consistent with the construction of the homotopy quotient\footnote{See also:\ \cite{Tu:2020} for a useful recapitulation of the logic of the descent in the action-groupoid setting.} and the descent principle for (2-)stacks,\ may appear somewhat abstract.\ This is to be contrasted with the straightforward conceptual approach to gauging in,\ {\it e.g.},\ the axiomatic setting of topological quantum field theory,\ in which gauging becomes essentially synonymous with the existence of distinguished defects in the `spacetime' of the field theory that implement the symmetry gauged through incorporation of symmetry-modelled discontinuities of its fields,\ see,\ {\it e.g.},\ \cite{Runkel:2008gr} for an early precursor in the 2$d$ (R)CFT setting,\ and \cite{Carqueville:2023jhb} for a recent overview of the purely topological approach,\ and an extensive bibliography.\ The penultimate  Sec.\,\ref{sec:gau-sigmod} of the present paper provides a complete reinterpretation of the previous results of the 2-stacky descent,\ based on the existence of the $\xcG$-equivariant structure \eqref{eq:Gext},\ in terms of a manifestly topological gauge-symmetry defect with simplicial induction.\ The reinterpretation conforms with the advocated presentation of the (coaugmented) PSM as the topological sector of the dynamical PAG $\si$-model and the pioneering treatment of defects in the latter field theory due to Runkel and the Author \cite{Runkel:2008gr,Suszek:2012ddg,Suszek:2013,Suszek:2022CSimpl} (see also \cite{Fuchs:2007fw} for some preliminary results).\ In this picture,\ building on the original proposal of \cite{Runkel:2008gr} (for discrete-group symmetries) and \cite{Suszek:2012ddg} (for their continuous counterparts),\ the 1-isomorphism $\Ups$ of the $\xcG$-equivariant structure is used to define cohomological data of a flat bi-brane (see:\ \cite{Fuchs:2007fw,Runkel:2008gr})---to be pulled back to the codimension-1 gauge-symmetry defect that separates trivialisation domains $\Si_i\subset O_i$ within the PSM worldsheet which host the local gerbes $\cG[\Theta_i]$ and across which the local matter fields $\xcF\t_i\circ\phi$ jump in a manner controlled by the transition 1-cocycle of $\xcP$---whereas the 2-isomorphism $\g$ determines the contribution to the relevant stratified differential character---the Dirac--Feynman amplitude of the coaugmented PSM with gauge symmetry defects---of junctions of such gauge-symmetry defects,\ both elementary (trivalent) and---through the homotopic induction mechanism discussed in \cite{Runkel:2008gr},\ exploiting the coherence of $\g$---also those of higher valence.\ The defect picture of the coaugmented PSM is the fourth---and last---key quantitative result of our investigation,\ which concludes the present work and opens avenues for future research,\ demarcated in the closing Sec.\,\ref{sec:summ-out}.\medskip

\begin{center}
*\hspace{2cm}*\hspace{2cm}*
\end{center}

The investigation presented in this work employs tools from a number of seemingly disparate mathematical domains.\ In order to facilitate the Reader's navigation through the paper and to make the latter essentially self-contained,\ the relevant background material---definitions,\ examples,\ propositions and theorems---has been collected in the Appendices.
\bigskip 
	
\noindent{\bf Acknowledgements:}\ The Author wishes to thank Janusz Grabowski,\ Noriaki Ikeda,\ Alexei Kotov,\ and Thomas Strobl for their sustained interest in and discussions about various aspects of the problems investigated in the present work.\ He has also profited from an exchange with Alexander Schmeding regarding the issue of the integrability of the algebra of compactly supported sections of a Lie algebroid.\ He gratefully acknowledges the hospitality of the Mathematical Research and Conference Center of the Institute of Mathematics of the Polish Academy of Sciences in B\c{e}dlewo,\ where the results of the work were first reported,\ in July 2026,\ in the form of a series of lectures given at the Summer School ``Differential-geometric and topological methods in modern physics'' for PhD and Master's students,\ coorganised by the Author jointly with Nils Carqueville,\ Janusz and Katarzyna Grabowscy,\ Thomas Strobl {\it et al}.\ The stimulating atmosphere of the Polish--Czech Workshop on Graded Geometry at the University of Hradec Kr\'alov\'e,\ organised by Alexei Kotov {\it et al.}\ in October 2026,\ at which the results were presented and discussed,\ is likewise noted with gratitude.\smallskip

In the course of this work,\ the Author was supported by the Excellence Initiative -- Research University (IDUB) programme at the University of Warsaw.\medskip

\noindent{\bf Disclaimer:}\ No \emph{Artificial} Intelligence---neither its tools nor its methods---has been employed in the research reported in this article.

\section{The classic Poisson $\si$-model and its flat directions --- reinterpreted}\label{sec:PsiM}

We shall be concerned with theories of classical fields over closed two-dimensional `spacetimes',\ taking values in cotangent Lie algebroids over integrable Poisson manifolds---the Poisson $\si$-models of Ikeda  \cite{Ikeda:1993fh} and Schaller and Strobl \cite{Schaller:1994es},\ and their generalisations with lagrangean gauge symmetry,\ first proposed by the Author in \cite{Suszek:2026col}.\ Our interest in this class of topological field theories is motivated by their key relevance to the deformation quantisation of Poisson structures in the sense of Kontsevich \cite{Kontsevich:1997vb},\ as established by Cattaneo and Felder in \cite{Cattaneo:1999fm,Cattaneo:2000iw}.\ For general definitions of the geometric structures employed,\ we refer the Reader to App.\,\ref{app:Lie}.

\subsection{The field theory}\label{sub:PSMasFT}
 
Let $(M,\Pi)$ be a Poisson manifold.\ Recall that the vertical vector-bundle morphism 
$\Pi^\sharp\colo\txT^*M\to\txT M$ induced by 
the Poisson bivector $\Pi\in\G(\wedge^2\txT M)$ endows the cotangent bundle $\pi_{\txT^*M}\colo\txT^*M\to M$ with the canonical structure of a Lie algebroid $(\txT^*M,\Pi^\sharp,[\cdot,\cdot]_{\rm K})$,\ with $\Pi^\sharp$ as the anchor,\ and with the Koszul bracket \cite{Koszul:1985bra} 
\qq\nn
[\cdot,\cdot]_{\rm K}\colo\Om^1(M)\x\Om^1(M)\too\Om^1(M),\ (\eta_1,\eta_2)\longmapsto\imath_{\Pi^\#(\eta_1)}\sfd\eta_2-\imath_{\Pi^\#(\eta_2)}\sfd\eta_1+\sfd\bigl(\eta_1\wedge\eta_2(\Pi)\bigr)
\qqq
as the Lie bracket.\ The latter is uniquely determined by the requirement of homomorphicity of the de Rham differential $\sfd\colo(C^\infty(M,\bR),\{\cdot,\cdot\}_\Pi)\to(\Om^1(M),[\cdot,\cdot]_{\rm K})$ in conjunction with the Leibniz rule,\ see:\ \cite{Coste:1987} for details.

The field theory that we want to discuss first is that of vector-bundle morphisms $(\vartheta,X)$ captured by commutative diagrams ($\pi_{\txT\Si}\colo\txT\Si\to\Si$ is the canonical base projection)
\qq\label{eq:PSM-field}
\alxydim{@C=1.25cm@R=1.cm}{ \txT\Si \ar[r]^{\vartheta\quad} \ar[d]_{\pi_{\txT\Si}} & \txT^*M \ar[d]^{\pi_{\txT^*M}} \\ \Si \ar[r]_X & M}\,.
\qqq
For these,\ we define the lagrangean density of the {\bf Poisson $\si$-model} 
\qq\label{eq:PSMAct}
\ceL_{\rm PSM}(X,\txT X;\vartheta)=\ev\circ\bigl(\vartheta\ox\txT X-\tfrac{1}{2}\,\vartheta\,\ox\bigl(\Pi^\sharp\circ\vartheta\bigr)\bigr)\circ{\rm Alt}\,,
\qqq
in which $\ev\colo \txT^*M\ox\txT M\to C^\infty(M,\bR)$ is the canonical evaluation map,\ and ${\rm Alt}\colo\txT\Si\ox\txT\Si\to\txT\Si\ox\txT\Si$ is the standard (un-weighted) alternator,\ ${\rm Alt}(v\ox w)=v\ox w-w\ox v$.\ Explicitly---in local coordiate charts:\ $\si^a,\ a\in\{1,2\}$ around $\si\in\Si$,\ and $x^\mu,\ \mu\in\ovl{1,\dim M}$ around $X(\si)\in M$,\ in which $\vartheta\underset{\tiny{{\rm loc.}}}{=}\vartheta_{a\mu}\,\sfd\si^a\ox\sfd(x^\mu\circ X)\equiv\vartheta_\mu\ox\sfd(x^\mu\circ X)$ and $\Pi\underset{\tiny{{\rm loc.}}}{=}\Pi^{\mu\nu}\,\p_\mu\wedge\p_\nu$---we have
\qq\nn
\ceL_{\rm PSM}(X,\txT X;\vartheta)\underset{\tiny{{\rm loc.}}}{=}\bigl(\vartheta_{a\mu}\,\p_b\bigl(x^\mu\circ X\bigr)+\tfrac{1}{2}\,\bigl(\Pi^{\mu\nu}\circ X\bigr)\,\vartheta_{a\mu}\,\vartheta_{b\nu}\bigr)\,\sfd\si^a\wedge\sfd\si^b\,.
\qqq
As demonstrated in \cite{Bojowald:2004wu},\ classical field configurations of this two-dimensional field theory,\ solving the Euler--Lagrange (EL) equations 
\qq\nn
\left\{ \barr{l}  \sfd\vartheta_\la+\tfrac{1}{2}\,\p_\la\Pi^{\mu\nu}(X)\,\vartheta_\mu\wedge\vartheta_\nu=0 \\ 
\sfd\bigl(x^\la\circ X\bigr)+\Pi^{\la\mu}(X)\,\vartheta_\mu=0 \earr \right.\,,\quad\la\in\ovl{1,\dim M}\,,
\qqq
are in a one-to-one correspondence with morphisms between the tangent Lie algebroid $(\txT\Si,\id_{\txT\Si},[\cdot,\cdot]_{\G(\txT\Si)})$ (with the trivial anchor $\id_{\txT\Si}$ and the commutator $[\cdot,\cdot]_{\G(\txT\Si)}$ of vector fields as the Lie bracket) and the cotangent Lie algebroid  $(\txT^*M,\Pi^\sharp,[\cdot,\cdot]_{\rm K})$.

The action functional 
\qq\label{eq:PsigM}
S_{\rm PSM}[X,\vartheta]=\int_\Si\,\ceL_{\rm PSM}(X,\txT X;\vartheta)
\qqq
of the Poisson $\si$-model  has flat directions---aka \emph{tangential} (or \emph{infinitesimal}) gauge symmetries---spanned on sections $\ep\in\G(X^*\txT^*M)$,\ with local presentations $\ep\underset{\tiny{{\rm loc.}}}{=}\ep_\mu\,\sfd(x^\mu\circ X)$.\ These model transformations of field variables
\qq\label{eq:tan-Pgauge}\qquad\qquad
x^\mu\circ X\longmapsto x^\mu\circ X-\bigl(\Pi^{\mu\nu}\circ X\bigr)\,\ep_\nu=:x^\mu\circ X^\ep\,,\qquad\qquad\vartheta_\mu\longmapsto\vartheta_\mu+\sfd\ep_\mu-\bigl(\p_\mu\Pi^{\la\nu}\circ X\bigr)\,\ep_\la\,\vartheta_\nu=:\vartheta^\ep_\mu\,,
\qqq
for which we obtain
\qq\label{eq:tanact-Pgauge}
S_{\rm PSM}[X^\vep,\vartheta^\vep]=S_{\rm PSM}[X,\vartheta]+\int_\Si\sfd\ep+\xcO(\ep^2)=S_{\rm PSM}[X,\vartheta]+\xcO(\ep^2)\,.
\qqq
The first of the two transformations admits a simple \emph{structural} interpretation:\ It is a linearisation of a smoothly $\Si$-indexed hamiltonian flow of $X$ with generator $\Pi^\sharp(\ep)$.\ The nature of the second one,\ on the other hand,\ is much less obvious,\ despite its familar affine structure,\ which led to its identification as a `gauge field' in \cite{Bojowald:2004wu}.\ The reason is that---unlike the first one---the second expression is neither geometric,\ {\it i.e.},\ it does not define a vector field on the space of vertical vector-bundle morphisms $\txT\Si\to\txT^*M$,\ nor straightforwardly algebraic,\ {\it i.e.},\ the commutator algebra of such transformations closes only on the space of solutions to the EL equations,\ see \cite[Sec.\,4]{Bojowald:2004wu} for details.\ Attempts at rectifying these shortcomings,\ undertraken in \cite{Bojowald:2004wu} and \cite{Kotov:2014iha},\ led to the emergence of two families of its (noncanonical) covariant algebraic extensions proportional to $F^\la(X,\vartheta)=\sfd\bigl(x^\la\circ X\bigr)+\Pi^{\la\mu}(X)\,\vartheta_\mu$.\ These extensions are by no means innocent,\ as they either call for a reintroduction of an explicit dependence of the $\ep_\mu$ on the section $X$,\ or require a choice of connection on the Koszul algebroid.\ And even then,\ the story does not end,\ as the demand that the extension define a flat direction of the action functional \eqref{eq:PsigM} imposes secondary constraints either on the gauge-symmetry data $\ep$ (with the dependence on $X$ reintroduced),\ or on the torsion of the connection used.\ It is to be emphasised that this substantial buildup of complexity,\ manifest in the analysis of \cite{Bojowald:2004wu},\ is not accompanied by any change of the status of the symmetry,\ which remains tangential after the extension.

In view of the above,\ we take as a firm basis of our subsequent considerations the \emph{structural} (geometric) tangential transformation of $X$,\ alongside the equally \emph{structural} property of the action functional,\ to wit:\ its strict invariance under the combined tangential symmetry of the two field variables in the \emph{linear} order in the functional transformation data $\ep_\mu$ for a \emph{closed} worldsheet $\Si\in\p^{-1}\emptyset$.

\brem\label{rem:tang-Pgauge}
The above characterisation of tangential gauge symmetries of the classic Poisson $\si$-model appears already in the original articles \cite{Ikeda:1993fh,Schaller:1994es}.\ They were also systematically determined in \cite{Falceto:2001eh} within the first-order formalism of Gaw\c{e}dzki,\ Kijowski,\ Tulczyjew and Szczyrba \cite{Gawedzki:1972ms,Kijowski:1973gi,Kijowski:1974mp,Kijowski:1976ze,Kijowski:1979dj}---as generators of the kernel of the presymplectic form of $\ceL_{\rm PSM}$ on the model's space of states.\ Their structure prompts us to regard them as smoothly $\Si$-indexed (and so compactly supported) \emph{closed} sections of $\txT^*M$,\ but the identification is obscured by the local nature of their definition,\ and the $\Si$-indexing itself (in which,\ in particular,\ a rewriting $\vep_\mu(\si,X(\si))=\ep_\mu(\si)$ would be justified,\ or even technically necessary,\ see:\ \cite{Bojowald:2004wu}).\ Hence,\ a precise identification of the subalgebra $\egt\subset\G_{\rm c}(\xcE)$ which is being \emph{gauged} here also seems somewhat unclear.\ The analysis of the enhancement of the classic formulation proposed in \cite{Suszek:2026col},\ which we carry out below,\ fixes $\egt$ in a structural manner.
\erem

\beg\label{ex:Lu-Weinstein}
Given a Poisson--Lie group $(\txG,\Pi)$ \cite{Drinfeld:1983,SemenovTianShansky:1985PL},\ there exists a canonical structure of a Lie algebroid $((\txT^*\txG\cong\txG\,\lx_{\Ad^*}\ggt^*,\txG,\bR^{\dim\,\txG},\pi_{\txT^*\txG}\equiv\pr_1),\Pi^\#,\{\cdot,\cdot\})$ on its cotangent bundle---the cotangent Lie algebroid,\ see:\ Ex.\,\ref{eg:cotang-Liealgd}.\ The dual $\ggt^*$ of $\ggt\equiv\Lie(\txG)$ is endowed with a Lie bracket $[\cdot,\cdot]^*$ induced from the Poisson bracket $\{\cdot,\cdot\}_\Pi$ on $C^\infty(\txG,\bR)\ni f_1,f_2$ through the identification $\ggt^*\equiv\txT^*_e\txG$ as
\qq\nn
[\txd f_1(e),\txd f_2(e)]^*:=\txd\{f_1,f_2\}_\Pi(e)\,,
\qqq 
and so $\ggt^*$ is canonically a Lie algebra.\ The corresponding Lie group $\txG^*$ is called the dual of $\txG$.\ The Poisson bracket on $C^\infty(\txG^*,\bR)$ is the unique one whose linearisation at the group unit $e^*\in\txG^*$ reproduces the Lie bracket $[\cdot,\cdot]$ on $\ggt\cong\ggt^*{}^*$.\ The Lie algebroid $\txG\,\lx_{\Ad^*}\ggt^*$ integrates to a symplectic Lie groupoid $\grpd{(\xcL\xcW}{\txG},\om_{\xcL\xcW})$---the {\bf Lu--Weinstein groupoid},\ introduced in \cite{Lu:1989sdg} (see also:\ \cite{Morgan:2019phd} for a detailed recent account).

The corresponding classic Poisson $\si$-model was discussed at length in relation to the maximally gauged WZNW $\si$-model with the dual Poisson--Lie group $(\txG^*,\Pi^*)$ as the target by Alekseev {\it et al.}\ in \cite{Alekseev:1995py},\ and subsequently by---{\it i.a.}---Falceto {\it et al.} in \cite{Falceto:2001eh} and \cite{Calvo:2003kv}.
\eeg

\subsection{An instructive toy example}\label{sub:toy-lPSM} 

Consider an $\bR$-linear space $V$ of dimension $D$,\ with a basis $\{\varphi^a\}^{a\in\ovl{1,D}}$ of $ V^*$ dual to a chosen basis $\{e_a\}_{a\in\ovl{1,D}}$ of $V$.\ The $\varphi^a$ are to be viewed as global coordinates on $V\ni v^b\,e_b\equiv v$,\ with $\varphi^a(v)=v^a$.\ A \emph{linear} Poisson structure on $V$ is then given by the formula
\qq\nn
\Pi_{\rm lin}(\varphi)=f^{ab}_{\ \ c}\,\varphi^c\,\tfrac{\p\ }{\p\varphi^a}\wedge\tfrac{\p\ }{\p\varphi^b}\,,\qquad f^{ab}_{\ \ c}\in\bR\,,
\qqq
and defines a Lie bracket
\qq\nn
[\varphi^a,\varphi^b]_\ggt=f^{ab}_{\ \ c}\,\varphi^c
\qqq
on $V^*\equiv\ggt$.\ Whenever the latter Lie algebra is quadratic,\ the corresponding Cartan--Killing form $K_\ggt\equiv K^{ab}\,\ev_{e_a}\ox\ev_{e_b}\colo V^*\ox V^*\too\bR$ enables us to identify $\ggt^*\equiv V^*{}^*\cong V$ with $\ggt$,\ and so also $C^\infty(\Si,V)$ with $C^\infty(\Si)\ox\ggt\ni X\equiv(\xi_a\circ X)\ox\varphi^a$ and $\G(\txT\Si\ox X^*\txT^*V)$ with $\Om^1(\Si)\ox\ggt\ni\vartheta\equiv\vartheta_a\ox\varphi^a$.\ The lagrangean density \eqref{eq:PSMAct} now rewrites as the {\bf linear Poisson $\si$-model}
\qq\nn
\ceL_{\rm lPSM}(X,\txT X;\vartheta)= K_\ggt\bigl(\vartheta\,\overset{\wedge}{,}\,\sfd X+\tfrac{1}{2}\,[\vartheta,X]_\ggt\bigr)\,.
\qqq
Its EL equations read
\qq\nn
\left\{ \barr{l}  \sfd\vartheta+\tfrac{1}{2}\,[\vartheta\,\overset{\wedge}{,}\,\vartheta]_\ggt=0 \\ 
\sfd X+[\vartheta,X]_\ggt=0 \earr \right.\,,
\qqq
and the tangential gauge symmetries $C^\infty(\Si,\bR)\ox\ggt\ni\ep$,\ identified in all generality in \eqref{eq:tan-Pgauge},\ are realised as 
\qq\nn
X\longmapsto X-[\ep,X]_\ggt\,,\qquad\qquad\vartheta\longmapsto\vartheta-[\ep,\vartheta]_\ggt+\sfd\ep\,.
\qqq

The above structure admits a natural geometric interpretation,\ which then paves the way to an obvious integration of the infinitesimal symmetries:\ The field theory is one of 
\bit
\item flat gauge fields $\vartheta$ (connection data) on the trivial bundle $\xcP_0:=\Si\x(\txG\x\ggt)\to\Si$,\ with the typical fibre given by the arrow manifold of the action groupoid $\txG\,\lx_{\txT_e\Ad}\ggt$ (with source map $s=\pr_2$ and target map $t=\txT_e\Ad$;\ see:\ Ex.\,\ref{eg:actgrpd}) associated with the adjoint action $\txT_e\Ad\colo\txG\x\ggt\to\ggt$ on $\ggt\cong V$ of the Lie group $\txG$ that integrates $\ggt$,\ and with the fibrewise action of the groupoid modelled on the right arrow adjunction:
\qq\nn
\xcP_0\,\,\fibx{s\circ\pr_2}{t}(\txG\x\ggt)\too\xcP_0,\ \bigl((\si,g,X),\bigl(h,\txT_e\Ad_{h^{-1}}(X)\bigr)\bigr)\longmapsto\bigl(\si,g\cdot h,\txT_e\Ad_{h^{-1}}(X)\bigr)\,;
\qqq
\item covariantly constant sections $X$ of the trivial quotient bundle $\xcF_0:=\xcP_0/\txG\,\lx_{\txT_e\Ad}\ggt\equiv\Si\x\ggt\to\Si$,\ with the typical fibre given by the object manifold of $\txG\,\lx_{\txT_e\Ad}\ggt$.
\eit
The tangential gauge transformations can now be understood as linearisations (around the identity diffeomorphism $\id_{\xcP_0}$) of the standard vertical $\txG\,\lx_{\txT_e\Ad}\ggt$-equivariant automorphisms
\qq\label{eq:int_gauge_lPSM}
\widetilde\la{}_h\colo \xcP_0\ni(\si,g,X)\longmapsto\bigl(\si,h(\si)\cdot g,X\bigr)\in\xcP_0\,,
\qqq
written in terms of arbitrary $h\in C^\infty(\Si,\txG)$.\ Indeed,\ write $h=\exp(-\ep)$ to obtain
\qq\nn
\txT_e\Ad_h\circ\vartheta-h^*\theta_{\rm R}=\vartheta
-[\ep,\vartheta]_\ggt+\sfd\ep+\xcO(\ep^2)\,,
\qqq
and 
\qq\nn
\txT_e\Ad_h\circ X=X-[\ep,X]_\ggt+\xcO(\ep^2)\,.
\qqq
In fact,\ it is easy to verify the invariance of the action functional \eqref{eq:PsigM} of the linear Poisson $\si$-model---to be denoted as $S_{\rm lPSM}$ below---under these large gauge transformations:\ We calculate---for $h\in C^\infty(\Si,\txG)$ arbitrary,\ and taking into account $\p\Si=\emptyset$---
\qq\nn
S_{\rm lPSM}[\txT_e\Ad_h\circ X,\txT_e\Ad_h\circ\vartheta-h^*\theta_{\rm R}]=S_{\rm lPSM}[X,\vartheta]+\int_\Si\sfd K_\ggt\bigl(X,h^*\theta_{\rm L}\bigr)=S_{\rm lPSM}[X,\vartheta]\,.
\qqq

In light of the findings of \cite{Strobl:2025},\ it is only natural to seek to recast the above integrated symmetry in the language of groupoid modules,\ see:\ Def.\,\ref{def:gr-mod}.\ To this end,\ we first endow the trivial adjoint bundle $\Ad(\xcP_0):=\Si\x(\txG\x\ggt)$ with the structure of a Lie groupoid with object manifold $\xcF_0$ by defining on the two the following structure maps:\ the source map $\widetilde s=\id_\Si\x s\colo \Ad(\xcP_0)\to\xcF_0,\ (\si,g,X)\mapsto(\si,X)$,\ the target map $\widetilde t=\id_\Si\x t\colo \Ad(\xcP_0)\to\xcF_0,\ (\si,g,X)\mapsto(\si,\txT_e\Ad_g(X))$,\ the identity map $\widetilde\Id\colo\xcF_0\to\Ad(\xcP_0),\ (\si,X)\mapsto(\si,e,X)$,\ the multiplication map $\widetilde\txm\colo\Ad(\xcP_0)\,\,\fibx{\widetilde s}{\widetilde t}\Ad(\xcP_0)\to\Ad(\xcP_0),\ ((\si,h,\txT_e\Ad_g(X)),(\si,g,X))\mapsto(\si,h\cdot g,X)$,\ and the inverse $\widetilde{\rm Inv}\colo\Ad(\xcP_0)\to\Ad(\xcP_0),\ (\si,g,X)\mapsto(\si,g^{-1},\txT_e\Ad_g(X))$.\ A natural left action of $\Ad\,\xcP_0$ on $\xcP_0$ \emph{along} $\xcF_0$ reads
\qq\nn
\la_{\xcP_0}\colo\Ad(\xcP_0)\,\,\fibx{\widetilde s}{\xcD_0}\xcP_0\too\xcP_0,\ \bigl((\si,h,\txT_e\Ad_g(X)),(\si,g,X)\bigr)\longmapsto(\si,h\cdot g,X)\,,
\qqq
where $\xcD_0:=\id_\Si\x t\colo\xcP_0\to\xcF_0$ is the quotient map.\ It is now clear that the gauge transformation \eqref{eq:int_gauge_lPSM} arises through the evaluation of the induced action  
\qq\nn
B\la_{\xcP_0}\colo\bB\bigl(\Ad(\xcP_0)\bigr)\x\xcP_0\too\xcP_0
\qqq
of the group $\bB(\Ad(\xcP_0))$ of bisections of the adjoint groupoid (see:\ Def.\,\ref{def:bisec}) on the left $\Ad(\xcP_0)$-module $\xcP_0$ on its specific elements
\qq\nn
\b_h=(\id_\Si,h)\x\id_\ggt\colo\xcF_0\too\Ad(\xcP_0),\ (\si,X)\longmapsto\bigl(\si,h(\si),X\bigr)\,,
\qqq
that is we have the identity
\qq\nn
\widetilde\la{}_h(\si,g,X)=B\la_{\xcP_0}\bigl(\b_h,(\si,g,X)\bigr)\,.
\qqq
Upon rewriting $\b_h$ as 
\qq\nn
\b_h(\si,X)=:\bigl(\si,\bigl(\g_h(\si)\bigr)(X)\bigr)\,,
\qqq
we then obtain a smooth family of bisections
\qq\nn
\g_h\colo\Si\too\bB\bigl(\txG\,\lx_{\txT_e\Ad}\ggt\bigr),\ \si\longmapsto\bigl(h(\si),\cdot\bigr)
\qqq
of the underlying action groupoid as a presentation of the gauge transformation.\ Note that the $\g_h$ map to the faithful image $\iota(\txG)$ of $\txG$ under the embedding $\iota\colo\txG\to\bB(\txG\,\lx_{\txT_e\Ad}\ggt),\ h\mapsto(h,\cdot)$ (see:\ Ex.\,\ref{eg:flatbisec}),\ in keeping with \cite[Ex.\,2.25]{Strobl:2025}.

In order to appreciate the exceptionality of the bisections just encountered,\ we need to delve deeper into the structure of the action groupoid,\ which \emph{integrates} the linear-Poisson target $(\ggt^*,\Pi_{\rm lin})$ under the aforementioned identification $\ggt^*\cong\ggt$.\ For that,\ we rewrite its arrow manifold equivalently as $\txG\x\ggt^*$,\ and proceed with its examination in the language of Poisson geometry in this more convenient presentation,\ in which it becomes the {\bf coadjoint-action groupoid} $\txG\,\lx_{\txT_e\Ad^*}\ggt^*$.\ The latter has the tangent Lie algebroid---the action algebroid $\xcE_\ggt=\ggt\,\lx_{\txT_e\Ad^*}\ggt^*$ (see:\ Ex.\,\ref{eg:actalgbrd})---canonically isomorphic with the cotangent Lie algebroid $\txT^*\ggt^*\cong\txT^*V\to V\cong\ggt^*$ with the Poisson anchor $\a_{\xcE_0}=\Pi_{\rm lin}^\sharp$.\  We begin by noting that 
the arrow manifold $\txG\x\ggt^*$ of that groupoid is canonically diffeomorphic with the total space of the cotangent bundle $\txT^*\txG\to\txG$,\ and---as such---it is endowed with the canonical symplectic structure \cite[Lem.\,6.20]{Crainic:2021}
\qq\nn
\om_{\rm lin}=-\sfd\theta_{\rm Liouville}\,,
\qqq
where $\theta_{\rm Liouville}=x\circ\theta_{\rm L}\equiv x^a\,\theta_{{\rm L}\,a}$ is the Liouville 1-form on $\txT^*\txG$,\ expressed in terms of the global coordinate functions $x=x^a\,\ev_{e_a}$ on $\ggt^*\equiv V^{**}\xleftarrow[\ev]{\cong}V$,\ for $\theta_{{\rm L}\,a}\ox\varphi^a=\theta_{\rm L}$.\ It is readily checked that the symplectic form $\om_{\rm lin}$ is multiplicative,\ in the sense of Def.\,\ref{def:mult-k-form},\ as a 2-form on the arrow manifold of the coadjoint-action groupoid.

Our first observation concerns the distinguished bisections $\g_h\equiv(h,\cdot)\in\bB(\txG\,\lx_{\txT_e\Ad^*}\ggt^*)$:\ Clearly,\ they annihilate the above symplectic form under pullback,
\qq\label{eq:omlin-holo}
\g_h^*\om_{\rm lin}=0\,.
\qqq
Such bisections were dubbed ($\om_{\rm lin}$-){\bf holonomic} by Crainic \cite{Crainic:2015msp},\ and that name,\ which was kept in \cite{Suszek:2026col},\ shall also be used in what follows.

Secondly,\ we note that the map $\kappa_{\xcE_\ggt}\colo\Id^*{\rm Ker}\,\txT s\to\txT^*\ggt^*$ canonically associated with the multiplicative 2-form $\om_{\rm lin}$ by Thm.\,\ref{thm:CSS} assumes a particularly simple form:\ Denote the right-invariant vector field on $\txG$ associated with the vector $\xi=\xi_a\,\varphi^a\in\ggt$ as $R_\xi\in\G(\txT\txG)$ to compute,\ at $v=v^b\,e_b V\cong\ggt^*$,
\qq\nn
\kappa_{\xcE_\ggt}(\xi)=\kappa_{\xcE_\ggt}\bigl(R_\xi(e),\brd0_{\txT\ggt^*}(v)\bigr)\equiv(e,\cdot)^*\bigl(\imath_{R_\xi}\om_{\rm lin}\bigr)(v)=\xi_a\,\sfd x^a\,,
\qqq
whence
\qq\nn
\kappa_{\xcE_\ggt}=\ev_{e_a}\ox\sfd x^a\,.
\qqq
Thus,\ upon expliciting the identification $K_\ggt^\flat\colo V^*\equiv\ggt\to\ggt^*\equiv V^{**}\cong V,\ \xi\mapsto K^{ab}\,\xi_a\,e_b=:x^b\,e_b$,\ we recover the first term in the lagrangean density of the linear Poisson $\si$-model as
\qq\nn
K_\ggt\bigl(\vartheta\,\overset{\wedge}{,}\,\sfd X\bigr)=\vartheta_a\wedge\sfd\bigl(K^{ab}\,\xi_a\circ X\bigr)\equiv-\sfd(x^a\circ X)\wedge\vartheta_a\,,
\qqq
or simply as
\qq\nn
K_\ggt\bigl(\vartheta\,\overset{\wedge}{,}\,\sfd X\bigr)=-X^*\kappa_{\xcE_\ggt}\overset{\wedge}{\circ}\vartheta\,,
\qqq
where $\overset{\wedge}{\circ}$ stands for the obvious composition of the wedge product in the differential-form factor with the standard composition of maps $\vartheta\colo\txT\Si\to\ggt$ and the $\ev_{e_a}\colo\ggt\to\bR$ in $\kappa_{\xcE_\ggt}$.\ Using the $\ad$-invariance of the Killing form,\ the other term in $\ceL_{\rm lPSM}$ is recovered similarly as
\qq\nn
\tfrac{1}{2}\,K_\ggt\bigl(\vartheta\,\overset{\wedge}{,}\,[\vartheta,X]_\ggt\bigr)&=&\tfrac{1}{2}\,K^{ab}\,(\xi_d\circ X)\,f^{cd}_{\ \ b}\,\vartheta_a\wedge\vartheta_c=-\tfrac{1}{2}\,K^{db}\,(\xi_d\circ X)\,f^{ca}_{\ \ b}\,\vartheta_a\wedge\vartheta_c\equiv-\tfrac{1}{2}\,\vartheta_a\wedge f^{ca}_{\ \ b}\,(x^b\circ X)\,\vartheta_c\cr\cr 
&\equiv&-\tfrac{1}{2}\,\corr{X^*\kappa_{\xcE_\ggt}\circ\vartheta\,\overset{\wedge}{,}\,X^*\a_{\xcE_\ggt}\circ\vartheta}\,.
\qqq
Altogether,\ then,\ we arrive at the formula (see also:\ Rem.\,\ref{rem:kappAi-expl})
\qq\label{eq:lPSM-rewrite}
-\tx{$\ceL$}_{\rm lPSM}(X,\txT X;\vartheta)\equiv X^*\kappa_{\xcE_\ggt}\overset{\wedge}{\circ}\vartheta+\tfrac{1}{2}\,\corr{X^*\kappa_{\xcE_\ggt}\circ\vartheta\,\overset{\wedge}{,}\,X^*\a_{\xcE_\ggt}\circ\vartheta}\,,
\qqq
which rephrases the lagrangean density of the Poisson $\sigma$-model under consideration in terms of the Spencer datum of the multiplicative symplectic form on the corresponding Lie groupoid (which happens to exist in this case).\ The formula constitutes the point of departure of the subsequent generalisation of our observations.

\subsection{A reinterpretation of the model}\label{sub:PSM-redef}

Let us assume\footnote{In the category of Lie algebroids,\ such an assumption is highly nontrivial as there is no counterpart of Lie's Third Theorem,\ see,\ {\it e.g.},\ \cite{Crainic:2003}.} henceforth that the target cotangent algebroid of the classic Poisson $\si$-model integrates to a Lie groupoid.\ The lagrangean density \eqref{eq:PSMAct} then admits an interesting reinterpretation,\ which leads to a symmetry enhancement and generalisation of the original field theory---advanced in \cite{Suszek:2026col}---with a direct link to the topological component of the (suitably specialised) dynamical Polyakov--Alvarez--Gaw\c{e}dzki (PAG) $\si$-model \cite{Deser:1976rb,Brink:1976sc,Polyakov:1981rd,Alvarez:1984es,Gawedzki:1987ak}.\ The first step towards the reinterpretation,\ proposed in \cite{Suszek:2026col},\ is the following rewriting of the two terms composing the lagrangian density $\ceL_{\rm PSM}(X,\txT X;\vartheta)$ in analogy with \eqref{eq:lPSM-rewrite}.\ The rewriting  employs the identity map $\id_{\txT^*M}$,\ viewed as a section of the tensor bundle $(\txT^*M)^*\ox\txT^*M$ with a local presentation $\id_{\txT^*M}\underset{\tiny{{\rm loc.}}}{=}\ev_{\p_\mu}\ox\sfd x^\mu$.\ Upon precomposing $\id_{\txT^*M}$ with the map $\vartheta$,\ we may then regard\footnote{The appropriate formal framework here is provided by the theory of foliated vector-valued differential forms.} the ensuing tensor $\id_{\txT^*M}\circ\vartheta\colo\txT\Si\to\txT^*M\rstr_{X(\Si)}$ as  a distinguished section of the tensor bundle $\txT^*(\Si\x M)\ox\txT^*(\Si\x M)\rstr_{{\rm graph}(X)}$,\ restricted to the graph $(\id_\Si,X)(\Si)\equiv{\rm graph}(X)$ of the field $X\in C^\infty(\Si,M)$.\ This uses the identification $\txT^*(\Si\x M)\cong\pr_1^*\txT^*\Si\oplus\pr_2^*\txT^*M$.\ The section may next be subjected independently to the following operations:
\bit
\item we antisymmetrise the tensor $\id_{\txT^*M}\circ\vartheta\in\G(\txT^*(\Si\x M)\ox\txT^*(\Si\x M)\rstr_{{\rm graph}(X)})$ and pull back the resulting 2-form from over ${\rm graph}(X)$---to be denoted suggestively as $(\id_{\txT^*M}\circ\vartheta)\circ{\rm Alt}$---along the section $\widetilde X\equiv(\id_\Si,X)$ of the trivial configuration bundle $\xcF_0\equiv\Si\x M\to M$;
\item we pair the contraction $\id_{\txT^*M}\circ\vartheta$ with the distinguished section of $\txT^*(\Si\x M)\ox\txT(\Si\x M)\rstr_{{\rm graph}(X)}$ induced by the map $\Pi^\sharp\circ\vartheta\colo\txT\Si\to\txT M\rstr_{X(\Si)}$,\ in the second tensor term (over $X(\Si)$),\ and antisymmetrise the result in the first tensor term (over $\Si$),\ whereby we obtain the composite object $\ev\circ((\id_{\txT^*M}\circ\vartheta)\ox(\Pi^\sharp\circ\vartheta))\circ{\rm Alt}$,\ also viewed as a 2-form over ${\rm graph}(X)$,\ and so to be pulled back to $\Si$ along $\widetilde X$.
\eit
(By a mild abuse of the notation,\ we use the same symbols:\ $\ev$ (for the evaluation) and ${\rm Alt}$ (for the un-weighted alternator) as before in the present setting.)\ The result of the rewriting reads
\qq\label{eq:PSM-pre-Spencer}
\ceL_{\rm PSM}(X,\txT X;\vartheta)\equiv-\widetilde X{}^*\bigl(\id_{\txT^*M}\circ\vartheta+\tfrac{1}{2}\,\ev\circ\bigl(\bigl(\id_{\txT^*M}\circ\vartheta\bigr)\ox\bigl(\Pi^\sharp\circ\vartheta\bigr)\bigr)\bigr)\circ{\rm Alt}\,.
\qqq

At this point,\ we may already take the first conceptual step towards the said reinterpretation by identifying in the above expression elements of the peculiar differential-geometric structure on the integrating Lie groupoid.\ The latter is a symplectic Lie groupoid (see:\ Ex.\,\ref{eg:symplgrpd}) $\grpd{(\xcG,\om)}{(M,\Pi)}$,\ with the tangent Lie algebroid $\xcE\equiv\Lie(\xcG)=(\Id^*\ker\,\txT s,\a_\xcE\equiv\txT t\circ\pr_2,[\cdot,\cdot]_\xcE)$ (see:\ Def.\,\ref{def:Liealgbrd}) canonically isomorphic to $(\txT^*M,\Pi^\sharp,[\cdot,\cdot]_{\rm K})$.\ The isomorphism,\ established in \cite{Weinstein:1987,Coste:1987},\ takes the (vertical) form
\qq\nn
\kappa_\xcE:=\Id^*\circ\om_\flat\colo\Id^*\ker\,\txT s\equiv M\,\fibx{\Id}{\pi_{\txT\xcG}}\ker\,\txT s\too\txT^*M,\ (m,\vep)\longmapsto\om(\vep,\cdot)\circ\txT_m\Id
\qqq
and recovers the Poisson structure on $C^\infty(M,\bR)$ as $\Pi=\kappa_\xcE\circ[\cdot,\cdot]_\xcE\circ(\kappa_\xcE^{-1}\x\kappa_\xcE^{-1})$.\ Thus,\ in particular,\ it relates the two anchors as
\qq\label{eq:kapE-anch}
\Pi^\sharp\circ\kappa_\xcE=\a_\xcE\,.
\qqq
\brem
Symplectic groupoids integrating---also in the formal sense (as singular spaces)---Poisson targets of the classic Poisson $\si$-model were studied at length by Cattaneo and Felder in \cite{Cattaneo:2000iw}.\ These arise in connection to the canonical deformation quatisation of the Poisson structure $(M,\Pi)$ \`a la Kontsevich \cite{Kontsevich:1997vb},\ whose relation to the path-integral quantisation of the Poisson $\si$-model was discovered by the same authors in their seminal work \cite{Cattaneo:1999fm}. 

In this framework,\ our assumption about the integrability of the Koszul algebroid means that the reduction of the phase space of the topological field theory $S_{\rm PSM}$ with respect to the gauge symmetries \eqref{eq:tan-Pgauge} is a non-singular finite dimensional manifold with a canonical structure of a symplectic Lie groupoid.
\erem

Properties of maps $\kappa_\xcE\colo\G(\xcE)\ni\vep\longmapsto\Id^*(\imath_\vep\rho)\in\Om^{k-1}(M)$ for arbitrary multiplicative $k$-forms on Lie groupoids were investigated at length by Crainic {\it et al.}\ in \cite{Crainic:2015msp} in the context of a Spencer-type presentation of such differential forms (see:\ Thm.\,\ref{thm:CSS}),\ of much relevance to the equivariantisation of the differential cohomology on $M$.\ These maps were dubbed (2-){\bf comomenta} in \cite{Suszek:2026col},\ and we shall stick to this denomination in what follows.\ The peculiarity of the symplectic case is the bijectivity of the comomentum $\kappa_\xcE$ engendered by the non-degenerate 2-form $\om$,\ which enables us to write $\Pi^\sharp=\a_\xcE\circ\kappa_\xcE^{-1}$.\ This leads to a further rewriting of the lagrangean density 
\qq\label{eq:PSM-Spencer}
\ceL_{\rm PSM}(X,\txT X;\vartheta)\equiv-\widetilde X{}^*\bigl(\kappa_\xcE\circ\widetilde\vartheta+\tfrac{1}{2}\,\ev\circ\bigl(\bigl(\kappa_\xcE\circ\widetilde\vartheta\bigr)\ox\bigl(\a_\xcE\circ\widetilde\vartheta\bigr)\bigr)\bigr)\circ{\rm Alt}\,,\qquad\qquad\widetilde\vartheta=\kappa_\xcE^{-1}\circ\vartheta\,,
\qqq
which prepares us for the last stage of the reformulation.\ In it,\ the question of the integrability of the tangential gauge transformations \eqref{eq:tan-Pgauge}  to the Lie-groupoidal level\footnote{The question was one of the key motivations for the construction of the principaloid bundle given in \cite{Strobl:2025}.} is addressed.

\brem\label{rem:kappAi-expl}
In view of the key r\^ole played by 2-forms of the above structure \eqref{eq:PSM-Spencer} in the remainder of the paper,\ we pause to write out their evaluation on pairs $((v_1,V_1),(v_2,V_2))\in(\txT_\si\Si\oplus\txT_m M)^{\x 2}\equiv \txT_{(\si,m)}(\Si\x M)^{\x 2}$.\ This is given by the formul\ae:
\qq\nn
\bigl(\bigl(\kappa_\xcE\circ\widetilde\vartheta\bigr)\circ{\rm Alt}\bigr)\bigl((v_1,V_1),(v_2,V_2)\bigr)=\bigl(\kappa_\xcE\bigl(\widetilde\vartheta(v_2)\bigr)\bigr)(V_1)-\bigl(\kappa_\xcE\bigl(\widetilde\vartheta(v_1)\bigr)\bigr)(V_2)\,,
\qqq
and
\qq\nn
\bigl(\ev\circ\bigl(\bigl(\kappa_\xcE\circ\widetilde\vartheta\bigr)\ox\bigl(\a_\xcE\circ \widetilde\vartheta\bigr)\bigr)\circ{\rm Alt}\bigr)\bigl((v_1,V_1),(v_2,V_2)\bigr)&=&\bigl(\kappa_\xcE\bigl(\widetilde\vartheta(v_1)\bigr)\bigr)\bigl(\a_\xcE\bigl(\widetilde\vartheta(v_2)\bigr)\bigr)-\bigl(\kappa_\xcE\bigl(\widetilde\vartheta(v_2)\bigr)\bigr)\bigl(\a_\xcE\bigl(\widetilde\vartheta(v_1)\bigr)\bigr)\cr\cr 
&=&2\bigl(\kappa_\xcE\bigl(\widetilde\vartheta(v_1)\bigr)\bigr)\bigl(\a_\xcE\bigl(\widetilde\vartheta(v_2)\bigr)\bigr)\,,
\qqq
where the last equality uses the basic property of the comomentum captured by \eqref{eq:comom-asymm}.
\erem

This last stage requires an initial conceptual leap:\ We need to replace the $\txT\Si$-foliated $\xcE$-valued 1-form $\widetilde\vartheta$ on ${\rm graph}(X)$---with an infinitesimal transformation law \eqref{eq:tan-Pgauge},\ giving rise to an `accidental' gauge symmetry of the lagrangean density \eqref{eq:PSM-pre-Spencer} ({\it i.e.},\ a symmetry which has not been built into the definition of $\widetilde\vartheta$ in an integrated form from the start,\ and only subsequently linearised)---by a restriction to ${\rm graph}(X)$ of a $\G_{\rm c}(\xcE)$-valued 1-form with an affine transformation behaviour under an \emph{intrinsically} defined action of the gauge group $\bB(\xcG)^\Si$ of the formerly identified gauge algebra $\G_{\rm c}(\xcE)^\Si\ni\ep$.\ Here,\ $\bB(\xcG)$ is the group of bisections of $\xcG$ (see:\ Def.\,\ref{def:bisec}),\ and its relation to $\G_{\rm c}(\xcE)$ is fixed by Thm.\,\ref{thm:LieBis}.\ Differential forms of this type arise naturally in the description of the tangent geometry of principaloid $\xcG$-bundles,\ introduced in \cite{Strobl:2025}.\ The latter were devised as geometric structures encoding---through their morphisms and compatible Ehresmann connections---the full physical content of the gauge principle for symmetries modelled on arbitrary Lie groupoids $\grpd{\xcG}{M}$,\ {\it i.e.},\ the gauge group,\ the gauge field,\ and a mechanism of the latter's canonical coupling to charged matter fields through the associated covariant derivative.\ This list of elementary objects was subsequently extended in \cite{Suszek:2026col} to include the equivariant-cohomological structure requisite for an effective stacky descent,\ to the (generically non-smooth) orbispace $M//\xcG$,\ of higher-geometric and -cohomological objects---such as the gerbe $\cG$ of the PAG $\si$-model---over the configuration fibre $M$ of the field theory with the categorified symmetry $\grpd{\xcG}{M}$ to be gauged.\ Thus,\ principaloid bundles seem tailored precisely to the task in hand.\ We now briefly recall the relevant  constructions from \cite{Strobl:2025,Suszek:2026col}. 

The fundamental object in \cite{Strobl:2025} is a {\bf principaloid $\xcG$-bundle},\ with a structure Lie groupoid $\grpd{\xcG}{M}$ (with structure maps $(s,t,\Id,\txm,\Inv)$),\ which is a fibre-bundle object in the category of right $\xcG$-module spaces (see:\ Def.\,\ref{def:gr-mod}),
\qq\nn
\alxydim{@C=1.25cm@R=1.cm}{ \cO\x\xcG & \cO\x_\Si\xcP \ar[l]^{\cong}_{\xcP\t} \ar[r] \ar[d] & \xcP \ar[d]_{\pi_\xcP} \ar[dr]^{\mu} & & \xcG \ar@{=>}[dl] \ar@/_1.pc/[ll]^{\xcR} \ar@/_2.25pc/[llll]_r \\ & \cO \ar[r]_{\jmath} & \Si & M & }\,,
\qqq
with a trivial $\xcG$-action on the base $\Si$,\ and that on the total space,\ $\xcR\colo \xcP\,\fibx{\mu}{t}\,\xcG\to\xcP$,\ defined for a moment map $\mu\colo\xcP\to M$ with local model $s$,\ and modelled---by means of $\xcG$-equivariant local trivialisations $\xcP\t$ over an open cover $\cO=\bigsqcup_{i\in I}\,O_i$ of $\Si$---on the canonical right-fibred action $r$ of $\grpd{\xcG}{M}$ on its typical fibre $\xcG$ through groupoid multiplication $r\colo\xcG\,\fibx{s}{t}\,\xcG\to\xcG,\ (g,h)\mapsto\txm(g,h)\equiv g.h\equiv r_h(g)$.\ As such,\ the bundle admits a local model 
\qq\label{eq:princ-clutch}
\xcP\cong\bigsqcup_{i\in I}\,(O_i\x\xcG)/\sim_{L_{\b_{\cdot\cdot}}},\qquad(x,g,j)\sim\bigl(x,L_{\b_{ij}(\si)}(g),i\bigr)\,,
\qqq
written in terms of a transition 1-cocycle $\b_{ij}\colo O_{ij}\equiv O_i\cap O_j\to\bB(\xcG)$, 
\qq\label{eq:tra-cocyc}
\b_{ij}(\si')\cdot\b_{kj}(\si')^{-1}\cdot\b_{ki}(\si')=\Id,\quad\si'\in O_i\cap O_j\cap O_k\equiv O_{ijk}\,,
\qqq
with values in the group of bisections of $\grpd{\xcG}{M}$,\ the latter acting on $\xcG$ by the canonical left-multiplication $L\colo\bB(\xcG)\to\Diff(\xcG)$ from Def.\,\ref{def:bisec-act}.

For the delooping groupoid $\bbB\txG\equiv\grpd{\txG}{\bullet}$ of a Lie group $\txG$,\ a principaloid $\bbB\txG$-bundle is a standard principal $\txG$-bundle $P$ with the free and proper (defining) right $\txG$-action $r\colo P\x\txG\to P$,\ whereas for the action groupoid $\txG\,\lx_\la M\equiv\grpd{\txG\x M}{M}$ associated with a smooth action $t\equiv\la\colo\txG\x M\to M$ of the Lie group we recover the extended bundle $P\x M$ \emph{if we co-restrict the transition 1-cocycle} to the subgroup $\iota(\txG)\subset\bB(\txG\,\lx_\la M)$ given by the image of the group homomorphism $\iota\colo\txG\to\bB(\txG\,\lx_\la M),\ g\longmapsto(g,\cdot)$ (see:\ Ex.\,\ref{eg:flatbisec}).\ The extended bundle admits a smooth quotient $P\x_\la M\equiv(P\x M)/\txG$ with respect to the free and proper diagonal $\txG$-action $\widetilde\la\colo(P\x M)\x\txG\to P\x M,\ ((p,m),g)\mapsto(r_g(p),\la_{g^{-1}}(m))$.\ The quotient is called the associated bundle,\ and plays a central r\^ole in the classic gauge principle for symmetries with the group model $\la$.\ Thus,\ the principaloid $\xcG$-bundle can be regarded as a natural (unified) generalisation of the two basic constructs of the classic gauge field theory.

The action $\varrho$ being principal,\ there exists a smooth Godement quotient $\xcF\equiv\xcP/\xcG$ (see:\ \cite{Serre:1964,Fernandes:2024}),\ termed the {\bf shadow bundle},\ which has the structure of a fibre bundle $\xcF\xrightarrow{\pi_\xcF}\Si$ with typical fibre $M$ and local model 
\qq\label{eq:shadow-clutch}
\xcF\cong\bigsqcup_{i\in I}\,(O_i\x M)/\sim_{t_*\b_{\cdot\cdot}},\qquad(x,m,j)\sim\bigl(x,t_*\bigl(\b_{ij}(\si))(m),i\bigr)\,,
\qqq
using the same transition 1-cocycle $\b_{ij}$ as before,\ but now realised by the canonical shadow action $t_*\colo\bB(\xcG)\to\Diff(M)$ from Def.\,\ref{def:bisec-act}.\ The quotient map $\xcD\colo\xcP\to\xcF$,\ termed the {\bf sitting-duck map},\ is a $\Si$-vertical surjective submersion locally modelled on $t$.\ The bundle reproduces the aforementioned associated bundle $P\x_\la M$ for the action groupoid and the structure group reduced to $\iota(\txG)$,\ and so the shadow bundle becomes a natural candidate for the r\^ole of the configuration bundle of the charged matter in the gauge field theory based on the construction recapitulated.

In the present picture,\ gauge transformations are identified as $\xcG$-equivariant vertical bundle automorphisms of the principaloid $\xcG$-bundle,\ composing the gauge group (of $\xcP$)
\qq\nn
{\rm Gau}(\xcP)={\rm Aut}_{\Bun(\Si)/\Si}(\xcP)\cap\Diff_{\varrho(\xcG)}(\xcP)\,.
\qqq
Locally,\ a gauge transformation $\Phi\in{\rm Gau}(\xcP)$ is presented by smooth maps $h_i\colo O_i\to\bB(\xcG),\ i\in I$ as
\qq\label{eq:gauge-trafo-P-loc}
\bigl(\xcP\t_i\circ\Phi\circ\xcP\t_i^{-1}\bigr)(\si,g)=\bigl(\si,L_{h_i(\si)}(g)\bigr),\quad(\si,g)\in O_i\x\xcG\,,
\qqq
the maps satisfying the usual gluing identities 
\qq\label{eq:gauge-glue}
h_i(\si')=\Ad_{\b_{ij}(\si')}\bigl(h_j(\si')\bigr)
\qqq
over intersections $O_{ij}\ni\si'$ of their respective domains.\ Automorphisms of $\xcP$ thus defined push down to distinguished automorphisms of the shadow bundle,\ {\it i.e.},\ we have a group homomorphism
\qq\nn
\xcF_*\colo{\rm Gau}(\xcP)\too{\rm Aut}_{\Bun(\Si)/\Si}(\xcF)\,,
\qqq
and the gauge-equivariance identity 
\qq\nn
\xcD\circ\Phi=\xcF_*(\Phi)\circ\xcD,\quad\Phi\in{\rm Gau}(\xcP)\,.
\qqq
Being vertical,\ the automorphisms map sections of $\xcF$ to new sections as
\qq\nn
\G(\xcF)\ni\phi\longmapsto\Phi\circ\phi=:\phi^\Phi\in\G(\xcF)\,,
\qqq
and we call $\phi^\Phi$ the {\bf gauge transform} of $\phi$.\ The image group 
\qq\nn
{\rm Gau}(\xcF)\equiv\xcF_*({\rm Gau}(\xcP)) 
\qqq
is termed the {\bf shadow gauge group}.\ Its elements are locally presented by smooth maps $t_*\circ h_i\colo O_i\to\Diff(M),\ i\in I$.

Consistently with our choice of $\xcG$-equivariance as the organising principle of the whole construction,\ we consider Ehresmann connections on principaloid $\xcG$-bundles which are represented by $\xcG$-equivariant connection 1-forms $\Theta\in\G(\txT^*\xcP\ox\txV\xcP)$,\ {\it i.e.},\ smoothly $\xcP$-indexed families of projectors from the tangent bundle $\txT\xcP\to\xcP$ to its vertical subbundle $\txV\xcP\equiv\ker\,\txT\pi_\xcP$ along the horizontal subbundle $\txH\xcP=\ker\,\Theta\subset\ker\,\txT\mu$,\ obeying the identities $\txT\varrho_g(\ker\,\Theta(p))=\ker\,\Theta(\varrho_g(p))$ at $p\in\mu^{-1}(\{t(g)\})$.\ These descend to (the tangent manifold of) the Godement quotient $\xcF$,\ and thus define shadow connection 1-forms $\Theta_\xcF\in\G(\txT^*\xcF\ox\txV\xcF)$,\ determined by the identity
\qq\nn
\Theta_\xcF\circ\txT\xcD=\txT\xcD\circ\Theta\,.
\qqq
A $\xcG$-equivariant connection 1-form becomes a source of a \emph{local} gauge field through trivialisation,\ but the structure of that field is subtler than in the classic case.\ This is only to be expected,\ if one takes into account the fact that the tangent Lie algebroid of the structure Lie groupoid,\ which replaces the tangent Lie algebra of the structure Lie group in its r\^ole of the field's codomain,\ is a vector bundle over the object manifold $M$.\ With this in mind,\ we write out---after \cite{Strobl:2025}---the local presentation of $\Theta$ over $O_i\x\xcG\ni(\si,g)$:
\qq\label{eq:Thet-loc}
\Theta_i(\si,g):=\txT\xcP\t_i\circ\Theta\circ\txT_{(\si,g)}\xcP\t_i^{-1}=\id_{\txT_g\xcG}-\txT_{\Id_{t(g)}}r_g\circ A_i\bigl(\si,t(g)\bigr)\,,
\qqq
from which the {\bf local gauge field} emerges as a ($\txT O_i$-foliated) $\xcE$-valued 1-form 
\qq\nn
A_i\in\G\bigl(\pr_1^*\txT^*O_i\ox\pr_2^*\xcE\bigr)
\qqq
over $O_i\x M$,\ subject to the affine gluing law
\qq\label{eq:glue-Aij}
A_i\bigl(\si,t_*\bigl(\b_{ij}(\si)\bigr)(m)\bigr)=\unl{\txT C}{}_{\b_{ij}(\si)}\circ A_j(\si,m)+\theta_{\rm R}\circ D_1\widetilde\b{}_{ij}(\si,m)\,,
\qqq
expressed in terms of the canonical action $\unl{\txT C}\colo\bB(\xcG)\to{\rm GL}(\xcE)$ of $\bB(\xcG)$ on $\xcE$ from Prop.\,\ref{prop:Bis-on-E} (covering the shadow action $t_*$ on the base),\ of the smooth maps 
\qq\label{eq:bij-tilde}
\widetilde\b{}_{ij}\colo O_{ij}\x M\too\xcG,\ (\si,m)\longmapsto\b_{ij}(\si)(m)\,,
\qqq
using the transition 1-cocycle $\{ \b_{ij}\}_{i,j\in I}$ of $\xcP$,\ with $D_1\widetilde\b{}_{ij}(\si,m)\equiv\txT_\si\widetilde\b{}_{ij}(\cdot,m)$,\ and of the right-invariant Maurer--Cartan form $\theta_{\rm R}$ on $\xcG$ (see:\ \cite{Fernandes:2014}).\ The latter is a $\ker\,\txT s$-foliated $\xcE$-valued 1-form on $\xcG$ given by
\qq\label{eq:MC-R}\qquad\qquad
\alxydim{@C=1.cm@R=1.cm}{ \ker\,\txT s \ar[r]^{\ \MCR} \ar[d]_{\pi_{\txT\xcG}\rstr_{\ker\,\txT s}} & \xcE \ar[d]^{\pi_\xcE\equiv\pr_1}\\ \xcG \ar[r]_{t\ } & M}\,,\qquad\quad\MCR(g)(w)=\bigl(t(g),\txT_g r_{g^{-1}}(w)\bigr)\,,\quad w\in(\ker\,\txT s)_g\,.
\qqq
Crucially for the present physical context,\ the local gauge field behaves in a simple (affine) manner under gauge transformations:\ For a gauge transformation $\Phi\in{\rm Gau}(\xcP)$ with local data $h_i$ as before,\ we find the identity
\qq\label{eq:gaugetrafo}
A_i^\Phi\bigl(\si,t_*\bigl(h_i(\si)\bigr)(m)\bigr)=\unl{\txT C}{}_{h_i(\si)}\circ A_i(\si,m)+\theta_{\rm R}\circ D_1\widetilde h{}_i(\si,m)\,,
\qqq
written,\ for arbitrary $\si\in O_i$ and $m\in M$,\ 
in terms of the smooth maps 
\qq\label{eq:gammi-tilde}
\widetilde h{}_i\colo O_i\x M\too\xcG,\ (\si,m)\longmapsto h_i(\si)(m)\,,
\qqq
with $D_1\widetilde h{}_i(\si,m)\equiv\txT_\si\widetilde h{}_i(\cdot,m)$.\ Here,\ $A_i^\Phi$ is the gauge-transformed local gauge field,\ {\it i.e.},\ the local datum of the {\bf gauge transform} of $\Theta$,
\qq\nn
\Theta^\Phi:=\Phi^{-1\,*}\Theta\equiv\txT\Phi\circ\Theta\circ\txT\Phi^{-1}\in\G\bigl(\txT^*\xcP\ox\txV\xcP\bigr)\,.
\qqq

The geometry of principaloid $\xcG$-bundles with the structure Lie groupoid given by the symplectic groupoid $\grpd{(\xcG,\om)}{(M,\Pi)}$ introduced before provides us with a natural candidate for a field $(X,\vartheta)$ of the classic Poisson $\si$-model,\ \emph{with an intrinsically defined action of the gauge group} $\bB(\xcG)^\Si$:\ We take a global section $\phi$ of the \emph{trivial} shadow $\xcF_{\rm triv}=\Si\x M$ of a trivial principaloid $\xcG$-bundle $\xcP_{\rm triv}=\Si\x\xcG$ endowed with an \emph{arbitrary} (trivial) connection 1-form $\Theta_{\rm triv}$,\ and subsequently localise the \emph{global} gauge field $A\in\G(\pr_1^*\txT^*\ox\pr_2^*\xcE)$ determined by the latter on that section $\phi(\Si)$---in this manner,\ we obtain the pair 
\qq\nn
\alxydim{@C=1.75cm@R=1.cm}{ \txT\Si \ar[r]^{\vartheta:=A(\phi)} \ar[d]_{\pi_{\txT\Si}} & \xcE \ar[d]^{\pi_\xcE} \\ \Si \ar[r]_{X:=\pr_2\circ\phi\ } & M}\,,
\qqq
and the ensuing redefinition of the lagrangean density of the classic Poisson $\si$-model\footnote{We insert the minus sign for pure notational convenience.}
\qq\label{eq:Psim-redef}
-\widetilde\ceL{}_{\rm PSM}(\phi,\txT\phi;A):=\ceL_{\rm PSM}\bigl(\pr_2\circ\phi,\txT(\pr_2\circ\phi);(\kappa_\xcE\circ A)(\phi)\bigr)\,.
\qqq

While the matching of the \emph{types} of the variables:\ $(X,\vartheta)$ in the Poisson $\si$-model and $(\phi,\txA)$ in the definition of the principaloid bundle with connection (and its shadow) is certainly a non-trivial indication of a potential relation between the two,\ it is desirable to solidify the latter by rewriting the lagrangean density $\ceL_{\rm PSM}$ further in such a manner that it appears less {\it ad hoc} and more \emph{natural} as an expression involving the (vector-valued) differential forms $\Theta_{\rm triv}$ and $\om$ on the total space and the typical fibre of the principaloid bundle $\xcP_{\rm triv}$,\ respectively.\ To this end,\ let us introduce---after \cite[Sec.\,4]{Strobl:2025}---the {\bf Atiyah--Ehresmann groupoid} ${\rm At}(\xcP)\equiv\grpd{(\xcP\fibx{\mu}{\mu}\xcP)/\xcG}{\xcF}$ of (a generic principaloid $\xcG$-bundle) $\xcP$,\ obtained as the Godement quotient of the submersion groupoid ${\rm Pair}_\mu(\xcP)$ (see:\ Ex.\,\ref{eg:subm-grpd}) with respect to the diagonal lift $(\xcR\x\xcR)\circ\D$ of the defining action $\xcR$ (here,\ $\D\colo\xcG\to\xcG\x\xcG,\ g\mapsto(g,g)$ is the canonical diagonal map),\ and equipped with structure maps induced from those of ${\rm Pair}_\mu(\xcP)$ under this equivariantisation as 
\qq\nn
&\sfS\colo(\xcP\fibx{\mu}{\mu}\xcP)/\xcG\too\xcF,\ [(p_1,p_2)]\longmapsto\xcD(p_2)\,,\qquad\qquad\sfT\colo(\xcP\fibx{\mu}{\mu}\xcP)/\xcG\too\xcF,\ [(p_1,p_2)]\longmapsto\xcD(p_1)\,,&\cr\cr 
&\sfI\colo\xcF\too(\xcP\fibx{\mu}{\mu}\xcP)/\xcG,\ f\longmapsto[(p,p)],\quad\tx{where}\ p\in\xcD^{-1}(f)\ \tx{is arbitrary}\,,&\cr\cr 
&\sfJ\colo(\xcP\fibx{\mu}{\mu}\xcP)/\xcG\too(\xcP\fibx{\mu}{\mu}\xcP)/\xcG,\ [(p_1,p_2)]\longmapsto[(p_2,p_1)]\,,&
\qqq
and
 \qq\nn
 \sfM\colo {\rm At}(\xcP)\,\,\fibx{\sfS}{\sfT}{\rm At}(\xcP)\too {\rm At}(\xcP),\ \bigl([(p_1,p_2)],[(p_2,p_3)]\bigr)\longmapsto[(p_1,p_3)]\,.
 \qqq
The groupoid fits into the short exact sequence of Lie groupoids (the {\bf Atiyah sequence}):
\qq\nn
\alxydim{@C=2.5cm@R=1.5cm}{ {\rm Ad}(\xcP) \ar@{^{(}->}[r]^{j_{{\rm Ad}(\xcP)}} \ar@<.5ex>[d]_{\sfT\vert\;\; } \ar@<-.5ex>[d]^{\;\; \sfS\vert} & {\rm At}(\xcP) \ar@{->>}[r]^{\pi} \ar@<.5ex>[d]_{\sfT \;\;} \ar@<-.5ex>[d]^{\;\; \sfS} & \Si\x\Si{} \ar@<.5ex>[d]_{\pr_1\;\; } \ar@<-.5ex>[d]^{\;\; \:\pr_2}\\ \xcF \ar@{=}[r] & \xcF \ar[r]_{\pi_\xcF} & \Si }\,,
\qqq
in which $\pi\colo{\rm At}(\xcP)\to\Si\x\Si,\ [(p_1,p_2)]\mapsto(\pi_\xcP(p_1),\pi_\xcP(p_2))$ and $j_{{\rm Ad}(\xcP)}$ is the embedding of the {\bf adjoint groupoid} ${\rm Ad}(\xcP) = \pi^{-1}(\Id(\Sigma))$.\ The latter is,\ in fact,\ a fibre bundle over $\Si$,\ with base projection $\pi_{\Ad(\xcP)}\colo\Ad(\xcP)\to\Si,\ [(p_1,p_2)]\mapsto\pi_\xcP(p_1)=\pi_\xcP(p_2)$,\ typical fibre $\xcG$,\ and the same transition 1-cocycle $\b_{ij}$ as $\xcP$,\ but realised through conjugation (see:\ Def.\,\ref{def:bisec-act}),
\qq\nn
\Ad(\xcP)\cong\bigsqcup_{i\in I}(O_i\x\xcG)/\sim_{C_{\b_{\cdot\cdot}}},\qquad(x,g,j)\sim\bigl(x,C_{\b_{ij}(\si)}(g),i\bigr)\,.
\qqq
The structure maps $\sfS\rstr,\sfT\rstr,\sfI\rstr$ and $\sfJ\rstr$ of the adjoint groupoid---induced from the respective structure maps of the Atiyah--Ehresmann groupoid through restriction---are readily seen to be vertical bundle morphisms.

To an arbitrary global section $\phi\in\G(\xcF)$ with presentation $\xcF\t_i\circ\phi=(\id_{O_i},X_i),\ X_i\colo O_i\to M$ in a local trivialisation $\xcF\t_i\colo\pi_\xcF^{-1}(O_i)\xrightarrow{\cong}O_i\x M$,\ we may now associate the $\phi$-{\bf unital section} 
\qq\nn
\Id_\phi:=\sfI\rstr\circ\phi\colo\Si\too\Ad(\xcP)\,,
\qqq 
with presentation
\qq\nn
\xcA\t_i\circ\Id_\phi=(\id_{O_i},\Id\circ X_i)\equiv(\id_{O_i}\x\Id)\circ(\id_{O_i},X_i)\colo O_i\too O_i\x\xcG
\qqq
in the corresponding local trivialisation $\xcA\t_i\colo\pi_{\Ad(\xcP)}^{-1}(O_i)\xrightarrow{\cong}O_i\x\xcG$.\ Note that for a principaloid $\bbB\txG$-bundle $\xcP=P$,\ with $\xcF=\Si$,\ we have $\phi=\id_\Si$,\ and then $\xcA\t_i\circ\Id_\phi=(\id_{O_i},e)$ is the standard unital section of the principal bundle $P$,\ which explains the name given to $\Id_\phi$.

In view of the identity of the local models of $\xcP$ and $\Ad(\xcP)$ over $O_i$,\ $\pi_\xcP^{-1}(O_i)\cong O_i\x\xcG\cong\pi_{\Ad(\xcP)}$,\ we may now evaluate the local projector fields $\Theta_i$ from \eqref{eq:Thet-loc} on the image of the $\xcA\t_i\circ\Id_\phi$,\ whereby we obtain the $\txT\xcG$-valued 1-forms
\qq\nn
\Theta_i\circ\txT_\si\bigl(\xcA\t_i\circ\Id_\phi\bigr)=\txT_{X_i(\si)}\Id\circ\txT_\si X_i-A_i\bigl(\si,X_i(\si)\bigr)\,.
\qqq
These take values in the domain of the multiplicative symplectic form $\om$ (or,\ more pedantically,\ of its pullback $\pr_2^*\om$ from the typical fibre),\ and so they yield expressions
\qq\nn
\pr_2^*\om\circ\bigl(\Theta_i\circ\txT\bigl(\xcA\t_i\circ\Id_\phi\bigr),\Theta_i\circ\txT\bigl(\xcA\t_i\circ\Id_\phi\bigr)\bigr)\rstr_{\txT O_i}\equiv(\id_{O_i},X_i)^*\om[\Theta_i]\,,
\qqq
with the local 2-forms
\qq\label{eq:omThe}
\om[\Theta_i]:=(\id_{O_i}\x\Id)^*\bigl(\pr_2^*\om\circ(\Theta_i,\Theta_i)\bigr)\in\Om^2(O_i\x M)
\qqq
which can be calculated directly,\ with the help of the identity $\Id^*\om=0$ and of Prop.\,\ref{prop:multi-anch},\ as
\qq
\om[\Theta_i](\si,m)&=&\om\circ\bigl(\txT_m\Id-A_i(\si,\Id_m),\txT_m\Id-A_i(\si,\Id_m)\bigr)\cr\cr 
&=&-\om\circ\bigl(\txT_m\Id, A_i(\si,\Id_m)\bigr)-\om\circ\bigl(A_i(\si,\Id_m),\txT_m\Id\bigr)+\om\circ\bigl(A_i(\si,\Id_m),A_i(\si,\Id_m)\bigr)\cr\cr 
&\equiv&\bigl(\kappa_\xcE(m)\circ A_i(\si,\Id_m)\bigr)\circ{\rm Alt}+\om\circ\bigl(A_i(\si,\Id_m),\txT_m\Id\circ\txT_{\Id_m}t\circ A_i(\si,\Id_m)\bigr)\cr\cr 
&\equiv&\bigl(\kappa_\xcE(m)\circ A_i(\si,\Id_m)+\tfrac{1}{2}\,\ev\circ\bigl(\bigl(\kappa_\xcE\circ A_i(\si,\Id_m)\bigr)\ox\bigl(\a_\xcE\circ A_i(\si,\Id_m)\bigr)\bigr)\bigr)\circ{\rm Alt}\,.\label{eq:omThe-guise}
\qqq
Thus,\ after the dust has cleared,\ we find,\ on the trivial principaloid $\xcG$-bundle with $I=\{i\},\ O_i\equiv\Si$ and $\xcA\t_i=\id_{\Si\x\xcG}=\xcP\t_i$,\ as well as $\phi=(\id_\Si,X)$ and $\Theta_{\rm triv}(\si,g)=\id_{\txT_g\xcG}-\txT_{\Id_{t(g)}}r_g\circ A(\si,t(g))$,\ the beautifully compact and natural expression
\qq\nn
\widetilde\ceL{}_{\rm PSM}(\phi,\txT\phi;A)=\phi^*\om[\Theta_{\rm triv}]
\qqq
involving the tensors $\om$ and $\Theta_{\rm triv}$,\ as desired.\ At this point---after the conceptual leap afforded by the assumed integration of the Poisson target to the symplectic Lie groupoid and the attendant principaloid $\xcG$-bundle---we may finally give a meaning to and answer the key question about the (large) gauge invariance of the action functional
\qq\label{eq:SPSM-beauty}
\widetilde S{}_{\rm PSM}[\phi,A]=\int_{\phi(\Si)}\om[\Theta_{\rm triv}]\,.
\qqq
And this is where we stumble upon a subtlety.

\subsection{The lagrangean anomaly}\label{sub:anom-kill-first}

Let us adopt the above \emph{Ansatz} for the field variables of the Poisson $\si$-model,\ and,\ accordingly,\ consider a gauge transformation of the action functional \eqref{eq:PSM-redef} modelled on an automorphism of the underlying principaloid $\xcG$-bundle $\xcP_{\rm triv}$.\ Thus,\ for $\Phi\in{\rm Gau}(\Si\x\xcG)$ presented by $h\colo\Si\to\bB(\xcG)$ as above,\ an adaptation of the master identity \eqref{eq:augs-glue} yields 
\qq\nn
&&\bigl(\xcF_*(\Phi)\circ\phi\bigr)^*\bigl(\kappa_\xcE\circ A^\Phi+\tfrac{1}{2}\,\ev\circ\bigl(\bigl(\kappa_\xcE\circ A^\Phi\bigr)\ox\bigl(\a_\xcE\circ A^\Phi\bigr)\bigr)\bigr)\circ{\rm Alt}\cr\cr
&=&\phi^*\bigl((t_*\circ h)^*\bigl(\kappa_\xcE\circ\unl{\txT C}{}_h\bigr)\circ A+\tfrac{1}{2}\,\ev\circ\bigl(\bigl((t_*\circ h)^*\bigl(\kappa_\xcE\circ\unl{\txT C}{}_h\bigr)\circ A\bigr)\ox\bigl(\a_\xcE\circ A\bigr)\bigr)\bigr)\circ{\rm Alt}+\phi^*\bigl(h\om-\widetilde h{}^*\om\bigr)\,,
\qqq
where $\widetilde h\colo\Si\x M\to\xcG,\ (\si,m)\mapsto h(\si)(m)$ as before,\ the pullback $(t_*\circ h)^*$ affects the second (codomain) tensor component in the morphism $\kappa_\xcE\circ\unl{\txT C}{}_h$ with the argument transformed by the automorphism \eqref{eq:Bis-on-E},\ and we have used the notation of App.\,\ref{app:lagr-anom} for $h\om$ and $\widetilde h{}^*\om$.\ In view of the arbitrariness of $(\phi,A)$,\ it is now clear that gauge invariance enforces upon us equalities:
\qq\nn
(t_*\circ h)^*\bigl(\kappa_\xcE\circ\unl{\txT C}{}_h\bigr)\circ A&\must& \kappa_\xcE\circ A\,,\cr\cr 
\ev\circ\bigl(\bigl((t_*\circ h)^*\bigl(\kappa_\xcE\circ\unl{\txT C}{}_h\bigr)\circ A\bigr)\ox\bigl(\a_\xcE\circ A\bigr)\bigr)&\must&\ev\bigl(\bigl(\kappa_\xcE\circ A\bigr)\ox\bigl(\a_\xcE\circ A\bigr)\bigr)
\qqq
\emph{and},\ separately,\ the cancellation of the second term,\ $\phi^*(h\om-\widetilde h{}^*\om)$,\ \emph{possibly after a correction} of the action functional $\widetilde S{}_{\rm PSM}$ independent of the gauge field $A$,\ and so using only some extra structure over $M$.\ An obvious additional requirement would then be that the correction do not affect the EL equations of the field theory.\ The relevance of the latter requirement is amplified by the structural interpretation of the solutions to the field equations,\ provided in \cite{Bojowald:2004wu}.

In the study of geometric consequences of the equivariance constraint
\qq\nn
(t_*\circ h)^*\bigl(\kappa_\xcE\circ\unl{\txT C}{}_h\bigr)\must\kappa_\xcE
\qqq
that solves the first of the two constraints,\ the identification of $\kappa_\xcE$ as the comomentum component of the Spencer data of the multiplicative 2-form $\om$ finally comes to the fore.\ As shown by Crainic {\it et al.} in \cite{Crainic:2015msp},\ the $\bB(\xcG)$-equivariance of the comomentum is,\ in general,\  obstructed,\ see:\ Thm.\,\ref{thm:CSS}.\ There is an exception to this rule,\ stated in Prop.\,\ref{prop:Spencer-on-holo}:\ The comomentum \emph{is} equivariant with respect to the action of the subgroup 
\qq\nn
\bB(\xcG,\om)\equiv\bB_\om(\xcG):=\{\ \b\in\bB(\xcG) \quad\vert\quad \b^*\om=0\ \}
\qqq
of $\om$-holonomic bisections (see:\ \eqref{eq:omlin-holo}),\ termed the {\bf lagrangean group} of $(\xcG,\om)$ in the symplectic setting \cite{Rybicki:2001}.\ It is a closed subgroup of the Lie group of bisections,\ with the tangent Lie algebra spanned on compactly supported de Rham 1-cocycles on $M$ (see:\ Thm.\,\ref{thm:LieBisOm}),\ and so the restriction to this subgroup in the gauging seems to be consistent with the tangential gauge-symmetry content,\ discussed in Rem.\,\ref{rem:tang-Pgauge}.\ Note also that these distinguished bisections descend to inner (symplectic leaf-preserving) Poisson automorphisms of the Poisson target $(M,\Pi)$ along the target map $t$.\ Hence,\ from now onwards,\ we assume the following {\bf lagrangean} (or {\bf $\om$-holonomic}) {\bf reduction of the structure group} of the symplectic principaloid $\xcG$-bundle $\xcP_{({\rm triv})}$:
\qq\label{eq:lagred}
\bB(\xcG)\searrow\bB(\xcG,\om)\,,
\qqq
and so,\ accordingly,\ denote the {\bf $\om$-holonomically reduced gauge group}
\qq\nn
{\rm Gau}_\om(\xcP_{\rm triv})=\bB(\xcG,\om)^\Si\,,
\qqq  
alongside its shadow
\qq\nn
{\rm Gau}_\om(\xcF_{\rm triv})=t_*\bigl(\bB(\xcG,\om)\bigr)^\Si\,.
\qqq
What remains,\ after this reduction,\ is to ensure the cancellation---for an arbitrary \emph{lagrangean} gauge tranformation $h\colo\Si\to\bB(\xcG,\om)$---of the $A$-independent term $\phi^*(h\om-\widetilde h{}^*\om)$.\ However,\ the reduction automatically nullifies its first summand,\ 
\qq\nn
h\om(\si,m)\equiv\bigl(h(\si)^*\om\bigr)(m)=0\,,
\qqq
and so we are left with the expression
\qq\label{eq:gtrafo-classic}
\widetilde S{}_{\rm PSM}[\phi^\Phi,A^\Phi]=\widetilde S{}_{\rm PSM}[\phi,A]-\int_\Si\phi^*\widetilde h{}^*\om\,.
\qqq

Thus,\ the classic Poisson $\si$-model,\ with the advocated reinterpretation in the language of symplectic principaloid $\xcG$-bundles,\ appears to \emph{exhibit a} {\bf \emph{lagrangean anomaly}} in general.\ We still need to convince ourselves that the anomaly derived above reproduces the original \emph{structural} results \eqref{eq:tan-Pgauge} and \eqref{eq:tanact-Pgauge} upon linearisation of $\Phi$.\ More specifically,\ we invoke Thm.\,\ref{thm:LieBisOm} and consider $h$ in the form of the exponential flow of a $\Si$-indexed Lie-algebra element $\kappa_\xcE^{-1}(\widetilde\ep)$ for $\widetilde\ep\in Z^1_{\rm c}(M)^\Si$ with a local (coordinate) presentation in a neighbourhood of the graph $\phi(\Si)$ of $X$ given by $\widetilde\ep(\si,m)=\ep_\mu(\si)\,\sfd x^\mu(m)$,\ {\it i.e.},\ we set
 \qq\nn
h(\si)=\exp\bigl(\kappa_\xcE^{-1}\bigl(\widetilde\ep(\si,\cdot)\bigr)\bigr)\,.
\qqq
Then---in keeping with the discussion presented at the end of Sec.\,\ref{sub:PSMasFT}---we seek to recover the tangential gauge-symmetry transformation of $X\equiv\pr_2\circ\phi$ from \eqref{eq:tan-Pgauge} and to prove the vanishing of the lagrangean anomaly in the linear order in $\widetilde\ep$ by expanding $h(\si)$ around the identity gauge transformation of $\phi(\Si)$ and keeping only terms at most linear in the `small' Lie-algebra element $\kappa_\xcE^{-1}(\widetilde\ep)$.\ Thus,\ we find (see:\ \eqref{eq:kapE-anch})
\qq\nn
\bigl(x^\mu\circ\pr_2\circ\xcF_*(\Phi)\circ\phi\bigr)(\si)&\equiv& x^\mu \bigl(t_*\bigl(h(\si)\bigr)\bigl(X(\si)\bigr)\bigr)=\bigl(x^\mu\circ X\bigr)(\si)+\bigl(\bigl(\txT t\circ\kappa_\xcE^{-1}\bigr)\bigl(\widetilde\ep\bigl(\si,X(\si)\bigr)\bigr)\bigr)^\mu+\xcO\bigl(\ep^2\bigr)\cr\cr 
&\equiv&\bigl(x^\mu\circ X\bigr)(\si)+\bigl(\bigl(\a_\xcE\circ\kappa_\xcE^{-1}\bigr)\bigl(\ep(\si)\bigr)\bigr)^\mu+\xcO\bigl(\ep^2\bigr)=\bigl(x^\mu\circ X\bigr)(\si)+\bigl(\Pi^\sharp\bigl(\ep(\si)\bigr)\bigr)^\mu+\xcO\bigl(\ep^2\bigr)\cr\cr 
&=&\bigl(x^\mu\circ X\bigr)(\si)+\Pi^{\nu\mu}\bigl(X(\si)\bigr)\ep_\nu(\si)+\xcO\bigl(\ep^2\bigr)
\qqq
for $\ep=\phi^*\widetilde\ep$,\ and
\qq\nn
\widetilde h{}^*\om(\si,m)&=&\om\bigl(\txT_{(\si,m)}\widetilde h,\txT_{(\si,m)}\widetilde h\bigr)=\om\bigl(\txT_m\bigl(h(\si)\bigr)+\txT_\si\widetilde h(\cdot,m)\bigr),\txT_m\bigl(h(\si)\bigr)+\txT_\si\widetilde h(\cdot,m)\bigr)\bigr)\cr\cr 
&=&\om\bigl(\txT_\si\widetilde h(\cdot,m)\bigr),\txT_m\bigl(h(\si)\bigr)\bigr)+\om\bigl(\txT_m\bigl(h(\si)\bigr),\txT_\si\widetilde h(\cdot,m)\bigr)\bigr)+\om\bigl(\txT_\si\widetilde h(\cdot,m)\bigr),\txT_\si\widetilde h(\cdot,m)\bigr)\bigr)\cr\cr 
&=&\om\bigl(\txT_\si\ep_\mu\ox\kappa_\xcE^{-1}\bigl(\sfd x^\mu(m)\bigr)\bigr),\txT_m\Id\bigr)+\om\bigl(\txT_m\Id,\txT_\si\ep_\mu\ox\kappa_\xcE^{-1}\bigl(\sfd x^\mu(m)\bigr)\bigr)+\xcO\bigl(\widetilde\ep{}^2\bigr)\cr\cr 
&=&\txT_\si\ep_\mu\ox\kappa_\xcE\bigl(\kappa_\xcE^{-1}\bigl(\sfd x^\mu(m)\bigr)\bigr)-\kappa_\xcE\bigl(\kappa_\xcE^{-1}\bigl(\sfd x^\mu(m)\bigr)\bigr)\ox\txT_\si\ep_\mu+\xcO\bigl(\widetilde\ep{}^2\bigr)\cr\cr 
&=&\txT_\si\ep_\mu\ox\sfd x^\mu(m)-\sfd x^\mu(m)\ox\txT_\si\ep_\mu+\xcO\bigl(\widetilde\ep{}^2\bigr)\,,
\qqq
whence also
\qq\nn
\phi^*\widetilde h{}^*\om=\sfd\ep_\mu\wedge\sfd\bigl(x^\mu\circ X)+\xcO\bigl(\ep^2\bigr)\equiv\sfd\ep+\xcO\bigl(\ep^2\bigr)\,,
\qqq
in perfect agreement with the original results.\ Having thus verified the validity of our reinterpretation of the classic PSM in the framework of symplectic principaloid $\xcG$-bundles,\ we conclude the present section with the discussion of the simplest conceivable scenario of the vanishing of the lagrangean anomaly,\ of which a more general variant shall be studied in Sec.\,\ref{sec:coaug-cancel}.

The embedded worldsheet $\phi(\Si)$ is closed,\ $\p\phi(\Si)=\emptyset$,\ and so the most direct solution to the problem in hand would be to demand that $\widetilde h{}^*\om\in\im\,\sfd$
for \emph{all} lagrangean gauge transformations $\Phi\in{\rm Gau}_\om(\xcP_{\rm triv})$.\ This,\ however,\ is not only unwieldy from the geometric point of view,\ but also---first and foremost---extremely restrictive,\ and in fact unnecessarily so.\ Indeed,\ upon adopting Dirac's quantum-mechanical interpretation of the classical lagrangian---put forward in his pioneering paper \cite{Dirac:1933pi},\ subsequently elaborated by Feynman in \cite{Feynman:1948pi},\ and much later adapted by Gaw\c{e}dzki in \cite{Gawedzki:1987ak} as the foundation of his successful higher-geometric/-cohomological approach to charged-path/loop dynamics (or string theory)---we should require a \emph{strict} invariance of the {\bf Dirac--Feynman} amplitude (we set $\hbar=1$)
\qq\label{eq:PSM-redef}
\cA_{\rm DF}^{\rm PSM}[\phi,A]:=\exp\bigl(\sfi\,\widetilde S{}_{\rm PSM}[\phi,A]\bigr)=\exp\left(\sfi\int_{\phi(\Si)}\om[\Theta_{\rm triv}]\right)\,.
\qqq
But then the condition for the vanishing of the anomaly,\ in its simplest variant,\ assumes the relaxed form
\qq\nn
\bigl\{\ {\rm Per}\bigl([\widetilde h{}^*\om]\bigr)\ |\ h\in{\rm Gau}_\om(\xcP_{\rm triv}) \ \bigr\}\overset{!}{\subset}2\pi\bZ\,.
\qqq 
This is still unwieldy because of the indexing by arbitrary lagrangean gauge transformations,\ and so we propose the stronger but also readily tractable and geometrically natural {\bf symplectic charge quantisation condition}
\qq\label{eq:PeromZ}
{\rm Per}\bigl([\om]\bigr)\overset{!}{\subset} 2\pi\bZ
\qqq
as the simplest structural solution to the problem of removal of the lagrangean anomaly in the classic Poisson $\si$-model.\ The condition admits a straightforward geometric interpretation:\ The symplectic form geometrises as a principal $\bC^\x$-bundle 
\qq\label{eq:preq-fram-om}
\alxydim{@C=1.25cm@R=1.cm}{ \cU\x\bC^\x & \cU\x_\xcG P_\om \ar[l]^{\cong}_{\t} \ar[r] \ar[d] & P_\om \ar[d]^{\pi_{P_\om}}  \\ & \cU \ar[r]_{\jmath} & \xcG }
\qqq
with local trivialisations trivialisations $\t=\{\t_\a\}_{\a\in A}$ over an open cover $\cU=\{U_\a\}_{\a\in A}$ of its base $\xcG$,\ and with a (principal) connection $\cA\in\Om^1(P_\om)$ of curvature $\om$.\ In the language of geometric quantisation \cite{Kostant:1970} (and with a view to a subsequent gerbe-theoretic generalisation),\ we witness here the emergence of the frame bundle of the prequantum bundle for $\om$.\ Its appearance is very natural from the point of view of quantisation of the target Poisson structure $(M,\Pi)$,\ see:\ \cite{Weinstein:1991}.

\beg
Being exact,\ the symplectic form $\om_{\rm lin}$ of the linear Poisson $\si$-model from Sec.\,\ref{sub:toy-lPSM} is automatically integral,\ and so the lagrangean anomaly vanishes identically in that model,\ in conformity with our analysis.
\eeg

\brem
It deserves to be pointed out that expressions of the form \eqref{eq:SPSM-beauty} are not at all unusal in the modelling of topological couplings of gauge fields on principaloid bundles to currents of probes charged under these fields,\ as represented by embedded worldvolumes.\ Indeed,\ in the classic 1$d$ $\si$-model describing the dynamics of a point-like probe charge $q$ in external gravitational and electromagnetic fields permeating a target manifold $M$,\ the topological coupling of the latter to the probe's worldline $X(\La)\subset M$ (for $\La\subset\bR$) can be written in the familiar manner
\qq\nn
q\,\int_{X(\La)}(\id_M\x\Id)^*\bigl(\theta\circ\Theta_{\rm triv}\bigr)
\qqq
in terms of a (global) connection 1-form $\Theta_{\rm triv}(m,u)=\id_{\txT_u\uj}-\txT_e r_u\circ A(m,\ast)$ on the (trivial) principaloid $\bbB\uj$-bundle $M\x\uj\ni(m,u)$ and the multiplicative (Maurer--Cartan) 1-form $\theta(u)=\frac{\sfd u}{\sfi u}$ on the arrow manifold $\uj$ of the (delooping) structure groupoid $\bbB\uj$ (with object manifold given by the singleton $\ast$).\ The latter has the constant identity map $\Id_\ast=e$,\ and so we recognise in the above integrand the usual pullback of the principal connection 1-form $\cA_{\rm triv}=\theta\circ\Theta_{\rm triv}\in\Om^1(M\x\uj)\ox\Lie\uj$ along a flat unital section $\id_M\x e\colo M\to M\x\uj$,\ evaluated on the image $X(\La)\subset M\equiv M\x\ast$ of the 1$d$ `spacetime' $\La$ in the shadow bundle $M\x\ast\equiv(M\x\uj)/\uj$ under a smooth embedding $X\in C^\infty(\La,M)$. 

Structurally related expressions have been encountered before in the setting of topological field theory,\ {\it e.g.},\ in the investigation of the AKSZ topological $\si$-model of \cite{Alexandrov:1995kv} carried out by Kotov and Strobl,\ and reported\footnote{The Author is grateful to Alexei Kotov for bringing this fact to his attention in a discussion subsequenbt to the completion of the work reported in the present article.}  in \cite{Kotov:2007nr}.
\erem

\section{Lagrangean-anomaly cancellation {\it via} gerbe coaugmentation}\label{sec:coaug-cancel}

Our reinetrpretation of the classic Poisson $\si$-model in Sec.\,\ref{sub:PSM-redef} is most concisely encapsulated in expression \eqref{eq:PSM-redef},\ which emphasises the topologicality of the field theory and immediately prompts a follow-up question as to an intrinsic geometric meaning---if any---of the 2-form \eqref{eq:omThe}.\ Differential forms with this general structure---a multiplicative 2-form $\rho\in\Om^2(\xcG)$ on the arrow manifold of a Lie groupoid $\grpd{\xcG}{M}$ evaluated as $\rho[\Theta_{(i)}]$ on the image of the (local) projector field $\Theta_{(i)}$ on the tangent bundle of a principaloid $\xcG$-bundle---have been encountered before \cite{Gawedzki:2010rn,Suszek:2026col},\ albeit invariably in the guise \eqref{eq:omThe-guise}.\ The geometric context of their appearance has been the gauging of a groupoidal symmetry modelled on $\grpd{\xcG}{M}$ in the topological sector of the dynamical 2$d$ Polyakov--Alvarez--Gaw\c{e}dzki (PAG) $\si$-model,\ the latter being captured by a Cheeger--Simons (ChS) differential character of degree $3$---the surface holonomy of a bundle gerbe \cite{Murray:1994db,Murray:1999ew}.\ The gauging has been demonstrated to hinge on two higher-algebraic mechanisms:\ a $\rho$-twisted equivariantisation of the gerbe $\cG$ over the simplicial nerve of the symmetry groupoid $\grpd{\xcG}{M}$ (see:\ Def.\,\ref{def:Gequiv-grb}) and a 2-stacky descent of the ensuing equivariantly augmented gerbe $\pr_2^*\cG\ox\cI_{\rho[\Theta_i]}$ to the total space of the shadow $\xcF$ of an arbitrary principaloid $\xcG$-bundle $\xcP$ from the shadow's local models $O_i\x M$.\ In what follows,\ we use the above contextualisation as a springboard for an essential generalisation of the basic scenario of the vanishing of the lagrangean anomaly \eqref{eq:PeromZ} proposed in the previous section,\ see:\ \cite{Suszek:2026col}.\medskip

The point of departure of our analysis is the transcription of the defining identities 
\qq\nn
\sfd\om=0\,,\qquad\qquad\bigl(\txm^*-\pr_1^*-\pr_2^*\bigr)\om=0\,,
\qqq
satisfied by the symplectic form $\om\in\Om^2(\xcG)$,\ in the language of the Bott--Schulman--Stasheff (BSS) cohomology of $\grpd{\xcG}{M}$ (see:\ Def.\,\ref{def:BSS-complex}):\ We find $\om$ in the intersection of the kernels of the two differentials of the BSS bicomplex.\ Assume given a de Rham 3-cocycle $H_M\in Z^3_{\rm dR}(M)$ which composes a {\bf BSS extension} $(H_M,\om,\eta,f)\in Z^3_{\rm BSS}(\grpd{\xcG}{M})\subset\Om^3(\xcG_0)\oplus\Om^2(\xcG_1)\oplus\Om^1(\xcG_2)\oplus\Om^0(\xcG_3)$ together with $\om$.\ The assumption forces $H_M$ to belong to the kernel of the Dupont operator $\D^{(3)}_{(0)}\equiv t^*-s^*$,\ and so an extension always exist.\ In the setting of the PAG $\si$-model with defects,\ defined in \cite{Runkel:2008gr},\ one usually considers extensions of the distinguished form\footnote{Such extensions have also been considered by many authors as twisted-multiplicative/(pre)symplectic differential forms on groupoids,\ see,\ {\it e.g.},\ \cite{Bursztyn:2004,Arias:2011,Bursztyn:2012},\ and also \cite{Severa:2001qm,Klimcik:2002,Cattaneo:2004} for a related discussion of twisted Poisson structures.} $(H_M,\om,0,0)$,\ which become naturally associated with defects implementing the rigid symmetry $\grpd{\xcG}{M}$.\ The restriction is in keeping with the multiplicativity of $\om$,\ and so we consistently impose it below. 

\brem\label{rem:Dupont}
The choice of the bicomplex in which we look for an extension is dictated by the identification of the geometric realisation of the simplicial manifold $N_\bullet(\grpd{\xcG}{M})$ (see:\ Def.\,\ref{def:fat-geom-real}) as the homotopy model
\qq\nn
\txB\xcG\equiv\Vert\xcG_\bullet\Vert
\qqq
of the orbispace $M//\xcG$,\ to which we want to effectively descend---through the gauging (see,\ {\it e.g.},\ \cite[Sec.\,9]{Gawedzki:2012fu})---the field theory of interest,\ and by the subsequent identification\footnote{See:\ Rem.\,\ref{rem:Dupont}.} of the $\xcG$-equivariant BSS cohomology as the (extended) de Rham cohomology of $\txB\xcG$,
\qq\label{eq:orbi-coho} 
H^\ast_{\rm dR}(\txB\xcG)\cong H^\ast_{\rm BSS}\bigl(\grpd{\xcG}{M}\bigr)\,,
\qqq
see:\ \cite{Bott:1976bss,Dupont:1976,Dupont:1988}.\ In view of the fundamental status of the latter in the chain of inferences followed in our discussion,\ we pause to recall its precise nature,\ after Dupont \cite{Dupont:1976}.

On one hand,\ we have the de Rham complex of the geometric realisation,\ 
\qq\nn
\Om^\bullet\bigl(\Vert\xcG_\bullet\Vert\bigr)=\bigoplus_{p\in\bN}\Om^p\bigl(\Vert\xcG_\bullet\Vert\bigr)\,,
\qqq 
with homogeneous components $\Om^p(\Vert\xcG_\bullet\Vert)$ of degree $p$ {\em defined},\ in an intuitively clear manner,\ as being generated over $\bR$ by families $\om^\bullet\equiv\{\om^{(n)}\}^{n\in\bN}$ of those restrictions $\om^{(n)}=\widetilde\om{}^{(n)}\rstr_{\xcG_n\x\triangle^n}$ of smooth $p$-forms $\widetilde\om^{(n)}\in\Om^p(\xcG_n\x\widetilde\triangle{}^n)$ over the smooth manifolds $\xcG_n\x\widetilde\triangle{}^n$ with (smooth) affine-hyperplane cartesian factors
\qq\nn
\widetilde\triangle^n:=\bigl\{ \ (t_1,t_2,\ldots,t_{n+1})\in\bR^{\x n+1} \quad\vert\quad \sum_{k=1}^{n+1}\,t_k=1 \ \bigr\}\subset\bR^{\x n+1}\,,
\qqq
which satisfy,\ for all $n\in\bN$,\ the `descent' identities
\qq\label{eq:hmtp-forms}
\bigl(d_k^{(n+1)}\x\id_{\triangle^n}\bigr)^*\om^{(n)}=
\bigl(\id_{\xcG_{n+1}}\x\widetilde\p{}_i^{(n+1)}\bigr)^*\widetilde\om{}^{(n+1)}\rstr_{\xcG_{n+1}\x\triangle^n}\,,
\qqq
expressed in terms of the smooth embeddings
\qq\nn
\widetilde\p{}_i^{(n+1)}\colo\widetilde\triangle{}^n\too\widetilde\triangle{}^{n+1},\ (t_1,t_2,\ldots,t_{n+1})\longmapsto(t_1,t_2,\ldots,t_i,0,t_{i+1},t_{i+2},\ldots,t_{n+1})\,,
\qqq
see:\ \eqref{eq:desc-geom-real}.\ Thus,
\qq\nn
\Om^p(\Vert\xcG_\bullet\Vert)=\prod_{n\in\bN}\Om^p(\xcG_n\x\widetilde\triangle{}^n)\rstr_{\xcG_n\x\triangle^n}/\sim\,,
\qqq
where $\sim$ stands for the `descent' identifications \eqref{eq:hmtp-forms}.\ The differential on $\Om^\bullet(\Vert\xcG_\bullet\Vert)$ is a family of operators
\qq\nn
\sfd_\bullet\colo\Om^p\bigl(\Vert\xcG_\bullet\Vert\bigr)\too\Om^{p+1}\bigl(\Vert\xcG_\bullet\Vert\bigr),\ \bigl(\om^{(n)}\bigr)^{n\in\bN}\longmapsto\bigl(\sfd\widetilde\om{}^{(n)}\rstr_{\xcG_n\x\triangle^n}\bigr)^{n\in\bN}
\qqq
determined by standard de Rham differential $\sfd$.\ The latter may be interpreted as the differential of the total complex of the bicomplex
\qq\nn
\bigl(\Om^{\bullet_1,\bullet_2}\bigl(\Vert\xcG_\bullet\Vert\bigr),\sfd^{(\bullet_1,\bullet_2)}_1,\sfd^{(\bullet_1,\bullet_2)}_2\bigr)
\qqq
with cochain groups
\qq\nn
\Om^{\bullet_1,\bullet_2}\bigl(\Vert\xcG_\bullet\Vert\bigr):=\prod_{n\in\bN}\Om^{\bullet_1}\bigl(\xcG_n\bigr)\ox_{C^\infty(\xcG_n\x\triangle^n,\bR)}\Om^{\bullet_2}\bigl(\triangle^n\bigr)/\sim\,,
\qqq
defined in terms of restrictions $\Om^{\bullet_2}(\triangle^n)\equiv\Om^{\bullet_2}(\widetilde\triangle^n)\rstr_{\xcG_n\x\triangle^n}$ and $C^\infty(\xcG_n\x\triangle^n,\bR)\equiv C^\infty(\xcG_n\x\widetilde\triangle{}^n,\bR)\rstr_{\xcG_n\x\triangle^n}$,\ and in terms of the obvious transcription of the previous `descent' identifications,\ and equipped with the $\id$-extended de Rham differentials:\ $\sfd^{(\bullet_1,\bullet_2)}_1\equiv\sfd_{(\xcG_\bullet)}\ox\id$ and $\sfd^{(\bullet_1,\bullet_2)}_2\equiv\id\ox\sfd_{(\widetilde\triangle{}^\bullet)}(\cdot)\rstr_{\triangle^\bullet}$.

On the other hand,\ we have the BSS bicomplex of $\grpd{\xcG}{M}$,
\qq\nn
\bigl(\Om^{\bullet_1}\bigl(\xcG_{\bullet_2}\bigr),\sfd^{(\bullet_1)}_{(\bullet_2)},\D_{(\bullet_2)}^{(\bullet_1)}\bigr)\,,
\qqq
with (the families of) the de Rham differentials $\sfd^{(\bullet_1)}_{(\bullet_2)}$ and the Dupont ones $\D^{(\bullet_1)}_{(\bullet_2)}$,\ as specified in Def.\,\ref{def:BSS-complex}.

An important conceptual result of \cite{Dupont:1976} was the determination of a chain-homotopy equivalence between the above two bicomplexes:
{\scriptsize\qq\nn
	\alxydim{@C=.5cm@R=1.5cm}{ & & & \cdots \ar[rrr]^(.5){\sfd_2^{(p+1,q-1)}} & & & \Om^{p+1,q}\bigl(\Vert\xcG_\bullet\Vert\bigr) \ar[rr]^(.5){\sfd_2^{(p+1,q)}} \ar@<.5ex>[dd]^(.25){\xcI^{(p+1,q)}}|(0.5)\hole|(0.61)\hole \ar@/^1pc/@<.25ex>[lll]^(.4){s^{p+1,q}} & & \Om^{p+1,q+1}\bigl(\Vert\xcG_\bullet\Vert\bigr) \ar@<.5ex>[dd]^(.25){\xcI^{(p+1,q+1)}}|(0.5)\hole|(0.61)\hole \ar[rrrr]^(.6){\sfd_2^{(p+1,q+1)}} \ar@/^1pc/@<.25ex>[ll]^(.4){s^{p+1,q+1}} & & & & \cdots \ar@/^1pc/@<.25ex>[llll]^(.4){s^{p+1,q+2}} \\ \cdots \ar[rrrrr]^(.65){\sfd_2^{(p,q-1)}} & & & & & \Om^{p,q}\bigl(\Vert\xcG_\bullet\Vert\bigr) \ar[rr]^(.6){\sfd_2^{(p,q)}} \ar@<.5ex>[dd]^(.25){\xcI^{(p,q)}} \ar@/^1pc/@<.25ex>[lllll]^(.4){s^{p,q}} \ar[ur]_{\sfd_1^{(p,q)}} & & \Om^{p,q+1}\bigl(\Vert\xcG_\bullet\Vert\bigr) \ar@<.5ex>[dd]^(.25){\xcI^{(p,q+1)}}  \ar[rr]^(.73){\sfd_2^{(p,q+1)}} \ar@/^1pc/@<.25ex>[ll]^(.4){s^{p,q+1}} \ar[ur]_{\sfd_1^{(p,q+1)}}  & & \cdots \ar@/^1pc/@<.25ex>[ll]^(.3){s^{p,q+2}} & & \\ & & & \cdots \ar'[rr]^(.7){\D^{(p+1)}_{(q-1)}}[rrr] & & & \Om^{p+1}(\xcG_q) \ar'[r]^(.75){\D^{(p+1)}_{(q)}}[rr] \ar@<.5ex>[uu]^(.25){\xcE^{(p+1)}_{(q)}}|(0.39)\hole|(0.5)\hole	& & \Om^{p+1}(\xcG_{q+1}) \ar[rrrr]^(.5){\D^{(p+1)}_{(q+1)}} \ar@<.5ex>[uu]^(.25){\xcE^{(p+1)}_{(q+1)}}|(0.39)\hole|(0.5)\hole & & & & \cdots \\ \cdots \ar[rrrrr]^(.45){\D^{(p)}_{(q-1)}} & & & & & \Om^p(\xcG_q) \ar[rr]^(.5){\D^{(p)}_{(q)}} \ar[ur]_(.6){\sfd^{(p)}_{(q)}} \ar@<.5ex>[uu]^(.25){\xcE^{(p)}_{(q)}} & & \Om^p(\xcG_{q+1}) \ar[rr]^(.55){\D^{(p)}_{(q+1)}} \ar[ur]_(.6){\sfd^{(p)}_{(q+1)}} \ar@<.5ex>[uu]^(.25){\xcE^{(p)}_{(q+1)}} & & \cdots & & }\,,
	\qqq}
\hspace{-5pt}{\it i.e.},\ of a chain homotopy $s^{(\bullet_1,\bullet_2)}\colo \Om^{\bullet_1,\bullet_2}(\Vert\xcG_\bullet\Vert)\to\Om^{\bullet_1,\bullet_2-1}(\Vert\xcG_\bullet\Vert)$ between chain maps $\xcE^{(\bullet_2)}_{(\bullet_1)}\circ\xcI^{(\bullet_1,\bullet_2)}$ and $\id_{\Om^{\bullet_1,\bullet_2}(\Vert\xcG_\bullet\Vert)}$,
\qq\nn
&\xcE^{(\bullet_2)}_{(\bullet_1)}\circ\xcI^{(\bullet_1,\bullet_2)}-\id_{\Om^{\bullet_1,\bullet_2}(\Vert\xcG_\bullet\Vert)}=s^{(\bullet_1,\bullet_2+1)}\circ\sfd_2^{(\bullet_1,\bullet_2)}+\sfd_2^{(\bullet_1,\bullet_2-1)}\circ s^{(\bullet_1,\bullet_2)}\,,&\cr\cr  &s^{(\bullet_1+1,\bullet_2)}\circ\sfd_1^{(\bullet_1,\bullet_2)}=\sfd_1^{(\bullet_1,\bullet_2-1)}\circ s^{(\bullet_1,\bullet_2)}\,,&
\qqq
alongside a trivial chain homotopy between $\xcI^{(\bullet_1,\bullet_2)}\circ\xcE^{(\bullet_2)}_{(\bullet_1)}$ and $\id_{\Om^{\bullet_1}(\xcG_{\bullet_2})}$,
\qq\nn
\xcI^{(\bullet_1,\bullet_2)}\circ\xcE^{(\bullet_2)}_{(\bullet_1)}=\id_{\Om^{\bullet_1}(\xcG_{\bullet_2})}\,.
\qqq
Morally,\ the chain maps $\xcE^{(\bullet_2)}_{(\bullet_1)}$ and $\xcI^{(\bullet_1,\bullet_2)}$---which intertwine $\sfd^{(\bullet_1,\bullet_2)}_1$ with $\sfd^{(\bullet_1)}_{(\bullet_2)}$ and,\ independently,\ $\sfd^{(\bullet_1,\bullet_2)}_2$ with $\D_{(\bullet_2)}^{(\bullet_1)}$---are (sums of) wedge products of pullbacks of forms on $\xcG_\bullet$ along suitable face maps with certain `elementary forms' introduced in \cite{Whitney:1957},\ and integrations over the simplex fibre(s),\ respectively---see:\ \cite{Dupont:1976} for technical details.\ The above chain-homotopy equivalence induces the desired isomorphism in cohomology:
\qq\nn
H^\ast_{\rm dR}(\txB\xcG):= H^\ast\bigl(\Om^{\bullet_1,\bullet_2}\bigl(\Vert\xcG_\bullet\Vert\bigr),\sfd^{(\bullet_1,\bullet_2)}_1,\sfd^{(\bullet_1,\bullet_2)}_2\bigr)\cong H^\ast\bigl(\Om^{\bullet_1}\bigl(\xcG_{\bullet_2}\bigr),\sfd^{(\bullet_1)}_{(\bullet_2)},\D_{(\bullet_2)}^{(\bullet_1)}\bigr)=:H^\ast_{\rm BSS}\bigl(\grpd{\xcG}{M}\bigr)\,.
\qqq
\erem

Suppose,\ next,\ that the 3-cocycle admits a \emph{partial} $\xcG$-equivariant prequantisation in the form of a gerbe $\cG$ of curvature $H_M$ together with a gerbe 1-isomorphism\footnote{Gerbe 1-isomorphisms of this special form were first considered by Fuchs {\it et al.}\ in \cite{Fuchs:2007fw} in relation to circular worldsheet defects.\ A fully fledged gerbe-theoretic description of arbitrary defect graphs (also those with intersections) was subsequently \emph{derived} and employed in an in-depth study of non-perturbative aspects of the Wess--Zumino--Novikov--Witten $\si$-model by Runkel and the Author in \cite{Runkel:2008gr},\ further elaborated in \cite{Runkel:2009sp} and,\ more recently,\ in \cite{Suszek:2022CSimpl}.}
\qq\label{eq:Ups-first}
\Upsilon\colo t^*\cG\xrightarrow{\ \cong\ }s^*\cG\ox\cI_\om\,,
\qqq
the latter providing a geometrisation of the BSS-cohomological condition
\qq\label{eq:BSS-1}
\D^{(3)}_{(0)}H_M=\sfd\om\,,
\qqq
in keeping with the identity $\curv(\cG_1\ox\cG_2)=\curv(\cG_1)+\curv(\cG_2)$,\ satisfied by arbitrary gerbes $\cG_a,\ a\in\{1,2\}$ over a common base.\ In virtue of Thm.\,\ref{thm:CH-BD} and the identities:\ $\Hol_{\cG_1\ox\cG_2}(\cdot)=\Hol_{\cG_1}(\cdot)\cdot\Hol_{\cG_2}(\cdot)$ and $\Hol_{f^*\cG}(\z)=\Hol_\cG(f_*\z)$ (written for $f\in C^\infty(X_1,X_2),\ \cG\in\bgrb_\nabla(X_2)_{(0)}$ and $\z\in Z_2(X_1)$),\ the surface holonomy of $\cG$ then satisfies the equality
\qq\label{eq:Upsalone}
t^*\Hol_\cG=s^*\Hol_\cG\cdot\Hol_{\cI_\om}\,,
\qqq
with
\qq\nn
\Hol_{\cI_\om}(c_2)=\exp\bigl(\sfi\int_{c_2}\om\bigr)\,,\qquad c_2\in Z_2(\xcG)\,,
\qqq
and so its pullback by a bisection $\b\in\bB(\xcG)$ yields
\qq\nn
(t_*\b)^*\Hol_\cG=\Hol_\cG\cdot\exp\bigl(\sfi\int_\cdot\b^*\om\bigr)\,.
\qqq
Lagrangean bisections thus become naturally distinguished as those whose shadows $t_*\b\in\Diff(M)$ preserve the holonomy ({\it i.e.},\ the isomorphism class) of the prequantisation $\cG$ of the 3-form component $H_M$ of the BSS 3-cocycle $(H_M,\om,0,0)$.\ From the point of view of the dynamical PAG $\si$-model with worldsheet $\Si\in\p^{-1}\emptyset$,\ target $(M,\txg_M,\cG)$ (for some metric $\txg_M\in\G(S^2\txT^*M)$),\ and the topological term in the DF amplitude given by
\qq\nn
\cA_{\rm DF}^{\si\,({\rm top})}[X]=\Hol_\cG\bigl(X(\Si)\bigr)\,,\qquad X\in C^\infty(\Si,M)\,,
\qqq
this condition acquires the physical interpretation of a rigid symmetry of the 2$d$ field theory,\ as long as the metric term is separately preserved\footnote{The most obvious possibility is $t_*\b\in{\rm Isom}(M,\txg_M)$,\ but more general scenarios could be conceived,\ see:\ \cite{Kotov:2014iha}.\ The Author is grateful to Thomas Strobl for pointing out this fact.} by $t_*\b$.

The 1-isomorphism \eqref{eq:Ups-first} is now recognised as the basic requisite for the gauging of the rigid groupoidal symmetry implemented  by $t_*(\bB(\xcG,\om))$ in the topological sector of the PAG $\si$-model, \ as it ensures the invariance of the topological DF amplitude for the {\bf connection-augmented gerbe}
\qq\label{eq:aug-grb}
\cG[\Theta_{\rm triv}]=\pr_2^*\cG\ox\cI_{\om[\Theta_{\rm triv}]}
\qqq
over the trivial shadow bundle $\xcF_{\rm triv}=\Si\x M$ under the corresponding gauge transformations ${\rm Gau}_\om(\xcF_{\rm triv})$.\ Indeed,\ for every $\Phi\in{\rm Gau}_\om(\xcP_{\rm triv})$ realised as $\Phi(\si,g)=(\si,L_{h(\si)}(g))$ by $h\colo\Si\to\bB(\xcG,\om)$ with  $\widetilde h\colo\Si\x M\to\xcG,\ (\si,m)\mapsto h(\si)(m)$ and $H\colo\Si\x M\to\Si\x M,\ (\si,m)\longmapsto(\si,t_*(h(\si))(m))$ as in App.\,\ref{app:lagr-anom},\ we obtain an induced gerbe 1-isomorphism
\qq\nn
\widetilde h{}^*\Upsilon\ox\id_{H^*\om[\Theta_{\rm triv}^\Phi]}\colo H^*\bigl(\cG[\Theta_{\rm triv}^\Phi]\bigr)\equiv\widetilde h{}^*t^*\cG\ox\cI_{H^*\om[\Theta_{\rm triv}^\Phi]}\xrightarrow{\ \cong\ }\widetilde h{}^*s^*\cG\ox\cI_{\widetilde h{}^*\om+H^*\om[\Theta_{\rm triv}^\Phi]}=\pr_2^*\cG\ox\cI_{\om[\Theta_{\rm triv}]}\equiv\cG[\Theta_{\rm triv}]\,,
\qqq
see:\ the specialisation \eqref{eq:augs-glue-red} of the master identity \eqref{eq:augs-glue}.\ In virtue of Thm.\,\ref{thm:CH-BD} (see also:\ Def.\,\ref{def:holG}),\ we infer the desired equality
\qq\nn
\Hol_{\cG[\Theta_{\rm triv}^\Phi]}\bigl((\xcF_*(\Phi)\circ\phi)(\Si)\bigr)=\Hol_{\cG[\Theta_{\rm triv}]}\bigl(\phi(\Si)\bigr)\,.
\qqq
Note that the previously derived lagrangean anomaly of the trivial term 
\qq\nn
\Hol_{\cI_{\om[\Theta_{\rm triv}]}}\bigl(\phi(\Si)\bigr)\equiv\cA_{\rm DF}^{\rm PSM}[\phi,A]
\qqq
in the topological DF amplitude $\Hol_{\cG[\Theta_{\rm triv}]}(\cdot(\Si))$ is now cancelled by that of the holonomy $\Hol_{\pr_2^*\cG}(\cdot(\Si))$ of the gerbe $\cG$ that geometrises $H_M$.\ This integrality of $\om$ is \emph{not} requisite in this situation.

The above considerations,\ inspired by the pioneering analysis of the action-groupoid case in \cite{Gawedzki:2010rn,Gawedzki:2012fu},\ lead us to an extension of the classic Poisson $\si$-model which we call---in view of the context in which it was first suggested in \cite{Suszek:2026col},\ and taking into account the topological triviality of the shadow bundle employed---the {\bf $\cG$-coaugmented trivial Poisson} ({\bf ctP}) {\bf $\si$-model}.\ It is a field theory
\qq\label{eq:ctP-sigmod}
\bigl(\Si,(\xcP_{\rm triv},\Theta_{\rm triv})\xrightarrow{\ \xcD_{\rm triv}\ }\xcF_{\rm triv},\bigl(\grpd{(\xcG,\om)}{(M,\Pi)},(\cG,\Upsilon)\bigr),\cA_{\rm DF}^{tP\si[\cG]}\bigr)\equiv P^{t[\cG]}_\si
\qqq
with the configuration bundle over the closed worldsheet $\Si$ given by the trivial shadow bundle $\xcF_{\rm triv}\equiv\Si\x M\to\Si$ of the trivial {\bf symplectic principaloid $(\xcG,\om)$-bundle} $\xcP_{\rm triv}\equiv\Si\x\xcG\to\Si$ (see:\ \cite[Ex.\,2.27]{Strobl:2025}) with the sitting duck $\xcD_{\rm triv}=\id_\Si\x t$ and the structure group reduced to the lagrangean group $\bB(\xcG,\om)$ of the symplectic structure Lie groupoid $\grpd{(\xcG,\om)}{(M,\Pi)}$.\ The typical fibre $M$ of the configuration bundle carries a Poisson bivector $\Pi\in\G(\wedge^2\txT M)$ and a gerbe $\cG$---termed the {\bf coaugmenting gerbe} in the present setting---alongside a gerbe 1-isomorphism $\Upsilon$ over the arrow manifold of the groupoid.\ The bundle $\xcP_{\rm triv}\ni(\si,g)$ comes with a $\xcG$-equivariant connection 1-form $\Theta_{\rm triv}$,\ admitting a global presentation $\Theta_{\rm triv}(\si,g)=\id_{\txT_g\xcG}-\txT_{\Id_{t(g)}}r_g\circ A(\si,t(g))$ as before.\ In it,\ $A\in\G(\pr_1^*\txT^*\Si\ox\pr_2^*\xcE)$ is the global gauge field,\ taking values in the tangent Lie algebroid $\xcE$ of the structure Lie groupoid.\ Classical configurations of this field theory are critical points of the DF amplitude---to be termed a {\bf trivial Poisson DF character}---
\qq\label{eq:ctP-sigmod-DF}
\cA_{\rm DF}^{tP\si[\cG]}\colo\G(\xcF_{\rm triv})\x{\rm Conn}(\xcP_{\rm triv})\too\uj,\ (\phi,\Theta_{\rm triv})\longmapsto\Hol_{\cG[\Theta_{\rm triv}]}\bigl(\phi(\Si)\bigr)\,,
\qqq
defined---in terms of the connection-augmented gerbe \eqref{eq:aug-grb}---on the product of the space ${\rm Conn}(\xcP_{\rm triv})$ of (compatible) connection 1-forms on $\xcP_{\rm triv}$ and the space $\G(\xcF_{\rm triv})\cong C^\infty(\Si,M)$ of global sections of $\xcF_{\rm triv}$.\ Just to reemphasise:\ The fundamental property of the trivial Poisson character is its invariance under simultaneous $\om$-\emph{holonomic} ({\it i.e.},\ lagrangean) gauge transformations of both field variables,
\qq\nn
\forall\ (\phi,\Theta_{\rm triv},\Phi)\in\G(\xcF_{\rm triv})\x{\rm Conn}(\xcP_{\rm triv})\x{\rm Gau}_\om(\xcP_{\rm triv})\colo\cA_{\rm DF}^{tP\si[\cG]}[\xcF_*(\Phi)\circ\phi,\Theta_{\rm triv}^\Phi]=\cA_{\rm DF}^{tP\si[\cG]}[\phi,\Theta_{\rm triv}]\,.
\qqq

An important property of the extension defined above is the tractability of the additive correction to the EL equations that it introduces.\ Indeed,\ in virtue of \eqref{eq:CS-char-bound},\ the correction depends linearly on the curvature $H_M$ of the gerbe $\cG$ as 
\qq\nn
\tfrac{1}{\sfi}\,\d\log\,\Hol_{\pr_2^*\cG}\bigl(\phi(\Si)\bigr)=\int_\Si\phi^*\bigl(\imath_{\d\phi}\pr_2^*H_M\bigr)\,.
\qqq
This fact distinguishes extensions which preserve the form---and so also the interpretation---of the classic EL equations,\ to wit:\ those with $H_M=0$.\ The corresponding gerbes represent classes in $H^2(M,\uj)$,\ see:\ Thm.\,\ref{thm:W3-tors}.\ Hence,\ for $M$ with $H_2(M)=0$,\ we have $\cG\cong\cI_0$.\ Whenever the gerbe $\cG$ is flat,\ {\it i.e.},\ $\curv(\cG)=0$,\ we call $P^{t[\cG]}_\si$ the {\bf flatly $\cG$-coaugmented trivial Poisson} ({\bf fctP}) {\bf $\si$-model}.\ In particular,\ for $\cG\equiv\cI_0$,\ we obtain the {\bf prequantum trivial Poisson $\si$-model}.

\beg
The exact symplectic form $\om_{\rm lin}$ of the linear Poisson $\si$-model from Sec.\,\ref{sub:toy-lPSM} gives rise to the zero gerbe $\cI_0$,\ which translates into a requirement that the coaugmentation be \emph{non}-twisted $\txG\,\lx_{\txT_e\Ad^*}\ggt^*$-equivariant,\ and hence entirely decoupled from the Poisson $\si$-model.\ However,\ the object manifold $\ggt^*$ is contractible,\ and therefore the zero gerbe $\cG=\cI_0$  is the unique (up to isomorphism) \emph{flat} coaugmentation.\ Thus,\ the zero coaugmentation can be viewed as a consistent---albeit trivial---prototype of the general one discussed in the present section.
\eeg

\beg
On a compact,\ simple and simply connected Lie group $\txG$,\ we have $H_2(\txG)=0$,\ and so there is a unique (up to a 2-gauge transformation) physically distinguished flat coaugmentation of the classic Poisson--Lie $\si$-model from Ex.\,\ref{eg:Lu-Weinstein},\ to wit:\ $\cG\equiv\cI_0$.
\eeg

\brem\label{rem:Lorentz}
The coaugmentation term in the trivial Poisson DF character can be expressed in terms of the Beilinson--Deligne (BD) data $(B_i,A_{ij},g_{ijk})\in A^2(\cO^M)$ of $\cG$ associated with a trivialising cover $\cO^M\equiv\{O^M_i\}_{i\in I}$ of $M$ (see:\ Def.\,\ref{def:Deligne} and Prop.\,\ref{prop:gerbes-as-H2}) as follows:\ Given a field $\phi=(\id_\Si,X)\in\G(\xcF_{\rm triv})$,\ choose a tessellation $\triangle_\Si$ of $\Si$---with the set $P_{\triangle_\Si}$ of plaquettes,\ the set $E_{\triangle_\Si}$ of edges between them,\ and the set $V_{\triangle_\Si}$ of vertices at which the latter join---subordinate to $\cO^M$ through the choice of an index map $\imath\colo\triangle_\Si\to I$ as $X(\t)\subset O^M_{\iota(\tau)},\ \tau\in\triangle_\Si$.\ Given a vertex $v\in V_{\triangle_\Si}$ on the boundary of an edge $e\in E_{\triangle_\Si}$ from the boundary of a plaquette $p\in P_{\triangle_\Si}$,\ define $\vep_{pev}\in\{-1,1\}$ to be $+1$ if $e$,\ when endowed with the orientation induced from $p$,\ is in-coming at $v$,\ and $-1$ otherwise.\ The holonomy of $\pr_2^*\cG$ can be written in the form
\qq
\Hol_{\pr_2^*\cG}\bigl(\phi(\Si)\bigr)=\prod_{p\in P_{\triangle_\Si}}\bigg(\exp\bigg(\sfi\int_{X(p)}\,B_{\imath(p)}\bigg)\cdot\prod_{e\in \p p}\bigg(\exp\bigg(\sfi\int_{X(e)}\,A_{\imath(p)\imath(e)}\bigg)\cdot\prod_{v\in \p e}g_{\imath(p)\imath(e)\imath(v)}\bigl(X(v)\bigr)^{\vep_{pev}}\bigg)\bigg)\,.\cr \label{eq:holo-gen}
\qqq

\begin{figure}[hbt]
\centering
\includegraphics[width=0.15\textwidth]{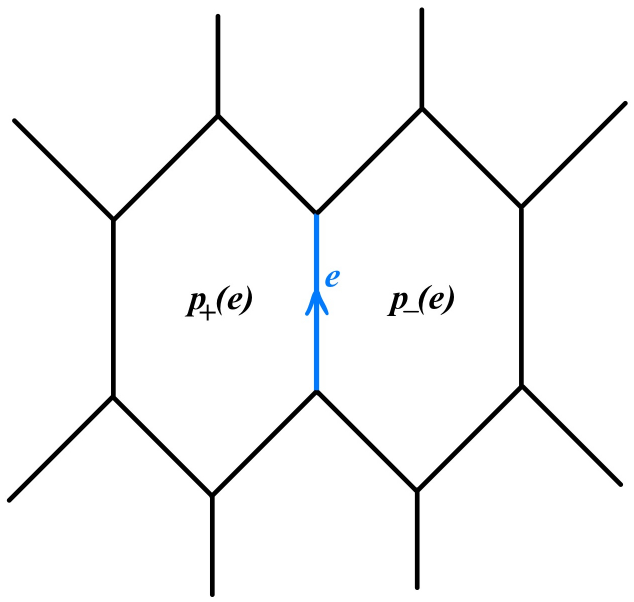}\hspace{3cm}\includegraphics[width=0.15\textwidth]{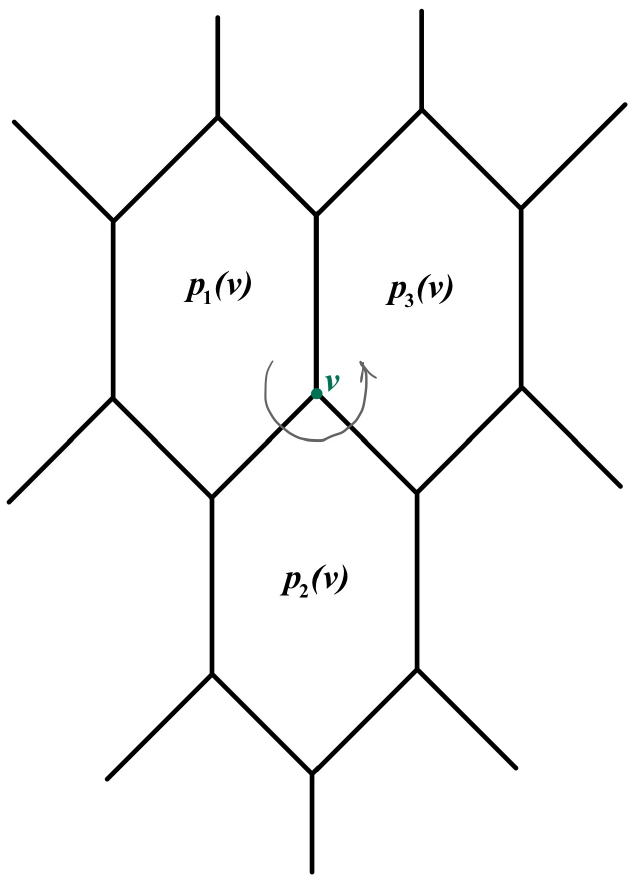}
\caption{Edges and vertices of the honeycomb tessellation.}\label{fig:honeycomb}
\end{figure}

The expression assumes a particularly neat form for trivalent tessellations,\ which can always be chosen,\ {\it e.g.},\ by taking the dual of a triangulation which refines a given tessellation.\ As every edge $e\in E_{\triangle_\Si}$ separates two plaquettes,\ a choice of orientation $v\in\G(\txT\Si\rstr_e)$ on $e$ identifies uniquely the plaquette $p_+(e)$ to the left of $e$ along $v$,\ and the plaquette $p_-(e)$ to the right of $e$ along $v$.\ Similarly,\ the orientation of $\Si$ (invariably assumed to be defined by the normal pointing towards the Reader in the figures attached) determines the 'anti-clockwise' cyclic order $(p_1(v),p_2(v),p_3(v))$ of plaquettes converging at the trivalent vertex $v\in V_{\triangle_\Si}$.\ The configuration is illustrated in Fig.\,\ref{fig:honeycomb}. 

\noindent The holonomy now factorises as
\qq
\Hol_{\pr_2^*\cG}\bigl(\phi(\Si)\bigr)=\prod_{p\in P_{\triangle_\Si}}\exp\bigg(\sfi\int_{X(p)}\,B_{\imath(p)}\bigg)\cdot\prod_{e\in E_{\triangle_\Si}}\exp\bigg(\sfi\int_{X(e)}\,A_{\imath(p_+(e))\imath(p_-(e))}\bigg)\cdot\prod_{v\in V_{\triangle_\Si}}g_{\imath(p_1(v))\imath(p_2(v))\imath(p_3(v))}\bigl(X(v)\bigr)\,.\cr \label{eq:holo}
\qqq
As a CS differential character of curvature $H_M$,\ the holonomy evaluates on 2-coboundaries $\p\Om\subset M$ as 
\qq\nn
\Hol_\cG\bigl(\p\Om\bigr)=\exp\left(\sfi\int_\Om H_M\right)\,,
\qqq
which explains its previously identified contribution to the EL equations. 

Taking into account the version of the holonomy for trivial gerbes,\ we arrive at the trivial Poisson DF character
\qq\nn
\cA_{\rm DF}^{tP\si[\cG]}[\phi,\Theta_{\rm triv}]=\exp\left(\sfi\int_{\phi(\Si)}\om[\Theta_{\rm triv}]\right)\cdot\Hol_{\pr_2^*\cG}\bigl(\phi(\Si)\bigr)\,.
\qqq
The formula simplifies considerably for flat coaugmentations.\ Indeed,\ the BD data of a flat gerbe can be reduced---through a suitable choice of the 2-gauge---to the form  $(B_i,A_{ij},g_{ijk})-D^{(1)}(\b_i,\a_{ij})=(0,0,c_{ijk})$ for some \v Cech 2-cocycle $(c_{ijk})\in\vZ^2(M,\uj)$,\ see:\ Thm.\,\ref{thm:W3-tors}.\ Using the invariance of the surface holonomy under 2-gauge transformations (see:\ Thm.\,\ref{thm:CH-BD}),\ we then obtain---for a trivalent tessellation,\ and in the notation of Rem.\,\ref{rem:Lorentz}---a factorised expression
\qq\nn
\cA_{\rm DF}^{tP\si[\cG]}[\phi,\Theta_{\rm triv}]=\exp\left(\sfi\int_{\varphi(\Si)}\om[\Theta]\right)\cdot\prod_{v\in V_{\triangle_\Si}}c_{\imath(p_1(v))\imath(p_2(v))\imath(p_3(v))}\,,
\qqq
in which the last term represents a constant-phase contribution of the coaugmenting gerbe $\cG$,\ \emph{encoding the latter's 2-gauge automorphisms and behaviour under (inner) Poisson automorphisms of $(M,\Pi)$ induced by large gauge transformations} from ${\rm Gau}_\om(\xcF_{\rm triv})$.
\erem

\subsection{Anomaly cancellation beyond prequantum bundles}\label{sub:killanom-beyond-preq}

The special status of flat coaugmentations,\ and the appearance of the \emph{trivial} ones $\cG\equiv\cI_0$ among them in physically relevant circumstances motivate a more detailed investigation of their structure,\ and a search for a direct relation with the scenario of the vanishing of the anomaly in the presence of a geometrisation of the symplectic structure $\om$,\ encountered at the end of Sec.\,\ref{sub:anom-kill-first}.\ We shall carry them out in the convenient formalism of the \v Cech--Deligne--Dupont (\v CDD) cohomology (see:\ Def.\,\ref{def:CDD}) over the simplicial nerve $N_\bullet(\grpd{\xcG}{M})\equiv\xcG_\bullet$ of the structure Lie groupoid (see:\ Ex.\,\ref{eg:face-nerve-LieGrpd}),\ with $\xcG_0\equiv M$,\ $\xcG_1=\xcG$,\ and $\xcG_n=\xcG\fibx{s\,}{t}\xcG\fibx{s\,}{t}\cdots\fibx{s\,}{t}\xcG \subset\xcG^{\x n},\ n>1$.\ The relevant \v CDD cochain bicomplex for the Deligne complex(es) $\xcD^\bullet(2;\xcG_n),\ n\in\bN$ of sheaves from Def.\,\ref{def:Deligne} is associated with a \emph{simplicial} open cover $\{\cO_{\xcG_n}\}_{n\in\bN}$ of this simplicial manifold (see:\ Def.\,\ref{def:simpl-cover}),\ whose existence follows from Thm.\,\ref{thm:simpl-ref}.\ In this formalism,\ the pair $(\cG,\Upsilon)$ is represented by local data (an incomplete 2-cochain) $(b,p)\in A^2(\cO_{\xcG_0})\oplus A^1(\cO_{\xcG_1})$ satisfying the hypercohomological identity
\qq\label{eq:CDD-tPcoaug}
\unl\D{}^{(2)}_{(0)}b+D^{(1)}_{(1)}p=\check\om\,,\qquad\qquad\check\om\equiv(\om\rstr_{O^{(1)}_{i_1}},0,1)\,.
\qqq
For $\cG=\cI_0$ \emph{in the trivial 2-gauge} $b=(0,0,1)$---which we \emph{fix} hencefort---the identity reduces to the simple one
\qq\nn
D^{(1)}_{(1)}p=\check\om\,.
\qqq
This identifies the 1-cochain $p=(\theta_{i_1},g_{i_1 j_1})\in A^1(\cO_{\xcG_1})$ on $\xcG_1$---through the Clutching Theorem for fibre bundles---as local BD data of a principal $\bC^\x$-bundle 
\qq\label{eq:preq-om-coaug}
P_\om\cong\bigsqcup_{i_1\in   I_1}\,\bigl(O^{(1)}_{i_1}\x\bC^\x\bigr)/\sim_{g_{\cdot\cdot}}\too\xcG_1\,,\qquad(g,z,j_1)\sim
\bigl(g,g_{i_1 j_1}(g)\cdot z,i_1\bigr)
\qqq
with the unitarily reduced structure group $\uj\swarrow\bC^\x$.\ Note,\ in particular,\ that the existence of $P_\om$ prerequires the integrality of periods of the symplectic form.\ The bundle then carries a (principal) connection 1-form $\cA_\om\in\Om(P_\om)$ with local presentations
\qq\label{eq:preq-conn}
\t_{i_1}^{-1*}\cA_\om(g,z)=\tfrac{\sfd z}{\sfi z}+\theta_{i_1}(g)\,,
\qqq
associated with local trivialisations $\t_{i_1}\colo\pi_{P_\om}^{-1}(O^{(1)}_{i_1})\xrightarrow{\cong}O^{(1)}_{i_1}\x\bC^\x,\ i_1\in I_1$.\ The local data $\theta_{i_1}$ satisfy the identities $\sfd\theta_{i_1}=\om\rstr_{O^{(1)}_{i_1}}$,\ and so they play the r\^ole of symplectic potentials for $\om$.\ The bundle $P_\om$,\ which is none other than the one stumbled upon in Sec.\,\ref{sub:anom-kill-first} and captured by diagram \eqref{eq:preq-fram-om},\ acquires the interpretation of the frame bundle of the prequantum (line) bundle $L_\om\cong P_\om\x_{\bC^\x}\bC$ (with (Koszul) connection $\nabla_\om$ induced from $\cA_\om$) over the symplectic arrow manifold $(\xcG,\om)$.\ It is a line bundle \emph{twisted} by the gerbe ({\it i.e.},\ it has an additional gauge symmetry inherited from it),\ but as long as we keep the trivial 2-gauge $b=(0,0,1)$,\ this non-standard status does not manifest.

On arrow manifolds with $H^2(\xcG,\uj)\neq 1$,\ there are choices of the flat coaugmenting gerbe $\cG$ other than $\cG=\cI_0$,\ which generalise the above structure in a cohomologically controllable manner.\ One way to realise this additional freedom is to keep the assumption \eqref{eq:PeromZ} of the prequantisability of $\om$---of independent significance emphasised in works \cite{Weinstein:1991,Crainic:2004,Bonechi:2005mw,Crainic:2007,Laurent-Gengoux:2007,Hawkins:2008} on the quantisation of Poisson manifolds---and use it to factorise the datum $\Upsilon$ of the tcP $\si$-model.\ Under the equivalence of categories $\Igt\colo\gt{Lin}_\om(\xcG)\cong\gt{Iso}_{\bgrb_\nabla(\xcG)}(\cI_0,\cI_\om)$ from \cite[Sec.\,3]{Waldorf:2007mm},\ the prequantum bundle $L_\om$ geometrising $\om$ gives rise to a gerbe 1-isomorphism $\Igt(L_\om)\colo\cI_0\xrightarrow{\cong}\cI_\om$,\ and so---altogether---we have a (flat) gerbe $\cG$ over $M$ in the kernel of $\unl\D{}^{(2)}_{(0)}$,\ by which we mean that there exists a 1-isomorphism
\qq\label{eq:untw-equiv-fcG}
\unl\Upsilon\colo t^*\cG\xrightarrow{\ \cong\ }s^*\cG\,,
\qqq
and---independently---the said prequantum bundle $L_\om$ over $(\xcG,\om)$.\ It is upon twisting $L_\om$ by $\unl\Upsilon$ that we recover the desired gerbe bimodule of curvature $\om$ (see:\ \cite{Fuchs:2007fw,Runkel:2008gr})---and so \emph{not} just a line bundle---as
\qq\label{eq:preq-and-prequivG}
\bigl(\id_{s^*\cG}\ox\Igt(L_\om)\bigr)\circ\unl\Upsilon\colo t^*\cG\xrightarrow{\ \cong\ }s^*\cG\equiv s^*\cG\ox\cI_0\xrightarrow{\ \cong\ }s^*\cG\ox\cI_\om\,.
\qqq
The existence of a 1-isomorphism $\unl\Ups$ is a necessary condition for a strict $\xcG$-equivariance of the coaugmenting gerbe,\ and---as such---it is a prerequisite of the gerbe's direct descent---in the sense of $\xcG$-equivariant cohomology---to the classifying space $\txB\xcG=\Vert N_\bullet(\grpd{\xcG}{M})\Vert$,\ understood as the homotopy model of the orbispace $M//\xcG$.\ Thus,\ a generic gerbe 1-isomorphism $\Upsilon$ is to be be regarded as a coherent---and physically sufficient---generalisation of the two rigid higher-geometric structures over $\grpd{\xcG}{M}$,\ in which their rigidity is lessened.\medskip

We conclude the present section with a statement---taken directly from \cite{Suszek:2026col}---about the topological obstruction against a flat coaugmentation of the classic Poisson $\si$-model.\ Note that the existence of the pair $(\cG,\Upsilon)$ is equivalent---in the \v CDD formalism introduced earlier---to the existence of a BD 2-cocycle $b_0\in A^2(\cO_{\xcG_0})$ such that the class of the 2-cocycle $\check\om-\unl\D{}^{(2)}_{(0)}b_0$ in $\bH^2(\xcG,\xcD^{\bullet}(2;\xcG))$ vanishes (for $\cO_{\xcG_1}$ sufficiently fine,\ see:\ Def.\,\ref{def:Deligne}).\ However,\ both $b_0$ and $\check\om$ have vanishing curvatures,\ and so they define (for $\cO_{\xcG_\bullet}$ sufficiently fine) cochains $[b_0]\in\cW^3(\xcG_0;0)\cong H^2(\xcG_0,\uj)$ and $[\check\om]\in\cW^3(\xcG_1;0)\cong H^2(\xcG_1,\uj)$ from the complex $(H^2(\xcG_\bullet,\uj),[\unl{\D}{}^{(2)}_{(\bullet)}]\rstr)\equiv(W_\xcG^\bullet,\widetilde\d{}_\xcG^{(\bullet)})$ defined in Prop.\,\ref{prop:ind-coc}. Accordingly,\ the triviality of the class 
\qq\nn
[\check\om-\unl\D{}^{(2)}_{(0)}b_0]=[\check\om]-\widetilde\d{}_\xcG^{(0)}[b_0]
\qqq
acquires an equivalent interpretation of the triviality of the class $[[\check\om]]_{\widetilde\d{}_\xcG}\in\widetilde h{}^1(\xcG,\uj)$ of the multiplicative symplectic form in the induced vertical cohomology $	\widetilde h{}^\bullet(\xcG,\uj)=H^\bullet(W_\xcG^\bullet,\widetilde\d{}_\xcG^{(\bullet)})$ from Prop.\,\ref{prop:ind-coc}.\ Thus,\ $\widetilde h{}^1(\xcG,\uj)$ becomes identified as the first obstruction cohomology group in the present field-theoretic setting.

\section{Extensions to the twisted sector through equivariance and 2-stacky descent}\label{sec:coaug-PSM-def}

In the previous section,\ we embedded the classic Poisson $\si$-model in the higher-geometric and -cohomological framework of the dynamical PAG $\si$-model by reinterpreting the former's DF amplitude as the surface holonomy of an augmentation of a gerbe $\cG$ over $M$ for which there exists a distinguished bimodule $\Upsilon$.\ Such augmentations were shown---first in \cite{Gawedzki:2010rn,Gawedzki:2012fu} (for action groupoids),\ and then in \cite{Suszek:2026col} (for arbitrary Lie groupoids)---to effect,\ under additional assumptions,\ the gauging of the Lie-groupoidal symmetry $\grpd{\xcG}{M}$---implemented by $\bB_\om(\xcG)$---of the differential character $\Hol_\cG$ in the topological DF amplitude of the original PAG $\si$-model with target $(M,\txg_M,\cG)$.\ Here,\ the gauging is understood as a configurational descent of the field theory from the \v Cech nerve $N_\bullet\cO_{\xcF\t}$ (see:\ Ex.\,\ref{eg:Cech-nerve}) of the trivialisation cover $\cO_{\xcF\t}\equiv\{O_i\x M\}_{i\in I}$ of the Godement shadow $\xcF\equiv\xcP/\xcG$ of a principaloid $\xcG$-bundle $\xcP$ of arbitrary topology---on which the (co)augmented differential characters for local variants of \eqref{eq:aug-grb} (with $\Theta$ replaced by $\Theta_i$) are well-defined and enjoy gauge invariance---to the total space of the bundle itself,\ $\xcF\cong N_0\cO_{\xcF\t}/\sim_{t_*\b_{\cdot\cdot}}$.\ The descent models an effective reduction of the configuration bundle $M\searrow M//\xcG$,\ with a deep mathematical meaning provided by the (universal) Cartan--Borel construction---reviewed in \cite[Sec.\,2]{Suszek:2026col},\ after \cite{Tu:2020} and \cite{Arias:2011}---which puts $\xcF$ in the r\^ole of a $\Si$-point in the aforementioned homotopy model $\txB\xcG=\Vert N_\bullet(\grpd{\xcG}{M})\Vert$ of the (generically non-smooth) orbispace $M//\xcG$.\ In this picture,\ the implementation of the gauge{\em d} symmetry ${\rm Gau}_\om(\xcF)$ on the \emph{trivial} configuration bundle $\xcF=\xcF_{\rm triv}$ represents the identification of \emph{smooth} field configurations in $M$ related by a smoothly $\Si$-indexed family of symmetry transformations.\ The incorporation of \emph{nontrivial} shadow bundles,\ on the other hand,\ accounts for the appearance of the symmetry-twisted sector (see:\ \cite{Dixon:1985jw} for a string-theoretic prototype) in the thus obtained effective model of dynamics on $M//\xcG$.\ This shall be expounded and concretised through the construction of gauge-symmetry defects in the next section.\ In the present one,\ we employ the mechanism of 2-stacky descent for the (co)augmented differential character on $N_0\cO_{\xcF\t}$,\ worked out in all generality in \cite{Suszek:2026col},\ to extend the original formulation of the Poisson $\si$-model to \emph{arbitrary} symplectic shadows $\xcF$.\ In this manner,\ we promote the model to the rank of a fully fledged topological gauge field theory,\ with the---by now integrated to the Lie-groupoidal level---gauge symmetry ${\rm Gau}_\om(\xcF)$,\ corresponding to inner Poisson automorphisms of $(M,\Pi)$.\ As we proceed,\ we use freely the notation and results from App.\,\ref{app:gerbarium},\ alongside those introduced heretofore.\medskip

Let $\xcP\to\Si$ be an arbitrary \emph{symplectic} principaloid $(\xcG,\om)$-bundle with the Godement shadow $\xcF\to\Si$,\ the two coming with the respective realisations:\ $L$ and $t_*$ of the $\om$-holonomically reduced structure group $\bB(\xcG,\om)$,\ see:\ \eqref{eq:lagred}.\ Moreover,\ let $\Theta$ be a (compatible) connection 1-form on $\xcP$.\ Given an open cover $\{O_i\}_{i\in I}$ of the common base $\Si$ of the two bundles,\ coming with the respective trivialisations:\ $\xcP\t_i\colo\pi_\xcP^{-1}(O_i)\xrightarrow{\cong}O_i\x\xcG$ and  $\xcF\t_i\colo\pi_\xcF^{-1}(O_i)\xrightarrow{\cong}O_i\x M$,\ the previous section defines the point of departure of our subsequent analysis as the $I$-indexed family of trivial Poisson DF characters $\{\Hol_{\cG[\Theta_i]}\}_{i\in I}$ associated to the family of local {\bf connection-augmented gerbes}
\qq\label{eq:conn-aug-grb-loc}\qquad\qquad
\cG[\Theta_\cdot]\in\prod_{i\in I}\,\bgrb_\nabla\bigl(\pi_\xcF^{-1}(O_i)\bigr)_{(0)}\,,\qquad\qquad \cG[\Theta_i]\equiv\cG[\Theta_\cdot](i):=\xcF\t_i^*\bigl(\pr_2^*\cG\ox\cI_{\om[\Theta_i]}\bigr)\,,\quad i\in I
\qqq
over the manifold $\check{Y}_\xcF:=\bigsqcup_{i\in I}\,\pi_\xcF^{-1}(O_i)$.\ The goal of the analysis is the identification of natural conditions under which $\cG[\Theta_\cdot]$ descends---in a sense to be made precise in what follows---along the surjective submersion $\check{\pi}_\xcF\colo\check{Y}_\xcF\twoheadrightarrow\xcF\,,\ (f,i)\longmapsto f$ (see:\ Ex.\,\ref{eg:Cech-nerve}) to a gerbe
\qq\label{eq:desc-Poisson-grb}
\unl{\cG[\Theta]}\in\bgrb_\nabla(\xcF)\qquad\tx{such that}\qquad \unl{\cG[\Theta]}\rstr_{\pi_\xcF^{-1}(O_i)}\cong\cG[\Theta_i],\ i\in I\,,
\qqq
which then defines the ($\xcP$-){\bf gauged Poisson DF character}
\qq\label{eq:gau-PoissonDF-char}
\cA_{\rm DF}^{P\si[\cG]/\xcP}\colo\G(\xcF)\x{\rm Conn}(\xcP)\too\uj,\ (\phi,\Theta)\longmapsto\Hol_{\unl{\cG[\Theta]}}\bigl(\phi(\Si)\bigr)
\qqq
for arbitrary global sections of the shadow $\xcF$ and arbitrary (compatible) connections on $\xcP$.\ Once the conditions have been met,\ we may ask if they ensure the invariance of the character under arbitrary lagrangean gauge transformations $\,{\rm Gau}_\om(\xcP)$.

\subsection{The descent of differential characters}\label{sub:diff-char-gauge}

The concept of descent of the \emph{local} trivial Poisson DF characters $\{\Hol_{\cG[\Theta_i]}\}_{i\in I}$ to a global character on the total space of a shadow bundle of arbitrary topology formalises neatly as an instantiation of the general idea of reconstruction of a gerbe---viewed as a 0-cell of the bicategory (in the sense of B\'enabou \cite{Benabou:1994bi}) of gerbes with connective structure $\bgrb_\nabla(M)$---from its local models,\ specified over an open cover of its base $M$.\ The idea found its concretisation in the work of Stevenson \cite{Stevenson:2000wj},\ subsequently elaborated by Nikolaus and Schweigert in \cite{Nikolaus:2011ehg},\ through the identification of the 2-stack structure over the smooth site\footnote{The abstract notion is to be understood in the sense of Breen \cite{Breen:1994}.} furnished by bundle gerbes,\ in direct analogy with the pioneering reconstruction of the 1-stack of principal $\bC^\x$-bundles,\ due to Brylinski \cite{Brylinski:1993ab}.\ The fundamental result of that work,\ which we employ hereunder,\ is the following
\bethe\cite{Stevenson:2000wj}\label{thm:descyk}
Given $Y,X\in\Man$,\ and a surjective submersion $\pi\colo Y\twoheadrightarrow X$,\ there exists a canonical equivalance
\qq\nn
\bgrb_\nabla(X)\cong\gt{Desc}(\pi)
\qqq
between the bicategory of (bundle) gerbes from Thm.\,\ref{thm:bgrb-bicat},\ and the descent bicategory from Def.\,\ref{def:desc-bicat}.
\ethe 
\noindent In the setting of interest,\ this is to be read in conjunction with Def.\,\ref{def:Cech-desc},\ in which the descent bicategory for the specific surjective submersions engendered by open covers---such as $\check{\pi}_\xcF$---are explicited.\ Accordingly,\ we are led to look for an extension of the family $\cG[\Theta_\cdot]$ to a {\bf \v Cech descent datum} $(\cG[\Theta_\cdot],\Phi_{\cdot\cdot},\varphi_{\cdot\cdot\cdot})\in\gt{Desc}(\check{\pi}_\xcF)_{(0)}$ which consists of {\bf transition 1-isomorphisms} 
\qq\nn
\Phi_{ij}\colo\cG[\Theta_i]\rstr_{\pi_\xcF^{-1}(O_{ij})}\xrightarrow{\ \cong\ }\cG[\Theta_j]\rstr_{\pi_\xcF^{-1}(O_{ij})}
\qqq
and {\bf composition 2-isomorphisms} 
\qq\nn
\varphi_{ijk}\colo\Phi_{jk}\circ\Phi_{ij}\rstr_{\pi_\xcF^{-1}(O_{ijk})}\xLongrightarrow{\cong}\Phi_{ik}\rstr_{\pi_\xcF^{-1}(O_{ijk})}\,,
\qqq
the latter satisfying the {\bf 2-cocycle identity}
\qq\nn
\varphi_{ikl}\bullet\bigl(\id_{\Phi_{kl}}\circ\varphi_{ijk}\bigr)\rstr_{\pi_\xcF^{-1}(O_{ijkl})}=\varphi_{ijl}\bullet\bigl(\varphi_{jkl}\circ\id_{\Phi_{ij}}\bigr)\rstr_{\pi_\xcF^{-1}(O_{ijkl})}\,.
\qqq

Upon noting the structural similarity between---on one hand---the gluing law \eqref{eq:glue-Aij} for the gauge-field components of the local presentations $\Theta_i$ and $\Theta_j$ of the connection 1-form in the local gerbes $\cG[\Theta_i]$ and $\cG[\Theta_j]$,\ respectively,\ and---on the other hand---the (gauge-)transformation law \eqref{eq:gaugetrafo} for the gauge-field components of the local presentations $\Theta^\Phi_i$ and $\Theta_i$ of the connection 1-forms $\Theta^\Phi$ and $\Theta$ in the local gerbes $\cG[\Theta^\Phi_i]\equiv\cG[\Theta^{h_i}_i]$ and $\cG[\Theta_i]$,\ respectively,\ we conclude that the transition 1-isomorphisms can be induced from the bimodule $\Upsilon$ as
\qq\label{eq:Phij-desc}
\Phi_{ij}:=\xcF\t_j^*\bigl(\widetilde\b{}_{ij}^*\Upsilon\ox\Id_{\cI_{B_{ij}^*\om[\Theta_i]}}\bigr)\rstr_{\pi_\xcF^{-1}(O_{ij})}
\qqq
for $\widetilde\b{}_{ij}$ as in \eqref{eq:bij-tilde},\ and 
\qq\nn
B_{ij}:=(\pr_1,t\circ\widetilde\b_{ij})\,.
\qqq 
This follows immediately from the identity $\xcF\t_i\circ\xcF\t_j^{-1}\rstr_{O_{ij}\x M}=B_{ij}$.\ Thus,\ the geometrisation $\Upsilon$ of the BSS-cohomological relation \eqref{eq:BSS-1} (itself presupposing the existence of the geometrisation $\cG$ of $H_M$) provides us with the second component of the desired \v Cech descent datum.\ This is a direct indication of the central r\^ole played by geometrisations of the BSS extensions $(H_M,\om,0,0)$ of the multiplicative 2-form $\om$ in the descent.\ Such geometrisations are known as $\om$-twisted $\xcG$-equivariant structures on $\cG$,\ and they consist,\ each,\ of a 1-isomorphism \eqref{eq:Ups-first} over $\xcG_1$,\ and a 2-isomorphism over $\xcG_2$
\qq\label{eq:gamma-2-iso}
\g\colo\bigl(\pr_2^*\Upsilon\ox\Id_{\cI_{\pr_1^*\om}}\bigr)\circ\pr_1^*\Upsilon\overset{\cong}{\Longrightarrow}\txm^*\Upsilon\,,
\qqq
the latter satisfying the coherence condition \eqref{eq:gamma-coh} over $\xcG_3$,\ see:\ Def.\,\ref{def:Gequiv-grb}.\ The existence of the geometrisation conforms with the physical interpretation of gerbe bimodules---such as $\Upsilon$---and distinguished 2-isomorphisms---such as $\g$---over stratified simplicial string backgrounds---such as $N_\bullet(\grpd{\xcG}{M})$---as higher-geometric data---bi-branes and inter-bi-branes,\ respectively---carried by prequantisable defects implementing symmetries---such as the aforementioned rigid groupoidal symmetries---and more general state correspondences (prequantisable \emph{local} symplectomorphisms of spaces of states) in the dynamical $\si$-model.\ These were originally worked out in \cite{Fuchs:2007fw,Runkel:2008gr,Runkel:2009sp},\ and elaborated in \cite{Suszek:2011hg,Suszek:2012ddg,Suszek:2013} and,\ more recently,\ in \cite{Suszek:2022CSimpl}.\ In particular,\ the multiplicativity of $\om$ can be viewed as a specialisation of the Defect-Junction Identity for the bi-brane curvature,\ derived in \cite[App.\,A.2]{Runkel:2008gr} and elucidated in \cite[Sec.\,5]{Suszek:2011hg}.\ Taking into account the reasoning recapitulated above,\ we assume henceforth---as a prerequisite of the descent under con\-sid\-er\-a\-tion---the existence of an $\om$-twisted $\xcG$-equivariant structure $(\cG,\Upsilon,\g)$ on the co-augmenting gerbe $\cG$.\ Further,\ {\it ex post} justification of the assumption shall be provided by the detailed construction of the gauge-symmetry defect in Sec.\,\ref{sub :gau-sym-def-top}.

As anticipated,\ the 2-isomorphism $\g$ gives rise to the sought-after composition 2-isomorphisms:
\qq\label{eq:varphijk-desc}
&&\varphi_{ijk}:=\xcF\t_k^*\bigl(\widetilde\b{}_{ijk}^*\g\ox\id_{\Id_{\cI_{B_{ik}^*\om[\Theta_i]}}}\bigr)\rstr_{\pi_\xcF^{-1}(O_{ijk})}\colo\Phi_{jk}\circ\Phi_{ij}\rstr_{\pi_\xcF^{-1}(O_{ijk})}\xLongrightarrow{\ \cong\ }\Phi_{ik}\rstr_{\pi_\xcF^{-1}(O_{ijk})}\,,
\qqq
expressed in terms of the smooth maps
\qq\label{eq:bijk-tilde}
\widetilde\b{}_{ijk}:=\bigl(\widetilde\b{}_{ij}\circ B_{jk},\widetilde\b{}_{jk}\bigr)\rstr_{O_{ijk}\x M}\colo O_{ijk}\x M\too\xcG_2\,.
\qqq
These,\ together with the transition 1-isomorphisms $\Phi_{ij}$,\ provide a bicategorial identification of the local gerbes over intersections of their respective domains,\ whose coherence is ensured by that of $\g$,\ as captured by \eqref{eq:gamma-coh},\ see \cite[App.\,E]{Suszek:2026col} for a straightforward proof.\ In virtue of Thm.\,\ref{thm:descyk},\ the existence of the \v Cech descent datum $(\{\cG[\Theta_i]\}_{i\in I},$ $\{\Phi_{ij}\}_{i,j\in I},\{\varphi_{ijk}\}_{i,j,k\in I})$ gives us a gerbe $\unl{\cG[\Theta]}$ as in \eqref{eq:desc-Poisson-grb},\ to be referred to as the {\bf descended Poisson gerbe} henceforth.

\subsection{The geometry and cohomology behind the descent}\label{sub:geom-cohom-equiv}

The construct underlying the descent of the local Poisson DF characters is the complete $\om$-twisted $\xcG$-equivariant structure $(\cG,\Upsilon,\gamma)$.\ The reasoning laid out in Sec.\,\ref{sub:killanom-beyond-preq} suggests that the structure be regarded as an \emph{essential} generalisation of a prequantum bundle over the symplectic groupoid $\grpd{(\xcG,\om)}{(M,\Pi)}$.\ The detailed analysis below substantiates this idea.

The structure of interest is conveniently described in the framework of the previously introduced \v CDD cohomology from Def.\,\ref{def:CDD}:\ A $\xcG$-equivariant gerbe $(\cG,\Upsilon,\g)$ is represented by a \v CDD 2-cochain $(b,p,f)\in A^2(\cO_{\xcG_0})\oplus A^1(\cO_{\xcG_1})\oplus A^0(\cO_{\xcG_2})$ satisfying the $\om$-twisted 2-cocycle identity (see:\ \eqref{eq:CDD-tPcoaug})
\qq\nn
\nabla_{(2)}(b,p,f)=\bigl(0,\check\om,0,1\bigr)\,.
\qqq
In the special case of the trivial flat coaugmenting gerbe $\cG=\cI_0$ in the trivial 2-gauge $b=(0,0,1)$,\ we rewrite the identity as the conjunction of the following coupled relations:
\qq\nn
D^{(1)}_{(1)}p=\check\om\,,\qquad\qquad\unl\D{}^{(1)}_{(1)} p=D^{(0)}_{(2)}f\,,\qquad\qquad\unl\D{}^{(0)}_{(2)}f=(1)\,,
\qqq
of which the first is a BD presentation of the prequantum bundle $L_\om$ introduced before.\ The second relation now translates into a statement of existence of a distinguished connection-preserving line-bundle isomorphism over $\xcG_2$
\qq\nn
\psi_\om\colo\pr_1^*L_\om\ox\pr_2^*L_\om\xrightarrow{\ \cong\ }\txm^*L_\om\,,
\qqq
and the last one encodes the latter's coherence (or associativity),\ expressed by the commutativity of the following diagram of connection-preserving line-bundle isomorphism over $\xcG_3$:
\qq\nn
\alxydim{@C=.75cm@R=1.cm}{ & \pr_1^*L_\om\ox\pr_2^*L_\om\ox\pr_3^*L_\om \ar[ld]_{\pr_{1,2}^*\psi_\om\ox\id_{\pr_3^*L_\om}\quad} \ar[rd]^{\ \id_{\pr_3^*L_\om}\ox\pr_{2,3}^*\psi_\om} & \\ (\txm\circ\pr_{1,2})^*L_\om\ox\pr_3^*L_\om \ar[rd]_{(\txm\x\id_\xcG)^*\psi_\om\quad} & & \pr_1^*L_\om\ox(\txm\circ\pr_{2,3})^*L_\om \ar[ld]^{\quad(\id_\xcG\x\txm)^*\psi_\om} \\ & \bigl(\txm\circ(\txm\x\id_\xcG)\bigr)^*L_\om\equiv\bigl(\txm\circ(\id_\xcG\x\txm)\bigr)^*L_\om& }\,.
\qqq
We thus arrive at a {\bf \emph{multiplicative} prequantum bundle} $((L_\om,\nabla_\om),\psi_\om)$ over $(\xcG,\om)$.\ Such and similar higher-geometric structures over symplectic Lie groupoids have attracted considerable interest in the past in the context of quantisation of Poisson structures.\ They were,\ {\it i.a.},\ discussed in \cite{Weinstein:1991,Crainic:2004,Crainic:2007,Laurent-Gengoux:2007,Hawkins:2008}.

It is instructive to see how the introduction of an \emph{arbitrary} flat gerbe $\cG$ lessens the rigidity and thereby enhances the scope of available completions of the classic Poisson $\si$-model whenever $H^2(\xcG,\uj)\neq 1$.\ Having already explained,\ in Sec.\,\ref{sub:killanom-beyond-preq},\ the way in which the 1-isomorphism $\Upsilon$ simultaneously generalises the structure of a prequantum bundle $L_\om$ and that of an untwisted (pre-)$\xcG$-equivariant gerbe,\ we shall now refine our analysis by taking as the point of departure the assumption of the prequantisability of $\om$.\ In other words,\ we consider given a prequantum bundle with BD data $p_\om\in A^1(\cO_{\xcG_1})$,\ subject to the relation
\qq\nn
D^{(1)}_{(1)}p_\om=\check\om\,.
\qqq
Denote $\unl p:=p-p_\om$\ and then rewrite the relations expressing the $\om$-twisted cocyclicity of $(b,p,f)$ in the form
\qq\nn
D^{(2)}_{(0)}b=0\,,\qquad\qquad
\unl\D{}^{(2)}_{(0)}b=-D^{(1)}_{(1)}\unl p\,,\qquad\qquad\unl\D{}^{(1)}_{(1)} (\unl p+p_\om)=D^{(0)}_{(2)}f\,,\qquad\qquad\unl\D{}^{(0)}_{(2)}f=(1)\,,
\qqq
which we may reinterpret as follows:\ The prequantum bundle $L_\om$ represented by $p_\om$ is \emph{not} multiplicative in general.\ It is only the gerbe bimodule $(\id_{s^*\cG}\ox\Igt(L_\om))\circ\unl\Upsilon$ of curvature $\om$ from \eqref{eq:preq-and-prequivG},\ encoded in the 1-cochain $\unl p+p_\om$,\ that exhibits (coherent) multiplicativity.\ Thus,\ not only does the generalisation affect the property of the prequantum bundle (its multiplicativity),\ but it also generalises the geometric category in which that property is internalised.\medskip

The existence of a fully fledged $\om$-twisted $\xcG$-equivariant gerbe $(\cG,\Upsilon,g)$ may be topologically obstructed---after \cite{Gawedzki:2010rn},\ we refer to the obstruction as the {\bf gauge anomaly}.\ Its top component was identified in Sec.\,\ref{sub:killanom-beyond-preq}.\ If the latter component vanishes,\ {\it i.e.},\ if the pair $(\cG,\Upsilon)$ exists,\ with the 1-isomorphism $\Upsilon$ represented by $p\in A^1(\cO_{\xcG_1})$,\ then the obstruction to the existence of the 2-isomorphism $\g$ is captured by the induced vertical cohomology $\widehat h{}^\bullet(\xcG,\uj)=H^\bullet(P_\xcG^\bullet,\widehat\d{}_\xcG^{(\bullet)})$ of the subcomplex $(P_\xcG^\bullet,\widehat\d{}_\xcG^{(\bullet)})\equiv(H^1(\xcG_\bullet,\uj),[\unl{\D}{}^{(1)}_{(\bullet)}])$ in the \v CDD bicomplex from Prop.\,\ref{prop:ind-coc}.\ Indeed,\ from relation \eqref{eq:CDD-tPcoaug},\ documenting the existence of $\Ups$,\ we read off a class 
\qq\nn
[\unl\D{}^{(1)}_{(1)}p]\in H^1\bigl(\xcG_2,\uj\bigr)\equiv P_\xcG^2\,,
\qqq
whose elements represent the 1-isomorphism $(\txm^*\Upsilon^{-1}\ox\Id)\circ(\pr_2^*\Upsilon\ox\Id)\circ\pr_1^*\Upsilon$.\ The existence of $\g$ means that---for $\cO_{\xcG_\bullet}$ sufficiently fine---
\qq\label{eq:gam-BD}
\unl\D{}^{(1)}_{(1)}p=-D^{(0)}_{(2)}f_\g
\qqq
for some 0-cochain $f_\g\in A^0(\cO_{\xcG_2})$.\ Hence,\ it is that class $[\unl\D{}^{(1)}_{(1)}p_\Ups]$ that quantifies the obstruction to the existence of $\g$.\ By construction,\ the class is a 2-cocycle,
\qq\nn
\widehat\d{}_\xcG^{(2)}[\unl\D{}^{(1)}_{(1)}p]\equiv[\unl{\D}{}^{(1)}_{(2)}\unl\D{}^{(1)}_{(1)}p]=0
\qqq
in the induced subcomplex $(P_\xcG^\bullet,\widehat\d{}_\xcG^{(\bullet)})$,\ and the freedom of redefining the 1-cochain $p$ as $p\mapsto p+q,\ q\in\ker\,D^{(1)}_{(1)}$ in conformity with the original relation \eqref{eq:CDD-tPcoaug} translates into a freedom of shifting the class $[\unl\D{}^{(1)}_{(1)}p]$ by an arbitrary coboundary $[\unl\D{}^{(1)}_{(1)}][q]\equiv\widehat\d{}_\xcG^{(1)}[q]$,\ with $[q]\in H^1(\xcG_1,\uj)\equiv P_\xcG^1$.\ Thus,\ altogether,\ the obstruction is---for $\cO_{\xcG_\bullet}$ sufficiently fine---given by the class
\qq\nn
[[\unl\D{}^{(1)}_{(1)}p]]_{\widehat\d{}_\xcG}\in \widehat h{}^2(\xcG,\uj)\,.
\qqq
Finally,\ if the triple $(\cG,\Upsilon,\g)$ exists,\ with the 2-isomorphism $\g$ represented by $f\in A^0(\cO_{\xcG_2})$,\ then the obstruction to the coherence of the 2-isomorphism $\g$ belongs to the induced vertical cohomology  $h^\bullet(\xcG,\uj)=H^\bullet(\G_\xcG^\bullet,\d_\xcG^{(\bullet)})$ of the subcomplex $(\G_\xcG^\bullet,\d_\xcG^{(\bullet)})\equiv(H^0\bigl(\xcG_\bullet,\uj),[\unl{\D}{}^{(0)}_{(\bullet)}])$ in the \v CDD bicomplex.\ In order to see this,\ note that the obstruction is measured by the cochain $\unl\D{}^{(0)}_{(2)}f$,\ and the latter satisfies
\qq\nn
D^{(0)}_{(3)}\unl\D{}^{(0)}_{(2)}f=\unl\D{}^{(1)}_{(2)}D^{(0)}_{(2)}f=-\unl\D{}^{(1)}_{(2)}\unl\D{}^{(1)}_{(1)}p=0\,.
\qqq
However,\ in the light of Prop.\,\ref{prop:BD-vs-Huj},\ $\ker\,D^{(0)}_{(n)}\equiv\bH^0(\cO_{\xcG_n},\xcD^{\bullet}(2;\xcG_n))=\bH^0(\xcG_n,\xcD^{\bullet}(2;\xcG_n))=H^0(\xcG_n,\uj)$ for $\cO_{\xcG_\bullet}$ sufficiently fine,\ and so $\unl\D{}^{(0)}_{(2)}f_\g\in\G^3_\xcG$.\ Moreover,\ $\d_\xcG^{(3)}\unl\D{}^{(0)}_{(2)}f\equiv\unl\D{}^{(0)}_{(3)}\unl\D{}^{(0)}_{(2)}f=0$,\ that is $\unl\D{}^{(0)}_{(2)}f$ is trivially a 3-cocycle in $(\G_\xcG^\bullet,\d_\xcG^{(\bullet)})$.\ At this stage,\ it remains to take into account the residual ambiguity in its definition built into the cohomological relation \eqref{eq:gam-BD} that documents the existence of $\g$.\ Changing the representative of $p$ within its cohomology class (or,\ equivalently,\ changing the representative of the 2-isomorphism class $[\Upsilon]$) as $p\mapsto p+D^{(0)}_{(1)}h,\ h\in A^0(\cO_{\xcG_1})$ produces a correction $f\mapsto f-\unl\D{}^{(0)}_{(1)} h$.\ On top of it,\ the 2-isomorphism data $f$ exhibit an intrinsic freedom of redefinition:\ $f\mapsto f+k,\ k\in\ker\,D^{(0)}_{(2)}$.\ We conclude that $f$ is determined up to (and so can be corrected by) a replacement
\qq\nn
f\mapsto f-\unl\D{}^{(0)}_{(1)} h+k,\ (h,k)\in A^0(\cO_{\xcG_1})\x\ker\,D^{(0)}_{(2)}\,,
\qqq
and so $\unl\D{}^{(0)}_{(2)}f\in\ker\,\d_\xcG^{(3)}$ is determined up to an arbitrary correction $\unl\D{}^{(0)}_{(2)}k$ with $k\in\ker\,D^{(0)}_{(2)}\equiv\G^2_\xcG$.\ Therefore,\ for $\cO_{\xcG_\bullet}$ sufficiently fine,\ the last obstruction is quantified by the class
\qq\nn
[\unl\D{}^{(0)}_{(2)}f]_{\d_\xcG}\in h^3\bigl(\xcG,\uj\bigr)\,.
\qqq
Note that the gauge anomaly may manifest only on Lie groupoids with a topology characterised by a non-vanishing degree-3 cohomology group $H^2(\xcG_1,\uj)\oplus H^1(\xcG_2,\uj)\oplus H^0(\xcG_3,\uj)$.

\brem
It is perhaps worth mentioning---for the sake of a more direct conceptualisation---that the trivial flat coaugmentation (using $\cG=\cI_0$),\ available for $\om$ with ${\rm Per}(\om)\subset 2\pi\bZ$,\ is,\ in essence,\ a case study of the higher-geometric construction of a bundle gerbe,\ originally worked out by Murray in \cite{Murray:1994db}.\ Indeed,\ we are dealing here with a trivial gerbe $\cI_B$ over the total space $Y\xcF\equiv\check{Y}_\xcF$ of the surjective submersion $\pi_{Y\xcF}\equiv\pi_{\check{Y}_\xcF}\colo Y\xcF\twoheadrightarrow\xcF$,\ with local components $\cI_B\rstr_{\pi_\xcF^{-1}(O_i)}=\cI_{\xcF\t_i^*\om[\Theta_i]},\ i\in I$.\ For the trivial gerbe to represent a gerbe on the base $\xcF$ of the submersion,\ there must exist a principal $\bC^\x$-bundle $\pi_L\colo L\to Y\xcF\x_\xcF Y\xcF$ over the $\xcF$-fibred square of $Y\xcF$,\ with a (principal) connection (form) $\cA_L\in\Om^1(L)$ of curvature $\sfd\cA_L=\pi_L^*(\pr_2^*-\pr_1^*)B$---in this r\^ole,\ we find the bundle with local components of the total space $P\rstr_{\pi_\xcF^{-1}(O_{ij})}=\xcF\t_j^*\widetilde\b{}_{ij}^*P_\om$ given by pullbacks of the principal $\bC^\x$-bundle $P_\om$ from \eqref{eq:preq-om-coaug},\ in conformity with \eqref{eq:Phij-desc},\ and with local components of the connection induced as the corresponding pullbacks of the principal connection $\cA_\om$ from \eqref{eq:preq-conn}.\ The $\xcF$-fibrewise groupoid structure on (fibres of) $P$,\ covering that of the submersion groupoid ${\rm Pair}_{\pi_{Y\xcF}}(Y\xcF)$ (see:\ Ex.\,\ref{eg:subm-grpd}),\ is now implemented by the connection-preserving principal-$\bC^\x$-bundle isomorphism $\mu_L\colo\pr_{1,2}^*L\ox\pr_{2,3}^*L\cong\pr_{1,3}^*L$ over $Y\xcF\x_\xcF Y\xcF\x_\xcF Y\xcF$ with local components $\mu_L\rstr_{\pi_\xcF^{-1}(O_{ijk})}=\xcF\t_k^*\widetilde\b{}_{ijk}^*\psi_\om$,\ in conformity with \eqref{eq:varphijk-desc}.\ Altogether,\ then,\ we arrive at Murray's (descended) bundle gerbe $(Y\xcF,B,L,\cA_L,\mu_L)$.

Clearly,\ the existence of more general coaugmentations (including non-trivial flat ones) reflects the existence of an additional {\it modulus} in this most elementary scheme.
\erem

\subsection{The coaugmented Poisson $\si$-model,\ and its gauge symmetry}\label{sub:coaug-PSM-def}

The descent of the local Poisson DF characters to the total space of the shadow $\xcF$ of an arbitrary symplectic principaloid $(\xcG,\om)$-bundle $\xcP$,\ whenever unobstructed,\ 
realises the first goal set at the beginning of the present section:\ It enables us to define a topological field theory
\qq\label{eq:coaug-PSM}
\bigl(\Si,(\xcP,\Theta)\xrightarrow{\ \xcD\ }\xcF,\bigl(\grpd{(\xcG,\om)}{(M,\Pi)},(\cG,\Upsilon,\g)\bigr),\cA_{\rm DF}^{P\si[\cG]/\xcP}\bigr)\equiv P_\si\,,
\qqq
dubbed the {\bf coaugmented Poisson $\si$-model} in \cite{Suszek:2026col},\ whose configuration bundle---the principal-$\xcG$-bundle object $\xcP\xrightarrow{\xcD}\xcF$ in $\Bun(\Si)$ with (compatible) connection $\Theta$,\ now no longer assumed trivial---carries the fully fledged $\om$-twisted $\xcG$-equivariant structure over its typical fibre $\xcG\xrightarrow{t}M$,\ and whose DF amplitude is given by the $\xcP$-gauged Poisson DF character \eqref{eq:gau-PoissonDF-char} for the descended Poisson gerbe $\unl{\cG[\Theta]}$.\ In the remainder of the present section,\ we shall convince ourselves that the cohomological mechanisms behind the descent---the $\om$-twisted $\xcG$-equivariance and the $\om$-holonomic ({\it i.e.},\ lagrangean) reduction---automatically ensure the attainment of the second goal,\ {\it i.e.},\ that the above topological field theory has a non-anomalous lagrangean gauge symmetry ${\rm Gau}_\om(\xcP)$.

First of all,\ let us clarify the very concept of gauge symmetry under the present circumstances.\ The configuration bundle $(\xcP,\Theta)\xrightarrow{\xcD}\xcF$ of the field theory under consideration carries the action of the group ${\rm Gau}_\om(\xcP)\ni\Phi$ of vertical $\xcG$-equivariant automorphisms of the symplectic principaloid $(\xcG,\om)$-bundle $\xcP$,\ locally modelled on $\bB(\xcG,\om)^\Si$.\ Due to its verticality,\ the action induces one on $\G(\xcF)$,\ and so gauge-transforming the first argument of the DF amplitude $\cA_{\rm DF}^{P\si[\cG]}$ is completely straightforward.\ The transformation of the connection 1-form $\Theta$,\ on the other hand,\ is subtler as the latter enters the action functional implicitly through the definition of the descended Poisson gerbe $\unl{\cG[\Theta]}$.\ Clearly,\ we need to understand the descent to $\xcF$ of the connection-augmented gerbes
\qq\nn\qquad\qquad
\cG[\Theta^\Phi_\cdot]\in\prod_{i\in I}\,\bgrb_\nabla\bigl(\pi_\xcF^{-1}(O_i)\bigr)_{(0)}\,,\qquad\qquad \cG[\Theta^\Phi_i]\equiv\cG[\Theta^\Phi_\cdot](i):=\xcF\t_i^*\bigl(\pr_2^*\cG\ox\cI_{\om[\Theta^\Phi_i]}\bigr)\,,\quad i\in I\,,
\qqq 
and subsequently verify the existence of a 1-isomorphism
\qq\label{eq:gauge-inv-descgrb}
\xcF_*(\Phi)^*\unl{\cG[\Theta^\Phi]}\cong\unl{\cG[\Theta]}
\qqq
for the descended Poisson gerbe $\unl{\cG[\Theta^\Phi]}$ such that $\unl{\cG[\Theta^\Phi]}\rstr_{\pi_\xcF^{-1}(O_i)}\cong\cG[\Theta^\Phi_i],\ i\in I$.\ Indeed,\ in virtue of Thm.\,\ref{thm:CH-BD},\ the above will imply the desired gauge invariance of the $\xcP$-gauged Poisson DF character,
\qq\nn
\Hol_{\unl{\cG[\Theta^\Phi]}}\bigl(\phi^\Phi(\Si)\bigr)=\Hol_{\unl{\cG[\Theta]}}\bigl(\phi(\Si)\bigr)\,.
\qqq

We begin by noting that due to the strict 2-functoriality of the gerbe-theoretic pullback (see:\ \cite{Waldorf:2007mm}) and the verticality of gauge transformations,\ the gauge-transformed gerbe $\xcF_*(\Phi)^*\unl{\cG[\Theta^\Phi]}$ is descended,\ in the previously discussed sense,\ from the \v Cech descent datum $(\xcF_*(\Phi)^*\cG[\Theta^\Phi_\cdot],\xcF_*(\Phi)^*\Phi^\Phi_{\cdot\cdot},\xcF_*(\Phi)^*\varphi^\Phi_{\cdot\cdot\cdot})\in\gt{Desc}(\check{\pi}_\xcF)_{(0)}$ obtained through pullback from the \v Cech descent datum $(\cG[\Theta^\Phi_\cdot],\Phi^\Phi_{\cdot\cdot},\varphi^\Phi_{\cdot\cdot\cdot})\in\gt{Desc}(\check{\pi}_\xcF)_{(0)}$,\ with the latter canonically induced from the $\om$-twisted $\xcG$-equivariant gerbe $(\cG,\Ups,\g)$ using---as before---the transition 1-cocycle of the gauge-transformed bundle $\Phi(\xcP)$.\ But that 1-cocycle is identical with the original one (see:\ \eqref{eq:gauge-glue}),\ and so we find---in the previously introduced notation---local gerbes 
\qq\nn
\cG[\Theta^\Phi_i]=\xcF\t_i^*\bigl(\pr_2^*\cG\ox\cI_{\om[\Theta^\Phi_i]}\bigr)\,,
\qqq
transition 1-isomorphisms
\qq\nn
\Phi^\Phi_{ij}=\xcF\t_j^*\bigl(\widetilde\b{}_{ij}^*\Upsilon\ox\Id_{\cI_{B_{ij}^*\om[\Theta^\Phi_i]}}\bigr)\rstr_{\pi_\xcF^{-1}(O_{ij})}\,,
\qqq
and composition 2-isomorphisms
\qq\nn
\varphi^\Phi_{ijk}=\xcF\t_k^*\bigl(\widetilde\b{}_{ijk}^*\g\ox\id_{\Id_{\cI_{B_{ik}^*\om[\Theta^\Phi_i]}}}\bigr)\rstr_{\pi_\xcF^{-1}(O_{ijk})}\,.
\qqq
In the light of Thm.\,\ref{thm:descyk},\ our task,\ at this stage,\ boils down to establishing a coherent 1-isomorphism (see:\ Def.\,\ref{def:Cech-desc})
\qq\nn
\bigl(\Psi_\cdot,\psi_{\cdot\cdot}\bigr)\colo\bigl(\xcF_*(\Phi)^*\cG[\Theta^\Phi_\cdot],\xcF_*(\Phi)^*\Phi^\Phi_{\cdot\cdot},\xcF_*(\Phi)^*\varphi^\Phi_{\cdot\cdot\cdot}\bigr)\xrightarrow{\ \cong\ }\bigl(\cG[\Theta_\cdot],\Phi_{\cdot\cdot},\varphi_{\cdot\cdot\cdot}\bigr)\,.
\qqq
It is not hard to check\footnote{See:\ \cite[App.\,E]{Suszek:2026col} for a lengthy but straightforward proof,\ invoking no more than identity \eqref{eq:augs-glue} and the coherence of $\g$.} that the 1-isomorphisms
\qq\nn
\Psi_i:=\xcF\t_i^*\bigl(\widetilde h{}_i^*\Upsilon\ox\Id_{\cI_{H_i^*\om[\Theta_i^\Phi]}}\bigr)
\qqq
(to be compared directly to the transition 1-isomorphisms \eqref{eq:Phij-desc}),\ together with the 2-isomorphisms
\qq\nn
\psi_{ij}:=\xcF\t_j^*\bigl(\widetilde{h\b}_{ij}^*\g^{-1}\bullet\widetilde{\b h}_{ij}^*\g\ox\id_{\Id_{\cI_{H_j^*B_{ij}^*\om[\Theta^\Phi_i]}}}\bigr)\,,
\qqq
defined in terms of smooth maps
\qq\nn
\widetilde{\b h}_{ij}:=\bigl(\widetilde\b{}_{ij}\circ H_j,\widetilde h{}_j\bigr)\rstr_{O_{ij}\x M}\colo O_{ij}\x M\too\xcG_2\,,\qquad\qquad\widetilde{h\b}_{ij}:=\bigl(\widetilde h{}_i\circ B_{ij},\widetilde\b{}_{ij}\bigr)\rstr_{O_{ij}\x M}\colo O_{ij}\x M\too\xcG_2\,,
\qqq
with $\widetilde h{}_i$ as before and
\qq\nn
H_i:=\xcF\t_i\circ\xcF_*(\Phi)\circ\xcF\t_i^{-1}\equiv\bigl( \pr_1,t\circ\widetilde h{}_i\bigr)\,,
\qqq
do satisfy the desired relations.\ This establishes the anticipated gauge symmetry of the coaugmented Poisson $\si$-model,
\qq\nn
\cA_{\rm DF}^{P\si[\cG]/\xcP}[\xcF_*(\Phi)\circ\phi,\Theta^\Phi]=\cA_{\rm DF}^{P\si[\cG]/\xcP}[\phi,\Theta]\,.
\qqq

\beg\label{eg:Lu-Weinstein}
Consider the Lu--Weinstein groupoid $(\xcL\xcW,\om_{\xcL\xcW})$ from Ex.\,\ref{ex:Lu-Weinstein}.\ We may conceive symplectic principaloid $(\xcL\xcW,\om_{\xcL\xcW})$-bundles $\xcP(\xcL\xcW,\om_{\xcL\xcW})$ over a given closed compact orientable 2-dimensional manifold $\Si$,\ with the $\om_{\xcL\xcW}$-holonomically reduced structure group $\bB(\xcL\xcW,\om_{\xcL\xcW})$,\ and their shadows $\xcF(\xcL\xcW,\om_{\xcL\xcW})\cong\xcP(\xcL\xcW,\om_{\xcL\xcW})/\xcL\xcW$.\ The general discussion above opens an avenue to a generalisation of the classic Poisson $\si$-model with target $(\txG,\Pi)$---through (flat) augmentation---to shadows of an arbitrary topology.\ A study thus oriented promises to shed new light on the long-known Poisson--Lie-type dualities among PAG $\si$-models and Poisson $\si$-models with Lie-group targets.
\eeg

\section{Topological gauge-symmetry defects}\label{sec:gau-sigmod}

Our hitherto considerations embed the Poisson $\si$-model in the setting of topological gauge field theory.\ In an abstract approach to topological field theory,\ {\it e.g.},\ in its functorial formulation in the quantum r\'egime,\ it is a well-established fact that the gauge\emph{-ing} of a symmetry can be \emph{realised} by topological defects carrying categorial data of the symmetry model---a group action,\ a groupoid,\ or a more general correspondence on the space of states,\ see:\ \cite{Frohlich:2006ch,Runkel:2008gr} for an early formalisation of the idea,\ and \cite{Carqueville:2023jhb} for a recent overview and an extensive bibliography.\ Here,\ defects are to be understood as self-intersecting codimension-1 {\it locii} of discontinuity of the configurations of the theory,\ separating its phases which inhabit connected components of spacetime bounded by defects.\ Their topologicality means that numerical values of observables (DF amplitudes,\ correlators {\it etc.}) do not change under homotopy moves of the defect embedded in the spacetime of the field theory,\ also those involving translations of intersections of its components of a non-zero dimension within the defect ({\it e.g.},\ the `homotopy moves' of pointlike junctions of defect lines in a defect graph embedded in the 2$d$ spacetime a 2$d$ field theory,\ considered in \cite[App.\,A.3]{Runkel:2008gr}).

In view of the above,\ it is natural to look for a reinterpretation of the coaugmented Poisson $\si$-model as a field theory with topological defects implementing the integrated groupoidal symmetry.\ That such reinterpretation is not merely plausible,\ but actually \emph{intrinsic} to the original definition of the $\si$-model in terms of the principal-$\xcG$-bundle object $(\xcP,\Theta)\xrightarrow{\xcD}(\xcF,\Theta_\xcF)$ and the bicategorial object $(\cG,\Ups,\g)$ over the nerve of its typical fibre $\xcG\xrightarrow{t}M$,\ can be inferred from the early studies of worldsheet defects and the attendant state correspondences\footnote{It is worth remarking at this point that the cross-breeding---initiated in \cite{Fuchs:2007fw,Runkel:2008gr}---between the functorial approach to the quantisation of the 2$d$ CFT with defefcts advanced in \cite{Frohlich:2006ch} and the higher-geometric formulation of the theory in the classical r\'egime advocated by Gaw\c{e}dzki resulted in the extraction---in \cite{Runkel:2008gr,Runkel:2009sp} and \cite{Suszek:2022CSimpl}---of highly nontrivial non-perturbative information on the quantised theory---such as elements of its Moore--Seiberg data---from the \emph{classical} gerbe-theoretic target structure of the PAG $\si$-model in the Wess--Zumino--Novikov--Witten (WZNW) setting,\ exceptionally tractable due to its extended symmetry.} in the dynamical 2$d$ PAG $\si$-model,\ reported in \cite{Fuchs:2007fw,Runkel:2008gr,Runkel:2009sp,Suszek:2011hg,Gawedzki:2012fu},\ and,\ in particular,\ in \cite{Suszek:2012ddg,Suszek:2022CSimpl}.\ Indeed,\ our reformulation of the classic Poisson $\si$-model in the bicategorial language of gerbe theory and its differential characters,\ given in Sec.\,\ref{sec:coaug-PSM-def} (following \cite{Suszek:2026col}),\ opens an avenue to the application of the concepts and techniques developed in \cite{Runkel:2008gr} (for the defect implementation of discrete symmetries of the 2$d$ field theory) and elaborated in \cite{Suszek:2012ddg} (for its continuous symmetries) in that very language.\ In it,\ defect lines are naturally associated with gerbe bimodules,\ and their intersections carry sheaf-cohomological data of distinguished gerbe 2-isomorphisms,\ and so the 2-stacky picture of descent worked out in Sec.\,\ref{sub:diff-char-gauge}---itself constituting a categorial enrichment of the underlying (1-)stacky structure of the configuration bundle object $\xcP\xrightarrow{\xcD}\xcF$,\ manifest in its local description \eqref{eq:princ-clutch} and \eqref{eq:shadow-clutch}---appears rather directly amenable to the desired reinterpretation.

In what follows,\ we carry out a detailed construction of the gauge-symmetry defect canonically encoded in the definition of the coaugmented Poisson $\si$-model \eqref{eq:coaug-PSM} and subsequently demonstrate how the various elements of the underlying $\om$-twisted $\xcG$-equivariant structure $(\cG,\Ups,\g)$ over the $\si$-model's simplicial target $\xcG_\bullet$ translate into non-generic properties of the defect,\ identifying it as a topological one in the sense of \cite{Runkel:2008gr}.\ The construction is readily seen to be rather tightly knitted,\ and so---conversely---the present reinterpretation of the TFT under consideration may be regarded as another piece of evidence for the adequacy and naturalness of its original definition,\ given---after \cite{Suszek:2026col}---in Sec.\,\ref{sec:coaug-PSM-def}.\medskip

By way of preparation for the construction of direct interest,\ let us adapt to the present purely topological setting the definition of the dynamical 2$d$ $\si$-model with a self-intersecting worldsheet defect formulated in all generality in \cite{Runkel:2008gr,Suszek:2011hg,Suszek:2022CSimpl} (see also:\ \cite{Gawedzki:2012fu}).\ In so doing,\ we keep the original terminology.\ The first elements to be recalled are the notions of {\bf $\cG_M$-bi-brane} $(Q,\iota_1,\iota_2,\varpi,\Psi)$ which augments a given {\bf bulk target} $(M,\cG_M)$ (from which we have removed the metric),\ and the corresponding {\bf circle-field configuration} $(\xi\,|\,\cD)$.\ These serve to give a formal meaning to the DF amplitude in the presence of a discontinuity---modelled by a collection of closed contours in $Q$---of the $\si$-model field $\phi\colo\Si\to M$ along a one-dimensional oriented submanifold $\cD\subset\Si$ with the topology of a disjoint union of a (finite) number of copies of $\bS^1$.\ Right from the outset,\ the construction calls for a generalisation of the concept of target space:\ In what follows,\ all manifolds in which $\si$-model fields take values are assumed to be \emph{stratified},\ by which we mean that they are disjoint unions of families of smooth manifolds (each of a definite dimension).\ For $\cD$ as described,\ we need two of these:\ the ({\bf bulk}) {\bf target space} $M$,\ and the {\bf bi-brane worldvolume} $Q$,\ the latter coming with two smooth maps $\iota_\a\colo Q\to M,\ \a\in\{1,2\}$,\ and a 2-form $\varpi\in\Om^2(Q)$.\ Given these,\ a circle-field configuration $(\xi\,\vert\,\cD)$ is a pair composed of the defect lines $\cD$ embedded in $\Si$,\ and a map
 \qq\nn
\xi\colo\Si\too M\sqcup Q
\qqq
such that 
\bit
\item  there exist smooth restrictions $\Si\setminus\cD\to M$ and $\cD\to Q$ of $\xi$;
\item for each oriented defect circle $\ell\subset\cD$,\ the map $\xi\rstr_{\Si\setminus\cD}$ extends from the patch to the left (resp.\ right) of $\ell$,\ in the direction defined by the orientation of $\ell$,\ to a smooth map $\xi_{|1}$ (resp.\ $\xi_{|2}$) $\colo(\Si\setminus\cD)\cup\ell\to M$ with the property
\qq\nn
\xi_{|\a}\rstr_\ell=\iota_\a\circ\xi\rstr_\ell\,.
\qqq
\eit

For such pairs $(\xi\,|\,\cD)$,\ we may define the {\bf topological $\si$-model}\footnote{Our specialisation of the definition of the 2$d$ PAG $\si$-model from \cite{Runkel:2008gr},\ from which the metric is removed,\ produces a stratified differential character,\ composed of a number of transport operators glued coherently.\ Nevertheless,\ in anticipation of the direct application of the resulting construction in the reinterpretation of the coaugmented Poisson $\si$-model,\ we regard the character as the DF amplitude of a purely topological $\si$-model.} with the circular defect $\cD$ in the presence of the remaining elements of the {\bf bi-brane string background} $((M,\cG_M),(Q,\iota_1,\iota_2,\varpi,\Psi))$,\ to wit:
\bit
\item a gerbe $\cG_M$ (of curvature $H_M$) on $M$;
\item a 2-form $\varpi\in\Om^2(Q)$,\ termed the {\bf bi-brane curvature},\ and a gerbe bimodule\footnote{Below,\ and in what follows,\ we use freely the standard gerbe-theoretic notation introduced in \cite{Murray:1994db,Murray:1999ew,Stevenson:2000wj} and \cite{Waldorf:2007mm},\ and recapitulated in App.\,\ref{app:gerbarium}.}
\qq\nn
\Psi\colo\iota_1^*\cG_M\xrightarrow{\ \cong\ }\iota_2^*\cG_M\ox\cI_\varpi\,.
\qqq
\eit
The DF amplitude $\cA^{t\si(M\sqcup Q)}_{\rm DF}$ of the topological $\si$-model for $(\xi\,|\,\cD)$, 
\qq\nn
\cA^{t\si(M\sqcup Q)}_{\rm DF}[(\xi\,|\,\cD)]=\Hol_{(\cG_M,\Psi)}(\xi\,|\,\cD)\,,
\qqq
is given by the ($\cD$-)decorated-surface holonomy for the circle-field configuration $(\xi\,|\,\cD)$ specified in \cite[Eq.\,(2.24)]{Runkel:2008gr}.\ The definition of $\cA^{t\si(M\sqcup Q)}_{\rm DF}$ uses one further important assumption about the target geometry,\ to wit---the existence of \v Cech extensions $\phi_\a\colo I^Q\to I^M$ of the respective bi-brane maps $\iota_\a,\ \a\in\{1,2\}$,\ which determine the subordination of the open cover $\cO^Q\equiv\{O^Q_i\}_{i\in I^Q}$ of $Q$ trivialising for $\Psi$ ({\it i.e.},\ such that $\Psi$ admits a corresponding BD presentation) to the formerly introduced open cover $\cO^M\equiv\{O^M_i\}_{i\in I^M}$ trivialising for $\cG_M$ captured by the relations
\qq\nn
\iota_\a\bigl(O^Q_i\bigr)\subset O^M_{\phi_\a(i)},\quad \a\in\{1,2\}\,.
\qqq

\brem\label{rem:D-deco-holo}
The $\cD$-decorated-surface holonomy can be written out in terms of the BD data:\ $(B_i,A_{ij},g_{ijk})\in A^2(\cO^M)$ of $\cG_M$ and $(P_i,K_{ij})\in A^1(\cO^Q)$ of $\Psi$ (associated with the open covers $\cO^M$ and $\cO^Q$ described above) as follows:\ We choose an arbitrary tessellation $\triangle_{\Si,\cD}$ of $\Si$ which restricts to a tessellation of the domains $\Si\setminus\cD$,\ and thus induces a tessellation $\triangle_\cD$ of $\cD$,\ with its edges $E_{\triangle_\cD}\subset E_{\triangle_{\Si,\cD}}$ and vertices $V_{\triangle_\cD}\subset V_{\triangle_{\Si,\cD}}$.\ Note that every edge $e\in E_{\triangle_\cD}$ separates two plaquettes:\ one to its left ($p^1(e)$),\ and one to its right ($p^2(e)$),\ in the direction of its orientation.\ Moreover,\ with every vertex $d\in V_{\triangle_\cD}$ from the boundary of an edge $e\in E_{\triangle_\cD}$,\ we may associate a sign $\vep_{ed}\in\{-1,1\}$ in analogy with the bulk case (see:\ Rem.\,\ref{rem:Lorentz}):\ $+1$ if $e$ is in-coming at $d$ when taken with the orientation inherited from $\cD$,\ and $-1$ otherwise.\ We denote the subset of $P_{\triangle_{\Si,\cD}}$ composed of plaquettes whose closures intersect $\cD$ and lie to the left (resp.\ to the right) of $\cD$ as $P_1$ (resp.\ $P_2$).\ Finally,\ we assume the tessellation to be subordinated to $\cO^M$ \emph{and} $\cO^Q$ along $\xi\colo\Si\to M\sqcup Q$ through a choice of an index map $\imath\colo\triangle_{\Si,\cD}\to I_M\cup I_Q$ such that $\imath(\triangle_{\Si,\cD}\setminus\triangle_\cD)\subset I_M$ and $\imath(\triangle_\cD)\subset I_Q$.\ The holonomy reads
\qq\nn
\Hol_{(\cG,\Psi)}(\xi\,|\,\cD)&=&\prod_{p\in P_{\triangle_{\Si,\cD}}}\bigg(\exp\bigg(\sfi\int_{\xi(p)}\,B_{\imath(p)}\bigg)\cdot\prod_{e\in \p p\setminus E_{\triangle_\cD}}\bigg(\exp\bigg(\sfi\int_{\xi(e)}\,A_{\imath(p)\imath(e)}\bigg)\cdot\prod_{v\in\p e\setminus V_{\triangle_\cD}}g_{\imath(p)\imath(e)\imath(v)}\bigl(\xi(v)\bigr)^{\vep_{pev}}\bigg)\bigg)\cr\cr   
&&\cdot\prod_{\a\in\{1,2\}}\prod_{p\in P_\a}\bigg(\bigg(\prod_{e\in\p p\setminus E_{\triangle_\cD}}\prod_{v\in\p e\cap V_{\triangle_\cD}}g_{\imath(p)\imath(e)\phi_\a(\imath(v))}\bigl(\xi_{|\a}(v)\bigr)^{\vep_{pev}}\bigg)\cr\cr 
&&\cdot\bigg(\prod_{e\in\p p\cap E_{\triangle_\cD}}\bigg(\exp\bigg(\sfi\int_{\xi_{|\a}(e)}\,A_{\imath(p)\phi_\a(\imath(e))}\bigg)\cdot\prod_{v\in\p e}g_{\imath(p)\phi_\a(\imath(e))\phi_\a(\imath(v))}\bigl(\xi_{|\a}(v)\bigr)^{\vep_{pev}}\bigg)\bigg)\bigg)\cr\cr 
&&\cdot\prod_{e\in E_{\triangle_\cD}}\bigg(\exp\bigg(\sfi\int_{\xi(e)}\,\bigl(P_{\imath(e)}\bigg)\cdot\prod_{d\in\p e}K_{\imath(e)\imath(d)}\bigl(\xi(d)\bigr)^{-\vep_{ed}}\bigg)\,.
\qqq
\erem

\begin{figure}[htb]
	\centering
	\includegraphics[width=0.5\textwidth]{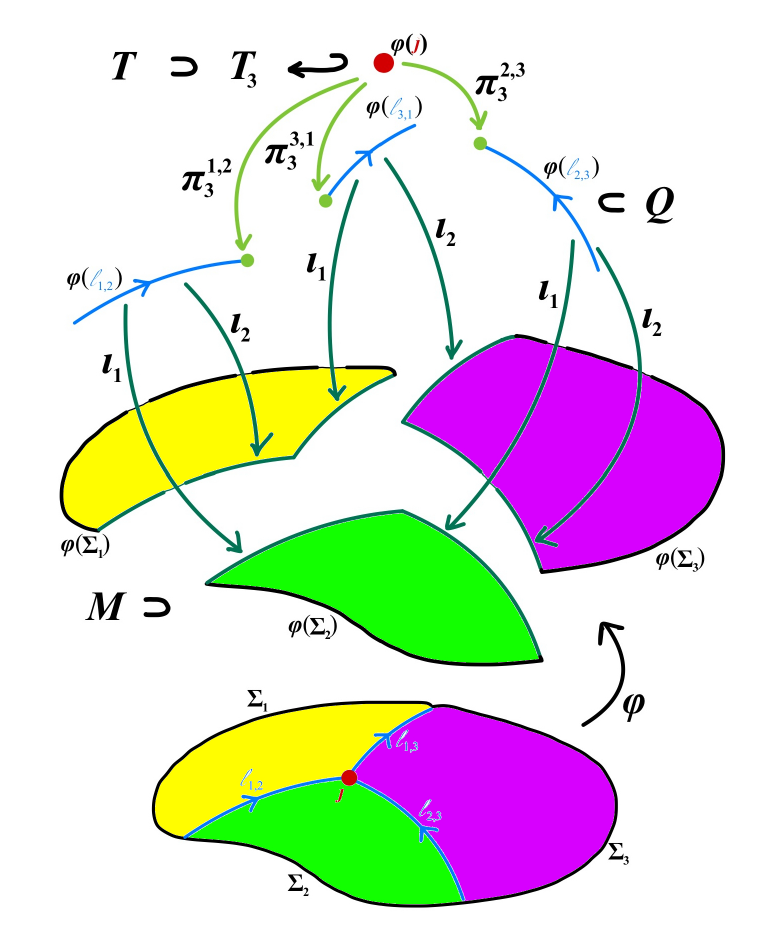}
	\caption{A network-field configuration $(\xi\,|\,\G)$ on a neighbourhood $\Si_1\cup\Si_2\cup\Si_3$ of a tri-valent defect junction $\jmath$ in a worldsheet $\Si$.}\label{fig:tri-net-field}
\end{figure}

Giving a rigorous meaning to the field theory in the presence of intersecting defect lines calls for further stratification of the $\si$-model data over the worldsheet and in the target,\ and adds another layer of gerbe-theoretic structure over the latter.\ The point of departure is the partitioning of $\Si$ into two-dimensional {\bf domains} separated by oriented edges of an embedded graph $\G\subset\Si$,\ termed the {\bf defect quiver}.\ The edges,\ called {\bf defect lines} and forming a set $\Egt_\G$,\ are either closed or bounded by vertices of arbitrary valence $n\geq 3$,\ called {\bf defect junctions}.\ Thus,\ the set $\Vgt_\G$ of defect junctions splits into fixed-valence subsets $\Vgt_\G^n\subset \bigcup_{m\geq 3}\,\Vgt_\G^m\equiv\Vgt_\G$).\ The field-theoretic construction now uses a tri-partite target space $\cM=M\sqcup Q\sqcup T$,\ formed by (stratified) manifolds:\ $M$ and $Q$ as before,\ augmented with the {\bf inter-bi-brane worldvolume} $T=\bigsqcup_{n\geq 3}\,T_n$ that is made up of smooth manifolds $T_n$.\ Each of the latter comes with a collection of maps,\ indexed by $\bZ/n\bZ\ni k-1$:\ $\pi_n^{k,k+1}\colo T_n\to Q$ (with the cyclic index identification $n+1\equiv 1$),\ and $\vep_n^{k,k+1}\colo T_n\to\{-1,1\}$ (the {\bf orientation maps},\ with the same cyclic index identification),\ jointly subject to the constraints:
\qq\label{eq:phase-maps_n}
\iota_2^{\vep_n^{k-1,k}}\circ\pi_n^{k-1,k}=\iota_1^{\vep_n^{k,k+1}}\circ\pi_n^{k,k+1}=:\pi_n^k\,,
\qqq 
in which $(\iota_1^{+1},\iota_2^{+1}):=(\iota_1,\iota_2)$ and $(\iota_1^{-1},\iota_2^{-1}):=(\iota_2,\iota_1)$.\ For these,\ we define a {\bf network-field configuration} $(\xi\,|\,\G)$ as a map
\qq\nn
\xi\colo\Si\too\cM
\qqq
such that
\bit
\item there exist smooth restrictions $\Si\setminus\G\to M,\ \G\setminus\Vgt_\G\to Q$ and $\Vgt_\G^n\to T_n$ of $\xi$;
\item for each defect line $\ell\in\Egt_\G$,\ the map $\xi\rstr_{\Si\setminus\G}$ extends from the domain to the left (resp.\ right) of $\ell$,\ in the direction defined by the orientation of $\ell$,\ to a smooth map $\xi_{|1}$ (resp.\ $\xi_{|2}$) $\colo(\Si\setminus\G)\cup(\ell\setminus\Vgt_\G)\to M$ with the property
\qq\nn
\xi_{|\a}\rstr_{\ell\setminus\Vgt_\G}=\iota_\a\circ\xi\rstr_{\ell\setminus\Vgt_\G }\,;
\qqq
\item for each $n$-valent defect junction $\jmath\in\Vgt_\G^n$ and defect lines $\ell_{k,k+1},\ k-1\in\bZ/n\bZ$ (with the cyclic index identification) meeting at $\jmath$ and ordered in a fixed cyclic way,\ the maps $\xi\rstr_{\ell_{k,k+1}\setminus\Vgt_\G}$ admit smooth extensions $\xi_{k,k+1}$ to $\jmath$,\ with the property
\qq\nn
\xi_{k,k+1}(\jmath)=\bigl(\pi_n^{k,k+1}\circ\xi\bigr)(\jmath)\,;
\qqq
\item for $\jmath$ and the $\ell_{k,k+1}$ as above,\ the orientation map $\vep_n^{k,k+1}$ takes the value $\vep_n^{k,k+1}(\xi(\jmath))=+1$ if $\ell_{k,k+1}$ is oriented towards $\jmath$,\ and the value $\vep_n^{k,k+1}(\xi(\jmath))=-1$ if $\ell_{k,k+1}$ is oriented away from $\jmath$ for junctions $\jmath$ with defect lines ordered counter-clockwise,\ and the opposite values for junctions with defect lines ordered clockwise.
\eit
The simple geometry behind the above complex description is illustrated in Fig.\,\ref{fig:tri-net-field}.

The definition of the topological $\si$-model with the defect quiver $\G$ can be formulated whenever we are given a family $\{\varphi_n\}_{n\in\bN_{\geq 3}}$ of gerbe 2-isomorphisms,\ over the corresponding inter-bi-brane worldvolumes $T_n$,
\qq\label{diag:2iso}
\xy (50,0)*{\bullet}="G3"+(5,4)*{\cG_{M\,n}^3\ox I_{\varpi_n^{1,2}+\varpi_n^{2,
			3}}}; (25,-20)*{\bullet}="G2"+(-10,0)*{\cG_{M\,n}^2\ox I_{\varpi_n^{1,2}}};
(75,-20)*{\ \vdots}="dots";
(35,-40)*{\bullet}="G1"+(0,-4)*{\cG_{M\,n}^1};
(65,-40)*{\bullet}="G1add"+(10.5,-4)*{\cG_{M\,n}^1\ox I_{\varpi_n^{1,2}+
		\varpi_n^{2,3}+\ldots+\varpi_n^{n,1}}}; (50,-40)*{}="id";
\ar@{->}|*+{\Phi_n^{2,3}\ox\id} "G2";"G3"
\ar@{->}|*+{\Phi_n^{3,4}\ox\id} "G3";"dots"
\ar@{->}|*+{\Phi_n^{1,2}} "G1";"G2"
\ar@{->}|*+{\Phi_n^{n,1}\ox\id} "dots";"G1add" \ar@{=}|*+{\id}
"G1"+(2,0);"G1add"+(-2,0) \ar@{=>}|*+{\varphi_n}
"G3"+(0,-3);"id"+(0,+3)
\endxy
\qqq
written in terms of 1-isomorphisms
$\,\Phi_n^{k,k+1}=(\pi_n^{k,k+1})^{*}\Phi^{\vep_n^{k,k+1}}\,$
between gerbes $\,\cG_{M\,n}^k= (\pi_n^{k})^{*}\cG_M$,\ and the trivial
gerbes with curvings
$\,\varpi_n^{k,k+1}=\vep_n^{k,k+1}\,(\pi_n^{k,k+1})^{*}\varpi$.\ We shall refer to the $\varphi_n$ as {\bf $n$-valent fusion 2-isomorphisms} in what follows.\ In the present setting,\ the ensuing fully fledged {\bf topological string background} 
\qq\nn
\bigl((M,\cG_M),(Q,\iota_1,\iota_2,\varpi,\Psi), (T_\bullet,\pi_\bullet^{\cdot,\cdot+1},\vep_\bullet^{\cdot,\cdot+1},\varphi_\bullet)\bigr)
\qqq
gives rise to the DF amplitude $\cA^{t\si(M\sqcup Q\sqcup T)}_{\rm DF}\equiv\cA^{t\si(\cM)}_{\rm DF}$ of the topological $\si$-model for $(\xi\,|\,\G)$,
\qq\label{eq:PAG-sigmodef}
\cA^{t\si(\cM)}_{\rm DF}[(\xi\,|\,\G)]=\Hol_{(\cG_M,\Psi,\varphi_\bullet)}(\xi\,|\,\G)\,,
\qqq
given by the ($\G$-)decorated-surface holonomy for the network-field configuration $(\xi\,|\,\G)$ detailed in \cite[Eq.\,(2.53)]{Runkel:2008gr}.\ As in the case of the topological $\si$-model with circular defects,\ the definition of $\cA^{t\si(\cM)}_{\rm DF}$ bases on the existence of a family of \v Cech extensions $\psi_n^{k,k+1}\colo I^{T_n}\to I^Q$ of the respective bi-brane maps $\pi_n^{k,k+1},\ k-1\in\bZ/n\bZ,\ n\in\bN_{\geq 3}$ (with the previous cyclic index identification),\ defined for the index sets of open covers:\ $\cO^{T_n}\equiv\{O^{T_n}_i\}_{i\in I^{T_n}}$ of $T_n$ trivialising for $\varphi_n$,\ and the previously defined $\cO^Q\equiv\{O^Q_i\}_{i\in I^Q}$.\ These enter the condition of subordination
\qq\nn
\pi_n^{k,k+1}\bigl(O^{T_n}_i\bigr)\subset O^Q_{\psi_n^{k,k+1}(i)},\quad k-1\in\bZ/n\bZ,\ n\in\bN_{\geq 3}\,.
\qqq
Together with the bi-brane \v Cech extensions $\phi_\a,\ \a\in\{1,2\}$ introduced before,\ the $\psi_n^{k,k+1}$  are subject to constraints which lift those imposed upon the corresponding manifold maps,\ \eqref{eq:phase-maps_n},\ and so they read
\qq\nn
\phi_2^{\vep_n^{k-1,k}}\circ\psi_n^{k-1,k}=\phi_1^{\vep_n^{k,k+1}}\circ\psi_n^{k,k+1}=:\psi_n^k\,,
\qqq 
where $(\phi_1^{+1},\phi_2^{+1}):=(\phi_1,\phi_2)$ and $(\phi_1^{-1},\phi_2^{-1}):=(\phi_2,\phi_1)$.\ 

\brem
In analogy with its predecessors,\ the $\G$-decorated-surface holonomy admits a presentation in terms of the BD data---those of the target gerbe $\cG_M$,\ $(B_i,A_{ij},g_{ijk})\in A^2(\cO^M)$,\ and of the bi-brane 1-isomorphism $\Psi$,\ $(P_i,K_{ij})\in A^1(\cO^Q)$,\ are augmented by those of the component inter-bi-brane 2-isomorphisms $\varphi_n$,\ $(f_{n,i})\in A^0(\cO^{T_n})$,\ the three families being associated with the open covers $\cO^M,\cO^Q$ and $\cO^{T_n}$ described above.\ The presentation calls for a choice of a tessellation $\triangle_{\Si,\G}$ of $\Si$ which restricts to a tessellation of the domains $\Si\setminus\G$,\ and thus induces a tessellation $\triangle_\G$ of $\G$,\ with edges $E_{\triangle_\G}\subset E_{\triangle_{\Si,\G}}$ and vertices $V_{\triangle_\G}\subset V_{\triangle_{\Si,\G}}$,\ the latter satisfying $V_{\triangle_\G}\supset\Vgt_\G$.\ Given $\triangle_{\Si,\G}$,\ we obtain two families of sets indexed by $\bN_{\geq 3}\ni n$:
\qq\nn
P_{\Vgt_\G}^n&:=&\bigl\{ \ p\in P_{\triangle_{\Si,\G}} \ \vert\ \exists\,e\in\p p\colo \p e\cap\Vgt_\G^n\neq\emptyset\ \bigr\}\,,\cr\cr 
E_{\Vgt_\G}^n&:=&\bigl\{ \ e\in E_{\triangle_\Si} \ \vert\ \p e\cap\Vgt_\G^n\neq\emptyset\ \bigr\}\,.
\qqq
Clearly,\ to every pair $(p,e)\in P_{\Vgt_\G}^n\x E_{\Vgt_\G}^n$ such that $e\in\p p$,\ we may now assign a side-label $\a(p,e)\in\{1,2\}$ determined uniquely by the identification of the side of the defect line $\ell\supset e$ on which $p$ lies (relative to the orientation of $\ell$),\ $p\in P_{\a(p,e)}$.\ Similarly,\ to every pair $(p,\jmath)\in P_{\Vgt_\G}^n\x\Vgt_\G^n$ such that there exists $e\in\p p$ with the property $\jmath\in\p e$,\ we may assign a domain-label $k(p,\jmath)-1\in\bZ/n\bZ$ determined uniquely by the identification of the domain,\ enclosed by the defect lines $\ell_{k(p,\jmath)-1,k(p,\jmath)}$ and $\ell_{k(p,\jmath),k(p,\jmath)+1}$ (for the fixed ordering of the defect lines converging at $\jmath$),\ in which $p$ is contained.\ Finally,\ to every pair $(e,\jmath)\in E_{\Vgt_\G}^n\x\Vgt_\G^n$ such that $\jmath\in\p e$,\ we may assign the defect-label $o(e,\jmath)-1\in\bZ/n\bZ$ determined uniquely by the inclusion $e\subset\ell_{o(e,\jmath),o(e,\jmath)+1}$ at $\jmath$ (for the aforementioned ordering).\ With all this structure predefined,\ we may finally give the formula for the $\G$-decorated-surface holonomy.\ Technically speaking,\ the latter differs from the one for the $\cD$-decorated-surface holonomy,\ written out in Rem.\,\ref{rem:D-deco-holo},\ only through the contribution of defect junctions,\ and so it is convenient to denote the expression for $\Hol_{(\cG_M,\Psi)}(\xi\,|\,\cD)$ \emph{with all contributions from $\Vgt_\G$ removed} as $\Hol_{(\cG_M,\Psi)}(\xi\,|\,\G\setminus\Vgt_\G)$,\ whereupon we may write
\qq\nn
\Hol_{(\cG_M,\Psi,\varphi_\bullet)}(\xi\,|\,\G)&=&\Hol_{(\cG_M,\Psi)}(\xi\,|\,\G\setminus\Vgt_\G)\cr\cr 
&&\cdot\prod_{n=3}^\infty\bigg(\prod_{p\in P_{\Vgt_\G}^n}\bigg(\bigg(\prod_{e\in\p p\setminus E_{\triangle_\G}}\prod_{\jmath\in\p e\cap\Vgt_\G^n}g_{\imath(p)\imath(e)\psi_n^{k(p,\jmath)}(\imath(\jmath))}\bigl(\pi_n^{k(p,\jmath)}\bigl(\xi(\jmath)\bigr)\bigr)^{\vep_{pe\jmath}}\bigg)\cr\cr &&\hspace{1.85cm}\cdot\bigg(\prod_{e\in\p p\cap E_{\triangle_\G}}\prod_{\jmath\in\p e\cap\Vgt_\G^n}g_{\imath(p)\phi_{\a(p,e)}(\imath(e))\psi_n^{k(p,\jmath)}(\imath(\jmath))}\bigl(\pi_n^{k(p,\jmath)}\bigl(\xi(\jmath)\bigr)\bigr)^{\vep_{pe\jmath}}\bigg)\bigg)\cr\cr 
&&\hspace{.8cm}\cdot\prod_{e\in E_{\Vgt_\G}^n}\prod_{\jmath\in\p e\cap\Vgt_\G^n}K_{\psi_n^{o(e,\jmath),o(e,\jmath)+1}(\imath(\jmath))}\bigl(\pi_n^{o(e,\jmath),o(e,\jmath)+1}\bigl(\xi(\jmath)\bigr)\bigr)^{-\vep_{e\jmath}}\bigg)\,.
\qqq
\erem

The formulation of the topological $\si$-model for network-field configurations over worldsheets with embedded defect quivers decorated by bicategorial data of gerbe bi-modules and fusion 2-isomorphisms,\ obtained above through specialisation of the original construction from \cite{Runkel:2008gr},\ provides us with a convenient framework in which to rigorously model field discontuities in the setting of the mono-phase topological $\si$-model such as the ctP $\si$-model \eqref{eq:ctP-sigmod-DF}.\ Indeed,\ upon specialising the general construction to the case of a \emph{single} component in the generically stratified bulk target $(M,\cG_M)$,\ we may conceive a topological string background implementing a prequantisable geometric symmetry on the configuration fibre $M$,\ see:\ \cite{Suszek:2011hg}.

\begin{figure}[htb]
\centering
\includegraphics[width=0.5\textwidth]{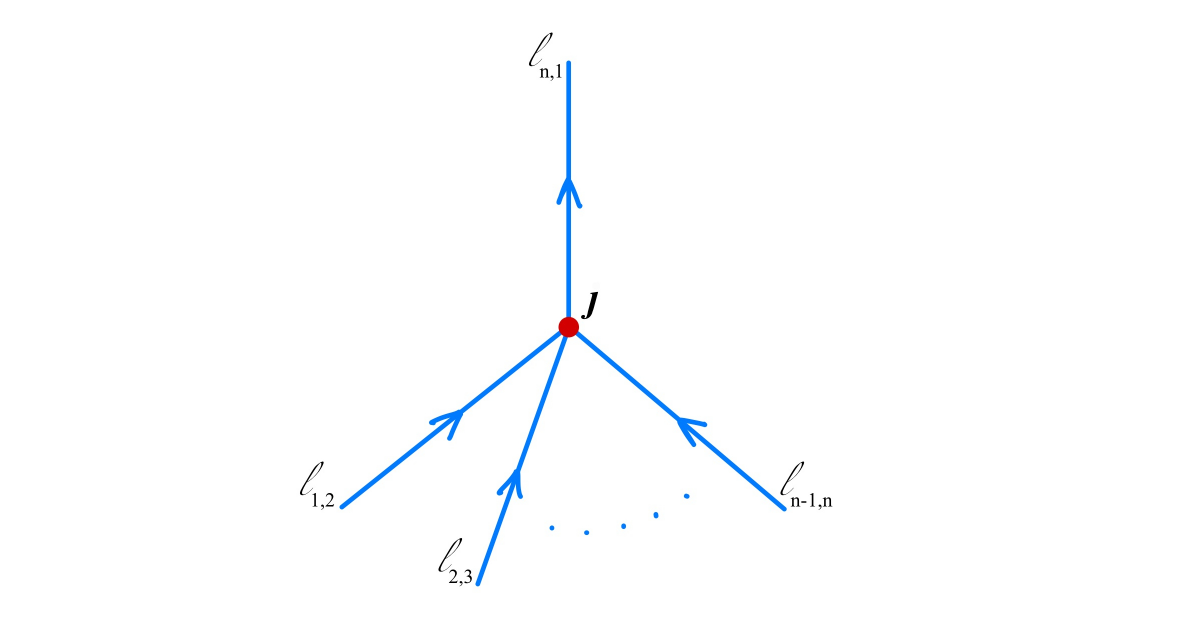}
\caption{An $n$-valent defect junction $\jmath$ with $n-1$ in-coming lines $\ell_{k,k+1},\ k\in\ovl{1,n-1}$ and 1 out-going defect line $\ell_{n,1}$.}\label{fig:many-2-one-junct}
\end{figure}

\beg \label{eg:grpd-rsymm-bckgrnd}
A very natural example,\ which---incidentally---paves the way to our subsequent considerations,\ is furnished by a multiplicatively $\om$-twisted $\xcG$-equivariant structure on the (flat) bulk gerbe $\cG$ from Sec.\,\ref{sub:diff-char-gauge}\ The corresponding bi-brane is the quintuple
\qq\nn
(\xcG,t,s,\om,\Upsilon)\,,
\qqq
with the worldvolume given by the arrow manifold $\xcG$ of a Lie groupoid $\grpd{\xcG}{M}$ over the bulk target space $M$---its target and source maps acting as bi-brane maps---and with the $\cG$-bi-module given by the 1-isomorphism $\Upsilon$ from \eqref{eq:Ups-first},\ which puts the multiplicative symplectic form $\om$ in the r\^ole of the bi-brane's curvature.\ As the distinguished component inter-bi-brane $(T_{n,++\cdots+-},\pi_n^{\cdot,\cdot+1},\vep_n^{\cdot,\cdot+1},\varphi_{n,++\cdots+-}),\ n>3$ with $\vep_n^{k,k+1}=(-1)^{\d_{k,n}}$,\ corresponding to the $(n-1)$-to-1 defect line configuration depicted in Fig.\,\ref{fig:many-2-one-junct},\ we take the manifold of composable $(n-1)$-tuples of arrows,
\qq\label{eq:many-2-one-ibb}
T_{n,++\cdots+-}=\xcG_{n-1}\equiv N_{n-1}(\grpd{\xcG}{M})\,,
\qqq
together with the superpositions 
\qq
&\pi_n^{1,2}=d^{(2)}_0\circ d^{(3)}_0\circ\cdots\circ d^{(n-1)}_0\,,\qquad \pi_n^{l,l+1}=d^{(2)}_2\circ d^{(3)}_2\circ\cdots\circ d^{(l)}_2\circ d^{(l+1)}_0\circ d^{(l+2)}_0\circ\cdots\circ d^{(n-1)}_0\,,\quad l\in\ovl{2,n-1}\,,&\cr\cr 
&\pi_n^{n,1}=d^{(2)}_1\circ d^{(3)}_1\circ\cdots\circ d^{(n-1)}_1& \label{eq:rig-pi-n}
x\qqq
of the face maps $d^{(\bullet+1)}_\cdot$ of the simplicial manifold $N_\bullet(\grpd{\xcG}{M})$,\ see:\ Ex.\,\ref{eg:face-nerve-LieGrpd},\ and 
\qq\nn
\varphi_{n,++\cdots+-}=\pi_n^{n,1\,*}d_\Upsilon\bullet\bigl(\id\circ\widehat\pi{}_n^{12\cdots n-2,n-1\,*}\g\bigr)\bullet\bigl(\id\circ\widehat\pi{}_n^{12\cdots n-3,n-2\,*}\g\bigr)\bullet\cdots\bullet\bigl(\id\circ\widehat\pi{}_n^{12,3\,*}\g\bigr)\bullet\bigl(\id\circ\widehat\pi{}_n^{1,2\,*}\g\bigr)\,,
\qqq
where
\qq
&\widehat\pi{}_n^{1,2}=d^{(3)}_0\circ d^{(4)}_0\circ\cdots\circ d^{(n-1)}_0\,,\qquad\widehat\pi{}_n^{12\cdots l,l+1}=d^{(3)}_2\circ d^{(4)}_2\circ\cdots\circ d^{(l+1)}_2\circ d^{(l+2)}_0\circ d^{(l+3)}_0\circ\cdots\circ d^{(n-1)}_0\,,\quad l\in\ovl{2,n-3}\,,&\cr\cr 
&\widehat\pi{}_n^{12\cdots n-2,n-1}=d^{(3)}_2\circ d^{(4)}_2\circ\cdots\circ d^{(n-1)}_2\,,&
\label{eq:rig-hatpi-n}
\qqq
and where $d_\Ups\colo(\Ups^{-1}\ox\id)\circ\Ups\overset{\cong}{\Rightarrow}\id$ is the death 2-isomorphism of $\Ups$,\ see:\ \cite[Sec.\,1.3]{Waldorf:2007mm} (where it was denoted by $i_l$).\ The tower of structures is completed by the elementary inter-bi-brane $(T_{3,++-},\pi_3^{\cdot,\cdot+1},\vep_3^{\cdot,\cdot+1},\varphi_{3,++-}),\ \vep_n^{1,2}=\vep_3^{2,3}=1=-\vep_3^{3,1}$ with worldvolume
\qq\nn
T_{n,++-}=\xcG_2\equiv N_2(\grpd{\xcG}{M})
\qqq
mapping to $\xcG$ through
\qq\nn
\pi_3^{1,2}=d^{(2)}_0\equiv\pr_1\,,\qquad\qquad\pi_3^{2,3}=d^{(2)}_2\equiv\pr_2\,,\qquad\qquad\pi_3^{3,1}=d^{(2)}_1\equiv\txm\,.
\qqq
On it,\ we take the elementary fusion 2-isomorphism
\qq\nn
\varphi_{3,++-}=\pi_3^{3,1\,*}d_\Upsilon\bullet\g\,.
\qqq
The remaining components of the inter-bi-brane are fairly straightforward to recover from the above data using the canonical structure of a monoidal bicategory with duality on $\bgrb_\nabla(\xcG_\bullet)$ (and strict pullback 2-functors),\ see:\ \cite[Sec.\,2.5.1]{Runkel:2008gr}.\ Importantly,\ the simplicial target space $\cM\equiv\bigsqcup_{n\in\bN}\, \xcG_n$ admits simplicial covers by Thm\,\ref{thm:simpl-ref},\ and so its bi-brane and inter-bi-brane maps are naturally \v Cech-extendable.\ This is an example of a topological string background with (simplicial) induction,\ first discussed in \cite[Sec.\,2.8]{Runkel:2008gr},\ and subsequently elaborated in \cite[Sec.\,1.2]{Suszek:2022CSimpl}.
\eeg

\subsection{The gauge-symmetry topological string background}\label{sub:gs-backgrbnd}

\begin{figure}[htb]
\centering
\includegraphics[width=0.6\textwidth]{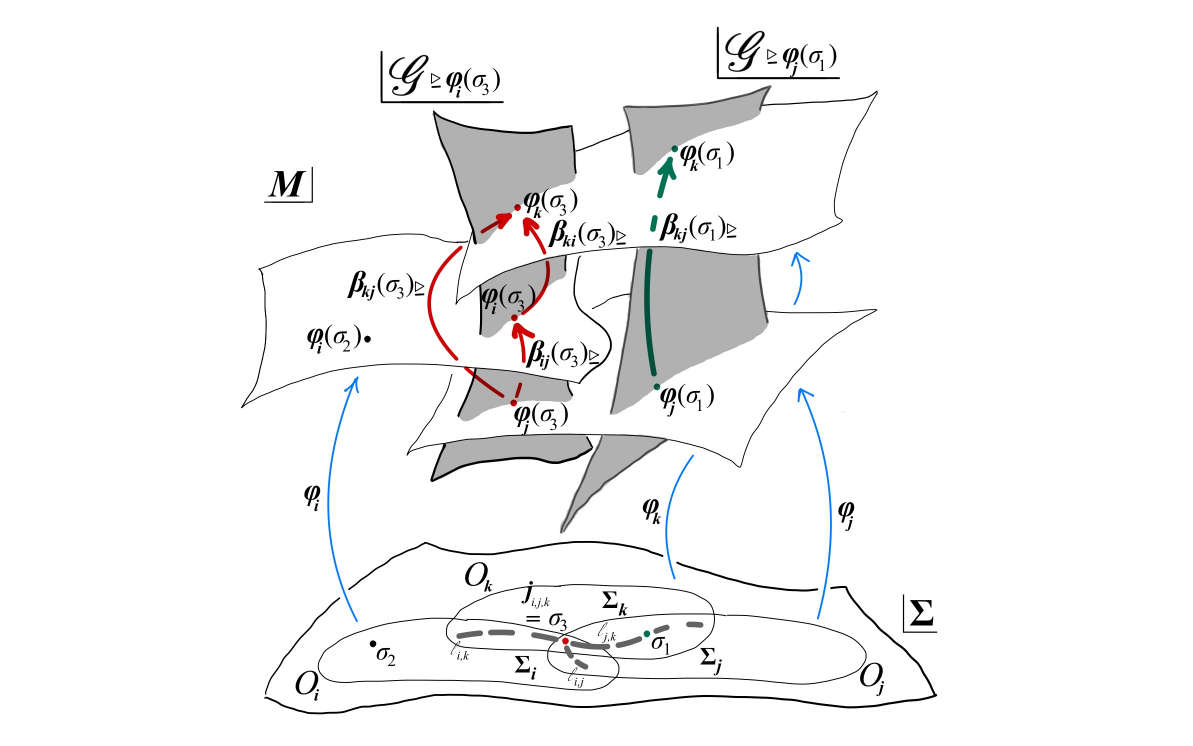}
\caption{A fragment of a gauge-symmetry defect graph $\G_{(\xcF)}$ inscribed in a trivialisation cover $\{O_i\}_{i\in I}$ of the worldsheet base $\Si$ of the shadow bundle $\xcF\to\Si$.\ The domains $\Si_i\subset O_i$ of local fields $\xi_i\colo O_i\to M$ are separated by defect lines $\ell_{i,j}\subset O_{ij}$ and join at defect junctions $\jmath_{i,j,k(\cdots)}\subset O_{ijk(\cdots)}$.\ The jump of the local field $\xi_i$ at $\ell_{i,j}$ within its symmetry orbit $\xcG\ulact\xi_i$ is effected by the shadow action of the transition 1-cocycle $\b_{ij}$ of $\xcF$.}\label{fig:gauge-sym-def}
\end{figure}

The gauging of a rigid symmetry $\grpd{\xcG}{M}$ in the coaugmented Poisson $\si$-model,\ as reviewed in Sec.\,\ref{sec:coaug-PSM-def},\ rather naturally gives rise to conditions in which symmetry-implementing defects emerge in the worldsheet.\ Indeed,\ when working with the Godement shadow $\xcF$ of a symplectic principaloid $(\xcG,\om)$-bundle $\xcP$ from Sec.\,\ref{sub:PSM-redef} as the configuration bundle of the $\si$-model,\ we encounter local fields $\phi_i\colo O_i\to M$ induced by global ones $\phi\in\G(\xcF)$ over the trivialising cover $\cO_\Si\equiv\{O_i\}_{i\in I}$ of the base $\Si$ of $\xcF$,\ with discontiuities over intersections $O_{ij}\ni\si$ of their respective domains determined by the $\bB(\xcG,\om)$-valued transition maps of $\xcF$ as
\qq\label{eq:loc-field-trans}
\phi_i(\si)=t_*\bigl(\b_{ij}(\si)\bigr)\bigl(\phi_j(\si)\bigr)\equiv\b_{ij}(\si)\ulact\phi_j(\si)\,.
\qqq
We may,\ accordingly,\ split $\Si$ into domains $\Si_i\subset O_i$ by inscribing in $\cO_\Si$ an oriented graph $\G_{(\xcF)}$ with edges $\ell_{i,j}\subset O_{ij}$ that meet at junctions $\jmath_{i_1,i_2,\ldots,i_n}\subset O_{i_1 i_2\cdots i_n}$---the ensuing worldsheet configuration is exemplified in Fig.\,\ref{fig:gauge-sym-def}.\ 
We subsequently consider the following topological string background:\ As the bulk target space,\ we take the disjoint union
\qq\nn
M_{(\xcF)}:=\bigsqcup_{i\in I}\,M_i\,,\quad M_i:=O_i\x M\,.
\qqq
On it,\ we put the target gerbe $\cG_{(\xcF)}$ given by the connection-augmented one from \eqref{eq:conn-aug-grb-loc},\ with restrictions
\qq\nn
\cG_{(\xcF)}\rstr_{M_i}:=\xcF\t_i^{-1\,*}\cG[\Theta_i]\equiv\pr_2^*\cG\ox\cI_{\om[\Theta_i]}\,.
\qqq 

Next,\ for every pair $(i,j)\in I^{\x 2}$ labelling an intersection $O_{ij}\neq\emptyset$,\ we write (see:\ \eqref{eq:bij-tilde})
\qq\label{eq:pro-defl-field}
\widehat\b{}_{ij}:=\bigl(\pr_1,\widetilde\b{}_{ij}\bigr)\colo O_{ij}\x M\too O_{ij}\x\xcG,\ (\si,m)\longmapsto\bigl(\si,\b_{ij}(\si)(m)\bigr)\,,
\qqq
and define the corresponding component bi-brane worldvolumes
\qq\nn 
Q_{ij}:=\widehat\b{}_{ij}\bigl(O_{ij}\x M\bigr)\,.
\qqq
Note that the latter can be viewed---each---as fibre bundles over $O_{ij}$,\ with typical fibre $M$.\ From these,\ the bi-brane worldvolume arises as
\qq\nn
Q_{(\xcF)}:=\bigsqcup_{(i,j)\in I^{\x 2}}\,Q_{ij}\,.
\qqq
The bi-brane comes with two maps to the target space:
\qq\nn
&\breve\iota{}_1:=\id_\Si\x t\x\pr_1\colo Q_{(\xcF)}\too M_{(\xcF)},\ \bigl(\si,\b_{ij}(\si)(m),i,j\bigr)\longmapsto\bigl(\si,\b_{ij}(\si)\ulact m,i\bigr)\,,&\cr\cr 
&\breve\iota{}_2:=\id_\Si\x s\x\pr_2\colo Q_{(\xcF)}\too M_{(\xcF)},\ \bigl(\si,\b_{ij}(\si)(m),i,j\bigr)\longmapsto\bigl(\si,m,j\bigr)\,,&
\qqq
and a $\cG_{(\xcF)}$-bimodule
\qq\nn
\Psi_{(\xcF)}:=\pr_2^*\Upsilon\ox\id\rstr_{Q_{(\xcF)}}\colo\breve\iota{}_1^*\cG_{(\xcF)}\xrightarrow{\ \cong\ }\breve\iota{}_2^*\cG_{(\xcF)}\ox\cI_{\varpi_{(\xcF)}}\,,
\qqq
whose curvature $\varpi$ restricts as
\qq\nn
\varpi_{ij}\equiv\varpi_{(\xcF)}\rstr_{Q_{ij}}=\bigl((\id_\Si\x t)^*\om[\Theta_i]-(\id_\Si\x s)^*\om[\Theta_j]+\pr_2^*\om\bigr)\rstr_{\txT Q_{ij}}\,.
\qqq
However,\ the tangent of the component bi-brane worldvolume $Q_{ij}$ is $\im\,\txT\widehat\b{}_{ij}$,\ and so the restrictions vanish identically,
\qq\label{eq:gauge-symm-bib-flat}
\varpi_{ij}\circ\txT\widehat\b{}_{ij}=B_{ij}^*\om[\Theta_i]-\om[\Theta_j]+\widetilde\b{}_{ij}^*\om\equiv 0\,,
\qqq 
see:\ \cite[Lem.\,E.1]{Suszek:2026col},\ or App.\,\ref{app:lagr-anom}.\ Hence,\ the bi-brane is actually flat,
\qq\nn
\Psi_{(\xcF)}\colo\breve\iota{}_1^*\cG_{(\xcF)}\xrightarrow{\ \cong\ }\breve\iota{}_2^*\cG_{(\xcF)}\,.
\qqq

Finally,\ we define the inter-bi-brane.\ We start with the elementary component of valence $3$,\ and then employ a variant of the induction mechanism that underlies the definition of the rigid-symmetry inter-bi-brane \eqref{eq:many-2-one-ibb}.\ As before,\ we restrict our discussion to components $T_{(\xcF)\,n,++\cdots+-}$ corresponding to junctions with $n-1$ in-coming defect lines and a single out-going one,\ leaving implicit the straightforward definition of the remaining components of the gauge-symmetry inter-bi-brane $(T_{(\xcF)\,\bullet},\breve\pi{}_\bullet^{\cdot,\cdot+1},\breve\vep{}_\bullet^{\cdot,\cdot+1},\varphi_{(\xcF)\,\bullet})$ over the stratified worldvolume $T_{(\xcF)}=\bigsqcup_{n\in\bN_{\geq 3}}\,T_{(\xcF)\,n}$.\ Thus,\ for every triple $(i,j,k)\in I^{\x 3}$ labelling an intersection $O_{ijk}\neq\emptyset$,\ we denote (see:\ \eqref{eq:bijk-tilde})
\qq\nn
\widehat\b{}_{ijk}:=\bigl(\pr_1,\widetilde\b{}_{ijk}\bigr)&\colo& O_{ijk}\x M\too O_{ijk}\x\xcG_2\cr\cr  &\colo&(\si,m)\longmapsto\bigl(\si,\b_{ij}(\si)\bigl( t_*\bigl(\b_{jk}(\si)\bigr)(m)\bigr),\b_{jk}(\si)(m)\bigr)\,,
\qqq
and define
\qq\nn
T_{ijk}:=\widehat\b{}_{ijk}\bigl(O_{ijk}\x M\bigr)\,,
\qqq
a fibre bundle over $O_{ijk}$,\ with typical fibre $M$.\ These compose the elementary inter-bi-brane worldvolume
\qq\nn
T_{(\xcF)\,3,++-}:=\bigsqcup_{(i,j,k)\in I^{\x 3}}\,T_{ijk}\,,
\qqq
endowed with the three maps to the bi-brane worldvolume:
\qq\nn
\breve\pi{}_3^{1,2}:=\id_\Si\x d^{(2)}_0\x\pr_{1,2}&\colo& T_{(\xcF)\,3,++-}\too Q_{(\xcF)}\cr\cr  
&\colo& \bigl(\si,\b_{ij}(\si)\bigl( t_*\bigl(\b_{jk}(\si)\bigr)(m)\bigr),\b_{jk}(\si)(m),i,j,k\bigr)\longmapsto\bigl(\si,\b_{ij}(\si)\bigl( t_*\bigl(\b_{jk}(\si)\bigr)(m)\bigr),i,j\bigr)\,,\cr\cr\cr
\breve\pi{}_3^{2,3}:=\id_\Si\x d^{(2)}_2\x\pr_{2,3}&\colo& T_{(\xcF)\,3,++-}\too Q_{(\xcF)}\cr\cr  
&\colo&\bigl(\si,\b_{ij}(\si)\bigl( t_*\bigl(\b_{jk}(\si)\bigr)(m)\bigr),\b_{jk}(\si)(m),i,j,k\bigr)\longmapsto\bigl(\si,\b_{jk}(\si)(m),j,k\bigr)\,,\cr\cr\cr 
\breve\pi{}_3^{3,1}:=\id_\Si\x d^{(2)}_1\x\pr_{1,3}&\colo& T_{(\xcF)\,3,++-}\too Q_{(\xcF)}\cr\cr  
&\colo&\bigl(\si,\b_{ij}(\si)\bigl( t_*\bigl(\b_{jk}(\si)\bigr)(m)\bigr),\b_{jk}(\si)(m),i,j,k\bigr)\longmapsto\bigl(\si,\b_{ik}(\si)(m),i,k\bigr)\,.
\qqq
Note that we have used the 1-cocycle condition \eqref{eq:tra-cocyc} when writing out $\breve\pi{}_3^{3,1}$.\ On $T_{(\xcF)\,3,++-}$,\ we put the fusion 2-isomorphism
\qq\nn
\varphi_{(\xcF)\,3,++-}:=\breve\pi{}_3^{3,1\,*}d_{\Psi_{(\xcF)}2}\bullet\g_2\,,
\qqq
with
\qq\nn
d_{\Psi_{(\xcF)}2}=\pr_2^*d_{\Psi_{(\xcF)}}\ox\id\,,\qquad\qquad\g_2=\pr_2^*\g\ox\id\,.
\qqq
The above structure generalises straightforwardly to arbitrary valence:\ The point of departure is the definition,\ for every $n$-tuple $(i_1,i_2,\ldots,i_n)\in I^{\x n}$ such that $O_{i_1 i_2\ldots i_n}\neq\emptyset$,\ of the map   
\qq\nn
&&\widehat\b{}_{i_1 i_2\cdots i_n}:=\bigl(\pr_1,\widetilde\b{}_{i_1 i_2}\circ B_{i_2 i_n},\widetilde\b{}_{i_2 i_3}\circ B_{i_3 i_n},\ldots,\widetilde\b{}_{i_{n-2} i_{n-1}}\circ B_{i_{n-1} i_n},\widetilde\b{}_{i_{n-1}i_n}\bigr)\colo O_{i_1 i_2\cdots i_n}\x M\too O_{i_1 i_2\cdots i_n}\x\xcG_{n-1}\,,
\qqq
and of the corresponding manifold
\qq\nn
T_{i_1 i_2\cdots i_n}:=\widehat\b{}_{i_1 i_2\cdots i_n}\bigl(O_{i_1 i_2\cdots i_n}\x M\bigr)\,,
\qqq
which enters the definition of the in ter-bi-brane worldvolume
\qq\nn
T_{(\xcF)\,n,++\ldots+-}:=\bigsqcup_{(i_1,i_2,\ldots,i_n)\in I^{\x n}}\,T_{i_1 i_2\cdots i_n}\,.
\qqq
The latter comes with maps
\qq\nn
\breve\pi_n^{k,k+1}:=\id_\Si\x\pi_n^{k,k+1}\x\pr_{k,k+1}\colo T_{(\xcF)\,n,++\ldots+-}\too Q_{(\xcF)}\,,\quad k\in\bZ/n\bZ\,,
\qqq
expressed in terms of their prototypes $\pi_n^{k,k+1},\ k-1\in\bZ/n\bZ$ from \eqref{eq:rig-pi-n},\ and the canonical projections $\pr_{l,l+1}\colo I^{\x n}\to I^{\x 2},\ (i_1,i_2,\ldots,i_n)\mapsto(i_l,i_{l+1}),\ l\nequiv n$ and $\pr_{n,1}\colo I^{\x n}\to I^{\x 2},\ (i_1,i_2,\ldots,i_n)\mapsto(i_1,i_n)$.\ Besides these,\ we have maps
\qq\nn
\mathring{\pi}_n^{1,2}=\id_\Si\x\widehat\pi{}_n^{1,2}\x\pr_{1,2,3}\,,\qquad\mathring{\pi}_n^{12\cdots l,l+1}=\id_\Si\x\widehat\pi{}_n^{12\cdots l,l+1}\x\pr_{1,l+1,l+2}\,,\quad l\in\ovl{2,n-2}
\qqq
from $T_{(\xcF)\,n,++\ldots+-}$ to $T_{(\xcF)\,3,++-}$,\ written in terms of the prototypes from \eqref{eq:rig-hatpi-n},\ and the canonical projections $\pr_{1,k+1,k+2}\colo I^{\x n}\to I^{\x 3},\ (i_1,i_2,\ldots,i_n)\mapsto(i_1,i_{k+1},i_{k+2}),\ k\in\ovl{1,n-2}$.\ On $T_{(\xcF)\,n,++\ldots+-}$,\ we now  set
\qq
\varphi_{(\xcF)\,n,++\cdots+-}=\breve\pi{}_n^{n,1\,*}d_{\Psi_{(\xcF)}2}\bullet\bigl(\id\circ\mathring{\pi}_n^{12\cdots n-2,n-1\,*}\g_2\bigr)\bullet\bigl(\id\circ\mathring{\pi}_n^{12\cdots n-3,n-2\,*}\g_2\bigr)\bullet\cdots\bullet\bigl(\id\circ\mathring{\pi}_n^{12,3\,*}\g_2\bigr)\bullet\bigl(\id\circ\mathring{\pi}_n^{1,2\,*}\g_2\bigr)\,.\cr 
\label{eq:fus-2-iso-n}
\qqq

With the topological string background for the ctP $\si$-model in the presence of the gauge-symmetry defect thus set up,\ we may finally describe field configurations 
\qq\nn
\xi\colo\Si\too M_{(\xcF)}\sqcup Q_{(\xcF)}\sqcup T_{(\xcF)}
\qqq
for which the $\si$-model is defined:\ As the component of the bulk $\si$-model field $\xi\rstr_{\Si\setminus\G_{(\xcF)}}$ over the domain (`phase') $\Si_i\subset O_i$,\ we take the corresponding restriction
\qq\nn
\xi\rstr_{\Si_i}:=\xcF\t_i\circ\phi\rstr_{\Si_i}\colo\Si_i\too\Si_i\x M\subset M_i
\qqq
of the local presentation $\xcF\t_i\circ\phi\equiv(\id_\Si,\phi_i)\colo O_i\to O_i\x M$ of a field $\phi\in\G(\xcF)$ of the coaugmented Poisson $\si$-model from Sec.\,\ref{sub:coaug-PSM-def}.

Next,\ as the component of the defect-line field $\xi\rstr_{\G_{(\xcF)}\setminus\Vgt_{\G_{(\xcF)}}}$ over the oriented defect line $\ell_{i,j}\subset O_{ij}$ separating the domains $\Si_i$ (on its left) and $\Si_j$ (on its right),\ we take the corresponding restriction
\qq\nn
\xi\rstr_{\ell_{i,j}}:=\widehat\b{}_{ij}\circ\xcF\t_j\circ\phi\rstr_{\ell_{i,j}}\colo\ell_{i,j}\too Q_{ij}
\qqq
of the map $\widehat\b{}_{ij}\circ\xcF\t_j\circ\phi\colo O_{ij}\to Q_{ij}$.\ The latter choice is consistent with the previous one in virtue of \eqref{eq:loc-field-trans},\ as we obtain,\ at every point $\si\in\ell_{i,j}\setminus\Vgt_{\G_{(\xcF)}}$,
\qq\nn
&(\breve\iota{}_1\circ\xi)(\si)=\bigl(\si,t_*\bigl(\b_{ij}(\si)\bigr)\bigl(\phi_j(\si)\bigr)\bigr)=\bigl(\si,\phi_i(\si)\bigr)\equiv\bigl(\xcF\t_i\circ\phi\bigr)(\si)\,,&\cr\cr 
&(\breve\iota{}_2\circ\xi)(\si)=\bigl(\si,s\bigl(\b_{ij}(\si)\bigr)\bigl(\phi_j(\si)\bigr)\bigr)=\bigl(\si,\phi_j(\si)\bigr)\equiv\bigl(\xcF\t_j\circ\phi\bigr)(\si)\,.&
\qqq
Here,\ the map $\xcF\t_i\circ\phi\rstr_{\Si_i\cup\ell_{i,j}\setminus\Vgt_{\G_{(\xcF)}}}$ (resp.\ $\xcF\t_j\circ\phi\rstr_{\Si_j\cup\ell_{i,j}\setminus\Vgt_{\G_{(\xcF)}}}$) is the obvious smooth local extension $\xi_{|1}\rstr_{\Si_j\cup\ell_{i,j}\setminus\Vgt_{\G_{(\xcF)}}}$ (resp.\ $\xi_{|2}\rstr_{\Si_j\cup\ell_{i,j}\setminus\Vgt_{\G_{(\xcF)}}}$) of the bulk field to the interior of the defect line $\ell_{i,j}$.

Finally,\ given a defect junction $\jmath_{i_1,i_2,\ldots,i_n}\in\Vgt_{\G_{(\xcF)}}$ at which the domains $\Si_{i_k},\ k\in\ovl{1,n}$ join---being ordered cyclically counterclockwise around the junction as reflected by their labelling,\ and separted pairwise by the defect lines $\ell_{k,k+1},\ k-1\in\bZ/n\bZ$ accordingly---we define the corresponding restriction of the defect-junction field $\xi\rstr_{\Vgt_{\G_{(\xcF)}}}$ as
\qq\nn
\xi\rstr_{\{\jmath_{i_1,i_2,\ldots,i_n}\}}:=\widehat\b{}_{i_1 i_2\cdots i_n}:\circ\xcF\t_{i_n}\circ\phi\rstr_{\{\jmath_{i_1,i_2,\ldots,i_n}\}}\colo\{\jmath_{i_1,i_2,\ldots,i_n}\}\too T_{(\xcF)\,n}\,.
\qqq
We readily check the compatibility of this choice with the previous ones for the distinguished case illustrated in Fig.\,\ref{fig:many-2-one-junct}.\ We find,\ at $\jmath\equiv\jmath_{i_1,i_2,\ldots,i_n}$ and for all $l\in\ovl{1,n-1}$,
\qq\nn
\bigl(\breve\pi{}_n^{l,l+1}\circ\xi\bigr)(\jmath)&=&\bigl(\jmath,\b_{i_l i_{l+1}}(\jmath)\bigl(t_*\bigl(\b_{i_{l+1} i_n}(\jmath)\bigr)\bigl(\phi_{i_n}(\jmath)\bigr)\bigr)\bigr)=\bigl(\jmath,\b_{i_l i_{l+1}}(\jmath)\bigl(\phi_{i_{l+1}}(\jmath)\bigr)\bigr)\cr\cr &\equiv&\bigl(\widehat\b{}_{i_l i_{l+1}}\circ\xcF\t_{i_{l+1}}\circ\phi\bigr)(\jmath)\,,\cr\cr\cr 
\bigl(\breve\pi{}_n^{n,1}\circ\xi\bigr)(\jmath)&=&\bigl(\jmath,\b_{i_1 i_2}(\jmath)\bigl(t_*\bigl(\b_{i_2 i_n}(\jmath)\bigr)\bigl(\phi_{i_n}(\jmath)\bigr)\bigr).\b_{i_2 i_3}(\jmath)\bigl(t_*\bigl(\b_{i_3 i_n}(\jmath)\bigr)\bigl(\phi_{i_n}(\jmath)\bigr)\bigr).\cdots\cr\cr 
&&\hspace{.5cm}.\b_{i_{n-2} i_{n-1}}(\jmath)\bigl(t_*\bigl(\b_{i_{n-1} i_n}(\jmath)\bigr)\bigl(\phi_{i_n}(\jmath)\bigr)\bigr).\b_{i_{n-1} i_n}(\jmath)\bigl(\phi_{i_n}(\jmath)\bigr)\bigr)\cr\cr 
&=&\bigl(\jmath,\b_{i_1 i_n}(\jmath)\bigl(\phi_{i_n}(\jmath)\bigr)\bigr)\equiv\bigl(\widehat\b{}_{i_1 i_n}\circ\xcF\t_{i_n}\circ\phi\bigr)(\jmath)\,,
\qqq
and so identify the maps $\widehat\b{}_{i_l i_{l+1}}\circ\xcF\t_{i_{l+1}}\circ\varphi\rstr_{\ell_{l,l+1}\cup\{\jmath_{i_1,i_2,\ldots,i_n}\}}$ as the smooth local extensions $\xi_{l,l+1}\rstr_{\ell_{l,l+1}\cup\{\jmath_{i_1,i_2,\ldots,i_n}\}}$ of the defect-line field to the half-capped defect line $\ell_{l,l+1}\cup\{\jmath_{i_1,i_2,\ldots,i_n}\}$,\ and---similarly---the maps $\widehat\b{}_{i_1 i_n}\circ\xcF\t_{i_n}\circ\varphi\rstr_{\ell_{1,n}\cup\{\jmath_{i_1,i_2,\ldots,i_n}\}}$ as the smooth local extensions $\xi_{n,1}\rstr_{\ell_{n,1}\cup\{\jmath_{i_1,i_2,\ldots,i_n}\}}$ of the defect-line field to the half-capped defect line $\ell_{n,1}\cup\{\jmath_{i_1,i_2,\ldots,i_n}\}$.\ The definition of the {\bf ctP $\si$-model for the gauge network-field configurations} $(\xi\,|\,\G)$ ({\bf GNP}) is now obtained through a specialisation of \eqref{eq:PAG-sigmodef},
\qq\label{eq:GNP-sigmodef}
\cA^{GNP}_{\rm DF}[(\xi\,|\,\G)]=\Hol_{(\cG_{(\xcF)},\Psi_{(\xcF)},\varphi_{(\xcF)\bullet})}(\xi\,|\,\G)\,.
\qqq

\subsection{Topologicality of the gauge-symmetry defect}\label{sub :gau-sym-def-top}

The introduction of defects into the framework of (classical\footnote{For the corresponding quantum field-theoretic discussion,\ the Reader is referred to \cite{Runkel:2008gr},\ where 2$d$ CFT with defect networks was defined.}) field theory gives rise to a number of novel problems,\ whose resolution is a prerequisite for a consistent application of the constructs reviewed in the opening paragraphs of the present Section \ref{sec:gau-sigmod} in the modelling of multi-phase resp.\ twisted-sector mechanics of charged (topological) loops through the principle of least action for the GNP $\si$-model for defect-network configurations,\ and the associated prequantisation in the sense of \cite{Gawedzki:1987ak,Suszek:2011hg}.\ Two of these problems become particularly prominent in the present context,\ in which we seek to model the groupoidal gauge symmetries of the coaugmented Poisson $\si$-model in terms of certain distinguished defect networks.\ The first one concerns the defect counterpart of the standard boundary condition (the latter being subsumed by the construction of the field theory with arbitrary defects,\ see:\ \cite[Secs.\,2.3 \& 2.4]{Runkel:2008gr}),\ to be obeyed by fields of the GNP $\si$-model at the defect network $\G$.\ The condition can,\ accordingly,\ be regarded as restricting the class of admissible field configurations in which the principle of least action is employed.\ The second one asks about the behaviour of the DF functional \eqref{eq:GNP-sigmodef} under homotopies of the embedding $\G\emb\Si$ of the defect quiver in a given worldsheet $\Si$,\ including translations of defect junctions along defect lines.\ There is an intimately related question concerning the compatibility of the definition of the DF functional with retractions of edges of the defect quiver which may result in changes of the power of the set of vertices of $\G$ and its stratification by the vertex valence.\ For the latter question to be well-posed,\ a homotopy of $\G$ has to be accompanied by a field redefinition allowed to be non-identity \emph{exclusively} inside the region---to be referred to as {\bf homotopy support} in what follows---in which the homotopy itself,\ when modelled as an ambient homotopy of $\Si$ with the embedded graph $\G$,\ differs from the identity map $\id_\Si$.\ This,\ then,\ entangles the three questions,\ because a redefined field has to satisfy the defect condition at each `instant' of the homotopy `time'.\ Such considerations naturally distinguish flat directions in the defect-position moduli space of the GNP $\si$-model along which the value of the DF amplitude remains constant---these are known as {\bf topological defects},\ see:\ \cite[Sec.\,2.9]{Runkel:2008gr} for details in the setting of the 2$d$ PAG $\si$-model.\ Their relevance in that setting stems from the fact that they intertwine the respective realisations of the full (bi-chiral) conformal (Virasoro) algebra in the `phases' of the field theory on either side of the defect,\ and thus correspond to fully fledged (invertible) prequantisable symmetries of the {\em conformal} field theory under consideration,\ see:\ \cite[Secs.\,3.2 \& 3.3]{Runkel:2008gr}. This fact underlies their direct inclusion in the very \emph{definition} of (symmetry-)orbifold topological quantum field theory and functorial quantisation of (symmetry-)orbifold 2$d$ conformal field theory \cite{Davydov:2011kb},\ which arises as a conceptually straightforward---albeit technically involved---categorial abstraction of the resummation of contributions of inequivalent associated/shadow bundles to the complete amplitude of the field theory with the corresponding symmetry model \emph{gauged},\ see:\ \cite[Sec.\,4.4]{Gawedzki:2010rn} (and the references given {\it ibidem},\ in particular:\ \cite{Hori:1994nc}) for a concrete example,\ and \cite[Sec.\,2]{Suszek:2026col} for a geometric/topological abstraction.\ Thus,\ it is the specialised question about the topologicality of the defect that comes to the fore in our case.\ All three questions mentioned above were addressed explicitly in \cite{Runkel:2008gr},\ and answers were found,\ on which our subsequent analysis can be built---these shall be recalled next.

As for the question concerning the defect counterpart of the boundary condition,\ the answer given in \cite[Sec.\,2.4 \& 2.7]{Runkel:2008gr} splits into parts:\ one for defect lines and another one for defect junctions.\ The former takes the form of an identity---dubbed the {\bf Defect Gluing Condition} ({\bf DGC}) in \cite{Runkel:2008gr}---which in the dynamical setting of the 2$d$ PAG $\si$-model relates the discontinuity of the (one-sided limits of) normal derivatives of the extended bulk field $\xi_{|\a}\rstr_{\Si\setminus\G},\ \a\in\{1,2\}$ on either side of the defect line $\ell\in\Egt_\G$ (the trans-defect momentum jump) to the value attained by the bi-brane curvature $\varpi$ on the tangent of the defect-line field $\xi\rstr_\ell$,\ but in the purely topological setting reduces to the equality:
\qq\label{eq:DGC}\qquad
\varpi\bigl(T_x\xi(\widehat t),\cdot\bigr)=\brd0_{T^*Q}\bigl(\xi (x)\bigr)\,,\quad x\in\ell\,,
\qqq
written in terms of the zero section $\brd0_{T^*Q}\colo Q\to T^*Q$,\ and $\widehat t\in\txT_x\ell$.\ The latter,\ baptised the {\bf Defect-Junction Identity},\ reads
\qq\nn
\sum_{k=1}^n\,\varpi_n^{k,k+1}=0\,,\quad n\in\bN_{\geq 3}\,,
\qqq
and so it reproduces the necessary condition for the existence of the relevant fusion 2-isomorphism $\varphi_n$.\ In particular,\ for the groupoidal-symmetry string background from Ex.\,\ref{eg:grpd-rsymm-bckgrnd} with $(M,Q,T_{3,++-})=(\xcG_0,\xcG_1,\xcG_2)$,\ it enforces the multiplicativity of the bi-brane curvature,\ see:\ (the comment below) \eqref{eq:gamma-2-iso}.\ Consequently,\ we shall not return to this condition in our subsequent considerations.

\begin{figure}[htb]
	\centering
	\includegraphics[width=0.6\textwidth]{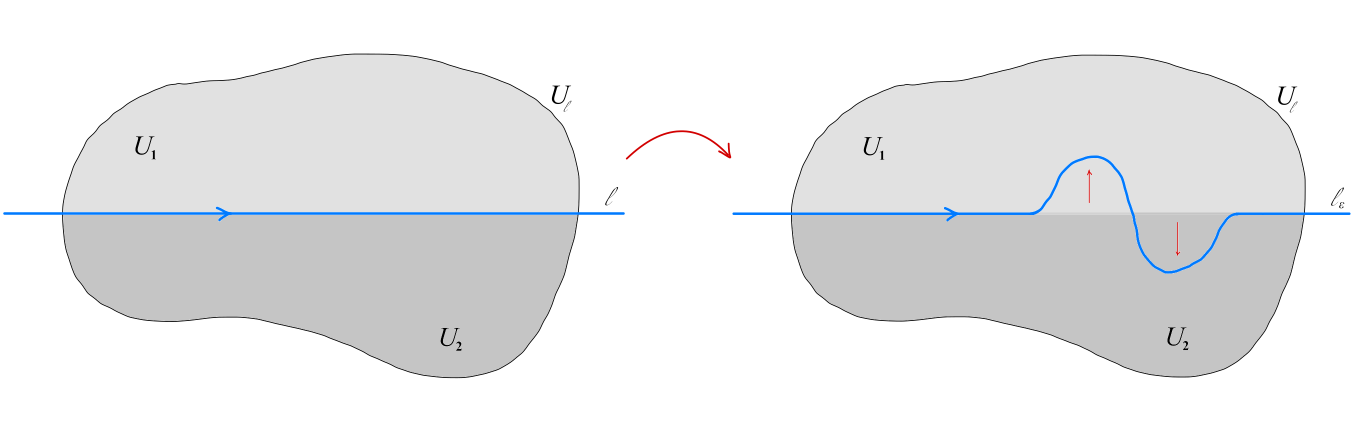}
	\caption{A homotopy deformation $\ell\to\ell_\vep$ of an edge $\ell$ of a defect quiver within its partial tubular neighbourhood $U_\ell=U_1\cup\widetilde\ell\cup U_2$,\ with $\widetilde\ell\subset\ell$.} 
	\label{fig:def-ho-lin}
\end{figure}

\begin{figure}[htb]
	\centering
	\includegraphics[width=0.6\textwidth]{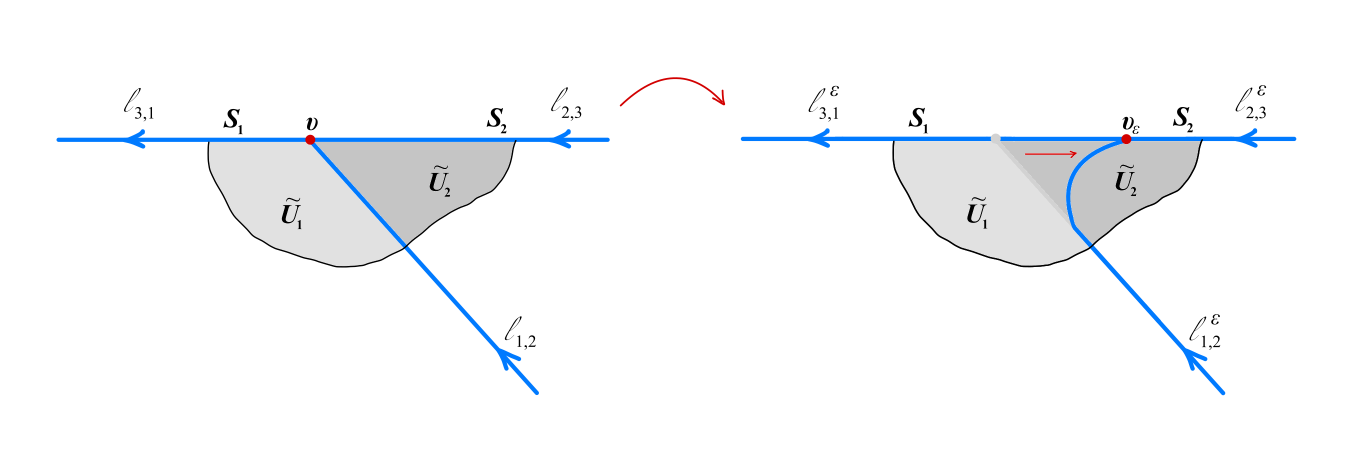}
	\caption{A homotopy move $v\to v_\vep$ of a 3-valent vertex $v$ of a defect quiver within its segment neighbourhood $S_1\cup\{v\}\cup S_2$ in the defect's edge $\ell_{2,3}\ni v$.\ The move entails a homotopy deformation $\ell_{1,2}\to\ell_{1,2}^\vep$ of the other edge $\ell_{1,2}\ni v$ within a partial tubular neighbourhood $\widetilde U{}_{\ell_{1,2}}=\widetilde U{}_1\cup\widetilde\ell{}_{1,2}\cup\widetilde U{}_2$ of $\ell_{1,2}$,\ with $\widetilde\ell{}_{1,2}\subset\ell_{1,2}$.} 
	\label{fig:def-ho-vert}
\end{figure}

As for the question about the topologicality of the defect,\ an important class of topological defects was identified in \cite[Sec.\,2.9]{Runkel:2008gr}:\ {\bf extendible defects},\ which we shall discuss next.\ As the name suggests,\ these are defects in the poly-phase $\si$-model with the following properties: 
\bit
\item defect fields $\xi\rstr_{\G\setminus\Vgt_\G}\colo\G\setminus\Vgt_\G\to Q$ and $\xi\rstr_{\Vgt_\G^n}\colo\Vgt_\G^n\to T_n,\ n\in\bN_{\geq 3}$ admit smooth \emph{extensions} to homotopy paths of their respective domains within homotopy supports for arbitrary homotopies;
\item the extensions determine network-field configurations at every `instant' of the homotopy `time';
\item the thus obtained network-field configurations satisfy the DGC \eqref{eq:DGC}.
\eit
Below,\ we shall unwrap the above concise characterisation of extendible defects for two elementary types of a defect-quiver homotopy in the topological setting:
\bit
\item[(i)] a homotopy deformation of an edge $\ell\in\Egt_\G$ which does not affect $\Vgt_\G$ (see:\ Fig.\,\ref{fig:def-ho-lin}):\ given a network-field configuration $\xi\colo\Si\to\cM$,\ we have an extension $\widehat\xi\colo U_\ell\to Q$ of $\xi\rstr_\ell\colo\ell\to Q$ to a (partial) tubular neighbourhood $U_\ell=U_1\cup \ell\cup U_2$ of $\ell$---with $U_\a\subset U_\ell\setminus\ell$ to the left ($\a=1$) resp.\ right ($\a=2$) of the edge,\ as seen in the direction of the orientation of $\ell$---which coincides with $\xi$ at $\ell$,\  $\widehat\xi\rstr_\ell=\xi\rstr_\ell$,\ and reproduces $\xi$ on either side of $\ell$ in $U_\ell$ as $\iota_\a\circ\widehat\xi\rstr_{U_\a}=\xi\rstr_{U_\a}$,\ and which satisfies the identity
\qq\nn
\varpi\bigl(T_x\widehat\xi(v),\cdot\bigr)=\brd0_{T^*Q}\bigl(\widehat\xi(x)\bigr)\,,
\qqq
written for an arbitrary vector $v\in T_x\Si$ at $x\in U_\ell$;
\item[(ii)] a homotopy move of a tri-valent defect junction $v\in\Vgt_\G$ along two of the edges of $\G$ joining at $v$,\ in which the remaining edge---say,\ $\ell_{1,2}$---is homotopically `dragged along' by $v$ (see:\ Fig.\,\ref{fig:def-ho-vert}):\ given a network-field configuration $\xi\colo\Si\to\cM$,\ we have \bit
\item an extension $\widehat\xi{}_v\colo S_v\to T_3$ of $\xi\rstr_v\colo\{v\}\to T_3$ to a segment neighbourhood $S_v=S_1\cup\{v\}\cup S_2\subset\G$ of $v$---with $S_\a\subset S_v\setminus\{v\}$ to the left ($\a=1$) resp.\ right ($\a=2$) of the vertex,\ as seen along the only edge $\ell_{1,2}\in\Egt_\G$ deformed by the homotopy,\ looking towards $v$---which coincides with $\xi$ at $v$,\ $\widehat\xi{}_v\rstr_v=\xi\rstr_v$,\ and reproduces $\xi$ on either side of $v$ in $S_v$ as $\pi_3^{3,1}\circ\widehat\xi{}_v\rstr_{S_1}=\xi\rstr_{S_1}$ and $\pi_3^{2,3}\circ\widehat\xi{}_v\rstr_{S_2}=\xi\rstr_{S_2 }$
\item an extension $\widehat\xi{}_{\ell_{1,2}}\colo\widetilde U{}_{\ell_{1,2}}\to Q$ of $\xi\rstr_{\ell_{1,2}}\colo\ell_{1,2}\to Q$ to a (partial) tubular neighbourhood $\widetilde U{}_{\ell_{1,2}}=\widetilde U{}_1\cup\ell_{1,2}\cup\widetilde U{}_2$ of $\ell_{1,2}$ containing $S_v$---with $\widetilde U{}_\a\subset\widetilde U{}_{\ell_{1,2}}\setminus\ell_{1,2}$ to the left ($\a=1$) resp.\ right ($\a=2$) of the edge,\ $\widetilde U{}_\a\supset S_\a$---which coincides with $\xi$ at $\ell_{1,2}$,\ $\widehat\xi{}_{\ell_{1,2}}\rstr_{\ell_{1,2}}=\xi\rstr_{\ell_{1,2}}$,\ reproduces $\xi$ on either side of $\ell_{1,2}$ as $\iota_\a^{\vep_3^{1,2}}\circ\widehat\xi{}_{\ell_{1,2}}\rstr_{\widetilde U{}_\a}=\xi\rstr_{\widetilde U{}_\a}$,\ and is reproduced by $\widehat\xi{}_v$ on $S_v$ as $\pi_3^{1,2}\circ\widehat\xi{}_v=\widehat\xi{}_{\ell_{1,2}}\rstr_{S_v}$,\smallskip
\eit
which satisfies the identities: 
\qq\nn
\varpi\bigl(T_x\widehat\xi{}_{\ell_{1,2}}(v),\cdot\bigr)=\brd0_{T^*Q}\bigl(\widehat\xi{}_{\ell_{1,2}}(x)\bigr)
\qqq
at $x\in\widetilde U{}_{\ell_{1,2}}$ and for $v,v^\perp\in T_x\Si$ as before,\ and 
\qq\nn
\varpi\bigl(T_{y_1}\bigl(\pi_3^{2,3}\circ\widehat\xi{}_v\bigr)\bigl(\widehat t{}_{2,3}\bigr),\cdot\bigr)=\brd0_{T^*Q}\bigl(\bigl(\pi_3^{2,3}\circ\widehat\xi{}_v\bigr)(y_1)\bigr)
\qqq
alongside
\qq\nn
\varpi\bigl(T_{y_2}\bigl(\pi_3^{3,1}\circ\widehat\xi{}_v\bigr)\bigl(\widehat t{}_{3,1}\bigr),\cdot\bigr)=\brd0_{T^*Q}\bigl(\bigl(\pi_3^{3,1}\circ\widehat\xi{}_v\bigr)(y_2)\bigr)
\qqq
at $y_\a\in S_\a,\ \a\in\{1,2\}$ and for an arbitrary vector $\widehat t{}_{2,3}$ (resp.\ $\widehat t{}_{3,1}$) tangent to $\ell_{2,3}$ (resp.\ $\ell_{3,1}$).
\eit

It is now entirely obvious that {\bf {\em the gauge-symmetry defect of the GNP $\si$-model is extendible,\ and hence topological}}.\ Indeed,\ the component defect-line fields $\xi\rstr_{\ell_{i,j}}$ and defect-junction fields $\xi\rstr_{\{\jmath_{i_1,i_2,\ldots,i_n}\}}$ are defined in terms of smooth maps with open neigbourhoods $O_{ij}\supset\ell_{i,j}$ and $O_{i_1 i_2\cdots i_n}$ of the respective components of the defect quiver as domains,\ and so are canonically extendible in terms of these maps.\ Moreover,\ the bi-brane $(Q_{(\xcF)},\breve\iota{}_1,\breve\iota{}_2,\varpi_{(\xcF)},\Psi_{(\xcF)})$ of the gauge-symmetry string background is flat,\ see:\ \eqref{eq:gauge-symm-bib-flat},\ which implies that the topological DGCs written out above are trivially satisfied by these extensions.\ This,\ though,\ is not the only peculiarity of the Poisson gauge-symmetry defect.\medskip

\begin{figure}[htb]
\centering
\includegraphics[width=0.6\textwidth]{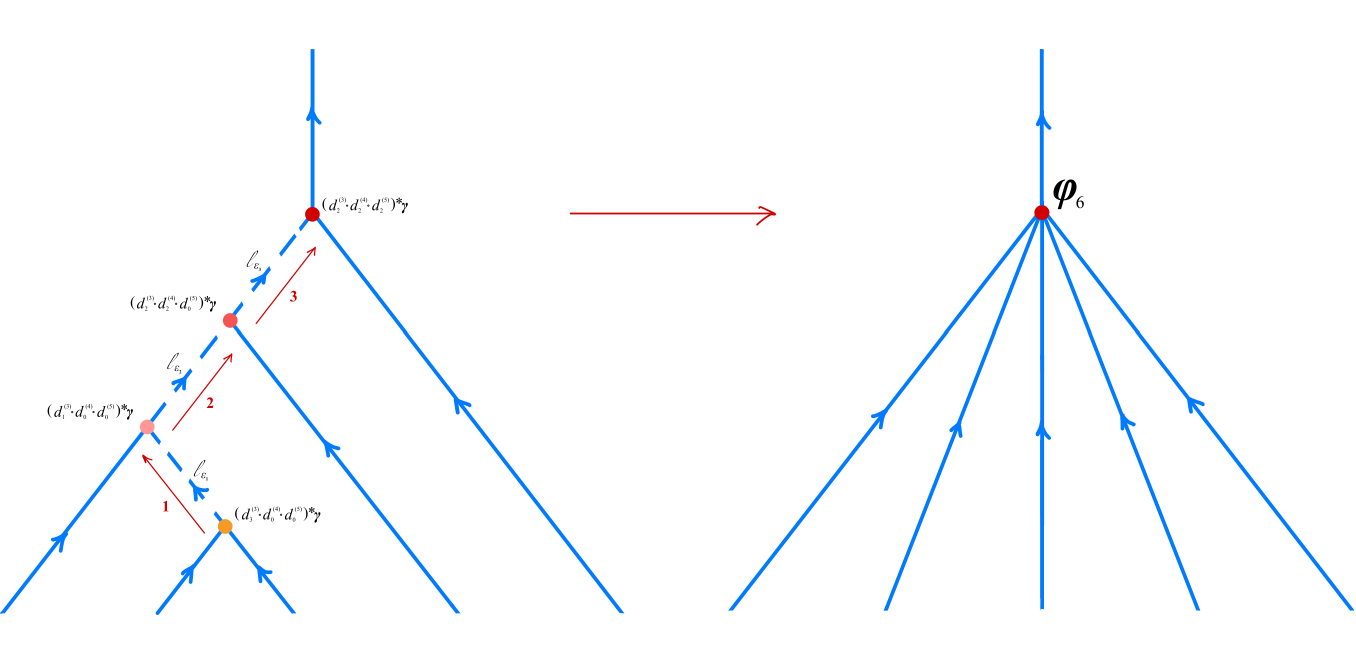}
\caption{Induction of a fusion 2-isomorphism $\varphi_6$ over a 5-to-1 defect junction from elementary fusion 2-isomorphisms $\g$ of the $\xcG$-equivariant structure over 2-to-1 defect junctions through a sequence of vertex homotopy moves.\ The latter represent vertical composition of the component 2-isomorphisms.} 
\label{fig:fu2iso-ind-5to1}
\end{figure}

\begin{figure}[htb]
	\centering
	\includegraphics[width=0.65\textwidth]{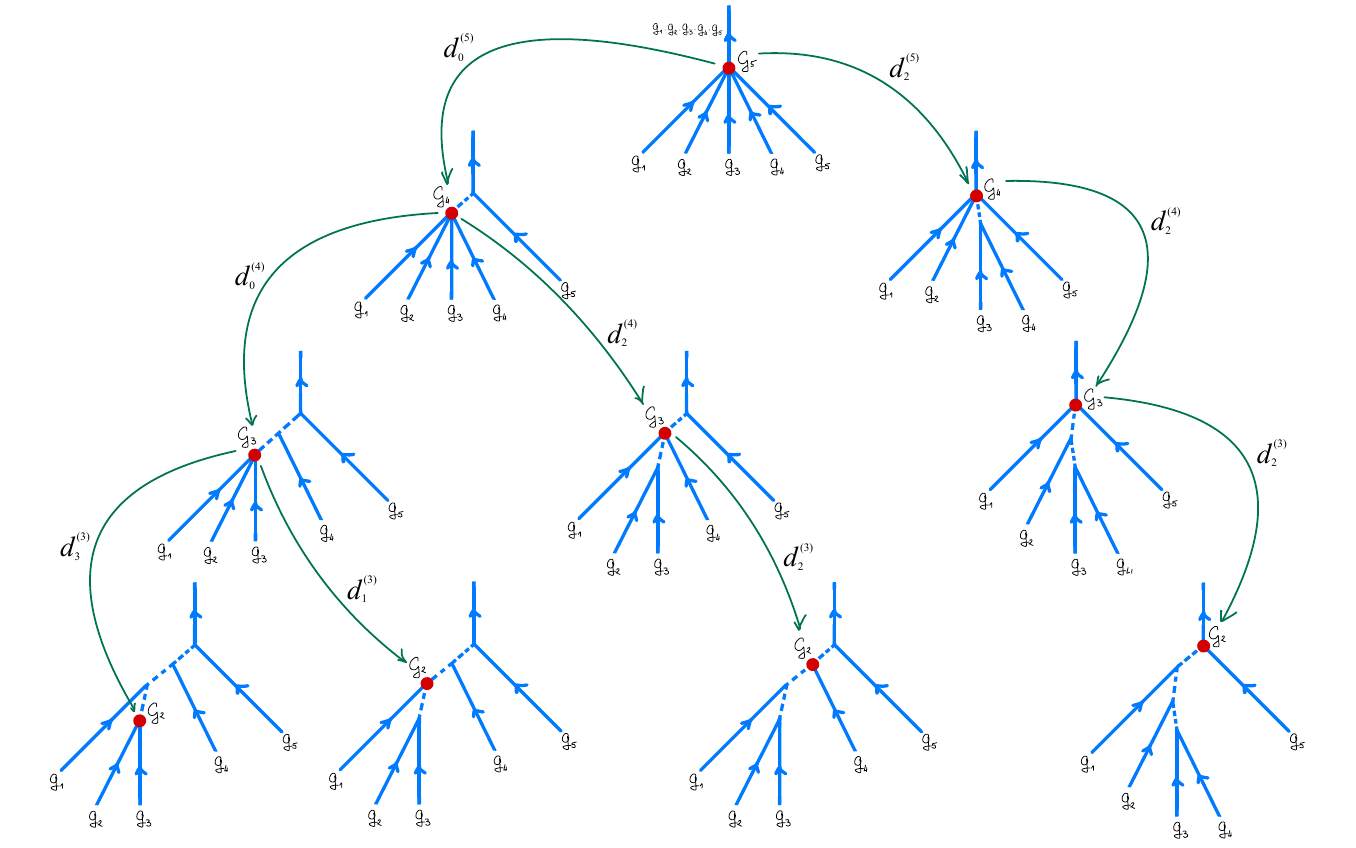}
	\caption{The four inequivalent binary-tree resolutions of the 6-valent vertex used in the induction procedure depicted in Fig.\,\ref{fig:fu2iso-ind-5to1}.} 
	\label{fig:bintree-res-val6}
\end{figure}

We conclude our study with the discussion of the compatibility of the definition of the GNP $\si$-model for arbitrary defect quivers,\ {\it i.e.},\ in its dependence on the second argument in $(\xi\,|\,\G)$,\ with retractions of edges of the defect quiver which reduce the number of its vertices and,\ in this process,\ produce vertices of valence higher than $3$ from the elementary 3-valent ones.\ Let us first explain carefully---following closely the logic of the original argument in \cite[Sec.\,2.9]{Runkel:2008gr}---the meaning of `compatibility' in this situation.\ First of all,\ note that the topologicality of the gauge-symmetry defect,\ established above,\ implies that we may obtain---without changing the value of the DF amplitude along the way---a defect vertex of arbitrary valence $n>3$ located at $\jmath_{i_1,i_2,\ldots,i_n}\in O_{i_1 i_2\cdots i_n}$ in a sequence of limiting elementary vertex-homotopy moves,\ of the type just considered,\ in each of which a 3-valent junction,\ separating phases $\Si_k,\ \Si_{k+1}$ and $\Si_{k+2}$ that converge at its location $\jmath^{k,k+1,k+2}_{i_1,i_2,\ldots,i_n}\in O_{i_1 i_2\cdots i_n}$,\ and linked by an arbitrarily short outgoing defect line $\ell_\vep\subset O_{i_1 i_2\cdots i_n}$ to a defect junction of valence $<n-2$ located at $\jmath_{i_1,i_2,\ldots,i_n}$,\ is shrunk to the latter point,\ $\ell_\vep\to\jmath_{i_1,i_2,\ldots,i_n}$.\ An example of such a multiple-retraction procedure is given in Fig.\,\ref{fig:fu2iso-ind-5to1}.\ 
It produces a particular composite sheaf-theoretic contribution to the DF amplitude which is pulled back,\ along $\xi$,\ to the final retract $\jmath_{i_1,i_2,\ldots,i_n}$ from the inter-bi-brane worldvolume $T_n$ and which represents the BD datum of a \emph{specific} 2-isomorphism $\Phi_{(\xcF)\,n}^{n,1}\circ\Phi_{(\xcF)\,n}^{n-1,n}\circ\cdots\circ\Phi_{(\xcF)\,n}^{1,2}\overset{\cong}{\Longrightarrow}\id$.\ It is clear,\ from the construction,\ that one of the sequences reproduces in this manner---through a \emph{complete} resolution of the $n$-valent vertex into a binary tree composed by 3-valent vertices for pairs of neighbouring defect lines\footnote{This self-explanatory concept was formalised in \cite{Suszek:2022CSimpl}.},\ as exemplified in Fig.\,\ref{fig:bintree-res-val6}---the contribution of the $n$-valent fusion 2-isomorphism \eqref{eq:fus-2-iso-n} from our definition of the gauge-symmetry topological string background.\ Thus,\ on one hand,\ the latter definition can be \emph{motivated} by the {\bf binary induction} procedure.\ On the other hand,\ for the logical consistency of this motivation,\ we must ensure that the result of the induction does \emph{not} depend on the choice of the complete resolution from which we generate the composite BD datum pulled back to a given defect junction---and this is exactly the formal meaning of the aforementioned `compatibility'.\ With this understanding,\ we may pass to a direct demonstration of the stated property of our construction.

\begin{figure}[htb]
	\centering
	\includegraphics[width=0.75\textwidth]{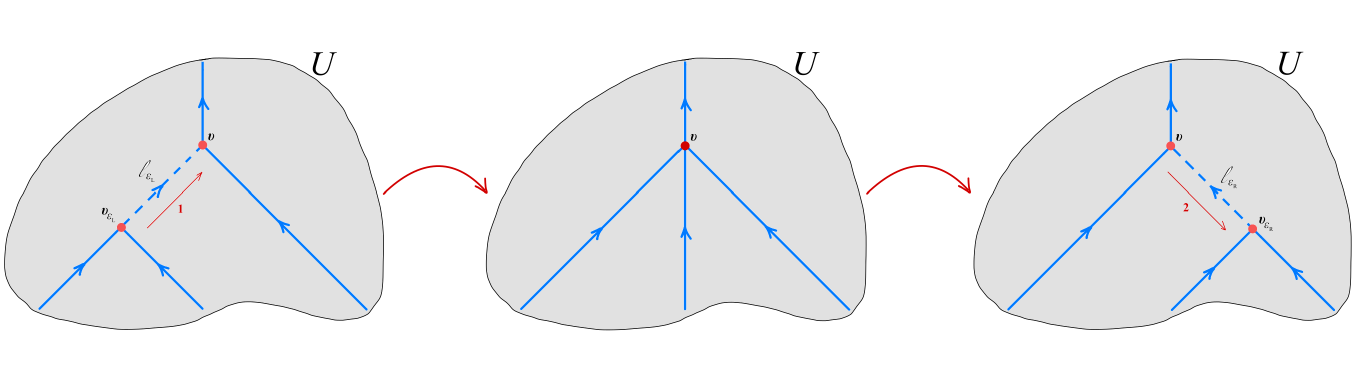}
	\caption{The associator homotopy move for a 4-valent vertex $v$ of the defect quiver,\ in which a neighbouring 3-valent vertex $v_{\vep_{\rm L}}$ is retracted,\ along an edge $\ell_{\vep_{\rm L}}$ of the defect quiver,\ to another 3-valent vertex,\ located at $v$,\ and the ensuing 4-valent vertex at $v$ is subsequently resolved,\ along an edge $\ell_{\vep_{\rm R}}$,\ into a pair of 3-valent vertices:\ one at $v$ and another one at a neighbouring point $v_{\vep_{\rm R}}$.\ The support of the homotopy is contained in a neighbourhood $U$ of the vertex.} 
	\label{fig:assoc-mov}
\end{figure}

Let us begin by providing a quantitative measure of the ambiguity in the definition of a 2-isomorphism $\Phi_{(\xcF)\,n}^{n,1}\circ\Phi_{(\xcF)\,n}^{n-1,n}\circ\cdots\circ\Phi_{(\xcF)\,n}^{1,2}\overset{\cong}{\Longrightarrow}\id$ in the convenient BD-cohomological language of Prop.\,\ref{prop:1iso-H1}.\ As follows from the last part of this proposition,\ captured by \eqref{eq:diff-2-iso-loconst},\ the ambiguity at a given defect junction is a (multiplicative) constant phase in general,\ and it is clear that phase differences between any two 2-isomorphisms induced from complete binary resolutions are products of phase factors coming from the generating {\bf associator moves} studied in \cite[Sec.\,2.8 \& 2.9]{Runkel:2008gr} in the present context\footnote{It is worth remarking that the study,\ upon specialisation to the setting of the WZNW PAG $\si$-model with a compact simple 1-connected Lie-group target $\txG$ and $Z(\txG)$-jump defects ({\it i.e.},\ gauge defects for the discrete symmetry group given by the centre $Z(\txG)$ of $\txG$),\ led to the reconstruction of the restriction of the Moore--Seiberg fusing matrix of the corresponding RCFT to the simple-current sector from the data of the WZNW gerbe,\ see:\ \cite[Sec.\,2.9.1]{Runkel:2008gr} and \cite[Sec.\,2]{Gawedzki:2008um}.},\ see:\ Fig.\,\ref{fig:assoc-mov}.\ The latter phase factors,\ in turn,\ are pullbacks of the 3-cocycle $\unl\D{}^{(0)}_{(2)}f$ determined by the BD data $f\in A^0(\cO_{\xcG_2})$ of the 2-isomorphism $\g$,\ whose non-triviality is an an obstruction against the coherence of the $\xcG$-equivariant structure $(\cG,\Ups,\g)$,\ discussed in Sec.\,\ref{sub:geom-cohom-equiv}.\ Thus,\ the existence of a \emph{coherent} such structure enables us to remove the ambiguity,\ whereupon the desired compatibility is established.\ As discussed at some length in \cite{Runkel:2008gr},\ this harmonises beautifully with the quantitative formal approach to gauging offered by abstract topological (quantum) field theory.

\section{Conclusions}\label{sec:summ-out}

In the present paper,\ the field and symmetry content of the classic Poisson $\si$-model (PSM) with a closed worldsheet $\Si$ was reinterpreted in terms of the geometric structure underlying the gauge principle for groupoidal symmetries,\ as laid out by Strobl and the Author in a recent work \cite{Strobl:2025}.\ The relevant Lie groupoid is the symplectic groupoid $\grpd{(\xcG,\om)}{(M,\Pi)}$ integrating the cotangent algebroid over the Poisson target $(M,\Pi)$---the existence of this groupoid was the base assumption of our analysis.\ In physical terms,\ the latter corresponds to the assumption of regularity of the physical phase space of the field theory,\ see:\ \cite{Cattaneo:2000iw}.\ The reinterpretation of the PSM as a theory of connections $\Theta_{\rm triv}$ on the trivial principaloid bundle $\xcP_{\rm triv}\equiv\Si\x\xcG\to\Si$ over the model's worldsheet,\ localised on sections of its Godement shadow $\xcF_{\rm triv}=\xcP_{\rm triv}/\xcG\equiv\Si\x M\to\Si$,\ allowed to explain the infinitesimal gauge symmetries of the PSM as linearisations of $\xcG$-equivariant automorphisms of $\xcP_{\rm triv}$ modelled on smoothly $\Si$-indexed families of global bisections of $\xcG$.\ The consistency of the reinterpretation then enforced the reduction of the codomain of the latter model to the \emph{lagrangean} subgroup $\bB(\xcG,\om)$ of the group $\bB(\xcG)$ of all bisections.\ Upon the reduction,\ the lagrangean anomaly of the (reinterpreted) PSM was derived,\ in which the multiplicativity of the symplectic form $\om$ and its presentation in terms of Spencer data \cite{Crainic:2015msp} played an instrumental r\^ole.\ The anomaly was shown to vanish identically for prequantisable Lie groupoids,\ {\it i.e.},\ those with ${\rm Per}([\om])\subset 2\pi\bZ$,\ geometrising as prequantum (frame) bundles $(P_\om,\cA)$ over $\xcG$ with connections $\cA\in\Om^1(P_\om)$ of curvature $\om$.

The PSM was reexpressed in a natural way in terms of the data of the symplectic principaloid bundle $\xcP_{\rm triv}$---the connection $\Theta_{\rm triv}$ on its total space and the symplectic form $\om$ on its typical fibre---as a holonomy of a trivial gerbe on $\xcF_{\rm triv}$,\ with the global curving given by a 2-form $\om[\Theta_{\rm triv}]\in\Om^2(\xcF_{\rm triv})$ encoding the coupling of the Lie algebroid-valued gauge field to the matter field $\phi\in\G(\xcF_{\rm triv})$ along which it is pulled back to the worldsheet.\ Drawing inspiration from an in-depth analysis \cite{Gawedzki:2010rn} of the gauging of group-modelled symmetries in the dynamical 2$d$ Polyakov--Alvarez--Gaw\c{e}dzki $\si$-model,\ and its generalisation to arbitrary Lie groupoids worked out in \cite{Suszek:2026col},\ in which similar 2-forms had been encountered,\ a general anomaly-cancellation mechanism was postulated.\ In it,\ the anomaly of the bare PSM is neutralised by an extra component of the target geometry,\ to wit:\ a background gerbe $\cG$ over $M$,\ endowed with an $\om$-twisted $\xcG$-equivariant structure $(\Ups,\g)$.\ The triple $(\cG,\Ups,\g)$ is to be viewed as a geometrisation of an extension of the de Rham 2-cocycle $\om\in Z^2_{\rm dR}(\xcG)$ from the kernel of the simplicial coboundary operator $\pr_1^*-\txm^*+\pr_2^*$ to a 3-cocycle $(H_M,\om,0,0)$ (with $H_M\equiv\curv(\cG)$) in the Bott--Shulman--Stasheff (BSS) bicomplex of the structure groupoid $\grpd{(\xcG,\om)}{(M,\Pi)}$,\ the latter's cohomology being regarded---after Dupont \cite{Dupont:1976}---as a model of the de Rham cohomology of the orbistack $M//\xcG$ (resp.\ $M//\bB(\xcG,\om)$),\ to which the gauging of the groupoidal symmetry is meant to descend the field theory.\ The gerbe $\cG$ was coupled to the PSM in a structured manner:\ Its pullback $\pr_2^*\cG$ to $\xcF_{\rm triv}$ was tensored with the trivial gerbe represented by $\om[\Theta_{\rm triv}]$,\ and the holonomy of the ensuing product gerbe $\cG[\Theta_{\rm triv}]=\pr_2^*\cG\ox\cI_{\om[\Theta_{\rm triv}]}$ was taken as the definition of the $\cG$-coaugmented PSM.\ The existence of the 1-isomorphism $\Ups$ of the $\xcG$-equivariant structure on $\cG$ was then demonstrated to ensure the vanishing of the lagrangean anomaly of the thus enhanced field theory.\ The contribution of the coaugmenting gerbe $\cG$ to the theory's Euler--Lagrange equations,\ being proportional to the curvature $H_M$,\ remains fully controllable.\ The structural interpretation of the critical points of the original PSM as Lie-algebroid morphisms distinguishes flat coaugmentations,\ $H_M=0$,\ which may,\ nevertheless,\ be nontrivial if $H^2(M,\uj)\neq 0$.

The coaugmentation postulated in the case of the trivial gauge bundle $\xcP_{\rm triv}$ was subsequently employed as a local model $\cG_i\equiv\xcF\t_i^*\cG[\Theta_i]$ of the higher-geometric structure on the shadow $\xcF$ of a symplectic principaloid $(\xcG,\om)$-bundle with connection  $(\xcP,\Theta)$ of arbitrary topology,\ associated with trivialisations  $\xcP\t_i\colo\pi_\xcP^{-1}(O_i)\cong O_i\x\xcG$ and $\xcF\t_i\colo\pi_\xcF^{-1}(O_i)\cong O_i\x M$,\ and coupled to the corresponding local connection $\Theta_i=\xcP\t_i^{-1\,*}\Theta$.\ The family $\{\cG_i\}_{i\in I}$ of local gerbes arising in this manner was shown to descend---in the sense of the structure of a 2-stack on the smooth site carried by gerbes \cite{Stevenson:2000wj}---to the total space of the shadow bundle.\ A key r\^ole in the descent was seen to be played by the fully fledged $\xcG$-equivariant structure $(\Ups,\g)$ on $\cG$,\ and so a complete cohomological description of topological obstructions against its existence---aka gauge anomalies---was worked out in terms of the induced simplicial (vertical) cohomology of the horizontal (Beilinson--Deligne) cohomology of the \v Cech--Deligne--Dupont resolution of the BSS bicomplex associated with a simplicial refinement of a good open cover of the nerve $\xcG_\bullet$ of $\xcG$.\ In the previously considered conservative scenario $\cG=\cI_0$,\ the $\xcG$-equivariant structure was demonstrated to reduce to a multiplicative structure on the prequantum (frame) bundle $(P_\om,\cA)$.\ The holonomy of the descended gerbe,\ coherently represented by the local gerbes $\cG_i$ over the covering open subsets $\pi_\xcF^{-1}(O_i)$,\ was then taken as the definition of the Dirac--Feynman amplitude of the generalised PSM with configuration bundle $(\xcP,\Theta)\to\xcF$ with the reduced structure group $\bB(\xcG,\om)$.\ The amplitude was verified to possess the desired gauge invariance---with respect to those automorphisms of $\xcP$ which are locally modelled on smoothly $O_i$-indexed families of lagrangean bisections of the symplectic groupoid $\grpd{(\xcG,\om)}{(M,\Pi)}$.

Finally,\ the generalised PSM was reformulated as a field theory with multiple phases represented by local trivialisations $\xcF\t_i\colo\pi_\xcF^{-1}(O_i)\cong O_i\x M$ of the configuration bundle,\ inhabited by local fields $(\xcF\t_i\circ\phi,\Theta_i)$,\ and separated by domain walls (edges) of a gauge-symmetry defect localised within intersections $O_i\cap O_j$ and carrying data of the $\xcG$-equivariant structure.\ The construction of the relevant stratified target space and of the attendant decorated character was developed along the lines of the precursor papers \cite{Runkel:2008gr,Suszek:2012ddg},\ in which the group-symmetry case had been treated at length.\ The gauge-symmetry defect was proven topological,\ that is amenable to canonical homotopic deformations within the worldsheet $\Si$ (also those involving its vertices),\ which do not affect the value of the Dirac--Feynman amplitude.\ The reformulation and its key structural properties conform with the standard approach to the gauging pursued in topological quantum field theory \cite{Runkel:2008gr,Carqueville:2023jhb}.\medskip

The novel interpretation and essential generalisation of the classic PSM advocated in the present paper hinge on a nontrivial interplay between the multiplicativity of the symplectic structure $(\xcG,\om)$ on the groupoid integrating the Poisson target $\txT^*M\to(M,\Pi)$,\ the lagrangean reduction $\bB(\xcG)\searrow\bB(\xcG,\om)$ of the structure group---and so,\ in particular,\ of the local model of gauge automorphisms---of the underlying gauge bundle $\xcP$,\ and the special form $\om[\Theta_i]$ of the basic coupling between the matter and gauge fields,\ mediated by the symplectic form $\om$.\ The relevance of the multiplicativity is further amplified by its embedding in the BSS model of the cohomology of the target orbistack $M//\xcG$,\ and subsequent geometrisation in the form of the coaugmenting gerbe $\cG$ endowed with an $\om$-twisted $\xcG$-equivariant structure $(\Ups,g)$,\ central to the descent of the local physics to the total space of the configuration bundle $\xcF$.\ The upshot is a tightly knitted higher-geometric/gauge construct bound by a natural mechanism of lagrangean-anomaly cancellation.\ The rather subtle internal coherence of this construct offers compelling evidence for the suitability of the two gauge systems employed in tandem:\ the (symplectic) principaloid $(\xcG,\om)$-bundles of \cite{Strobl:2025} and the $\xcG$-equivariant gerbes of \cite{Gawedzki:2010rn,Suszek:2026col}.

The proposed definition of the topological gauge field theory opens up avenues for further investigation of its internal structure and related issues:
\bit
\item The question---of central relevance to the application of the theory in Kontsevich's programme of deformation quantisation---of a systematic identification of quantum-mechanically consistent target backgrounds for the boundary variant of the theory (with $\p\Si\neq\emptyset$) can now be posed rigorously in the framework of the relative $\xcG$-equivariant cohomology underlying the gauging of lagrangean symmetries in the presence of symmetric gerbe modules,\ with the relevant (general) structures and intuitions worked out in \cite{Gawedzki:2012fu,Suszek:2012ddg,Suszek:2022CSimpl}.
\item The study of twisted Poisson structures,\ initiated in \cite{Severa:2001qm,Klimcik:2002},\ lifts straightforwardly to and so can be pursued naturally in the setting of gerbe theory,\ and the attendant hypercohomology\footnote{The Author is grateful to Alexei Kotov for suggesting this line of research.}.
\item Much of the discussion of the gauging in the topologically trivial sector $(\xcP_{\rm triv},\Theta_{\rm triv})\to\xcF_{\rm triv}$ can be formulated purely in terms of an action of lagrangean bisections on the cotangent algebroid over the Poisson target $(M,\Pi)$,\ without any reference to principaloid bundles (and---under additional assumptions---to gerbes).\ One may then try to better understand the algebroidal aspect of the gauging procedure,\ and its projection to the Poisson base\footnote{The Author is grateful to Janusz Grabowski for pointing out this direction of further study.}.
\item In search of examples of symplectic groupoids of direct relevance to physics,\ one stumbles upon the Lu--Weinstein groupoid that integrates the cotangent algebroid over a Poisson--Lie group \cite{Lu:1989sdg}.\ Recall that the latter arises naturally in the symplectic analysis of the Wess--Zumino--Novikov--Witten (WZNW) $\si$-model of RCFT \cite{Gawedzki:2001rm,Falceto:2001eh}.\ The distinguished structure of a \emph{double} symplectic Lie groupoid exhibited by the Lu--Weinstein groupoid---submersing both:\ the original Poisson--Lie group $(\txG,\Pi)$ and its dual $(\txG^*,\Pi^*)$ as object manifolds---invites a gerbe-theoretic study of various dualities formerly considered in the literature in the present gerbe-coaugmented field-theoretic (and hence prequantisable,\ see:\ \cite{Gawedzki:1987ak}) setting,\ such as the duality between the Poisson $\si$-model with target $(\txG^*,\Pi^*)$ and the fully gauged WZNW $\si$-model with target $\txG$ \cite{Alekseev:1995py,Falceto:2001eh},\ the open-closed duality between Poisson $\si$-models for the dual targets $(\txG,\Pi)$ and $(\txG^*,\Pi^*)$ \cite{Calvo:2003kv},\ or the intricate Poisson $T$-duality between the respective dynamical (PAG) $\si$-models \cite{Klimcik:1995ux,Alekseev:1995ym}.
\eit  
We hope to return to these topics in a future work.

\appendix

\section{Essential facts about Lie groupoids and algebroids}\label{app:Lie}
	
The present appendix contains elements of the theory of Lie groupoids and algebroids---definitions with examples,\ propositions and theorems---which constitute the conceptual core of the physically oriented considerations in the main text.
	
\bedef\cite{Brandt:1926}\label{def:grpd}
A \textbf{groupoid} is a small category with all morphisms invertible.\ Thus,\ it is a septuple $\Gr=(\obj\Gr\equiv M,\morf\Gr\equiv\xcG,s, t,\Id,\txm\equiv.,\Inv\equiv(\cdot)^{-1})$---to be abbreviated as $\grpd{\xcG}{M}$---which is composed of a pair of sets: 
\bit
\item the \textbf{object set} $M$; 
\item the \textbf{arrow set} $\xcG$,
\eit
and a quintuple of \textbf{structure maps}: 
\bit
\item the \textbf{source} ({\bf map}) $ s\colo \xcG\to M$;
\item the \textbf{target} ({\bf map}) $  t\colo \xcG\to M$; 
\item the \textbf{unit} ({\bf map}) $\Id\colo M\to\xcG,\ m\mapsto\Id_m$;
\item the \textbf{inverse} ({\bf map}) $\Inv\colo \xcG\to\xcG,\  g\mapsto \Inv(g)\equiv g^{-1}$;
\item the \textbf{multiplication} ({\bf map}) $\txm\colo \xcG\fibx{s}{t}\xcG\to\xcG,\ ( g, h)\mapsto \txm( g, h)\equiv g.h$,
\eit 
subject to the conditions (in force whenever the expressions are well-defined):
\bit
\item[(i)] $ s( g. h)= s(h),\   t( g. h)=  t(g)$;
\item[(ii)] $( g. h). k= g.( h.k)$;
\item[(iii)] $\Id_{  t( g)}. g=g= g.\Id_{ s( g)}$;
\item[(iv)] $ s( g^{-1})=  t( g),\  t(g^{-1})= s( g),\  g. g^{-1}=\Id_{  t( g)},\
	g^{-1}. g=\Id_{ s(g)}$.
\eit
	
A \textbf{morphism} between two groupoids $\Gr_a,\ a\in\{1,2\}$ is a functor $\Phi\colo \Gr_1\to\Gr_2$.
	
A \textbf{Lie groupoid} is a groupoid whose object and arrow sets are smooth manifolds,\ whose structure maps are smooth,\ and whose source and target maps are surjective submersions.\ A morphism between two Lie groupoids is a functor between them with smooth object and morphism components. 
	
A Lie groupoid is called 
\bit
\item {\bf transitive} if the {\bf anchor} $(t,s)\colo\xcG\to M\x M$ is surjective;
\item {\bf locally trivial} if $(t,s)\colo\xcG\to M\x M$ is a surjective submersion.
\eit
\exdef
	
	\beg\label{eg:deloop} 
	Every Lie group $(\txG,\txm_\txG,\Inv_\txG,e)$ gives rise to a Lie groupoid 
	\qq\nn
	\bbB\txG\equiv\bigl(\{\bullet\},\txG,\bullet,\bullet,e,\txm_\txG,\Inv_\txG\bigr)\,, 
	\qqq
	with the singleton $\{\bullet\}$ as the object manifold.\ It is termed the {\bf delooping} ({\bf groupoid}) of $\txG$.
	\eeg
	
	\beg\label{eg:pairgrpd}
	Every manifold $M\in\Man$ gives rise to a Lie groupoid 
	\qq\nn
	{\rm Pair}(M)\equiv\bigl(M,M\x M,\pr_2,\pr_1,\D,\pr_{1,4},\t\bigr)\,,
	\qqq 
	with unit $\D\colo M\to M\x M,\ m\mapsto(m,m)$,\ multiplication $\pr_{1,4}\colo(M\x M)\ \fibx{\pr_2\,\,}{\pr_1}(M\x M)\to M\x M,\ ((m_4,m_2),$ $(m_2,m_1))\mapsto(m_4,m_1)$,\ and inverse $\t\colo M\x M\to M\x M,\ (m_1,m_2)\mapsto(m_2,m_1)$.\ It is called the {\bf pair groupoid} of $M$.
	\eeg
	
	\beg\label{eg:subm-grpd}
	Given $M,Y\in\Man$ and a surjective submersion $\pi\colo Y\twoheadrightarrow M$,\ there arises a Lie groupoid
	\qq\nn
	{\rm Pair}_\pi(Y)\equiv\bigl(Y,Y\fibx{\pi}{\pi}Y,\pr_2,\pr_1,\D,\pr_{1,4},\t\bigr)\subset{\rm Pair}(Y)\,,
	\qqq 
	defined in the notation of Ex.\,\ref{eg:pairgrpd}.\ It is called the {\bf submersion groupoid} of $\pi$.
	\eeg
	
	\beg\label{eg:actgrpd}
	Given a Lie group $\txG$ and a manifold $M$ on which $\txG$ acts smoothly as $\la\colo\txG\to\Diff(M)$,\ there arises a Lie groupoid 
	\qq\nn
	\txG\,\lx_\la M\equiv\bigl(M,\txG\x M,\pr_2,\widetilde\la\equiv\ev\circ(\la\x\id_M),(e,\cdot),(\txm_\txG\circ\pr_{1,3},\pr_4),(\Inv_\txG\circ\pr_1,\widetilde\la)\bigr)\,,
	\qqq
	defined in terms of the evaluation map $\ev\colo\Diff(M)\x M\to M,\ (f,m)\mapsto f(m)$.\ It is called the {\bf action groupoid} of $\la$.\ By a mild abuse of the notation,\ we identify $\la$ and $\widetilde\la$ in this work.
	\eeg
	
	\beg\cite{Weinstein:1987,Coste:1987}\label{eg:symplgrpd}
	A Lie groupoid $(M,\xcG,s,t,\Id,\txm,\Inv)$ is termed {\bf symplectic} if there exists a closed non-degenerate 2-form $\om\in\Om^2(\xcG)$,\ which makes $(\xcG,\om)$ a symplectic manifold,\ and such that the graph of $\txm$ is a lagrangean submanifold in $(\xcG^{\x 3},(\pr_1^*+\pr_2^*-\pr_3^*)\om)$.\ As shown in \cite[Thm.\,1.1,\ Chap.\,II]{Coste:1987},\ the object manifold $M$ then carries a canonical Poisson bivector $\Pi\in\G(\bigwedge^2\txT M)$,\ relative to which $s$ is Poisson,\ and $t$ is anti-Poisson.\ Furthermore,\ $\Id(M)$ is lagrangean.
	\eeg
	
	\bedef[\cite{Moerdijk:2003mm}]\label{def:bisec}
	Adopt the notation of Def.\,\ref{def:grpd}.\ A ({\bf global}) {\bf bisection of} $\Gr$ is a section $\b\colo M\to\xcG$ of the surjective submersion $ s\colo \xcG\to M$ such that the induced map 
	\qq\nn
	t_*\b\equiv  t\circ\b\colo M\too M
	\qqq
	is a diffeomorphism.\ Equivalently,\ it is a submanifold $S\subset\xcG$ with the property that both restrictions:\ $ s\rstr_S$ and $t\rstr_S$ are diffeomorphisms.\ We shall denote the set of bisections as 
	\qq\nn
	\BisGr\equiv\bB(\xcG)\,.
	\qqq
	
	A {\bf local bisection of} $\Gr$ is a local section $\b\colo O\to\xcG$ of $ s$ over an open subset $O\subset M$,\ such that the induced map 
	\qq\nn
	t_*\b\equiv  t\circ\b\colo O\too  (t\circ\b)(O)
	\qqq
	is a diffeomorphism.\ We shall denote the set of local bisections as 
	\qq\nn
	{\rm Bisec}_{\rm loc}(\Gr)\equiv\bB_{\rm loc}(\xcG)\,.
	\qqq
	\exdef
	
	\bedef
	Adopt the notation of Def.\,\ref{def:bisec}.\ The {\bf group of bisections} of a Lie groupoid $\grpd{\xcG}{M}$ is the canonical structure of a group on $\bB(\xcG)$.\ Its binary operation is defined as
	\qq\nn
	\cdot\colo \bB(\xcG)\x\bB(\xcG)\too\bB(\xcG),\ (\b_2,\b_1)\longmapsto\b_2\bigl(t_*\b_1(\cdot)\bigr).\b_1(\cdot)\equiv\b_2\cdot\b_1\,.
	\qqq
	The neutral element is $\Id$,\ also termed the {\bf unit bisection},\ and the corresponding inverse is 
	\qq\nn
	\Inv\colo \bB(\xcG)\too\bB(\xcG),\ \b\longmapsto\Inv\circ\b\circ\bigl(  t_*\b\bigr)^{-1}\equiv\b^{-1}\,.
	\qqq
	
	Similarly,\ the {\bf group of local bisections} of $\grpd{\xcG}{M}$ is the canonical structure of a group on $\bB_{\rm loc}(\xcG)$.\ Its binary operation is defined as
	\qq\nn
	\cdot\colo \bB_{\rm loc}(\xcG)\x\bB_{\rm loc}(\xcG)\too\bB_{\rm loc}(\xcG),\ (\b_2,\b_1)\longmapsto\b_2\bigl(t_*\b_1(\cdot)\bigr).\b_1(\cdot)\rstr_{(t_*\b_1)^{-1}({\rm dom}(\b_2))\cap{\rm dom}(\b_1)}\equiv\b_2\cdot\b_1\,.
	\qqq
	The neutral element is $\Id$,\ and the corresponding inverse is 
	\qq\nn
	\Inv\colo \bB_{\rm loc}(\xcG)\too\bB_{\rm loc}(\xcG),\ \b\longmapsto\Inv\circ\b\circ\bigl( t_*\b\bigr)^{-1}\rstr_{t_*\b({\rm dom}(\b))}\equiv\b^{-1}\,.
	\qqq
	\exdef
	
\beg
${\rm Bisec}(\bbB\txG)\cong\txG$ (see:\ Ex.\,\ref{eg:deloop}).
\eeg
	
\beg
${\rm Bisec}({\rm Pair}(M))\cong\Diff(M)$ (see:\ Ex.\,\ref{eg:pairgrpd}).
\eeg
	
\beg\label{eg:flatbisec}
The group ${\rm Bisec}(\txG\,\lx_\la M)$ (see:\ Ex.\,\ref{eg:actgrpd}) contains a subgroup 
\qq\nn
\txG\cong\iota(\txG)\subset{\rm Bisec}(\txG\,\lx_\la M)\,,\qquad\tx{where}\qquad\iota\colo\txG\too{\rm Bisec}(\txG\,\lx_\la M),\ g\longmapsto(g,\cdot)\,.
\qqq
We refer to bisections from $\iota(\txG)$ as {\bf flat}.
\eeg
	
\beg\cite{Coste:1987,Rybicki:2001}\label{eg:lagrbisec} 
The group $\bB(\xcG)$ of the symplectic groupoid from from Ex.\,\ref{eg:symplgrpd} contains a subgroup 
\qq\nn
\bB(\xcG,\om):=\bigl\{ \ \b\in\bB(\xcG) \quad\vert\quad \b(M)\subset\xcG \tx{ is lagrangean}\ \bigr\}\equiv\bigl\{ \ \b\in\bB(\xcG) \quad\vert\quad \b^*\om=0\ \bigr\}\subset\bB(\xcG)\,.
\qqq
These bisections are called {\bf lagrangean},\ and $\bB(\xcG,\om)$ is termed the {\bf lagrangean group} of $\xcG$.
\eeg
	
\bedef\label{def:B-compl}
A Lie groupoid $\grpd{\xcG}{M}$ with the property:
\qq\nn
\forall\ g\in\xcG\ \exists\ \b\in\bB(\xcG)\colo\b\bigl(s(g)\bigr)=g
\qqq 
is termed {\bf $\bB(\xcG)$-complete}.
\exdef
	
\beg
All source- resp.\ target-connected Lie groupoids are $\bB(\xcG)$-complete (see:\ \cite[Thm.\,3.1]{Zhong:2009}),\ and so is also every action groupoid from Ex.\,\ref{eg:actgrpd}. 
\eeg
	
\begin{propanition}\label{prop:beta-split}
Every bisection $\b\in\bB_{\rm loc}(\xcG)$ (see:\ Def.\,\ref{def:bisec}) determines,\ over each point $m\in{\rm dom}(\b)$,\ a direct-sum decomposition
\qq\nn
\txT_{\b(m)}\xcG\cong\txT_{\b(m)}\bigl(\b({\rm dom}(\b))\bigr)\oplus\bigl(\ker\,\txT s\bigr)_{\b(m)}\,.
\qqq
We call the corresponding pair $(\chi_\b,\upsilon_\b)$ of complementary projectors,
\qq\nn
\chi_\b(m)=\txT_{\b(m)}(\b\circ s)\,,\qquad\qquad\upsilon_\b(m)=\id_{\txT_{\b(m)}\xcG}-\chi_\b(m)\,,
\qqq
the {\bf splitting} of $\txT\xcG\rstr_{\b({\rm dom}(\b))}$ {\bf induced by} $\b$.
\end{propanition}
	
\bedef\cite[Sec.\,3.1]{Crainic:2015msp}\label{def:J1B}
Adopt the notation of Def.\,\ref{def:bisec}.\ The {\bf first-jet groupoid} of $\xcG$ is the groupoid
\qq\nn
\bigl(M,J^1\xcG,\pi_{J^1},\t,j^1\Id,j^1\txm,j^1\Inv\bigr)
\qqq
with the arrow manifold $J^1\xcG$ consisting of first jets $j^1_m\b$ of local bisections $\b\in\bB_{\rm loc}(\xcG)$ (${\rm dom}(\b)\ni m$) over $M$,\ and with source $\pi_{J^1}(j^1_m\b)=m$ and target $\t(j^1_m\b)=t_*\b(m)$,\ identity $j^1\Id(m)=j^1_m\Id$,\ multiplication $j^1\txm(j^1_{t_*\b_1(m)}\b_2,j^1_m\b_1)=j^1_m(\b_2\cdot\b_1)$,\ and inverse $j^1\Inv(j^1_m\b)=j^1_{t_*\b(m)}(\b^{-1})$. 
\exdef
	
\berop\cite[Sec.\,3.1]{Crainic:2015msp}
There exists a group homomorphism 
\qq\label{eq:j1-hom}
j^1\colo\bB(\xcG)\too\bB(J^1\xcG),\ \b\longmapsto j^1_\cdot\b\,.
\qqq
\eerop
	
\bedef\label{def:tstar}
The {\bf shadow action of} $\bB(\xcG)$ {\bf on} $M$ is 
\qq\nn
t_*\colo \bB(\xcG)\x M\too M,\ (\b,m)\longmapsto t\bigl(\b(m)\bigr)\equiv\b\ulact m\,.
\qqq
We refer by the same name to and use the same symbol for the group homomorphism 
\qq\label{eq:BisGr-act-obj}
t_*\colo \bB(\xcG)\too{\rm Diff}(M)\,.
\qqq 
\exdef
	
\bedef\label{def:bisec-act}
The {\bf left-multiplication of} $\xcG$ {\bf by} $\bB(\xcG)$ is the left action
\qq\nn
L\colo \bB(\xcG)\x\xcG\too\xcG,\ \bigl(\b, g\bigr)\longmapsto\b\bigl(  t\bigl( g\bigr)\bigr). g\equiv L_\b\bigl( g\bigr)\equiv \b\lact g\,.
\qqq
The {\bf right-multiplication of} $\xcG$ {\bf by} $\bB(\xcG)$ is the right action
\qq\nn
R\colo  \xcG\x\bB(\xcG)\too\xcG,\ \bigl( g,\b\bigr)\longmapsto g.\bigl(\b^{-1}\bigl(s\bigl( g\bigr)\bigr)\bigr)^{-1}\equiv R{}_\b\bigl( g\bigr)\equiv g\ract\b\,.
\qqq
The {\bf conjugation of} $\xcG$ {\bf by} $\bB(\xcG)$ is the \emph{left} action
\qq\nn
C\colo \bB(\xcG)\x\xcG\too\xcG,\ (\b, g)\longmapsto\b\bigl(  t( g)\bigr). g.\b\bigl( s( g)\bigr)^{-1}\equiv C_\b( g)\equiv \b\lact g\ract\b^{-1}\,.
\qqq
\exdef
	
\bedef\cite{Ehresmann:1959}\label{def:gr-mod}
A \textbf{right-$\xcG$-module space},\ or a \textbf{right-$\xcG$-module} for short,\ is a triple $(X,\mu,\varrho)$ composed of 
\bit
\item a smooth manifold $X$;
\item a smooth map $\mu\colo X\to M$,\ called the (\textbf{right}) \textbf{moment map};
\item a smooth map
\qq\nn
\varrho\colo  X\,\fibx{\mu}{t}\xcG\too
X,\ (x, g)\longmapsto\varrho(x, g)\equiv\varrho_{ g}(x)\equiv x\mact g\,,
\qqq
termed the (\textbf{right}) \textbf{action} ({\bf map}),
\eit 
subject to the relations (in force whenever the expressions are well-defined):
\bit
\item[(GrM1)] $\mu(x\mact g)= s( g)$;
\item[(GrM2)] $x\mact\Id_{\mu(x)}=x$;
\item[(GrM3)] $(x\mact g)\mact h=x\mact( g. h)$.
\eit
	
Similarly,\ a {\bf left-$\xcG$-module} ({\bf space}) is a triple $(X,\mu,\la)$ composed of 
\bit
\item a smooth manifold $X$;
\item a smooth map $\mu\colo X\to M$,\ called the (\textbf{left}) \textbf{moment map};
\item a smooth map
\qq\nn
\la\colo \xcG\,\fibx{s}{\mu}X\too
X,\ (g,x)\longmapsto\la(g,x)\equiv\la_g(x)\equiv g\mlact x\,, 
\qqq
termed the (\textbf{left}) \textbf{action} ({\bf map}),
\eit 
subject to the relations (in force whenever the expressions are well-defined):
\bit
\item[(GlM1)] $\mu(g\mlact x)= t(g)$;
\item[(GlM2)] $\Id_{\mu(x)}\mlact x=x$;
\item[(GlM3)] $h\mlact(g\mlact x)=(h.g)\mlact x$.
\eit
	
A right action $\varrho$ is termed \textbf{free} if the following implication obtains:
\qq\nn
x\mact g=x\qquad\Longrightarrow\qquad g=
\Id_{\mu(x)}\,,
\qqq
so that,\ in particular,\ the \textbf{isotropy group} $\xcG_m= s^{-1}(\{m\})\cap  t^{-1}(\{m\})$ of $m\in M$ acts freely (in the usual sense) on the fibre $\mu^{-1}(\{m\})$.\ A free left action is defined analogously.
	
A right action $\varrho$ is termed \textbf{transitive} if for any two points $x,x'\in X$ there exists an arrow $ g\in\xcG$ such that $x'=x\mact g$.\ A transitive left action is defined analogously.
	
The subset 
\qq\nn
\xcG\mlact x:=\bigl\{\ g\mlact x \quad\vert\quad g\in s^{-1}\bigl(\mu(x)\bigr) \ \bigr\}
\qqq
is called the {\bf left $\xcG$-orbit} of $x\in X$.\ {\bf Right $\xcG$-orbits} are defined analogously.
	
Let $\Gr_a\equiv\grpd{\xcG_a}{M_a},\ a\in\{1,2\}$ be a pair of Lie groupoids,\ and let $(X_a,\mu_a,\varrho_a)$ be the respective
right-$\xcG_a$-modules.\ A \textbf{morphism} between the latter is a pair $(\Theta,\Phi)$ consisting of a smooth manifold map $\Theta\colo  X_1\to X_2$ together with a functor $\Phi\colo \Gr_1\to\Gr_2$ for which the following diagrams commute
\qq
&\alxydim{@C=1.cm@R=1.cm}{ X_1 \ar[r]^{\Theta} \ar[d]_{\mu_1} &
X_2 \ar[d]^{\mu_2}\cr M_1 \ar[r]_{\Phi} &
M_2}\,,& \label{diag:mu-Th-mu}\\\cr\cr
&\alxydim{@C=1.5cm@R=1.cm}{ X_1\ \fibx{\mu_1}{t_1}\,\xcG_1
\ar[r]^{\Theta\x\Phi} \ar[d]_{\varrho_1} &  X_2\
\fibx{\mu_2}{t_2}\,\xcG_2 \ar[d]^{\varrho_2}\cr  X_1
\ar[r]_{\Theta} &  X_2}\,.& \label{diag:Th-intertw}
\qqq
For every Lie groupoid $\Gr\equiv\grpd{\xcG}{M}$,\ right $\xcG$-modules together with the corresponding morphisms with $\Phi=\id_\Gr$ form a {\bf category of right $\xcG$-modules} $\Man_\xcG$.\ Its $\hom$-sets shall be denoted as
\qq\nn
C^\infty_\xcG(X_1,X_2)\equiv\Hom_{\Man_\xcG}\bigl(\bigl(X_1,\mu_1,\varrho_1\bigr),\bigl(X_2,\mu_2,\varrho_2\bigr)\bigr)\,,
\qqq
with elements referred to as {\bf smooth $\xcG$-equivariant maps} between $X_1$ and $X_2$.\ In the particular case of $(X_2,\varrho_2)=(X_1,\varrho_1)\equiv(X,\varrho)$ and smoothly invertible maps,\ we employ the notation
\qq\nn
\Diff_{\varrho(\xcG)}(X)=\{\ f\in C^\infty_\xcG(X,X) \quad\vert\quad \exists\ f^{-1}\in C^\infty_\xcG(X,X) \ \}
\qqq
\exdef
	
	\beg\label{eg:st-act}
	There exists canonical $\xcG$-module structures on the object manifold $M$:\ the left-$\xcG$-module structure
	\qq\nn
	\bigl(M,\id_M,t\circ\pr_1\bigr)\,,
	\qqq
	with the {\bf target action}
	\qq\nn
	(t\circ\pr_1)(g,m)=t(g)\,,
	\qqq
	and the right-$\xcG$-module structure
	\qq\nn
	\bigl(M,\id_M,s\circ\pr_2\bigr)\,,
	\qqq
	with the {\bf source action}
	\qq\nn
	(s\circ\pr_2)(m,g)=s(g)\,.
	\qqq
	\eeg
	
	\beg\label{eg:fibred-act}
	There exist two commuting canonical $\xcG$-module structures on the arrow manifold $\xcG$:\ the left-$\xcG$-module structure 
	\qq\nn
	(\xcG,  t,l\equiv.)\,,
	\qqq
	with the {\bf left-fibred action}
	\qq\label{eq:Lgrpdact}
	l\bigl( h, g\bigr)\equiv l_{ h}\bigl( g\bigr)= h. g\,,
	\qqq
	and the right-$\xcG$-module structure 
	\qq\label{eq:can-R-Gr-mod}
	(\xcG, s,r\equiv.)\,,
	\qqq
	with the {\bf right-fibred action}
	\qq\label{eq:Rgrpdact}
	r\bigl( g, h\bigr)\equiv r_{ h}\bigl( g\bigr)= g. h\,.
	\qqq
	\eeg
	
	\berop\label{prop:G-in-B_act}
	Adopt the notation of Defs.\,\ref{def:bisec} and \ref{def:gr-mod}.\ The action $\la\colo \xcG\,\fibx{s}{\mu}X\to X$ of $\xcG$ canonically induces a left action $B\la\colo\bB(\xcG)\x X\to X$ of $\bB(\xcG)$ determined by the commutative diagram
	\qq\nn
	\alxydim{@C=1.75cm@R=1.25cm}{ \bB(\xcG)\x M\,\fibx{\id_M}{\mu}X \ar[r]^{\qquad\ev\x\id_X} & \xcG\,\fibx{s}{\mu}X \ar[d]^{\la} \\ \bB(\xcG)\x X \ar[u]^{\id_{\bB(\xcG)}\x(\mu,\id_X)} \ar[r]_{\quad B\la} & X }\,,
	\qqq
	in which $\ev\colo \bB(\xcG)\x M\to\xcG,\ (\b,m)\mapsto\b(m)$ is the {\bf evaluation map}.\ An analogous statement holds true for right $\xcG$-modules.
	\eerop
	
	\bedef\cite{Ehresmann:1959,Moerdijk:1991}\label{def:princ-gr-bun} 
	Adopt the notation of Def.\,\ref{def:gr-mod}.\ A \textbf{right principal $\xcG$-bundle} is a quintuple $(\breP,\Si,\pi_{\breP},\mu,\varrho)$ composed of a pair of smooth manifolds: 
	\bit
	\item the \textbf{total space} $\breP$;
	\item the \textbf{base} $\Si$,
	\eit 
	and a triple of smooth maps: 
	\bit
	\item a surjective submersion $\pi_{\breP}\colo \breP\to\Si$,\
	termed the \textbf{bundle projection}; 
	\item the \textbf{moment map} $\mu\colo \breP\to M$;
	\item the \textbf{action} ({\bf map}) $\varrho\colo \breP\fibx{\mu}{t}\xcG\to\breP$,
	\eit
	with the following properties:
	\bit
	\item[(PGr1)] $(\breP,\mu,\varrho)$ is a right $\xcG$-module;
	\item[(PGr2)] $\pi_{\breP}$ is $\xcG$-invariant in the sense made precise by the commutative diagram
	\qq\nn
	\alxydim{@C=1.cm@R=1.cm}{\breP\fibx{\mu}{t}
		\xcG \ar[r]^{\quad\varrho} \ar[d]_{\pr_1} & \breP
		\ar[d]^{\pi_{\breP}}\cr \breP \ar[r]_{\pi_{\breP}} & \Si}\,;
	\qqq
	\item[(PGr3)] the map
	\qq\nn
	(\pr_1,\varrho)\colo \breP\fibx{\mu}{t}\xcG\too\breP\fibx{\pi_{\breP}}{\pi_{\breP}}\breP,\ (p,g)\longmapsto(p,p\mact g)
	\qqq
	is a diffeomorphism,\ so that $\xcG$ acts freely and transitively on $\pi_{\breP}$-fibres.\ The smooth inverse of $(\pr_1,\varrho)$ takes the form
	\qq\label{eq:div-map}
	(\pr_1,\varrho)^{-1}=:(\pr_1,\phi_{\breP})\,,\qquad\qquad\phi_{\breP}\colo \breP\fibx{\pi_{\breP}}{\pi_{\breP}}\breP\too\xcG\,,
	\qqq 
	and $\phi_{\breP}$ is called the \textbf{division map}.
	\eit
	\noindent The Lie groupoid $\xcG$ is termed the {\bf structure groupoid} of $\breP$ in this setting.
	
	We represent a right principal $\xcG$-bundle by the simplified diagram 
	\qq\label{diag:princGrdiag}\qquad\qquad
	\alxydim{@C=.75cm@R=1cm}{ \breP \ar@{->>}[d]_{\pi_\breP} \ar[rd]^{\mu} & & \xcG \ar@{=>}[ld] \ar@/_1.pc/[ll]_{\varrho} \\ \Si & M & }\,,
	\qqq
	in which the remaining structure is implicit.
	
	Let $\Gr_a\equiv\grpd{\xcG_a}{M_a},\ a\in\{1,2\}$ be a pair of Lie groupoids and let $(\breP{}_a,\Si_a,\pi_{\breP{}_a},\mu_a,\varrho_a)$ be the respective right principal $\xcG_a$-bundles.\ A \textbf{morphism} between the two bundles is a triple $(\Theta,f,\Phi)$ composed of a map $f\colo\Si_1\to\Si_2$ between the bases,\ and a morphism $(\Theta,\Phi)$ between the corresponding right $\xcG_a$-modules $(\breP{}_a,\mu_a,\varrho_a)$ which covers $f$ in the sense expressed by the commutative diagram 
	\qq\nn
	\alxydim{@C=1.cm@R=1.cm}{ \breP{}_1 \ar[r]^{\Theta} \ar[d]_{\pi_{\breP{}_1}} & \breP{}_2
		\ar[d]^{\pi_{\breP{}_2}}\cr \Si_1 \ar[r]_{f} & \Si_2 }\,.
	\qqq
	
	{\bf Left principal $\xcG$-bundles} $(\breP,\Si,\pi_\breP,\mu,\la)$ (and morphisms between them) are defined analogously.\ The corresponding diagrams take the self-explanatory form
	\qq\nn
	\alxydim{@C=.75cm@R=1cm}{ \xcG \ar@{=>}[rd] \ar@/^1.pc/[rr]^{\la} & & \breP \ar@{->>}[d]^{\pi_\breP} \ar[ld]_{\mu} \\ & M & \Si}\,,
	\qqq
	\exdef
	
	\beg\cite[Remark 5.34(1)]{Moerdijk:2003mm}\label{eg:triv-grpd-bndle}
	There exists a canonical structure of a right principal $\xcG$-bundle on $\grpd{\xcG}{M}$,\ given by 
	\qq\nn
	U_\xcG:=(\xcG,M,t,s,r)\,,
	\qqq 
	where $r$ is the right-fibred action \eqref{eq:Rgrpdact}.\ We call this structure the \textbf{unit bundle of} $\grpd{\xcG}{M}$.
	\eeg
	
	\bedef\cite{Pradines:1967}\label{def:Liealgbrd}
	Let $X$ be a smooth manifold.\ A ({\bf real}) {\bf Lie algebroid over} $X$ ({\bf of rank} $N\in\bN^\x$) is a triple $((\cE,X,\bR^N,\pi_\cE),\a_\cE,[\cdot,\cdot]_\cE)$ composed of
	\bit
	\item a vector bundle $(\cE,X,\bR^N,\pi_\cE)$,\ with total space $\xcE$,\ base $X$,\ typical fibre $\bR^N$,\ and base projection $\pi_\cE\colo\cE\to X$;
	\item a (vertical) vector-bundle morphism 
	\qq\nn
	\alxydim{@C=1.cm@R=1.cm}{ \cE \ar[r]^{\a_\cE\quad} \ar[d]_{\pi_\cE} & \txT X \ar[d]^{\pi_{\txT X}}\\ X \ar@{=}[r] & X}\,,
	\qqq
	termed the {\bf anchor} ({\bf map});
	\item a binary operation $[\cdot,\cdot]_\cE\colo \G(\cE)\x\G(\cE)\to\G(\cE)$,
	\eit
	satisfying the following conditions:
	\bit
	\item $[\cdot,\cdot]_\cE$ is a Lie bracket;
	\item $\forall \vep_1,\vep_2\in\G(\cE)\ \forall f\in C^\infty(X,\bR)\colo [\vep_1,f\lact\vep_2]_\cE=f\lact[\vep_1,\vep_2]_\cE+\a_\cE(\vep_1)(f)\lact\vep_2$ (the {\bf Leibniz rule}).
	\eit
	\exdef
	
\beg\label{eg:tanLiealgbrd}
Fix a Lie groupoid $\grpd{\xcG}{M}$.\ Let 
\qq\nn
\G(\txT\xcG)_{\rm R}=\bigl\{\  \xcV\in\G(\ker\,\txT s) \quad\vert\quad \forall g\in\xcG\ \forall h\in s^{-1}(\{t(g)\})\colo \txT_h r_{g}\bigl(\xcV(h)\bigr)=\xcV(h.g) \ \bigr\}
\qqq
be the set of {\bf right-invariant vector fields on} $\xcG$,\ defined in terms of the right-fibred action from Ex.\,\ref{eq:Rgrpdact}.\ The ({\bf right}) {\bf tangent algebroid of} $\xcG$ is the vector bundle 
\qq\nn
\pr_1\colo\Lie(\xcG)\equiv\xcE:=\Id^*\ker\,\txT s=M\,\fibx{\Id}{\pi_{\txT\xcG}}\ker\,\txT s\too M
\qqq 
with anchor 
\qq\nn
\a_\xcE=\txT t\circ\pr_2\,,
\qqq 
and Lie bracket 
\qq\nn
[\cdot,\cdot]_\xcE=\iota_{\rm R}\circ[\cdot,\cdot]_{\txT\xcG}\circ\bigl(\iota_{\rm R}^{-1}\x\iota_{\rm R}^{-1}\bigr)
\qqq
induced by the canonical isomorphism $\iota_{\rm R}\colo \G(\txT\xcG)_{\rm R}\to\G(\xcE),\ \xcV\mapsto(\cdot,\xcV(\Id(\cdot)))$.
\eeg
	
\beg
The Lie algebroid $\Lie(\bbB\txG)$ (see:\ Ex.\,\ref{eg:deloop}) is the tangent Lie algebra $\ggt\equiv{\rm Lie}(\txG)$ of $\txG$.
\eeg
	
\beg
The Lie algebroid $\Lie({\rm Pair}(M))$ (see:\ Ex.\,\ref{eg:pairgrpd}) is the tangent bundle $\txT M$,\ with anchor $\id_{\txT M}$,\ and with the Lie bracket given by the commutator of vector fields on $M$.
\eeg
	
\beg\label{eg:actalgbrd}
The Lie algebroid $\ggt\,\lx_\la M\equiv\Lie(\txG\,\lx_\la M)$ (see:\ Ex.\,\ref{eg:actgrpd}) is a vector bundle canonically isomorphic to the trivial bundle
\qq\nn
\alxydim{@C=1.5cm@R=1.5cm}{ \ggt \ar@{^{(}~>}[r] & \ggt\x M \ar[d]_{\pr_2} \\ & M }\,,
\qqq
with the anchor canonically induced by the fundamental vector field $\cK\equiv-\txT\la\circ(\cdot,\brd0_{\txT M})\colo\ggt\to\G(\txT M)$ of $\la$,\ and ---for a choice $\{t_a\}_{a\in\ovl{1,\dim\,\txG}}$ of a basis in $\ggt$---evaluating on $(f^a(\cdot)\, t_a,\cdot)\in\G(\ggt\x M)$ as
\qq\nn
\a_{\ggt\,\lx_\la M}\bigl(f^a(\cdot)\, t_a,\cdot\bigr)=-f^a(\cdot)\,\cK_{t_a}(\cdot)\,,
\qqq
and with the Lie bracket canonically induced by that on $\ggt$---with basis presentation $[t_a,t_b]_\ggt=f_{ab}^{\ \ c}\,t_c$---and the commutator on $\G(\txT M)$ as
\qq\nn
[f_1^a\,t_a,f_2^b\,t_b]_{\ggt\,\lx_\la M}=\bigl(f_2^b\,\cK_{t_b}\bigl(f_1^a\bigr)-f_1^b\,\cK_{t_b}\bigl( f_2^a\bigr)-f_1^b\,f_2^c\,f_{bc}^{\ \ a}\bigr)\,t_a\,.
\qqq
\eeg
	
\beg\cite{Coste:1987}\label{eg:cotang-Liealgd}
The tangent algebroid of the symplectic Lie groupoid from Ex.\,\ref{eg:symplgrpd} is canonically isomorphic to the {\bf cotangent Lie algebroid}
\qq\nn
\bigl(\bigl(\txT^*M,M,\bR^{\dim\,M},\pi_{\txT^*M}\bigr),\Pi^\#,\{\cdot,\cdot\}\bigr)\,,
\qqq
with anchor 
\qq\label{eq:anchor-Poisson}
\Pi^\#\colo\txT^*M\to\txT M\,,
\qqq
canonically induced by the Poisson bivector $\Pi$,\ and the {\bf Koszul bracket} \cite{Koszul:1985bra}
\qq\nn
\{\cdot,\cdot\}\colo\G(\txT^*M)\x\G(\txT^*M)\too\G(\txT^*M),\ (\eta_1,\eta_2)\longmapsto\imath_{\Pi^\#(\eta_1)}\sfd\eta_2-\imath_{\Pi^\#(\eta_2)}\sfd\eta_1+\sfd\bigl(\eta_1\wedge\eta_2(\Pi)\bigr)\,,
\qqq 
as the Lie bracket.
\eeg
	
\bethe\cite{Rybicki:2002ALG}\label{thm:LieBis}
Adopt the notation of Defs.\,\ref{def:bisec},\ and of Ex.\,\ref{eg:tanLiealgbrd}.\ The Lie algebra of the Lie group $\bB(\xcG)$ is isomorphic to the Lie algebra $\G_{\rm c}(\xcE)$ of smooth compactly supported sections of $\xcE\equiv\Lie(\xcG)$.
\ethe

\bethe\cite{Rybicki:2001}\label{thm:LieBisOm}
Adopt the notation of Exs.\,\ref{eg:flatbisec} and \ref{eg:cotang-Liealgd},\ The Lie algebra of the Lie group $\bB(\xcG,\om)$ is isomorphic to the Lie algebra $Z^1_{\rm c}(M)$ of smooth compactly supported de Rham 1-cycles on  $M$.
\ethe
	
\berop\label{prop:Bis-on-E}
Adopt the notation of Defs.\,\ref{def:tstar} and \ref{def:bisec-act},\ and of Ex.\,\ref{eg:tanLiealgbrd}.\ There is a canonical action $\unl{\txT C}\colo\bB(\xcG)\to{\rm GL}(\xcE)$ of the group $\bB(\xcG)$ on the tangent algebroid $\xcE\equiv\Lie(\xcG)$ by vector-bundle isomorphisms.\ It lifts the restriction of the conjugation $C$ to the identity bisection $\Id(M)\subset\xcG$,\ and so covers the shadow action $t_*$ in the sense captured by the commutative diagram,\ written for $\b\in\bB(\xcG)$,
\qq\label{eq:Bis-on-E}
\alxydim{@C=1.cm@R=1.cm}{ \xcE \ar[r]^{\unl{\txT C}_\b} \ar[d]_{\pi_\xcE} & \xcE \ar[d]^{\pi_\xcE}\\ M \ar[r]_{t_*\b} & M}\,.
\qqq
\eerop
	
\berop\cite[Sec.\,3.1]{Crainic:2015msp}\label{prop:J1GonMandE}
Adopt the notation of Ex.\,\ref{def:J1B}.\ There is a canonical structure of (left) $J^1\xcG$-module on $\txT M$,\ with moment map $\pi_{\txT M}\colo\txT M\to M$,\ and action map
\qq\nn
\widetilde t\colo J^1\xcG\fibx{\pi_{J^1}}{\pi_{\txT M}}\txT M\too\txT M,\ \bigl(j^1_m\b,v\bigr)\longmapsto\bigl(\txT_{\b(m)}t\circ \txT_m\b\bigr)(v)\equiv\widetilde t{}_{j^1_m\b}(v)\,,
\qqq
written in terms of an arbitrary representative $\b\in\bB_{\rm loc}(\xcG)$ of $j^1_m\b$.\ There is a similar structure on the tangent algebroid $\xcE\equiv\Lie(\xcG)$,\ with moment map $\pi_\xcE\colo\xcE\to M$,\ and action
\qq\nn
\widetilde C\colo J^1\xcG\fibx{\pi_{J^1}}{\pi_\xcE}\xcE\too\xcE,\ \bigl(j^1_m\b,\vep\bigr)\longmapsto\bigl(\txT_{\b(m)}r_{\b(m)^{-1}}\circ \txT_{(\b(m),\Id_m)}\txm\bigr)\bigl(\bigl(\txT_m\b\circ \txT_{\Id_m}t\bigr)(\vep),\vep\bigr)\equiv\widetilde C{}_{j^1_m\b}(\vep)\,.
\qqq
	
For every $j^1_m\unl\b\in J^1\xcG$ such that there exists $\b\in\bB(\xcG)$ with the property $j^1(\b)(m)=j^1_m\unl\b$,\ written in terms of the group homomorphism \eqref{eq:j1-hom},\ the following identities hold true
\qq\label{eq:tildetC-4-glob}
\widetilde t{}_{j^1_m\unl\b}=\txT_m(t_*\b)\,,\qquad\qquad\qquad\widetilde C{}_{j^1_m\unl\b}=\unl{\txT C}_\b\rstr_{\xcE_m}\,.
\qqq
\eerop

\section{Simplicial tools,\ and the Bott--Schulman--Stasheff cohomology}\label{app:simplicial}
	
Below,\ we list the requisites of the simplicial model of equivariant cohomology of Lie groupoids and its sheaf-theoretic resolution employed in the discussion of multiplicative BSS extensions of de Rham cohomology classes and their geometrisations.
	
	\becon
	Let $\triangle$ be the standard simplex category,\ with objects $[n]=\{0,1,2,\ldots,n\}\subset\bN$ (see,\ {\it e.g.}:\ \cite{Kan:1957}).\ We denote the standard generators of its $\hom$-sets (of order-preserving maps) as
	\qq\nn
	\d^{(n+1)}_i\colo[n]\too[n+1],\ k\longmapsto\left\{ \barr{cl} k & {\rm if}\ k<i \\ k+1 & {\rm if}\ k\geq i \earr \right.\,,\qquad i\in\ovl{0,n}
	\qqq
	({\bf coface maps}),\ and
	\qq\nn
	\si^{(n)}_j\colo[n+1]\too[n],\ k\longmapsto\left\{ \barr{cl} k & {\rm if}\ k\leq j \\ k-1 & {\rm if}\ k>j \earr \right.\,,\qquad j\in\ovl{0,n}
	\qqq
	({\bf codegeneracy maps}).
	\econ
	
	\bedef
	For $n\in\bN$,\ the {\bf topological $n$-simplex} is the topological space
	\qq\nn
	\triangle^n:=\bigl\{ \ (t_1,t_2,\ldots,t_{n+1})\in\bR_{\geq 0}^{\x n+1} \quad\vert\quad \sum_{k=1}^{n+1}\,t_k=1 \ \bigr\}\subset\bR^{\x n+1}\,.
	\qqq
	For $i\in\ovl{0,n+1}$,\ the {\bf $i^{\rm th}$ $n$-face inclusion} is the map
	\qq\nn
	\p_i^{(n+1)}\colo\triangle^n\too\triangle^{n+1},\ (t_1,t_2,\ldots,t_{n+1})\longmapsto(t_1,t_2,\ldots,t_i,0,t_{i+1},t_{i+2},\ldots,t_{n+1})\,.
	\qqq
	\exdef
	
	\bedef\label{def:simpl-obj}
	Let $\cC$ be a category.\ A {\bf simplicial object} in $\cC$ is a presheaf $C_\bullet\colo\triangle^{\rm op}\to\cC$.\ The functorial images of the generators of the $\hom$-sets $\triangle([k],[l]),\ k,l\in\bN$ are called {\bf face maps} $d^{(n+1)}_i=C_\bullet(\d^{(n+1)}_i),\ i\in\ovl{0,n}$ and {\bf degeneracy maps} $s^{(n)}_j=C_\bullet(\si^{(n)}_j),\ j\in\ovl{0,n}$ (defined for all $n\in\bN$),\ respectively.\ Given simplicial objects $C^a_\bullet,\ a\in\{1,2\}$,\ a {\bf simplicial morphism} between them is a natural transformation $\eta\colo C^1_\bullet\Rightarrow C^2_\bullet$. 
	
	In particular,\ a simplicial object $M_\bullet\colo\triangle^{\rm op}\to\Man$ in the category $\Man$ of smooth manifolds is termed a {\bf simplicial manifold}.\ Its (smooth) face maps determine the {\bf Dupont operators}
	\qq\nn
	\D^{(\bullet)}_{(n)}:=\sum_{k=0}^{n+1}\,(-1)^k\,d_k^{(n+1)*}\colo\Om^\bullet(M_n)\too\Om^\bullet(M_{n+1})
	\qqq
	on differential forms on its component objects $M_n=M_\bullet([n])\in\Man$.
	\exdef
	
	\beg\cite{Segal:1968}\label{eg:face-nerve-LieGrpd}
	Adopt the notation of Defs.\,\ref{def:grpd} and \ref{def:simpl-obj}.\ The {\bf nerve} of a Lie groupoid $\grpd{\xcG}{M}$ is a simplicial manifold $\xcG_\bullet\equiv N_\bullet(\grpd{\xcG}{M})$ with component objects $\xcG_0\equiv N_0(\grpd{\xcG}{M})=M$ and $\xcG_n\equiv N_n(\grpd{\xcG}{M})=\xcG\fibx{s}{t}\xcG\fibx{s}{t}\cdots\fibx{s}{t}\xcG\subset\xcG^{\x n}$ ($n>0$),\ face maps ($m>0$)
	\qq\nn
	&d^{(1)}_0=t\,,\qquad\qquad d^{(1)}_1=s\,,&\cr\cr
	&d^{(m+1)}_i(g_{m+1},g_m,\ldots,g_1)=\left\{ \barr{cl} (g_{m+1},g_m,\ldots,g_2) & {\rm if}\ i=0 \\ (g_{m+1},g_m,\ldots,g_{i+2},g_{i+1}\circ g_i,g_{i-1},g_{i-2},\ldots,g_1) & {\rm if}\ i\in\ovl{1,m} \\ 
	(g_m,g_{m-1},\ldots,g_1) & {\rm if}\ i=m+1 \earr \right.\,,&
	\qqq
	and degeneracy maps
	\qq\nn
	&s^{(0)}_0=\Id\,,&\cr\cr
	&s^{(m)}_j(g_m,g_{m-1},\ldots,g_1)=\left\{ \barr{cl} (g_m,g_{m-1},\ldots,g_1,\Id_{s(g_1)}) & {\rm if}\ j=0 \\ (g_m,g_{m-1},\ldots,g_{j+1},\Id_{t(g_j)},g_j,g_{j-1},\ldots,g_1) & {\rm if}\ j\in\ovl{1,m} \earr \right.\,.&
	\qqq
	\eeg
	
	\beg\label{eg:Cech-nerve}
	Let $M\in\Man$,\ and let $\cO=\{O_i\}_{i\in I}$ be an open cover of $M$.\ The manifold $\check{Y}_\cO:=\bigsqcup_{i\in I}\,O_i$ (endowed with the standard disjoint-sum topology and smooth structure inherited from the submanifolds $O_i\subset M$) submerses $M$ as
	\qq\nn
	\check{\pi}\colo\check{Y}_\cO\twoheadrightarrow M\,,\ (x,i)\longmapsto x\,.
	\qqq
	The nerve $N_\bullet({\rm Pair}_{\check\pi}(\check Y{}_\cO))\equiv N_\bullet\cO$ of the submersion groupoid ${\rm Pair}_{\check\pi}(\check Y{}_\cO)$ (see:\ Ex.\,\ref{eg:subm-grpd}),\ with $N_n\cO\cong\bigsqcup_{i_0,i_1,\ldots,i_n\in I}\,O_{i_0 i_1\cdots i_n}$,\ where $O_{i_0 i_1\cdots i_n}\equiv O_{i_0}\cap O_{i_1}\cap\cdots\cap O_{i_n}$,\ is termed the {\bf \v Cech nerve} of $\cO$.\ The corresponding Dupont operators are known as the {\bf \v Cech coboundary operators}.
	\eeg
	
	\bedef\cite{Segal:1974}\label{def:fat-geom-real}
	Adopt the notation of Def.\,\ref{def:simpl-obj}.\ Let $X_\bullet\colo\triangle^{\rm op}\to\Top$ be a simplicial object in the category $\Top$ of topological spaces,\ with objects $X_n\equiv X_\bullet([n])$ and face maps $\xi^{(n+1)}_i=X_\bullet(\d^{(n+1)}_i),\ i\in\ovl{0,n}$.\ A ({\bf fat}) {\bf geometric realisation} of $X_\bullet$ is the quotient topological space
	\qq\nn
	\Vert X_\bullet\Vert:=\bigsqcup_{n\in\bN}\,\bigl(X_n\x\triangle^n\bigr)/\sim
	\qqq
	obtained through identifications---written for $(x,t)\in X_{n+1}\x\triangle^n$---
	\qq\label{eq:desc-geom-real}
	\bigl(\xi^{(n+1)}_i(x),t\bigr)\sim\bigl(x,\p_i^{(n+1)}(t)\bigr)\,.
	\qqq
	\exdef
	
	\bedef\cite{Bott:1976bss,Dupont:1976,Dupont:1988}\label{def:BSS-complex}
	Adopt the notation of Ex.\,\ref{eg:face-nerve-LieGrpd}.\ The {\bf Bott--Schulman--Stasheff} ({\bf BSS}) {\bf bicomplex} of a Lie groupoid $\grpd{\xcG}{M}$ is the cochain bicomplex 
	\qq\nn
	\bigl(\Om^{\bullet_1}\bigl(\xcG_{\bullet_2}\bigr),\sfd^{(\bullet_1)}_{(\bullet_2)},\D_{(\bullet_2)}^{(\bullet_1)}\bigr)
	\qqq
	with the two commuting coboundary operators:\ the de Rham differentials $\sfd^{(\bullet_1)}_{(\bullet_2)}\colo\Om^{\bullet_1}(\xcG_{\bullet_2})\to\Om^{\bullet_1+1}(\xcG_{\bullet_2})$,\ and the Dupont operators $\D_{(\bullet_2)}^{(\bullet_1)}\colo\Om^{\bullet_1}(\xcG_{\bullet_2})\to\Om^{\bullet_1}(\xcG_{\bullet_2+1})$ (see:\ Def.\,\ref{def:simpl-obj}).\ Its total cohomology,\ {\it i.e.},\ the cohomology of the diagonal complex ${\rm Tot}^\bullet(\Om^{\bullet_1}(\xcG_{\bullet_2}),\sfd^{(\bullet_1)}_{(\bullet_2)},\D^{(\bullet_1)}_{(\bullet_2)})$ with components
	\qq\nn
	{\rm Tot}^r\bigl(\Om^{\bullet_1}(\xcG_{\bullet_2}),\sfd^{(\bullet_1)}_{(\bullet_2)},\D^{(\bullet_1)}_{(\bullet_2)}\bigr)=\bigoplus_{p+q=r}\,\Om^p(\xcG_q)\,,
	\qqq
	and the {\bf BSS coboundary operator}
	\qq\nn
	D_{(r)}\rstr_{\Om^p(\xcG_q)}=\D^{(p)}_{(q)}+(-1)^q\,\sfd^{(p)}_{(q)}\,,
	\qqq
	is called the {\bf BSS cohomology} of $\grpd{\xcG}{M}$.\ We denote its groups as
	\qq\nn
	H^r_{\rm BSS}(\grpd{\xcG}{M})=\tfrac{\ker\,D_{(r)}}{\im\,D_{(r-1)}}\,.
	\qqq
	\exdef
	
	\bedef\label{def:mult-k-form}
	Adopt the notation of Def.\,\ref{def:BSS-complex}.\ A $k$-form $\rho\in\Om^k(\cG_1)$ is called {\bf multiplicative} if $\rho\in\ker\,\D^{(k)}_{(1)}$.
	\exdef
	
\berop\cite[Lem.\,4.2]{Crainic:2015msp}\label{prop:multi-anch}
Adopt the notation of Def.\,\ref{def:mult-k-form} and \Reqref{eq:MC-R},\ and let $\rho\in\Om^k(\xcG)$ be multiplicative.\ Then,\ the following identity holds true at $g\in\xcG$ for all $\vep\in(\ker\,\txT s)_g$ and $\xi_l\in\txT_g\xcG,\ l\in\ovl{1,k-1}$:
\qq\nn
\rho(g)(\vep,X_1,X_2,\ldots,X_{k-1})=\rho\bigl(\Id_{t(g)}\bigr)\bigl(\theta_{\rm R}(\vep),\txT_g(\Id\circ t)(X_1),\txT_g(\Id\circ t)(X_2),\ldots,\txT_g(\Id\circ t)(X_{k-1})\bigr)\,.
\qqq
\eerop
	
\bethe\cite[Thm.\,1 \& Prop.\,4.1]{Crainic:2015msp}\label{thm:CSS}
Adopt the notation of Defs.\,\ref{def:J1B} and \ref{def:mult-k-form},\  of Ex.\,\ref{eg:tanLiealgbrd},\ and of \Reqref{eq:MC-R}.\ Let $\grpd{\xcG}{M}$ be a Lie groupoid with tangent algebroid $\xcE$,\ and let $\grpd{J^1\xcG}{M}$ be the first-jet groupoid of $\xcG$.\ For every multiplicative $k$-form $\rho\in\Om^k(\xcG)$,\ there exists a pair $(\kappa,S)$ composed of a vector-bundle morphism 
\qq\label{eq:comom-expl}
\alxydim{@C=0.75cm@R=1.cm}{ \xcE \ar[rr]^{\kappa\qquad} \ar[dr]_{\pi_\xcE} & & \wedge^{k-1}\txT^*M \ar[dl]^{\pi_{\wedge^{k-1}\txT^*M}} \\ & M & }\,,\qquad\qquad \kappa(\vep)=\Id^*\bigl(\imath_\vep\rho\bigr)\,,
\qqq
and a map 
\qq\label{eq:Spencer-expl}
\alxydim{@C=0.75cm@R=1.cm}{ J^1\xcG \ar[rr]^{S\quad} \ar[dr]_{\t} & & \wedge^k\txT^*M \ar[dl]^{\pi_{\wedge^k\txT^*M}} \\ & M & }\,,\qquad\qquad S\bigl(j^1_m\b\bigr)=\rho_{\b(m)}\circ\wedge^k\bigl(\txT_m\b\circ\bigl(\widetilde t{}_{j^1_m\b}\bigr)^{-1}\bigr)\,,
\qqq
with the following properties,\ written---in terms of the canonical actions of $J^1\xcG$:\ $\widetilde t{}_\cdot$ on $\txT M$ and $\widetilde C$ on $\xcE$ described in Prop.\,\ref{prop:J1GonMandE}---for all $g\in\xcG$ with $s(g)=m$ and arbitrary $j^1_m\b\in J^1\xcG$ with $\b(m)=g$,\ $\vep,\vep_1,\vep_2\in\xcE_m$,\ and $v_1,v_2,\ldots,v_{k-1}\in\txT_m M$:
\qq\label{eq:comom-equiv}
&&\kappa\bigl(\widetilde C{}_{j^1_m\b}(\vep)\bigr)\bigl(\widetilde t{}_{j^1_m\b}(v_1),\widetilde t{}_{j^1_m\b}(v_2),\ldots,\widetilde t{}_{j^1_m\b}(v_{k-1})\bigr)-\kappa(\vep)(v_1,v_2,\ldots,v_{k-1})\\ \cr
&=&S\bigl(j^1_m\b\bigr)\bigl(\widetilde t{}_{j^1_m\b}\bigl(\a_\xcE(\vep)\bigr),\widetilde t{}_{j^1_m\b}(v_1),\widetilde t{}_{j^1_m\b}(v_2),\ldots,\widetilde t{}_{j^1_m\b}(v_{k-1})\bigr)\,,\nn
\qqq
and
\qq\label{eq:comom-asymm}
\imath_{\a_\xcE(\vep_1)}\kappa(\vep_2)+\imath_{\a_\xcE(\vep_2)}\kappa(\vep_1)=0\,.
\qqq
The $k$-form $\rho$ evaluates on an arbitrary $k$-tuple $(X_1,X_2,\ldots,X_k)\in\txT_g\xcG$ at $g\in s^{-1}(\{m\})$ as
\qq\label{eq:rho-as-Skappa}
\rho(X_1,X_2,\ldots,X_k)=S\bigl(j^1_m\b\bigr)\bigl(\widetilde t{}_{j^1_m\b}(\unl{X}_1),\widetilde t{}_{j^1_m\b}(\unl{X}_2),\ldots,\widetilde t{}_{j^1_m\b}(\unl{X}_k)\bigr)\\ \cr 
+\sum_{p=1}^k\,\sum_{\sigma\in{\rm Shuffle}(p,k-p)}\,(-1)^{|\si|}\,\bigl(\kappa\circ\theta_{\rm R}\bigr)(\vep_{\si(1)})\bigl(\a_\xcE(\vep_{\si(2)}),\ldots,\a_\xcE(\vep_{\si(p)}),\widetilde t{}_{j^1_m\b}(\unl{X}_{\si(p+1)}),\ldots,\widetilde t{}_{j^1_m\b}(\unl{X}_{\si(k)})\bigr)\,,\nn
\qqq
where $\unl{X}_i:=\txT_g s(X_i),\ i\in\ovl{1,k}$,\ and---for $j^1_m\b\in J^1\xcG$ with $\b(m)=g$ chosen arbitrarily---
\qq\nn
\vep_j:=\bigl(\id_{\txT_g\xcG}-\txT_m\b\circ\txT_g s\bigr)(X_j)\,,\quad j\in\ovl{1,k}\,.
\qqq
Above,\ ${\rm Shuffle}(p,k-p)\subset S_k$ denotes the set of $(p,k-p)$-shuffles of $\ovl{1,k}$,\ and $(-1)^{|\cdot|}\colo S_k\to\{-1,1\}$ is the sign function.
\ethe
	
\bedef\label{def:Spencer-pair}
Adopt the notation of Thm.\,\ref{thm:CSS}.\ The bundle map $\kappa$ and the map $S$ is called ($k${\bf -}){\bf co\-mo\-men\-tum} and---after \cite{Crainic:2015msp}---the {\bf scalar} ($k${\bf -}){\bf Spencer operator} of $\rho$,\ respectively.\ The pair $(\kappa,S)$ shall be referred to as the {\bf Spencer pair} of $\rho$.
\exdef

\begin{propanition}\label{prop:rholo-grp}
Adopt the notation of Defs.\,\ref{def:bisec} and \ref{def:mult-k-form}.\ Let $\rho\in\Om^k(\xcG)$ be multiplicative.\ The set 
\qq\nn
\bB_\rho(\xcG):=\{\ \b\in\bB(\xcG) \quad\vert\quad \b^*\rho=0\ \}
\qqq
is a subgroup of the group $\bB(\xcG)$ of bisections of $\xcG$.\ We shall call it the {\bf $\rho$-holonomic group} of $\xcG$.
\end{propanition}
	
\berop\label{prop:Spencer-on-holo}
Adopt the notation of Defs.\,\ref{def:Spencer-pair},\ \ref{prop:rholo-grp} and \ref{def:J1B}.\ Let $(\kappa,S)$ be the Spencer pair of a multiplicative $k$-form $\rho\in\Om^k(\xcG)$ on a Lie groupoid $\grpd{\xcG}{M}$,\ and let
	\qq\nn
	\bB_\rho(J^1\xcG):=j^1\bigl(\bB_\rho(\xcG)\bigr)
	\qqq
	be the homomorphic image of the $\rho$-holonomic group of $\xcG$ in the group $\bB(J^1\xcG)$ of bisections of the first-jet groupoid $\grpd{J^1\xcG}{M}$ of $\xcG$ along the group homomorphism $j^1$ of \Reqref{eq:j1-hom}.\ The restriction of the scalar $k$-Spencer operator $S$ to $\bB_\rho(J^1\xcG)$ vanishes identically,\ {\it i.e.},
	\qq\nn
	\forall\ \b\in\bB_\rho(\xcG)\ \forall\ m\in M\colo S\bigl(j^1_m\b\bigr)=0\,.
	\qqq
	Hence,\ the $k$-comomentum $\kappa$ is equivariant with respect to the natural action of $\bB_\rho(J^1\xcG)$ on the tangent algebroid $\xcE$ of $\xcG$ and the exterior bundle $\wedge^\bullet\txT^*M$ over its object manifold $M$.
	\eerop
	
\bedef\label{def:simpl-cover}
Adopt the notation of Def.\,\ref{def:simpl-obj}.\ Let $M_\bullet\colo\triangle^{\rm op}\to\Man$ be a simplicial manifold with face maps $\{d^{(n+1)}_i\ \vert\ i\in\ovl{0,n}\}_{n\in\bN}$.\ A family $\{\cO_{M_n}\}_{n\in\bN}$ of open covers $\cO_{M_n}=\{O^{(n)}_{i_n}\}_{i_n\in I_n}$ of the respective objects $M_n\in\Man$ is called {\bf simplicial} if the index sets $I_n,\ n\in\bN$ compose a simplicial set $I_\bullet\colo\triangle^{\rm op}\to\Set$ with face maps $\iota^{(n+1)}_i\equiv I_\bullet(\d^{(n+1)}_i)\colo I_{n+1}\to I_n$ such that the relation
\qq\nn
d^{(n+1)}_j\bigl(O^{(n+1)}_{i_n}\bigr)\subset O^{(n)}_{\iota^{(n+1)}_j(i_n)}
\qqq
holds true for all $j\in\ovl{0,n+1}$ and $n\in\bN^\x$.
\exdef
	
\bethe\cite{Tu:2006}\label{thm:simpl-ref}
For every simplicial manifold,\ and every collection of open covers of its objects,\ there exists a refinement of the latter which is simplicial in the sense of Def.\,\ref{def:simpl-cover}.
\ethe
	
\berop\label{prop:tricompl}
Adopt the notation of Def.\,\ref{def:simpl-cover}.\ Let $M_\bullet\colo\triangle^{\rm op}\to\Man$ be a simplicial manifold with face maps $\{d^{(n+1)}_i\ \vert\ i\in\ovl{0,n}\}_{n\in\bN}$.\ Given a simplicial family $\{\cO_{M_n}\}_{n\in\bN}$ of open covers $\cO_{M_n}=\{O^{(n)}_{i_n}\}_{i_n\in I_n}$ of the respective objects $M_n\in\Man$,\ the Dupont operators $\D^{(\bullet_1)}_{(\bullet_2)}\colo\Om^{\bullet_1}(M_{\bullet_2})\to\Om^{\bullet_1}(M_{\bullet_2+1})$ canonically extend to \v{C}ech cochains with values in the sheaves $\unl{\Om^{\bullet_1}}(M_{\bullet_2})$ of locally smooth $\bullet_1$-forms on $M_{\bullet_2}$ as
\qq\nn
\D^{(p,q)}_{(r)}&\colo&\vC^p\bigl(\cO_{M_r},\unl{\Om^q}\bigr)\too\vC^p\bigl(\cO_{M_{r+1}},\unl{\Om^q}\bigr)\cr\cr 
&\colo&\om\equiv(\om_{j_0 j_1\ldots j_p})\longmapsto\left(\sum_{k=0}^{r+1}\,(-1)^k\,d_k^{(r+1)*}\om_{\iota^{(r+1)}_k(i_0) \iota^{(r+1)}_k(i_1)\ldots \iota^{(r+1)}_k(i_p)}\equiv\bigl(\D^{(p,q)}_{(r)}\om\bigr)_{i_0 i_1\ldots i_p}\right)\,,
\qqq 
and thus give rise to a cochain tricomplex  
\qq\nn
\bigl(\vC^{\bullet_1}\bigl(\cO_{M_{\bullet_3}},\unl{\Om^{\bullet_2}}\bigr),\vd^{(\bullet_1,\bullet_2)}_{(\bullet_3)},\sfd^{(\bullet_1,\bullet_2)}_{(\bullet_3)},\D^{(\bullet_1,\bullet_2)}_{(\bullet_3)}\bigr)\,.
\qqq
\eerop

\section{The lagrangean anomaly---a derivation}\label{app:lagr-anom}

We present an adaptation of the result \cite[Lem.\,E.2]{Suszek:2026col} which computes the gauge transformation of the action functional of the classic Poisson $\si$-model \eqref{eq:PSMAct} subject to the reinterpretation \eqref{eq:PSM-redef} from Sec.\,\ref{sub:PSM-redef}.\ In our discussion,\ we use the notation of Sec.\,\ref{sub:PSM-redef}.

Let $\grpd{(\xcG,\om)}{(M,\Pi)}$ be a symplectic Lie groupoid,\ and let $\xcP\to\Si$ be a principaloid $\xcG$-bundle endowed with connection $\Theta$ with local presentation \eqref{eq:Thet-loc} over a trivialising cover $\{O_i\}_{i\in I}$ of $\Si$.\ Given an arbitrary gauge transformation $\Phi\in{\rm Gau}(\xcP)$ with local data $h_i\colo O_i\to\bB(\xcG)$,\ denote
\qq\nn
\widetilde h{}_i\colo O_i\x M\too\xcG,\ (\si,m)\longmapsto h_i(\si)(m)
\qqq 
alongside
\qq\nn
H_i=\bigl(\pr_1,t\circ\widetilde h{}_i\bigr)\colo O_i\x M\too O_i\x M,\ (\si,m)\longmapsto\bigl(\si,t_*\bigl(h_i(\si)\bigr)(m)\bigr)\,,
\qqq
where $t_*$ is the shadow action from Def.\,\ref{def:tstar}.

The 2-form
\qq\nn
\om[\Theta_i]:=\bigl(\kappa_\xcE\circ A_i+\tfrac{1}{2}\,\ev\circ\bigl(\bigl(\kappa_\xcE\circ A_i\bigr)\ox\bigl(\a_\xcE\circ A_i\bigr)\bigr)\bigr)\circ{\rm Alt}\in\Om^2(O_i\x M)\,,
\qqq
satisfies the {\bf master identity}
\qq\label{eq:augs-glue}
\om^{h_i}[\Theta_i]-H_i^*\om[\Theta_i^\Phi]=\widetilde h{}_i^*\om-h_i\om\,,
\qqq
in which
\qq\nn
\om^{h_i}[\Theta_i]=\bigl(\kappa_\xcE^{h_i}\circ A_i+\tfrac{1}{2}\,\ev\circ\bigl(\bigl(\kappa_\xcE^{h_i}\circ A_i\bigr)\ox\bigl(\a_\xcE\circ A_i\bigr)\bigr)\bigr)\circ{\rm Alt}\,,\qquad\qquad\kappa_\xcE^{h_i}(\si,m)=\bigl(t_*\bigl(h_i(\si)\bigr)\bigr)^*\bigl(\kappa_\xcE\circ\unl{\txT C}_{h_i(\si)}\rstr_{\xcE_m}\bigr)
\qqq
and 
\qq\nn
h_i\om(\si,m)=\om\circ\bigl(\txT_m\bigl(h_i(\si)\bigr),\txT_m\bigl(h_i(\si)\bigr)\bigr)\,.
\qqq
Accordingly,\ whenever $h_i$ takes values in the \emph{lagrangean} group of $(\xcG,\om)$ from Ex.\,\ref{eg:lagrbisec},\ we obtain the reduced identity
\qq\label{eq:augs-glue-red}
\om[\Theta_i]-H_i^*\om[\Theta_i^\Phi]=\widetilde h{}_i^*\om\,,
\qqq

The proof of the identity departs from the observation that the pullback  
\qq\nn
\widetilde h{}_i^*\om(\si,m)\equiv\om\circ\bigl(\txT_{(\si,m)}\widetilde h{}_i,\txT_{(\si,m)}\widetilde h{}_i\bigr)\,,
\qqq
evaluated at $(\si,m)\in O_i\x M$,\ localises the multiplicative 2-form $\om$ on the bisection $h_i(\si)(M)\subset\xcG$,\ whose tangent at $\widetilde h{}_i(\si,m)$ admits a splitting (see:\ Prop.\,\ref{prop:beta-split}) engendered by the decomposition 
\qq\nn
\txT_{(\si,m)}\widetilde h{}_i=\txT_m\bigl(h_i(\si)\bigr)+\txT_\si\widetilde h{}_i(\cdot,m)\,,
\qqq
with $\im\,\txT_\si\widetilde h{}_i(\cdot,m)\subset(\ker\,\txT s)_{\widetilde h{}_i(\si,m)}$.\ Here,\ it is to be understood that $\txT_\si\widetilde h{}_i(\cdot,m)$ (resp.\ $\txT_m(h_i(\si))$) acts on the first (resp.\ the second) summand in $\txT_\si O_i\oplus\txT_m M\equiv\txT_{(\si,m)}(O_i\x M)$.\ Upon invoking \eqref{eq:rho-as-Skappa},\ we then obtain
\qq\nn
\widetilde h{}_i^*\om(\si,m)&=&\bigl(\kappa_\xcE\circ\theta_{\rm R}\circ\txT_\si\widetilde h{}_i(\cdot,m)\circ\pr_1\bigr)\bigl(\txT_m\bigl(t_*h_i(\si)\bigr)\circ\pr_2\bigr)-\bigl(\kappa_\xcE\circ\theta_{\rm R}\circ \txT_\si\widetilde h{}_i(\cdot,m)\circ\pr_2\bigr)\bigl(\txT_m\bigl(t_*h_i(\si)\bigr)\circ\pr_1\bigr)\cr\cr 
&&+\bigl(\kappa_\xcE\circ\theta_{\rm R}\circ \txT_\si\widetilde h{}_i(\cdot,m)\circ\pr_1\bigr)\bigl(\txT_{\widetilde h{}_i(\si,m)}t\circ\txT_\si\widetilde h{}_i(\cdot,m)\circ\pr_2\bigr)+h_i\om(\si,m)\,,
\qqq
with $\pr_1$ and $\pr_2$ projecting canonically onto the first and second vector argument of the 2-form,\ respectively.\ The $\bB(\xcG)$-equivariance of $\a_\xcE$ (see:\ \cite[Cor.\,E.3\,(ii)]{Strobl:2025}) enables us to rewrite the above as 
\qq\nn
\widetilde h{}_i^*\om(\si,m)&=&\bigl(\kappa^{h_i}_\xcE\circ\widetilde h{}_i^*\vartheta_{\rm L}(\si,m)\circ\pr_1\bigr)(\pr_2)-\bigl(\kappa^{h_i}_\xcE\circ\widetilde h{}_i^*\vartheta_{\rm L}(\si,m)\circ\pr_2\bigr)(\pr_1)\cr\cr 
&&+\bigl(\kappa^{h_i}_\xcE\circ\widetilde h{}_i^*\vartheta_{\rm L}(\si,m)\circ\pr_1\bigr)\bigl(\a_\xcE\circ\widetilde h{}_i^*\vartheta_{\rm L}(\si,m)\circ\pr_2\bigr)+h_i\om(\si,m)
\qqq
using the shorthand notation (see:\ Prop.\,\ref{prop:J1GonMandE})
\qq\nn
\widetilde h{}_i^*\vartheta_{\rm L}(\si,m):=\unl{\txT C}_{h_i(\si)^{-1}}\circ\theta_{\rm R}\circ \txT_\si\widetilde h{}_i(\cdot,m)=\txT_{\Id_{t_*(h_i(\si))(m)}}C_{h_i(\si)^{-1}}\circ\theta_{\rm R}\circ\txT_\si\widetilde h{}_i(\cdot,m)\,.
\qqq
Identity \eqref{eq:comom-asymm} leads to the final rewriting:
\qq\nn
\widetilde h{}_i^*\om(\si,m)-h_i\om(\si,m)&=&\bigl(\kappa_\xcE\circ\widetilde h{}_i^*\vartheta_{\rm L}(\si,m)\circ\pr_1\bigr)(\pr_2)-\bigl(\kappa_\xcE\circ\widetilde h{}_i^*\vartheta_{\rm L}(\si,m)\circ\pr_2\bigr)(\pr_1)\cr\\
&&+\tfrac{1}{2}\,\bigl(\bigl(\kappa_\xcE\circ\widetilde h{}_i^*\vartheta_{\rm L}(\si,m)\circ\pr_1\bigr)\bigl(\a_\xcE\circ\widetilde h{}_i^*\vartheta_{\rm L}(\si,m)\circ\pr_2\bigr) \label{eq:rho-pull-bij} \\ \cr 
&&-\bigl(\kappa_\xcE\circ\widetilde h{}_i^*\vartheta_{\rm L}(\si,m)\circ\pr_2\bigr)\bigl(\a_\xcE\circ\widetilde h{}_i^*\vartheta_{\rm L}(\si,m)\circ\pr_1\bigr)\bigr)\,.\nn
\qqq
	
On the other hand,\ we have,\ as before,
\qq\nn
\txT_{(\si,m)}H_i&=&\bigl(\pr_1,\a_\xcE\circ\theta_{\rm R}\circ \txT_\si\widetilde h{}_i(\cdot,m)+\txT_m\bigl(t_*h_i(\si)\bigr)\bigr)\cr\cr 
&=&\bigl(\id_{\txT_\si\Si}\x\txT_m\bigl(t_*h_i(\si)\bigr)\bigr)\circ\bigl(\id_{\txT_{(\si,m)(O_i\x M)}}+\a_\xcE\circ\widetilde h{}_i^*\vartheta_{\rm L}(\si,m)\bigr)\bigr)\,,
\qqq
and so we begin the calculation of $H_i^*\om[\Theta_i^\Phi](\si,m)$ from
\qq\nn
&&\om[\Theta_i^\Phi]\circ\bigl(\id_{\txT_\si\Si}\x \txT_m\bigl(t_*h_i(\si)\bigr),\id_{\txT_\si\Si}\x \txT_m\bigl(t_*h_i(\si)\bigr)\bigr)\cr\cr 
&\equiv&\bigl(\bigl(t_*\bigl(h_i(\si)\bigr)\bigr)^*\kappa_\xcE\circ A_i^\Phi\bigl(\si,\bigl(t_*\bigl(h_i(\si)\bigr)\bigr)(m)\bigr)\cr\cr &&+\tfrac{1}{2}\,\ev\bigl(\bigl(\kappa_\xcE\circ A_i^\Phi\bigl(\si,\bigl(t_*\bigl(h_i(\si)\bigr)\bigr)(m)\bigr)\bigr)\ox\bigl(\a_\xcE\circ A_i^\Phi\bigl(\si,\bigl(t_*\bigl(h_i(\si)\bigr)\bigr)(m)\bigr)\bigr)\bigr)\bigr)\circ{\rm Alt}\cr\cr 
&=&\bigl(\bigl(t_*\bigl(h_i(\si)\bigr)\bigr)^*\kappa_\xcE\circ\unl{\txT C}{}_{h_i(\si)}\circ\bigl(A_i+\widetilde h{}_i^*\vartheta_{\rm L}\bigr)(\si,m)\cr\cr 
&&+\tfrac{1}{2}\,\ev\bigl(\bigl(\kappa_\xcE\circ \unl{\txT C}{}_{h_i(\si)}\circ\bigl(A_i+\widetilde h{}_i^*\vartheta_{\rm L}\bigr)(\si,m)\bigr)\ox\bigl(\a_\xcE\circ\unl{\txT C}{}_{h_i(\si)}\circ\bigl(A_i+\widetilde h{}_i^*\vartheta_{\rm L}\bigr)(\si,m)\bigr)\bigr)\bigr)\circ{\rm Alt}\cr\cr 
&=&\bigl(\kappa^{h_i}_\xcE\circ A_i+\kappa^{h_i}_\xcE\circ\widetilde h{}_i^*\vartheta_{\rm L}+\tfrac{1}{2}\,\ev\bigl(\bigl(\kappa^{h_i}_\xcE\circ\bigl(A_i+\widetilde h{}_i^*\vartheta_{\rm L}\bigr)\bigr)\ox\bigl(\a_\xcE\circ\bigl(A_i+\widetilde h{}_i^*\vartheta_{\rm L}\bigr)\bigr)\bigr)\bigr)(\si,m)\circ{\rm Alt}\cr\cr 
&\equiv&\om^{h_i}[\Theta_i](\si,m)+\bigl(\kappa^{h_i}_\xcE\circ\widetilde h{}_i^*\vartheta_{\rm L}+\tfrac{1}{2}\,\ev\bigl(\bigl(\kappa^{h_i}_\xcE\circ A_i\bigr)\ox\bigl(\a_\xcE\circ\widetilde h{}_i^*\vartheta_{\rm L}\bigr)\bigr)+\tfrac{1}{2}\,\ev\bigl(\bigl(\kappa^{h_i}_\xcE\circ\widetilde h{}_i^*\vartheta_{\rm L}\bigr)\ox\bigl(\a_\xcE\circ A_i\bigr)\bigr)\cr\cr &&+\tfrac{1}{2}\,\ev\bigl(\bigl(\kappa^{h_i}_\xcE\circ\widetilde h{}_i^*\vartheta_{\rm L}\bigr)\ox\bigl(\a_\xcE\circ\widetilde h{}_i^*\vartheta_{\rm L}\bigr)\bigr)\bigr)(\si,m)\circ{\rm Alt}\,,
\qqq
substituting \eqref{eq:gaugetrafo} and once more invoking the $\bB(\xcG)$-equivariance of $\a_\xcE$ along the way.

This we subsequently reduce to the form 
\qq\nn
&&\om[\Theta_i^\Phi]\circ\bigl(\id_{\txT_\si\Si}\x \txT_m\bigl(t_*h_i(\si)\bigr),\id_{\txT_\si\Si}\x \txT_m\bigl(t_*h_i(\si)\bigr)\bigr)\cr\cr 
&=&\om^{h_i}[\Theta_i](\si,m)+\bigl(\kappa^{h_i}_\xcE\circ\widetilde h{}_i^*\vartheta_{\rm L}+\ev\bigl(\bigl(\kappa^{h_i}_\xcE\circ A_i\bigr)\ox\bigl(\a_\xcE\circ\widetilde h{}_i^*\vartheta_{\rm L}\bigr)\bigr)+\tfrac{1}{2}\,\ev\bigl(\bigl(\kappa^{h_i}_\xcE\circ\widetilde h{}_i^*\vartheta_{\rm L}\bigr)\ox\bigl(\a_\xcE\circ\widetilde h{}_i^*\vartheta_{\rm L}\bigr)\bigr)\bigr)(\si,m)\circ{\rm Alt}
\qqq
with the help of identity \eqref{eq:comom-asymm}.\ At this stage,\ it remains to precompose the above with the cartesian square of the linear operator $\id_{\txT_{(\si,m)}(O_i\x M)}+\a_\xcE\circ\widetilde h{}_i^*\vartheta_{\rm L}(\si,m)\equiv\id_\txT+\a_\xcE\circ\widetilde h{}_i^*\vartheta_{\rm L}(\si,m)$ to obtain the final expression for $H_i^*\om[\Theta_i^\Phi](\si,m)$.\ To this end,\ we calculate---invoking the foliated nature of the various forms involved and Rem.\,\ref{rem:kappAi-expl} along the way,\ and dropping $(\si,m)$ for the sake of transparency---
\qq\nn
&&\bigl(\om^{h_i}[\Theta_i]+\ev\circ\bigl(\bigl(\kappa^{h_i}_\xcE\circ A_i\bigr)\ox\bigl(\a_\xcE\circ\widetilde h{}_i^*\vartheta_{\rm L}\bigr)\bigr)\circ{\rm Alt}\bigr)\circ\bigl(\id_\txT+\a_\xcE\circ\widetilde h{}_i^*\vartheta_{\rm L},\id_\txT+\a_\xcE\circ\widetilde h{}_i^*\vartheta_{\rm L}\bigr)\cr\cr 
&=&\om^{h_i}[\Theta_i]\circ\bigl(\bigl(\id_\txT,\id_\txT\bigr)+(\id_\txT,\a_\xcE\circ\widetilde h{}_i^*\vartheta_{\rm L})+(\a_\xcE\circ\widetilde h{}_i^*\vartheta_{\rm L},\id_\txT\bigr)\bigr)+\ev\circ\bigl(\bigl(\kappa^{h_i}_\xcE\circ A_i\bigr)\ox\bigl(\a_\xcE\circ\widetilde h{}_i^*\vartheta_{\rm L}\bigr)\bigr)\circ\bigl(\id_\txT,\id_\txT\bigr)\circ{\rm Alt}\cr\cr 
&=&\om^{h_i}[\Theta_i]+\bigl(\bigl(\kappa^{h_i}_\xcE\circ A_i\bigr)\circ\bigl(\bigl(\id_\txT,\a_\xcE\circ\widetilde h{}_i^*\vartheta_{\rm L}\bigr)+\bigl(\a_\xcE\circ\widetilde h{}_i^*\vartheta_{\rm L},\id_\txT\bigr)\bigr)+\ev\circ\bigl(\bigl(\kappa^{h_i}_\xcE\circ A_i\bigr)\ox\bigl(\a_\xcE\circ\widetilde h{}_i^*\vartheta_{\rm L}\bigr)\bigr)\bigr)\circ{\rm Alt}\cr\cr 
&=&\om^{h_i}[\Theta_i]\,.
\qqq
Similarly,\ we obtain
\qq\nn
&&\bigl(\kappa^{h_i}_\xcE\circ\widetilde h{}_i^*\vartheta_{\rm L}+\tfrac{1}{2}\,\ev\circ\bigl(\bigl(\kappa^{h_i}_\xcE\circ\widetilde h{}_i^*\vartheta_{\rm L}\bigr)\ox\bigl(\a_\xcE\circ\widetilde h{}_i^*\vartheta_{\rm L}\bigr)\bigr)\bigr)\circ\bigl(\id_\txT+\a_\xcE\circ\widetilde h{}_i^*\vartheta_{\rm L},\id_\txT+\a_\xcE\circ\widetilde h{}_i^*\vartheta_{\rm L}\bigr)\circ{\rm Alt}\cr\cr 
&=&\bigl(\kappa^{h_i}_\xcE\circ\widetilde h{}_i^*\vartheta_{\rm L}\bigr)\circ\bigl(\bigl(\id_\txT,\id_\txT\bigr)+\bigl(\id_\txT,\a_\xcE\circ\widetilde h{}_i^*\vartheta_{\rm L}\bigr)+\bigl(\a_\xcE\circ\widetilde h{}_i^*\vartheta_{\rm L},\id_\txT\bigr)\bigr)\circ{\rm Alt}\cr\cr 
&&+\tfrac{1}{2}\,\ev\circ\bigl(\bigl(\kappa^{h_i}_\xcE\circ\widetilde h{}_i^*\vartheta_{\rm L}\bigr)\ox\bigl(\a_\xcE\circ\widetilde h{}_i^*\vartheta_{\rm L}\bigr)\bigr)\circ\bigl(\id_\txT,\id_\txT\bigr)\circ{\rm Alt}\cr\cr &=&\bigl(\kappa^{h_i}_\xcE\circ\widetilde h{}_i^*\vartheta_{\rm L}-\tfrac{1}{2}\,\ev\circ\bigl(\bigl(\kappa^{h_i}_\xcE\circ\widetilde h{}_i^*\vartheta_{\rm L}\bigr)\ox\bigl(\a_\xcE\circ\widetilde h{}_i^*\vartheta_{\rm L}\bigr)\bigr)\bigr)\circ{\rm Alt}\,.
\qqq
Thus,\ altogether,\ we arrive at the formula
\qq\nn
H_i^*\om[\Theta_i^\Phi]=\om^{h_i}[\Theta_i]+\bigl(\kappa^{h_i}_\xcE\circ\widetilde h{}_i^*\vartheta_{\rm L}-\tfrac{1}{2}\,\ev\circ\bigl(\bigl(\kappa^{h_i}_\xcE\circ\widetilde h{}_i^*\vartheta_{\rm L}\bigr)\ox\bigl(\a_\xcE\circ\widetilde h{}_i^*\vartheta_{\rm L}\bigr)\bigr)\bigr)\circ{\rm Alt}\,,
\qqq
from which the desired equality \eqref{eq:augs-glue} ensues through a direct comparison with \eqref{eq:rho-pull-bij} .

\section{Gerbes,\ characters,\ equivariance and descent}\label{app:gerbarium}
	
In this appendix,\ elements of the theory of gerbes are introduced,\ which are instrumental in the discussion of the gauging of rigid Lie-groupoidal symmetries of the PAG $\si$-model.\ A good general reference for the theory is provided by Brylinski's monograph \cite{Brylinski:1993ab} and Ph.D.\ theses:\ \cite{Stevenson:2000wj} and \cite{Waldorf:2007phd}.
	
\bedef\label{def:Deligne}
Let $X\in\Man$.\ The {\bf degree-2 Deligne complex} of $X$ is the cochain complex
\qq\nn
\xcD^{\bullet}(2;X)\qquad\colo\qquad\brd0\xrightarrow{\hspace{45pt}}\unl{\uj}{}_X\xrightarrow{\ \, \sfd^{(0)}=-\sfi\sfd\log\ \, }\unl{\Om^1}(X)\xrightarrow{\quad\ \sfd^{(1)}=\sfd\quad\ }\unl{\Om^2}(X)
\qqq
of sheaves of abelian groups:\ $\unl{\uj}{}_X\equiv\unl{\uj}$ (locally smooth $\uj$-valued functions on $X$),\ $\unl{\Om^1}(X)\equiv\unl{\Om^1}$ (locally smooth real 1-forms on $X$),\ and $\unl{\Om^2}(X)\equiv\unl{\Om^2}$ (locally smooth real 2-forms on $X$).\ Given an open cover $\cO=\{O_i\}_{i\in I}$ of $X$,\ the corresponding {\bf \v{C}ech--Deligne bicomplex} is the semi-bounded bicomplex
{\small\qq\nn
\alxydim{@C=1.5cm@R=1.cm}{ 0 \ar@{-->}[r] & 0 \ar@{-->}[r] & 0 \ar@{-->}[r] & 0 \ar@{-->}[r] & \cdots \\ \vC^0\bigl(\cO,\unl{\Om^2}\bigr) \ar[r]_{\vd^{(0,2)}} \ar@{-->}[u] & \vC^1\bigl(\cO,\unl{\Om^2}\bigr) \ar[r]_{\vd^{(1,2)}} \ar@{-->}[u] & \vC^2\bigl(\cO,\unl{\Om^2}\bigr) \ar[r]_{\vd^{(2,2)}} \ar@{-->}[u] & \vC^3\bigl(\cO,\unl{\Om^2}\bigr) \ar[r]_{\qquad\vd^{(3,2)}} \ar@{-->}[u] & \cdots \\ \vC^0\bigl(\cO,\unl{\Om^1}\bigr) \ar[r]_{\vd^{(0,1)}} \ar[u]^{\sfd^{(0,1)}} & \vC^1\bigl(\cO,\unl{\Om^1}\bigr) \ar[r]_{\vd^{(1,1)}} \ar[u]^{\sfd^{(1,1)}} & \vC^2\bigl(\cO,\unl{\Om^1}\bigr) \ar[r]_{\vd^{(2,1)}} \ar[u]^{\sfd^{(2,1)}} & \vC^3\bigl(\cO,\unl{\Om^1}\bigr) \ar[r]_{\qquad\vd^{(3,1)}} \ar[u]^{\sfd^{(3,1)}} & \cdots \\ \vC^0\bigl(\cO,\unl{\uj}\bigr) \ar[r]_{\vd^{(0,0)}} \ar[u]^{\sfd^{(0,0)}} & \vC^1\bigl(\cO,\unl{\uj}\bigr) \ar[r]_{\vd^{(1,0)}} \ar[u]^{\sfd^{(1,0)}} & \vC^2\bigl(\cO,\unl{\uj}\bigr) \ar[r]_{\vd^{(2,0)}} \ar[u]^{\sfd^{(2,0)}} & \vC^3\bigl(\cO,\unl{\uj}\bigr) \ar[r]_{\qquad\vd^{(3,0)}} \ar[u]^{\sfd^{(3,0)}} & \cdots}
\qqq}
\hspace{-5pt}with the horizontal differentials $\vd^{(p,q)}=\vd^{(p)}_{\cS_q}$ given by the \v{C}ech coboundary operators (see:\ Ex.\,\ref{eg:Cech-nerve}) of the \v{C}ech complexes with values in the respective sheaves $(\cS_0,\cS_1,\cS_2)=(\unl{\uj},\unl{\Om^1},\unl{\Om^2})$,\ and with the vertical differentials $\sfd^{(p,q)}$ given by the obvious extensions of the differentials $\sfd^{(q)}$.
	
Its diagonal complex
\qq\nn
A^{\bullet}(\cO)\qquad\colo\qquad\alxydim{@C=1.25cm@R=1.25cm}{ \brd0 \ar[r] & A^0(\cO) \ar[r]^{D^{(0)}} & A^1(\cO) \ar[r]^{D^{(1)}} & A^2(\cO) \ar[r]^{D^{(2)}} & A^3(\cO) \ar[r]^{D^{(3)}} & \cdots}
\qqq
has groups of cochains 
\qq\nn
A^0(\cO)&=&\vC^0\bigl(\cO,\unl{\uj}\bigr)\cr\cr 
A^1(\cO)&=&\vC^0\bigl(\cO,\unl{\Om^1})\bigr)\oplus\vC^1\bigl(\cO,\unl{\uj}\bigr)\cr\cr 
A^2(\cO)&=&\vC^0\bigl(\cO,\unl{\Om^2}\bigr)\oplus\vC^1\bigl(\cO,\unl{\Om^1}\bigr)\oplus\vC^2\bigl(\cO,\unl{\uj}\bigr)\cr\cr 
A^3(\cO)&=&\vC^1\bigl(\cO,\unl{\Om^2}\bigr)\oplus\vC^2\bigl(\cO,\unl{\Om^1}\bigr)\oplus\vC^3\bigl(\cO,\unl{\uj}\bigr)\qquad\ldots
\qqq
and the {\bf Deligne differentials}
\qq\nn
D^{(0)}(f_i)&=&\bigl(-\sfi\,f_i^{-1}\,\sfd f_i,f_j^{-1}\cdot f_i\rstr_{O_{i j}}\bigr)\cr\cr 
D^{(1)}(\Pi_i,\chi_{ij})&=&\bigl(\sfd\Pi_i,-\sfi\,\chi_{ij}^{-1}\,\sfd\chi_{ij}+(\Pi_j-\Pi_i)\rstr_{O_{ij}},\chi_{jk}^{-1}\cdot\chi_{ik}\cdot\chi_{ij}^{-1}\rstr_{O_{ijk}}\bigr)\cr\cr 
D^{(2)}(B_i,A_{ij},g_{ijk})&=&\bigl(\sfd A_{ij}-(B_j-B_i)\rstr_{O_{ij}},-\sfi\,g_{ijk}^{-1}\,\sfd g_{ijk}+(A_{jk}-A_{ik}+A_{ij})\rstr_{O_{ijk}},g_{jkl}^{-1}\cdot g_{ikl}\cdot g_{ijl}^{-1}\cdot g_{ijk}\rstr_{O_{ijkl}}\bigr)\cr\cr
{\it etc.}
\qqq
Its cohomology 
\qq\nn
\bH^\bullet\bigl(\cO,\xcD^{\bullet}(2;X)\bigr)\equiv H^\bullet\bigl(A^\bullet(\cO),D^{(\bullet)}\bigr)
\qqq
is termed the ({\bf real}) {\bf degree-2 Beilinson--Deligne cohomology} of $X$ associated to $\cO$.\ Under refinement of the cover ({\it i.e.},\ for $\cO$ sufficiently fine---{\it e.g.},\ good),\ it yields the ({\bf real}) {\bf degree-2 Beilinson--Deligne cohomology} of $X$
\qq\nn
\bH^\bullet\bigl(X,\xcD^{\bullet}(2;X)\bigr)=\underset{\scriptsize{\cO}}{\varinjlim}\,\bH^\bullet\bigl(\cO,\xcD^{\bullet}(2;X)\bigr)
\qqq
\exdef
	
	\berop\label{prop:BD-vs-Huj}
	Adopt the notation of Def.\,\ref{def:Deligne}.\ There exist canonical isomorphisms
	\qq\nn
	\bH^0\bigl(X,\xcD^{\bullet}(2;X)\bigr)\cong H^0\bigl(X,\uj\bigr)\,,\qquad\qquad\bH^1\bigl(X,\xcD^{\bullet}(2;X)\bigr)\cong H^1\bigl(X,\uj\bigr)\,.
	\qqq
	\eerop
	
	\bethe\label{thm:W3-tors}
	Adopt the notation of Def.\,\ref{def:Deligne}.\ Every class $[(B_i,A_{ij},g_{ijk})]\in\bH^2(X,\xcD^{\bullet}(2;X))$ defines a 3-form $H\in Z^3_{\rm dR}(X)$ with restrictions $H\rstr_{O_i}=\sfd B_i$,\ called the {\bf curvature} of $[(B_i,A_{ij},g_{ijk})]$.\ The set $\cW^3(X;H)$ of classes of curvature $H$ is a torsor under an action of $\cW^3(X;0)$ given by
	\qq\nn
	&&\cW^3(X;0)\x\cW^3(X;H)\too\cW^3(X;H)\cr\cr  &&\bigl([(\b_i,\a_{ij},\g_{ijk})],[(B_i,A_{ij},g_{ijk})]\bigr)\longmapsto[(B_i+\b_i,A_{ij}+\a_{ij},g_{ijk}\cdot\g_{ijk})]\,.
	\qqq
	There is a canonical isomorphism
	\qq\nn
	\cW^3(X;0)\cong H^2\bigl(X,\uj\bigr)\,.
	\qqq
	\ethe
	
	\bedef\label{def:Grb}
	Let $X\in\Man$,\ and let $H\in\Om^3(X)$ be a de Rham 3-cocycle with periods in $2\pi\bZ$.\ A {\bf gerbe} of {\bf curvature} $H\equiv\curv(\cG)$ over $X$ is a sextuple $\cG\equiv(Y,B,L,\cA_L,\mu_L)$ composed of
	\bit
	\item a surjective submersion $Y\xrightarrow{\pi_Y}X$;
	\item a 2-form $B\in\Om^2(Y)$,\ termed the {\bf curving} of $\cG$,\ satisfying the identity $\sfd B=\pi_Y^*H$;
	\item a principal $\bC^\x$-bundle $L\xrightarrow{\pi_L}Y\x_X Y$  with principal connection 1-form $\cA_L\in\Om^1(L)$ satisfing the identity $\sfd\cA_L=\pi_L^*(\pr_2^*-\pr_1^*)B$;
	\item a connection-preserving isomorphism $\mu_L\colo\pr_{1,2}^*L\ox\pr_{2,3}^*L\xrightarrow{\cong}\pr_{1,3}^*L$ of principal $\bC^\x$-bundles over $Y\x_X Y\x_X Y$ (the tensor product is Brylinski's contracted product from \cite[Sec.\,2.1]{Brylinski:1993ab}),\ satisfying the coherence condition 
	\qq\nn
	\pr_{1,3,4}^*\mu_L\circ\bigl(\pr_{1,2,3}^*\mu_L\ox\id\bigr)=\pr_{1,2,4}^*\mu_L\circ\bigl(\id\ox\pr_{2,3,4}^*\mu_L\bigr)
	\qqq
	over $Y\x_X Y\x_X Y\x_X Y$.
	\eit
	
	Given gerbes $\cG_a,\ a\in\{1,2\}$ as above,\ a {\bf 1-isomorphism} $\Phi\colo\cG_1\xrightarrow{\cong}\cG_2$ is a quadruple $\Phi\equiv(Z,E,\cA_E,\a_E)$ composed of
	\bit
	\item a surjective submersion $Z\xrightarrow{\pi_Z}Y_{1,2}\equiv Y_1\x_X Y_2$;
	\item a principal $\bC^\x$-bundle $E\xrightarrow{\pi_E}Z$  with principal connection 1-form $\cA_E\in\Om^1(E)$ satisfing the identity $\sfd\cA_E=\pi_E^*\pi_Z^*(\pr_2^*B_2-\pr_1^*B_1)$;
	\item a connection-preserving isomorphism $\a_E\colo \pi_Z^{\x 2\,*}\pr_{1,3}^*L_1\ox\pr_2^*E\xrightarrow{\cong}\pr_1^*E\ox\pi_Z^{\x 2\,*}\pr_{2,4}^*L_2$ of principal $\bC^\x$-bundles over $Z\x_X Z$,\ satisfying the coherence condition expressed by the commutative diagram
	\qq\nn
	\alxydim{@C=3.75cm@R=1.cm}{ \pi_Z^{\x 3\,*}\pr_{1,3}^*L_1\ox\pi_Z^{\x 3\,*}\pr_{3,5}^*L_1\ox\pr_3^*E \ar[r]^{\hspace{35pt}\pi_Z^{\x 3\,*}\pr_{1,3,5}^*\mu_{L_1}\ox\id} \ar[d]_{\id\ox\pr_{2,3}^*\a_E} & \pi_Z^{\x 3\,*}\pr_{1,5}^*L_1\ox\pr_3^*E \ar[dd]^{\pr_{1,3}^*\a_E} \\ \pi_Z^{\x 3\,*}\pr_{1,3}^*L_1\ox\pr_2^*E\ox\pi_Z^{\x 3\,*}\pr_{4,6}^*L_2 \ar[d]_{\pr_{1,2}^*\a_E\ox\id} & \\ \pr_1^*E\ox\pi_Z^{\x 3\,*}\pr_{2,4}^*L_2\ox\pi_Z^{\x 3\,*}\pr_{4,6}^*L_2 \ar[r]_{\hspace{35pt}\id\ox\pi_Z^{\x 3\,*}\pr_{2,4,6}^*\mu_{L_2}} & \pr_1^*E\ox\pi_Z^{\x 3\,*}\pr_{2,6}^*L_2 }
	\qqq
	over $Z\x_X Z\x_X Z$.
	\eit
	
	Given 1-isomorphisms $\Phi_b\colo\cG_1\xrightarrow{\cong}\cG_2,\ b\in\{1,2\}$ as above,\ a {\bf 2-isomorphism} $\psi\colo\Phi_1\overset{\cong}{\Longrightarrow}\Phi_2$ is an equivalence class of triples $(W,\b)$ composed of
	\bit
	\item a surjective submersion $W\xrightarrow{\pi_W}Z_1\x_{Y_{1,2}}Z_2$;
	\item a connection-preserving isomorphism $\b\colo\pi_W^*\pr_1^*E_1\xrightarrow{\cong}\pi_W^*\pr_2^*E_2$ of principal $\bC^\x$-bundles over $W$,\ subject to the coherence condition expressed by the commutative diagram
	\qq\nn
	\alxydim{@C=2.5cm@R=1.cm}{ \pi_W^{\x 2}{}^*\pr_{1,3}^*\bigl(\pi_{Z_1}^{\x 2\,*}\pr_{1,3}^*L_1\ox\pr_2^*E_1\bigr) \ar[r]^{\pi_W^{\x 2\,*}\pr_{1,3}^*\a_{E_1}} \ar[d]_{\id\ox\pr_2^*\b} & \pi_W^{\x 2\,*}\pr_{1,3}^*\bigl(\pr_1^*E_1\ox\pi_{Z_1}^{\x 2\,*}\pr_{2,4}^*L_2\bigr) \ar[d]^{\pr_1^*\b\ox\id} \\ \pi_W^{\x 2\,*}\pr_{2,4}^*\bigl(\pi_{Z_2}^{\x 2\,*}\pr_{1,3}^*L_1\ox\pr_2^*E_2\bigr) \ar[r]_{\pi_W^{\x 2\,*}\pr_{2,4}^*\a_{E_2}} & \pi_W^{\x 2\,*}\pr_{2,4}^*\bigl(\pr_1^*E_2\ox\pi_{Z_2}^{\x 2\,*}\pr_{2,4}^*L_2\bigr) }
	\qqq
	over $W\x_X W$.
	\eit
	We consider pairs $(W_c,\b_c),\ c\in\{1,2\}$ as above equivalent if there exists a manifold $\widetilde W$ together with surjective submersions $\pi_c\colo\widetilde W\to W_c$ obeying $\pi_{W_2}\circ\pi_2=\pi_{W_1}\circ\pi_1$ and such that $\pi_2^*\b_2=\pi_1^*\b_1$.
	\exdef
	
\bethe\cite{Stevenson:2000wj,Waldorf:2007mm}\label{thm:bgrb-bicat}
Adopt the notation of Def.\,\ref{def:Grb}.\ Gerbes over a given base $X\in\Man$,\ together with gerbe 1-isomorphisms between them,\ and gerbe 2-isomorphisms between the latter,\ form a bicategory $\bgrb_\nabla(X)$ (in the sense of \cite{Benabou:1994bi}) with monoidal structure $\ox$ and duality.\ Moreover,\ every smooth map $f\in C^\infty(X_1,X_2)$ canonically induces a (strict monoidal) 2-functor $f^*\colo\bgrb_\nabla(X_2)\to\bgrb_\nabla(X_1)$.
\ethe
	
\berop\label{prop:gerbes-as-H2}
Adopt the notation of Defs.\,\ref{def:Deligne} and \ref{def:Grb},\ and of Thm.\,\ref{thm:W3-tors}.\ There is a one-to-one correspondence between 1-isomorphism classes of gerbes over $X$ and classes in $\bH^2(X,\xcD^{\bullet}(2;X))$.\ Under this correspondence,\ 1-isomorphism classes of gerbes of curvature $H$ are identified with elements in $\cW^3(X;H)$.\ In particular,\ such gerbes are classified by the cohomology group $H^2(X,\uj)$.
\eerop
	
\berop\label{prop:1iso-H1}
Adopt the notation of Def.\,\ref{def:Deligne}.\ Let the 2-cocycles $b_a\in\ker\,D^{(2)},\ a\in\{1,2\}$ represent the respective gerbes $\cG_a\in\bgrb_\nabla(X)$ under the correspondence from Prop.\,\ref{prop:gerbes-as-H2}.\ A 1-isomorphism $\Phi\colo\cG_1\xrightarrow{\cong}\cG_2$ is then represented by a 1-cochain $p\in A^1(\cO)$ such that 
\qq\nn
b_2=b_1+D^{(1)}p\,,
\qqq
and a 2-isomorphism $\varphi\colo\Phi_1\overset{\cong}{\Rightarrow}\Phi_2$ between two such 1-isomorphisms $\Phi_b\colo\cG_1\xrightarrow{\cong}\cG_2,\ b\in\{1,2\}$,\ given by the respective 1-cochains $p_b\in A^1(\cO)$,\ is represented by a 0-cochain $f\in A^0(\cO)$ such that 
\qq\nn
p_2=p_1+D^{(0)}f\,.
\qqq
Accordingly,\ 2-isomorphism classes of 1-isomorphisms between given gerbes $\cG_a,\ a\in\{1,2\}$ are classified by the cohomology group $H^1(X,\uj)$.
	
Finally,\ the difference between two 0-cochains $f_c\in A^0(\cO),\ c\in\{1,2\}$ representing 2-isomorphisms between a given pair of 1-isomorphisms $\Phi_b,\ b\in\{1,2\}$ as above is a local constant,\ {\it i.e.},\ it satisfies (for $\cO$ sufficiently fine)
\qq\label{eq:diff-2-iso-loconst}
f_2-f_1\in\ker\,D^{(0)}\cong\uj^{|\pi_0(X)|}.
\qqq
\eerop
	
\bedef\cite{Cheeger:1985}\label{def:CS-char}
Let $X\in\Man$,\ and let $C_k(X)$ and $Z_k(X)$,\ for $k\in\bN$,\ denote the abelian groups of (smooth) $k$-chains and $k$-cycles in $X$,\ respectively.\ For $n\in\bN$,\ a {\bf differential character} of {\bf degree} $n$ on $X$ is a pair $(\chi,\om)$ composed of an abelian-group homomorphism
\qq\nn
\chi\colo Z_{n-1}(X)\too\uj
\qqq
and a closed $n$-form $\om_\chi$,\ termed the {\bf curvature} of $\chi$,\ which satisfies the identity
\qq\label{eq:CS-char-bound}
\chi\circ\p(\cdot)=\exp\left(\sfi\int_\cdot\,\om_\chi\right)\,.
\qqq
The group of differential characters of degree $n$ is denoted as
\qq\nn
\widehat H{}^n\bigl(X,\uj\bigr)\,.
\qqq 
\exdef
	
\bethe\cite[Prop.\,1.5.7]{Brylinski:1993ab}\label{thm:CH-BD}
Adopt the notation of Defs.\,\ref{def:Deligne} and \ref{def:CS-char}.\ There exists a canonical bijection
\qq\nn
\bH^2\bigl(X,\xcD^{\bullet}(2;X)\bigr)\cong\widehat H{}^3\bigl(X,\uj\bigr)\,.
\qqq
\ethe
	
	\bedef\label{def:holG}
	Adopt the notation of Def.\,\ref{def:Grb}.\ The image of the hypercohomology class representing a gerbe $\cG$ under the bijection from Thm.\,\ref{thm:CH-BD} is termed the {\bf holonomy} of $\cG$,\ and denoted as $\Hol_\cG$.
	\exdef
	
	\bedef\cite{Stevenson:2000wj,Gawedzki:2010rn}\label{def:desc-bicat}
	Adopt the notation of Thm.\,\ref{thm:bgrb-bicat}.\ Let $Y,X\in\Man$,\ and let $\pi\colo Y\twoheadrightarrow X$ be a surjective submersion.\ Denote $Y^{[n]}:=\{ (y_1,y_2,\ldots,y_n)\in Y^{\x n}\,\vert\,\forall_{i,j\in\ovl{1,n}}\ \pi(y_i)=\pi(y_j)\}$,\ and---for every $k\in\ovl{1,n-1}$---$\pr_{i_1 i_2\cdots i_k}=(\pr_{i_1},\pr_{i_2},\ldots,\pr_{i_k})$,\ where $\pr_l\colo Y^{[n]}\to Y$ is the canonical projection onto the $l$-th cartesian factor.\ The {\bf descent bicategory} $\gt{Desc}(\pi)$ for $\pi$ is the bicategory in which 
	\bit
	\item 0-cells are triples $(\cG,\Phi,\varphi)$ composed of a gerbe $\cG\in\bgrb_\nabla(Y)$,\ a gerbe 1-isomorphism 
	\qq\nn
	\Phi\colo\pr_1^*\cG\xrightarrow{\ \cong\ }\pr_2^*\cG
	\qqq 
	over $Y^{[2]}$,\ and a gerbe 2-isomorphism 
	\qq\nn
	\varphi\colo\pr_{2,3}^*\Phi\circ\pr_{1,2}^*\Phi\xLongrightarrow{\cong}\pr_{1,3}^*\Phi
	\qqq 
	over $Y^{[3]}$,\ the latter satisfying the coherence identity
	\qq\nn
	\pr_{1,3,4}^*\varphi\bullet(\Id_{\pr_{3,4}^*\Phi}\circ\pr_{1,2,3}^*\varphi)=\pr_{1,2,4}^*\varphi\bullet(\pr_{2,3,4}^*\varphi\circ\Id_{\pr_{1,2}^*\Phi})
	\qqq
	over $Y^{[4]}$;
	\item 1-cells are pairs $(\Psi,\psi)\colo(\cG_1,\Phi_1,\varphi_1)\xrightarrow{\cong}(\cG_2,\Phi_2,\varphi_2)$ composed of a gerbe 1-isomorphism
	\qq\nn
	\Psi\colo\cG_1\xrightarrow{\ \cong\ }\cG_2
	\qqq
	over $Y$,\ and a gerbe 2-isomorphism 
	\qq\nn
	\psi\colo\pr_2^*\Psi\circ\Phi_1\xLongrightarrow{\cong}\Phi_2\circ\pr_1^*\Psi
	\qqq
	over $Y^{[2]}$,\ the latter satisfying the coherence identity
	\qq\nn
	\pr_{1,3}^*\psi\bullet\bigl(\id_{\pr_3^*\Psi}\circ\varphi_1\bigr)=\bigl(\varphi_2\circ\id_{\pr_1^*\Psi}\bigr)\bullet\bigl(\id_{\pr_{2,3}^*\Phi_2}\circ\pr_{1,2}^*\psi\bigr)\bullet\bigl(\pr_{2,3}^*\psi\circ\id_{\pr_{1,2}^*\Phi_1}\bigr)
	\qqq
	over $Y^{[3]}$;
	\item 2-cells $\xi\colo(\Psi_1,\psi_1)\overset{\cong}{\Rightarrow}(\Psi_2,\psi_2)$ in $\gt{Desc}(\pi)((\cG_1,\Phi_1,\varphi_1),(\cG_2,\Phi_2,\varphi_2))\ni(\Psi_1,\psi_1),(\Psi_2,\psi_2)$ are given by gerbe 2-isomorphisms 
	\qq\nn
	\xi\colo\Psi_1\xLongrightarrow{\ \cong\ }\Psi_2
	\qqq
	satisfying the coherence identity
	\qq\nn
	\bigl(\id_{\Phi_2}\circ\pr_1^*\xi\bigr)\bullet\psi_1=\psi_2\bullet\bigl(\pr_2^*\xi\circ\id_{\Phi_1}\bigr)\,.
	\qqq
	\eit
	\exdef
	
	\bedef\cite{Gawedzki:2010rn}\label{def:CDD}
	Adopt the notation of Def.\,\ref{def:Deligne} and Ex.\,\ref{eg:face-nerve-LieGrpd}.\ Consider the family of \v{C}ech--Deligne complexes
	$(A^{\bullet_1}(2;\xcG_{\bullet_2}),D^{(\bullet_1)}_{(\bullet_2)})$ associated with a {\em simplicial} family $\cO_{\xcG_\bullet}\equiv\{\cO_{\xcG_n}\}_{n\in\bN}$ of open covers $\cO_{\xcG_n}=\{O^{(n)}_{i_n}\}_{i_n\in I_n}$---in the sense of Def.\,\ref{def:simpl-cover}---of the components $\xcG_n\in\Man$ of the nerve $\xcG_\bullet\equiv N_\bullet(\grpd{\xcG}{M})$.\ We shall call the bicomplex 
	\qq\nn
	\left(A^{\bullet_1}\bigl(\cO_{\xcG_{\bullet_2}}\bigr),D^{(\bullet_1)}_{(\bullet_2)},\unl\D{}^{(\bullet_1)}_{(\bullet_2)}\right)
	\qqq
	with the vertical differentials
	\qq\nn
	&\unl\D{}^{(0)}_{(\bullet)}=\D^{(0,0)}_{(\bullet)}\,,\qquad\qquad\unl\D{}^{(1)}_{(\bullet)}=\D^{(0,1)}_{(\bullet)}\oplus\D^{(1,0)}_{(\bullet)}\,,&\cr\cr &\unl\D{}^{(2)}_{(\bullet)}=\D^{(0,2)}_{(\bullet)}\oplus\D^{(1,1)}_{(\bullet)}\oplus\D^{(2,0)}_{(\bullet)}\,,\qquad\qquad\unl\D{}^{(3)}_{(\bullet)}=\D^{(1,2)}_{(\bullet)}\oplus\D^{(2,1)}_{(\bullet)}\oplus\D^{(3,0)}_{(\bullet)}\qquad\qquad{\it etc.}&
	\qqq
	given in terms of the extensions of Dupont operators from\footnote{We implicitly pass from the additive notation to the multiplicative one for sections of $\unl{\uj}$.} Prop.\,\ref{prop:tricompl},\ the {\bf \v{C}ech--Deligne--Dupont} ({\bf \v{C}DD}) {\bf bicomplex} of $\grpd{\xcG}{M}$ associated to $\cO_{\xcG_\bullet}$.\ Its total cohomology,\ {\it i.e.},\ the cohomology of the diagonal complex 
	\qq\nn
	{\rm Tot}^\bullet\left(A^{\bullet_1}\bigl(\cO_{\xcG_{\bullet_2}}\bigr),D^{(\bullet_1)}_{(\bullet_2)},\unl\D{}^{(\bullet_1)}_{(\bullet_2)}\right)
	\qqq 
	with components
	\qq\nn
	{\rm Tot}^r\left(A^{\bullet_1}\bigl(\cO_{\xcG_{\bullet_2}}\bigr),D^{(\bullet_1)}_{(\bullet_2)},\unl\D{}^{(\bullet_1)}_{(\bullet_2)}\right)=\bigoplus_{p+q=r}\,A^p\bigl(\cO_{\xcG_q}\bigr)\,,
	\qqq
	and the {\bf \v{C}DD differentials}
	\qq\nn
	\nabla_{(r)}\rstr_{A^p(\cO_{\xcG_q})}=D^{(p)}_{(q)}+(-1)^p\,\unl\D{}^{(p)}_{(q)}\,,
	\qqq
	shall be referred to as the {\bf \v{C}DD cohomology} of $\grpd{\xcG}{M}$ associated to $\cO_{\xcG_\bullet}$.
	\exdef
	
	\berop\label{prop:ind-coc}
	Adopt the notation of Def.\,\ref{def:CDD}.\ The \v{C}DD bicomplex canonically determines---for a sufficiently fine (simplicial) $\cO_{\xcG_\bullet}$---cochain complexes:
	\qq\nn
	\bigl(\G_\xcG^\bullet,\d_\xcG^{(\bullet)}\bigr)&:=&\bigl(H^0\bigl(\xcG_\bullet,\uj\bigr),[\unl{\D}{}^{(0)}_{(\bullet)}]\bigr)\,,\cr\cr 
	\bigl(P_\xcG^\bullet,\widehat\d{}_\xcG^{(\bullet)}\bigr)&:=&\bigl(H^1\bigl(\xcG_\bullet,\uj\bigr),[\unl{\D}{}^{(1)}_{(\bullet)}]\bigr)\,,
	\qqq
	and
	\qq\nn
	\bigl(W_\xcG^\bullet,\widetilde\d{}_\xcG^{(\bullet)}\bigr):=\bigl(H^2\bigl(\xcG_\bullet,\uj\bigr),[\unl{\D}{}^{(2)}_{(\bullet)}]\rstr\bigr)\,,
	\qqq
	with the respective coboundary operators induced by the vertical differentials as
	\qq\nn
	[\unl{\D}{}^{(k)}_{(n)}]\colo H^k\bigl(\xcG_n,\uj\bigr)\cong\bH^k\bigl(\xcG_n,\xcD^{\bullet}(2;\xcG_n)\bigr)\ni[\xi]\longmapsto[\unl{\D}{}^{(k)}_{(n)}\xi]\in\bH^k\bigl(\xcG_{n+1},\xcD^{\bullet}(2;\xcG_{n+1})\bigr)\cong H^k\bigl(\xcG_{n+1},\uj\bigr)
	\qqq
	for $k\in\{0,1\}$,\ and
	\qq\nn
	[\unl{\D}{}^{(2)}_{(n)}]\colo H^2\bigl(\xcG_n,\uj\bigr)\cong\cW^3(\xcG_n;0)\ni[b]\longmapsto[\unl{\D}{}^{(2)}_{(n)}b]\in\cW^3(\xcG_{n+1};0)\cong H^2\bigl(\xcG_{n+1},\uj\bigr)\,,
	\qqq
	and with cohomology groups to be denoted as
	\qq\nn
	h^\bullet\bigl(\xcG,\uj\bigr)=H^\bullet\bigl(\G_\xcG^\bullet,\d_\xcG^{(\bullet)}\bigr)\,,\qquad\qquad\widehat h{}^\bullet\bigl(\xcG,\uj\bigr)=H^\bullet\bigl(P_\xcG^\bullet,\widehat\d{}_\xcG^{(\bullet)}\bigr)\,,
	\qqq
	and
	\qq\nn
	\widetilde h{}^\bullet\bigl(\xcG,\uj\bigr)=H^\bullet\bigl(W_\xcG^\bullet,\widetilde\d{}_\xcG^{(\bullet)}\bigr)\,.
	\qqq
	\eerop
	\beroof
	Obvious.\ (See:\ Prop.\,\ref{prop:BD-vs-Huj} and Thm.\,\ref{thm:W3-tors}.)
	\eroof
	
\bedef\cite{Gawedzki:2010rn,Gawedzki:2012fu}\label{def:Gequiv-grb}
Adopt the notation of Def.\,\ref{def:mult-k-form} and Thm.\,\ref{thm:bgrb-bicat},\ and let $\rho\in\Om^2(\xcG)$ be multiplicative.\ A {\bf $\rho$-twisted $\xcG$-equivariant gerbe} is a triple $(\cG,\Upsilon,\g)$ composed of
\bit
\item a gerbe $\cG\in\bgrb_\nabla(M)$;
\item a gerbe 1-isomorphism $\Upsilon\colo d^{(1)*}_0\cG\xrightarrow{\cong}d^{(1)*}_1\cG\ox\cI_\rho$ over $\xcG$;
\item a gerbe 2-isomorphism $\g\colo(d^{(2)*}_2\Upsilon\ox\Id_{\cI_{d^{(2)*}_0\rho}})\circ d^{(2)*}_0\Upsilon\overset{\cong}{\Longrightarrow}d^{(2)*}_1\Upsilon$ over $\xcG_2$\,,
\eit
satisfying the coherence identity (with the self-explanatory subscripts on the identity 2-isomorphisms dropped for the sake of clarity)
\qq\label{eq:gamma-coh}
d^{(3)*}_2\g\bullet\bigl(\id\circ d^{(3)*}_0\g\bigr)=d^{(3)*}_1\g\bullet\bigl(\bigl(d^{(3)*}_3\g\ox\id\bigr)\circ\id\bigr)
\qqq
over $\xcG_3$.\ The pair $(\Upsilon,\g)$ is called a {\bf $\rho$-twisted $\xcG$-equivariant structure on} $\cG$.\ Whenever $\rho=0$,\ we call the corresponding $0$-twisted $\xcG$-equivariant gerbes and $0$-twisted $\xcG$-equivariant structures on them simply $\xcG${\bf -equivariant}.
	
A $\xcG${\bf -equivariant 1-isomorphism} between a $\rho_1$-twisted $\xcG$-equivariant gerbe $(\cG_1,\Upsilon_1,\g_1)$ and a $\rho_2$-twisted $\xcG$-equivariant gerbe $(\cG_2,\Upsilon_2,\g_2)$ is a pair $(\Phi,\varphi)$ composed of
\bit
\item a gerbe 1-isomorphism $\Phi\colo\cG_1\xrightarrow{\cong}\cG_2$ over $M$;
\item a gerbe 2-isomorphism $\varphi\colo(d^{(1)*}_1\Phi\ox\Id_{\cI_{\rho_1}})\circ\Upsilon_1\overset{\cong}{\Longrightarrow}\Upsilon_2\circ d^{(1)*}_0\Phi$ over $\xcG$,
\eit 
satisfying the coherence identity (written in the shorthand notation as above)
\qq\label{eq:Gequiv-iso-coh}
d^{(2)*}_1\varphi\bullet\bigl(\id\circ\g_1\bigr)=\bigl(\g_2\circ\id\bigr)\bullet\bigl(\id\circ d^{(2)*}_0\varphi\bigr)\bullet\bigl(\bigl(d^{(2)*}_2\varphi\ox\id\bigr)\circ\id\bigr)
\qqq
over $\xcG_2$.
	
A $\xcG${\bf -equivariant 2-isomorphism} between $\xcG$-equivariant 1-isomorphisms $(\Phi_c,\varphi_c),\ c\in\{1,2\}$ as above is a gerbe 2-isomorphism $\psi\colo\Phi_1\overset{\cong}{\Longrightarrow}\Phi_2$ over $M$,\ satisfying the coherence identity (written as the ones above)
\qq\nn
\varphi_2\bullet\bigl(\bigl(d^{(1)*}_1\psi\ox\id\bigr)\circ\id\bigr)=\bigl(\id\circ d^{(1)*}_0\psi\bigr)\bullet\varphi_1
\qqq
over $\xcG$.
	
Given a multiplicative 2-form $\rho\in\Om^2(\xcG)$,\ the corresponding $\rho$-twisted $\xcG$-equivariant gerbes together with $\xcG$-equivariant 1-isomorphisms between them,\ and $\xcG$-equivariant 2-isomorphisms between the latter compose the {\bf bicategory of $\rho$-twisted $\xcG$-equivariant gerbes} 
\qq\nn
\bgrb_\nabla(M)^{(\xcG,\rho)}\,.
\qqq 
Within the latter,\ there exists a full sub-bicategory 
\qq\nn
\bgrb_\nabla(M)^{(\xcG,0)}\equiv\bgrb_\nabla(M)^\xcG\,,
\qqq
termed the {\bf bicategory of $\xcG$-equivariant gerbes}.
\exdef
	
\bedef\cite{Stevenson:2000wj}\label{def:Cech-desc}
Adopt the notation of Def.\,\ref{eg:Cech-nerve} and Thm.\,\ref{thm:bgrb-bicat}.\ Let $X\in\Man$,\ and let $\cU=\{U_i\}_{i\in I}$ be an open cover of $X$.\ Over the manifold $\check{Y}_\cU:=\bigsqcup_{i\in I}\,U_i$,\ there arises the {\bf \v Cech descent bicategory} $\gt{Desc}(\check{\pi})$:\ Its 0-cells---{\bf \v Cech descent data}---are triples 
\qq\nn
\check{\cG}\equiv\bigl(\{\cG_i\}_{i\in I},\{\Phi_{ij}\}_{i,j\in I},\{\varphi_{ijk}\}_{i,j,k\in I}\bigr)
\qqq
composed of $I$-indexed families of
\bit
\item gerbes $\cG_i\in\bgrb_\nabla(U_i)$;
\item gerbe 1-isomorphisms $\Phi_{ij}\colo\cG_i\rstr_{U_{ij}}\xrightarrow{\cong}\cG_j\rstr_{U_{ij}}$;
\item gerbe 2-isomorphisms $\varphi_{ijk}\colo\Phi_{jk}\circ\Phi_{ij}\rstr_{U_{ijk}}\xLongrightarrow{\cong}\Phi_{ik}\rstr_{U_{ijk}}$
\eit 
satisfying the coherence identities
\qq\label{eq:coh-Cech-0}
\varphi_{ikl}\bullet\bigl(\id_{\Phi_{kl}}\circ\varphi_{ijk}\bigr)\rstr_{U_{ijkl}}=\varphi_{ijl}\bullet\bigl(\varphi_{jkl}\circ\id_{\Phi_{ij}}\bigr)\rstr_{U_{ijkl}}\,.
\qqq
Its 1-cells---{\bf coherent 1-isomorphisms between \Cv ech descent data} $\check{\cG}^a,\ a\in\{1,2\}$---are pairs 
\qq\nn
\check{\Psi}\equiv\bigl(\{\Psi_i\}_{i\in I},\{\psi_{ij}\}_{i,j\in I}\bigr)
\qqq
composed of $I$-indexed families of
\bit
\item gerbe 1-isomorphisms $\Psi_i\colo\cG_i^1\xrightarrow{\cong}\cG_i^2$;
\item gerbe 2-isomorphisms $\psi_{ij}\colo\Psi_j\circ\Phi_{ij}^1\rstr_{U_{ij}}\xLongrightarrow{\cong}\Phi_{ij}^2\circ\Psi_i\rstr_{U_{ij}}$,
\eit
satisfying the coherence identities
\qq\label{eq:psi-phi-coh}
\psi_{ik}\bullet\bigl(\id_{\Psi_k}\circ\varphi_{ijk}^1\bigr)\rstr_{U_{ijk}}=\bigl(\varphi_{ijk}^2\circ\id_{\Psi_i}\bigr)\bullet\bigl(\id_{\Phi_{jk}^2}\circ\psi_{ij}\bigr)\bullet\bigl(\psi_{jk}\circ\id_{\Phi_{ij}^1}\bigr)\rstr_{U_{ijk}}\,.
\qqq
Finally,\ its 2-cells---{\bf coherent 2-isomorphisms between} ({\bf coherent}) {\bf 1-isomorphisms} $\check{\xi}\colo\check{\Psi}{}^1\overset{\cong}{\Rightarrow}\check{\Psi}{}^2$ in $\gt{Desc}(\check\pi)(\check{\cG}{}^1,\check{\cG}{}^2)\ni\check{\Psi}{}^1,\check{\Psi}{}^2$---are $I$-indexed families gerbe 2-isomorphisms 
\qq\nn
\check{\xi}\equiv\{\ \xi_i\colo\Psi^1_i\xLongrightarrow{\ \cong\ }\Psi^2_i\ \}_{i\in I}
\qqq
satisfying the coherence identities
\qq\nn
\bigl(\id_{\Phi^2_{ij}}\circ\pr_1^*\xi_i\bigr)\bullet\psi^1_{ij}\rstr_{U_{ij}}=\psi^2_{ij}\bullet\bigl(\pr_2^*\xi_j\circ\id_{\Phi^1_{ij}}\bigr)\rstr_{U_{ij}}\,.
\qqq
\exdef

\end{document}